\documentclass[10pt,doctor]{is-thesis}
\usepackage{amssymb}
\usepackage{amsmath}
\usepackage{amsfonts}
\usepackage{graphics,graphicx}
\usepackage{latexsym,epsfig} 
\usepackage{revsymb} 
\usepackage{ulem}
\usepackage{arydshln}
\usepackage{fn2end}
\usepackage{makeidx}
\usepackage{enumerate}
\usepackage{theorem}
\theoremstyle{break}
\newtheorem{Theorem}{Theorem}[chapter]
\newtheorem{Notation}[Theorem]{Notation}
\newtheorem{Definition}[Theorem]{Definition}
\newtheorem{Lemma}[Theorem]{Lemma}
\newtheorem{Proposition}[Theorem]{Proposition}
\newtheorem{Corollary}[Theorem]{Corollary}

\newtheorem{Example}[Theorem]{Example}
\newtheorem{Fact}[Theorem]{Fact}
\newcommand{\f}{\frac}
\newcommand{\C}{{\mathbb{C}}}

\newcommand\HH{\mathcal{H}}  
\newcommand\BB{\mathcal{B}}
\newcommand\KK{\mathcal{K}}
\newcommand\OO{\mathcal{O}}

\newcommand\Hi{\mathcal{H}_\mathtt{in}}
\newcommand\Ho{\mathcal{H}_\mathtt{out}}
\newcommand\Ha{\mathcal{H}_\mathtt{aux}}
\newcommand\BHi{\BB(\Hi)}
\newcommand\BHo{\BB(\Ho)}
\newcommand\BHa{\BB(\Ha)}
\def\Tr{\mathop{\mathrm{Tr}}}
\renewcommand\Tr{\mathop{\mathrm{Tr}}\nolimits}
\newcommand\bra[1]{{\langle{#1}|}}
\newcommand\ket[1]{{|{#1}\rangle}}
\newcommand\ketbra[1]{\ket{#1}\!\bra{#1}}
\newcommand\braket[1]{\langle{#1}|{#1}\rangle}
\newcommand\qdv[2]{H\left({#1}\,||\,{#2}\right)}
\def\id{\mathop{\mathrm{id}}}
\def\Id{\mathop{\mathrm{Id}}}
\def\oId{\mathop{\mathrm{Id}^\#}}
\def\Range{\mathop{\mathrm{Range}}}
\def\MM{{\Lambda}}
\def\dM{{\MM^{\!(\dagger)}}}
\def\fE{{\mathfrak E}}
\def\fM{{\mathfrak M}}
\def\fX{{\mathfrak{X}}}
\def\tM{\tilde{\MM}}
\def\oN{{\otimes n}}
\def\d{{\otimes 2}}

\newcommand\ignore[1]{}

\renewcommand\Range{\mathop{\mathrm{Range}}\nolimits}
\newcommand\ox[1]{{\otimes{#1}}} 

\newcommand\density[1]{\BB({#1})}
\newcommand{\qed}{\hfill $\Box$}
\newcommand{\1}{{\openone}}

\newcommand{\tr}{{\operatorname{Tr}}}

\def\av{^{\rm Av}}

\def\pmx{\begin{pmatrix}}
\def\emx{\end{pmatrix}}

\newenvironment{proof}%
{{\noindent{\bf Proof\ }}}{\hfill $\Box$\vskip 0.5\baselineskip}
\def\emph{\it}
\newcommand{\I}{\sqrt{-1}}
\newcommand{\AB}{^{\>\sf A:B}}
\newcommand{\OA}{^{\>\tt out:aux}}
\newcommand{\oa}{^{\>\tt o:a}}
\def\T{\Lambda}
\def\Tq{{\Lambda_{\tt 4}}}
\def\@cite#1#2{$^{\hbox{\scriptsize{#1\if@tempswa , $2\fi})}}$}

\newcommand{\objhsw}{S(\sum_{i=1}^n p_i \, \T \rho_i) - \sum_{i=1}^n p_i \, S( \T \rho_i)}

\makeindex
\begin{document}
{

\vspace*{0mm}
    {\thispagestyle{empty}
    \huge
     \noindent This document contains \\ \\
      \indent \indent the doctoral dissertation (94 pages)\\ \\
      \indent \indent \indent submitted to \\ \\ 
      \indent \indent \indent \indent the Graduate School \\ \\
      \indent \indent \indent \indent of the University of Tokyo\\ \\
     \indent and \\ \\
     \indent \indent the slides (mainly Japanese, 14 pages herein, \\ \\
     \indent \indent \indent \indent \indent originally a PowerPoint file with 28 slides)\\ \\
     \indent \indent \indent used at the presentation \\ \\
     \indent \indent \indent \indent for the degree of the doctor \\ \\ 
     \indent \indent \indent in Octover 1995. \\ \\ \\   
     \indent \indent \indent \indent \indent \indent  
     \indent \indent \indent \indent 
     \\ \\
     Dr. of Information Science and Technology, \vspace*{-4mm} \\ \\ \\
     \indent \indent \indent \indent \indent \indent  
     \indent \indent \indent \indent \indent \indent       
    \Huge 
     Toshiyuki Shimono \vspace*{1mm}
     \\
\hspace*{95mm}\includegraphics[width=60mm]{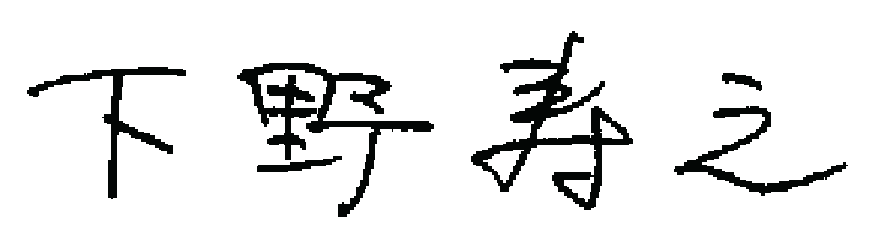} \\
\indent\indent\includegraphics[width=85mm]{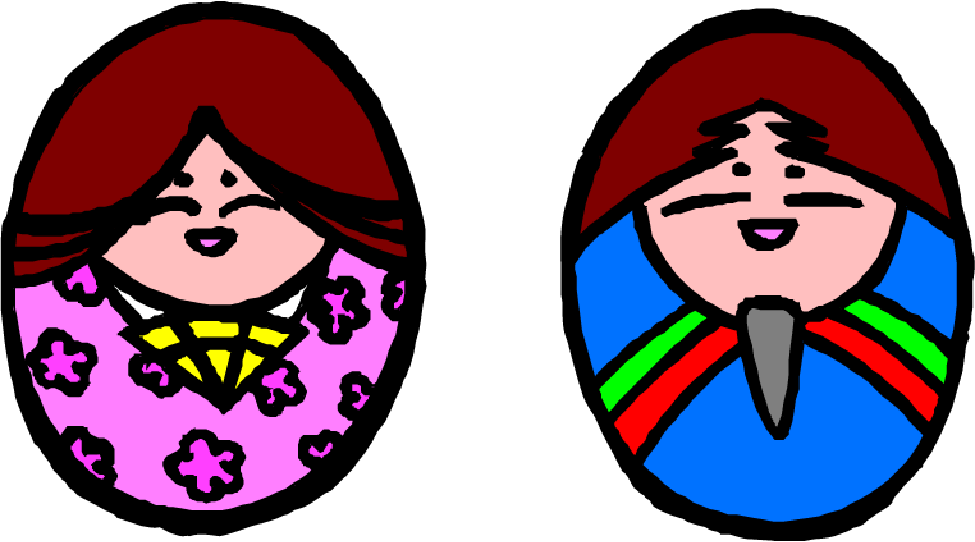}\\    } 
}%
\newpage 
\thispagestyle{empty}
\begin{center}
\begin{minipage}{160mm}{
\Large\sl
\hfil ``Beam me over,'' Alice: A cricket's quantum journey
\medskip \medskip 

\Large
This thesis addresses two known quantities in quantum information science: (1) entanglement cost, and (2) Holevo capacity. These quantities will be crucial values when teleportation becomes common in daily life, perhaps centuries from now.
\bigskip

Assume that Alice desires to send a singing Japanese cricket to her friend Bob in America, and that Alice and Bob already share a quantum entanglement.  First, Alice sends Bob a mass of information bits resulting from the interaction between the cricket she holds in her hand and half of the entanglement. Subsequently, Bob receives the information bits and manipulates the other half of the entanglement, transforming them back into the original cricket. Examining this situation from an instrumental engineering viewpoint, quantifying the amount of the quantum entanglement and the number of information bits is crucial for this transmission.  If both values are enough, Alice could even send herself to Bob's place instead of the tiny cricket.
\bigskip

The topics of this thesis therefore are: (1) the mathematical properties of the entanglement cost, such as whether it is an additive measure similar to normal length or weight; and (2) how to calculate the Holevo capacity, an ultimately achievable limit of the information conveyance capacity of an information channel, such as of a single photon passing through an optical fiber or space.  These two distinct quantities are magically tied together by several ``additive or not'' hypotheses, which await mathematical proof. 

}
\end{minipage}
\end{center}

\vspace*{15mm}\noindent
{\LARGE \hfill $>>$\ \ The submitted thesis is from the next page. $>>$ \hfill }

\newpage
\thispagestyle{empty}
\vspace*{-29mm}\hspace*{-32.3mm}%
\noindent\includegraphics[width=210mm,height=297mm,bb=0 0 594 841]{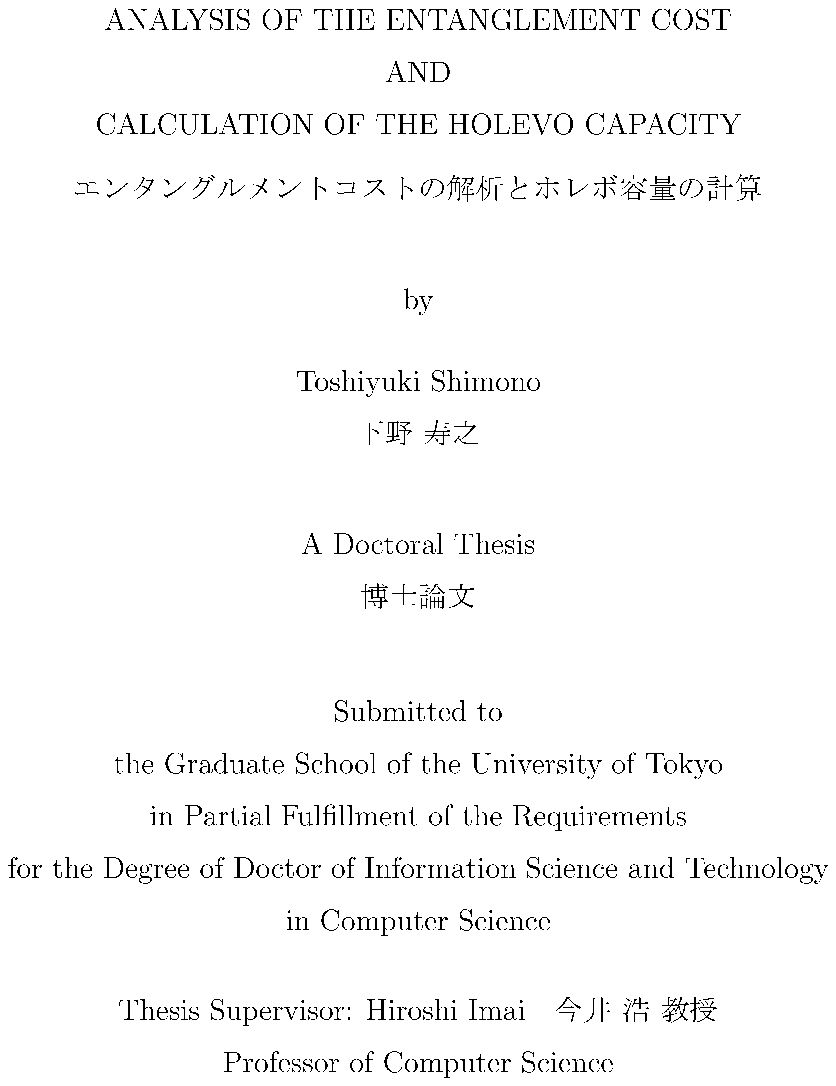}
\newpage
\thispagestyle{empty}
\vspace*{-29mm}\hspace*{-32.3mm}%
\noindent\includegraphics[width=210mm,height=297mm,bb=0 0 594 841]{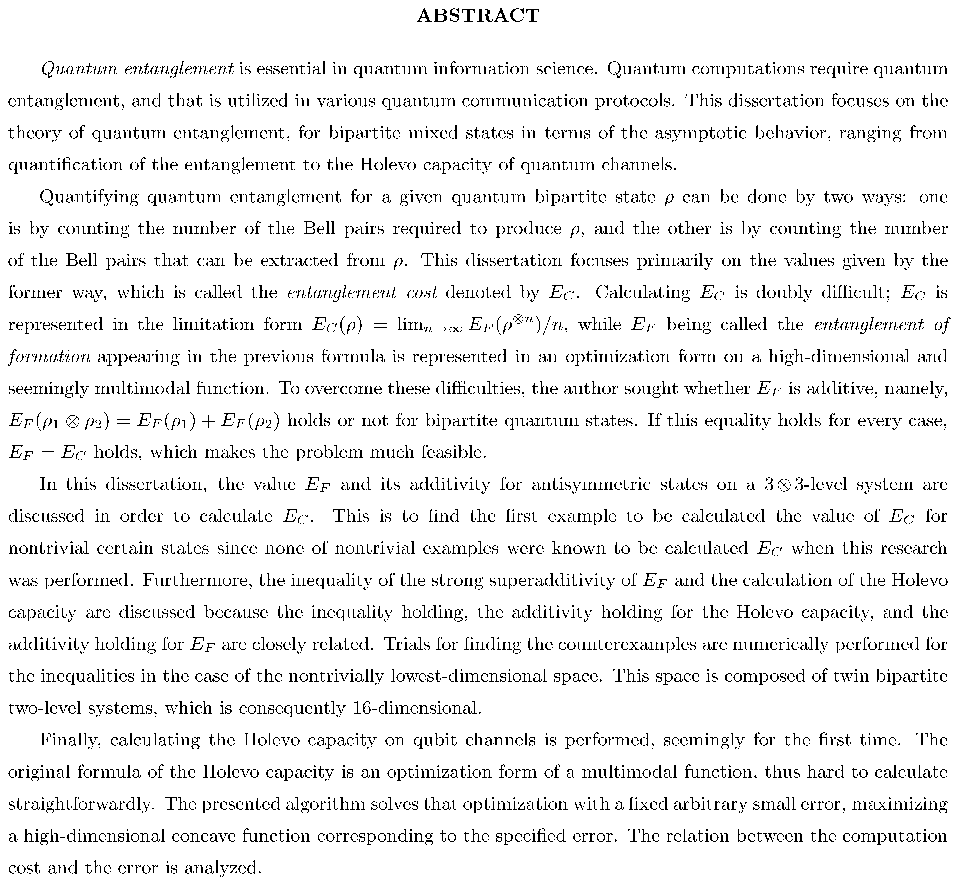}
\newpage
\thispagestyle{empty}
\vspace*{-29mm}\hspace*{-32.3mm}%
\noindent\includegraphics[width=210mm,height=297mm,bb=0 0 594 841]{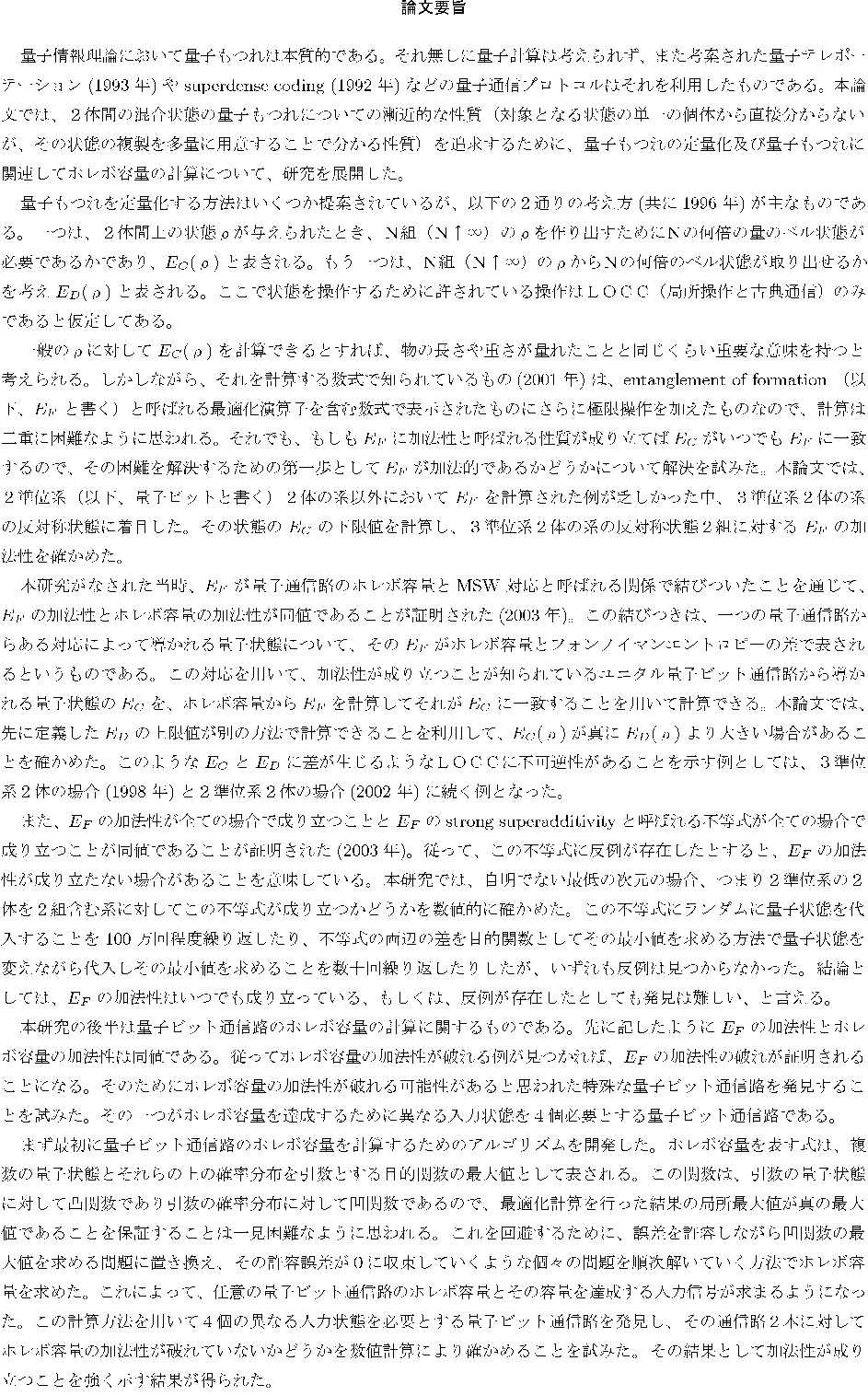}

\newpage

%
	\begin{acknowledge}%
I would like to say thank you to all the following people:  
 Prof.~Hiroshi Imai, my supervisor; Prof.~Keiji Matsumoto, the former
 head of the Tokyo branch of the Imai project; Prof.~Masahito Hayashi, the present head.
Jozef Gruska, the advisor of the project; Hwang Won Young, Wang Xian Bing, and Masahiro Hachimori, 
 with whom I talked over many issues; Andreas Winter, Fan Heng, and Mary Beth Ruskai, 
 the coauthors of our articles, who offered me many experiences; Jumpei Niwa;
 Takako Sakuragi; Takashi Yamada; Jun Hasegawa; Fran\c{c}ois Le Gall; Sonoko Moriyama, and Mayumi Oto. 
I do not have enough time nor memory to recall all those who have supported me
--- thus please forgive me for any gross omission.

 I have conducted research in my way to  hopefully be the root of a new methodology 
 in the realm of research,
 rather than imitating others.
 Nonetheless I will not forget that this dissertation could not have been completed without a lot of 
 support from those  mentioned above. 
 Their help was immense, and I hope to continue to utilize it beyond this work.

\bigskip
Toshiyuki Shimono, December 2004.

\end{acknowledge}

%
%
%
	\tableofcontents
	\listoffigures
	\listoftables
%
	\newpage\pagenumbering{arabic}

\makeendnotes

\begin{minipage}{\linewidth}
\vspace*{4cm}
\part{Prologue {} 	--~Introduction~and~Background~--}
\end{minipage}
{
\begin{minipage}{\linewidth}
\vspace*{4cm}
\begin{center}
\includegraphics[width=8cm]{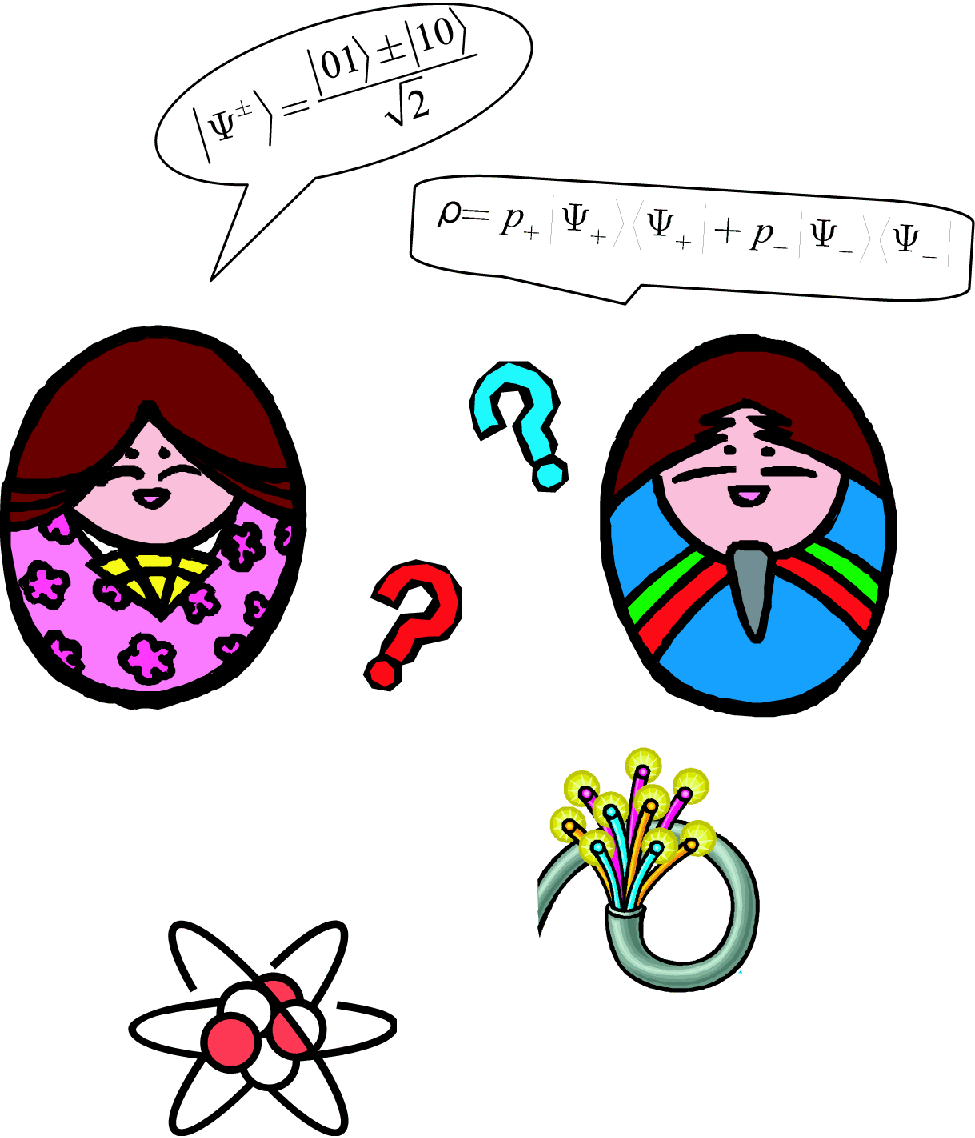}
\index{Alice and Bob}
\end{center}
\end{minipage}
}

\makeendnotes
	%


\chapter{Quantum Information Science}\label{ch:01}

	Quantum information science,  the consolidation of the rules and the intelligence 
	of this physical world, is the emerging essence of science. We need tools to 
	understand and investigate this developing science. 

	\medskip

	Prior to the chapter of the introduction, we present here what 
	the {\it quantum information science} is in this chapter.
	Concepts used in  this dissertation are explained, compactly, neatly, 
	and hopefully coherently. 
	%

	\section{What is a quantum state?}

		 What is a quantum state?  It specifies a specific physical state at a given time
		 in the framework of quantum physics. Quantum physics attempts  to explain most everything
		 in the world, especially the micro-world.  Generally, a quantum state is the ``superposition'' of 
		 multiple specified states, as often depicted by Schr\"odinger's cat, that is the superposition
		 of a living cat and a dead cat.\index{superposition}
		\index{Schr\"odinger!---'s cat}
		  One needs to understand such peculiar phenomena
		 because information devices are becoming smaller and smaller toward  nanometer world. Predictions 
		 say that fifteen or twenty years from now, each bit of information will 
		  be contained in as few as one atom as technology continues to grow exponentially, as it has 
		for more than forty years. 

		 In this dissertation, we deal solely with static quantum states; 
		 we do not consider physical continuous time transition on states that might involve the 
		 \index{Schr\"odinger!--- equation}
		 Schr\"odinger equation, that is, $H(t)\ket{\psi(t)}=\sqrt{-1}\hbar\frac{\partial}{\partial t}\ket{\psi(t)}$. Still, in the static framework, there are many interesting 
		 phenomena such as {\it quantum entanglement}.  It is an anomalous correlation on two sites or 
		 more, which is not depicted in the framework of classical views of physics. Einstein et 
		 al.~opposed quantum physics because quantum entanglement seemed to raise the issue of superluminal 
		 communication \cite{EPR}. 
		 Quantum entanglement is, however, being experimentally confirmed, and many protocols utilizing it are
		 proposed and being tested experimentally. 
		 
		 Here, again, we ask ``What is a quantum state?''  
		 The definitions, mainly from mathematical viewpoints, follows. 

	\subsection{Pure state and its tensor product} 
	\subsubsection{Pure states}

	 A {\it pure} state $\left| \psi  \right\rangle$ \index{pure state}\index{state!pure ---} is a vector of which the length is one, 
	dwelling on a specified vector space, or a Hilbert space, of a complex number field. 
	\index{Hilbert space}
	One can 
	represent it by a column vector of which the elements are complex numbers as,
	\[
	\left| \psi \right\rangle = \left( 
	\begin{array}{c}
	x_1 \\ x_2 \\ \vdots \\ x_d
	\end{array}
	\right)
	\] 
	with the conditions $x_1,x_2,\ldots ,x_d\in\mathbb{C}$ and $ |x_1|^2 + |x_2|^2 + \ldots + |x_d|^2 = 1	$. 
	Note that this is just a representation, and a vector treated in quantum information science
	is an element of a vector space, more precisely a {\it Hilbert space} over {\it complex number field}.
	The space is spanned by some bases, such as $\{\ket{\leftrightarrow},\ket{\updownarrow}\}$, 
	 $\{\ket{\uparrow},\ket{\downarrow}\}$, or simply $\{\ket{1},\ket{2},\ldots,\ket{d}\}$,
	as exemplified in the next.  

	\smallskip\noindent
	{\bf The polarization of a single photon:}\index{polarization!--- of a single photon}
	\begin{figure}[ht]
		\begin{center}
\unitlength 0.1in
\begin{picture}( 56.3000, 19.8000)(  1.3000,-22.8000)
%
\special{pn 8}%
\special{pa 1320 2280}%
\special{pa 1344 2250}%
\special{pa 1368 2220}%
\special{pa 1390 2190}%
\special{pa 1414 2160}%
\special{pa 1436 2130}%
\special{pa 1458 2100}%
\special{pa 1478 2068}%
\special{pa 1498 2038}%
\special{pa 1516 2008}%
\special{pa 1532 1978}%
\special{pa 1548 1948}%
\special{pa 1562 1918}%
\special{pa 1574 1888}%
\special{pa 1586 1858}%
\special{pa 1594 1828}%
\special{pa 1600 1798}%
\special{pa 1604 1768}%
\special{pa 1606 1738}%
\special{pa 1604 1708}%
\special{pa 1600 1678}%
\special{pa 1592 1648}%
\special{pa 1582 1618}%
\special{pa 1570 1588}%
\special{pa 1554 1558}%
\special{pa 1538 1528}%
\special{pa 1524 1498}%
\special{pa 1518 1468}%
\special{pa 1522 1436}%
\special{pa 1536 1406}%
\special{pa 1558 1378}%
\special{pa 1586 1358}%
\special{pa 1618 1350}%
\special{pa 1648 1358}%
\special{pa 1676 1378}%
\special{pa 1694 1408}%
\special{pa 1700 1446}%
\special{pa 1694 1488}%
\special{pa 1678 1522}%
\special{pa 1658 1540}%
\special{pa 1636 1532}%
\special{pa 1620 1504}%
\special{pa 1610 1466}%
\special{pa 1614 1428}%
\special{pa 1630 1398}%
\special{pa 1652 1374}%
\special{pa 1682 1356}%
\special{pa 1714 1348}%
\special{pa 1746 1352}%
\special{pa 1770 1370}%
\special{pa 1786 1400}%
\special{pa 1790 1440}%
\special{pa 1784 1482}%
\special{pa 1770 1514}%
\special{pa 1758 1518}%
\special{pa 1746 1492}%
\special{pa 1740 1450}%
\special{pa 1738 1408}%
\special{pa 1746 1374}%
\special{pa 1764 1354}%
\special{pa 1794 1352}%
\special{pa 1822 1366}%
\special{pa 1844 1392}%
\special{pa 1858 1424}%
\special{pa 1864 1456}%
\special{pa 1862 1486}%
\special{pa 1852 1516}%
\special{pa 1838 1546}%
\special{pa 1820 1574}%
\special{pa 1798 1602}%
\special{pa 1778 1630}%
\special{pa 1758 1656}%
\special{pa 1740 1684}%
\special{pa 1724 1712}%
\special{pa 1714 1740}%
\special{pa 1708 1768}%
\special{pa 1706 1796}%
\special{pa 1708 1824}%
\special{pa 1714 1852}%
\special{pa 1722 1882}%
\special{pa 1732 1910}%
\special{pa 1746 1938}%
\special{pa 1762 1968}%
\special{pa 1782 1996}%
\special{pa 1802 2026}%
\special{pa 1824 2056}%
\special{pa 1848 2084}%
\special{pa 1874 2114}%
\special{pa 1900 2142}%
\special{pa 1926 2172}%
\special{pa 1954 2202}%
\special{pa 1982 2232}%
\special{pa 1990 2240}%
\special{sp}%
%
\special{pn 8}%
\special{ar 1720 1460 368 368  0.0000000 6.2831853}%
%
\special{pn 8}%
\special{pa 2200 1430}%
\special{pa 2970 1430}%
\special{fp}%
\special{sh 1}%
\special{pa 2970 1430}%
\special{pa 2904 1410}%
\special{pa 2918 1430}%
\special{pa 2904 1450}%
\special{pa 2970 1430}%
\special{fp}%
\special{pa 2100 1120}%
\special{pa 2540 810}%
\special{fp}%
\special{sh 1}%
\special{pa 2540 810}%
\special{pa 2474 832}%
\special{pa 2496 842}%
\special{pa 2498 866}%
\special{pa 2540 810}%
\special{fp}%
\special{pa 1740 970}%
\special{pa 1780 560}%
\special{fp}%
\special{sh 1}%
\special{pa 1780 560}%
\special{pa 1754 624}%
\special{pa 1776 614}%
\special{pa 1794 628}%
\special{pa 1780 560}%
\special{fp}%
\special{pa 1350 990}%
\special{pa 1160 700}%
\special{fp}%
\special{sh 1}%
\special{pa 1160 700}%
\special{pa 1180 768}%
\special{pa 1190 746}%
\special{pa 1214 746}%
\special{pa 1160 700}%
\special{fp}%
\special{pa 1180 1370}%
\special{pa 950 1330}%
\special{fp}%
\special{sh 1}%
\special{pa 950 1330}%
\special{pa 1012 1362}%
\special{pa 1004 1340}%
\special{pa 1020 1322}%
\special{pa 950 1330}%
\special{fp}%
\special{pa 1260 1660}%
\special{pa 970 1810}%
\special{fp}%
\special{sh 1}%
\special{pa 970 1810}%
\special{pa 1038 1798}%
\special{pa 1018 1786}%
\special{pa 1020 1762}%
\special{pa 970 1810}%
\special{fp}%
\special{pa 2170 1670}%
\special{pa 2400 1920}%
\special{fp}%
\special{sh 1}%
\special{pa 2400 1920}%
\special{pa 2370 1858}%
\special{pa 2364 1882}%
\special{pa 2340 1884}%
\special{pa 2400 1920}%
\special{fp}%
%
\special{pn 8}%
\special{ar 3340 1430 220 680  0.0000000 6.2831853}%
%
\special{pn 4}%
\special{pa 3530 1090}%
\special{pa 3120 1500}%
\special{fp}%
\special{pa 3540 1140}%
\special{pa 3120 1560}%
\special{fp}%
\special{pa 3550 1190}%
\special{pa 3130 1610}%
\special{fp}%
\special{pa 3550 1250}%
\special{pa 3130 1670}%
\special{fp}%
\special{pa 3560 1300}%
\special{pa 3140 1720}%
\special{fp}%
\special{pa 3560 1360}%
\special{pa 3150 1770}%
\special{fp}%
\special{pa 3560 1420}%
\special{pa 3160 1820}%
\special{fp}%
\special{pa 3560 1480}%
\special{pa 3170 1870}%
\special{fp}%
\special{pa 3560 1540}%
\special{pa 3190 1910}%
\special{fp}%
\special{pa 3550 1610}%
\special{pa 3200 1960}%
\special{fp}%
\special{pa 3540 1680}%
\special{pa 3220 2000}%
\special{fp}%
\special{pa 3530 1750}%
\special{pa 3240 2040}%
\special{fp}%
\special{pa 3520 1820}%
\special{pa 3270 2070}%
\special{fp}%
\special{pa 3500 1900}%
\special{pa 3300 2100}%
\special{fp}%
\special{pa 3450 2010}%
\special{pa 3360 2100}%
\special{fp}%
\special{pa 3520 1040}%
\special{pa 3120 1440}%
\special{fp}%
\special{pa 3510 990}%
\special{pa 3120 1380}%
\special{fp}%
\special{pa 3490 950}%
\special{pa 3120 1320}%
\special{fp}%
\special{pa 3480 900}%
\special{pa 3130 1250}%
\special{fp}%
\special{pa 3460 860}%
\special{pa 3140 1180}%
\special{fp}%
\special{pa 3440 820}%
\special{pa 3150 1110}%
\special{fp}%
\special{pa 3410 790}%
\special{pa 3160 1040}%
\special{fp}%
\special{pa 3380 760}%
\special{pa 3180 960}%
\special{fp}%
\special{pa 3320 760}%
\special{pa 3230 850}%
\special{fp}%
%
\special{pn 13}%
\special{ar 4540 1430 114 114  0.0000000 6.2831853}%
%
\special{pn 8}%
\special{pa 4670 1430}%
\special{pa 4702 1394}%
\special{pa 4734 1358}%
\special{pa 4762 1326}%
\special{pa 4786 1298}%
\special{pa 4806 1276}%
\special{pa 4822 1264}%
\special{pa 4830 1260}%
\special{pa 4830 1266}%
\special{pa 4822 1284}%
\special{pa 4810 1310}%
\special{pa 4792 1344}%
\special{pa 4770 1380}%
\special{pa 4748 1420}%
\special{pa 4724 1460}%
\special{pa 4702 1498}%
\special{pa 4682 1534}%
\special{pa 4666 1562}%
\special{pa 4656 1586}%
\special{pa 4650 1598}%
\special{pa 4654 1600}%
\special{pa 4664 1592}%
\special{pa 4682 1574}%
\special{pa 4704 1550}%
\special{pa 4730 1520}%
\special{pa 4758 1484}%
\special{pa 4790 1448}%
\special{pa 4820 1410}%
\special{pa 4852 1372}%
\special{pa 4880 1338}%
\special{pa 4908 1308}%
\special{pa 4930 1284}%
\special{pa 4946 1266}%
\special{pa 4958 1260}%
\special{pa 4960 1262}%
\special{pa 4956 1278}%
\special{pa 4946 1302}%
\special{pa 4930 1332}%
\special{pa 4910 1368}%
\special{pa 4888 1408}%
\special{pa 4864 1448}%
\special{pa 4842 1488}%
\special{pa 4820 1526}%
\special{pa 4802 1556}%
\special{pa 4788 1582}%
\special{pa 4782 1596}%
\special{pa 4782 1600}%
\special{pa 4788 1592}%
\special{pa 4804 1574}%
\special{pa 4822 1548}%
\special{pa 4846 1516}%
\special{pa 4874 1478}%
\special{pa 4902 1440}%
\special{pa 4932 1400}%
\special{pa 4960 1362}%
\special{pa 4988 1328}%
\special{pa 5010 1300}%
\special{pa 5030 1280}%
\special{pa 5044 1270}%
\special{pa 5050 1270}%
\special{pa 5050 1284}%
\special{pa 5042 1310}%
\special{pa 5030 1342}%
\special{pa 5014 1382}%
\special{pa 4996 1424}%
\special{pa 4978 1464}%
\special{pa 4960 1504}%
\special{pa 4946 1538}%
\special{pa 4936 1564}%
\special{pa 4930 1578}%
\special{pa 4932 1580}%
\special{pa 4942 1568}%
\special{pa 4958 1544}%
\special{pa 4978 1512}%
\special{pa 5002 1476}%
\special{pa 5028 1436}%
\special{pa 5054 1394}%
\special{pa 5078 1356}%
\special{pa 5102 1320}%
\special{pa 5122 1294}%
\special{pa 5138 1276}%
\special{pa 5148 1270}%
\special{pa 5152 1278}%
\special{pa 5148 1300}%
\special{pa 5140 1332}%
\special{pa 5128 1372}%
\special{pa 5114 1414}%
\special{pa 5098 1458}%
\special{pa 5084 1500}%
\special{pa 5072 1536}%
\special{pa 5064 1562}%
\special{pa 5060 1578}%
\special{pa 5064 1580}%
\special{pa 5072 1566}%
\special{pa 5088 1540}%
\special{pa 5106 1506}%
\special{pa 5128 1468}%
\special{pa 5150 1424}%
\special{pa 5172 1382}%
\special{pa 5194 1344}%
\special{pa 5212 1310}%
\special{pa 5228 1284}%
\special{pa 5238 1272}%
\special{pa 5242 1272}%
\special{pa 5238 1288}%
\special{pa 5230 1316}%
\special{pa 5220 1354}%
\special{pa 5206 1396}%
\special{pa 5190 1440}%
\special{pa 5176 1482}%
\special{pa 5164 1522}%
\special{pa 5154 1552}%
\special{pa 5150 1574}%
\special{pa 5152 1582}%
\special{pa 5160 1574}%
\special{pa 5174 1554}%
\special{pa 5190 1524}%
\special{pa 5212 1488}%
\special{pa 5234 1446}%
\special{pa 5258 1404}%
\special{pa 5280 1362}%
\special{pa 5300 1324}%
\special{pa 5318 1294}%
\special{pa 5332 1272}%
\special{pa 5340 1260}%
\special{pa 5340 1264}%
\special{pa 5336 1282}%
\special{pa 5326 1310}%
\special{pa 5312 1346}%
\special{pa 5296 1388}%
\special{pa 5280 1430}%
\special{sp}%
%
\special{pn 8}%
\special{pa 5280 1430}%
\special{pa 5760 1430}%
\special{fp}%
\special{sh 1}%
\special{pa 5760 1430}%
\special{pa 5694 1410}%
\special{pa 5708 1430}%
\special{pa 5694 1450}%
\special{pa 5760 1430}%
\special{fp}%
%
\special{pn 8}%
\special{pa 3700 1430}%
\special{pa 3732 1394}%
\special{pa 3764 1358}%
\special{pa 3792 1326}%
\special{pa 3816 1298}%
\special{pa 3836 1276}%
\special{pa 3852 1264}%
\special{pa 3860 1260}%
\special{pa 3860 1266}%
\special{pa 3852 1284}%
\special{pa 3840 1310}%
\special{pa 3822 1344}%
\special{pa 3800 1380}%
\special{pa 3778 1420}%
\special{pa 3754 1460}%
\special{pa 3732 1498}%
\special{pa 3712 1534}%
\special{pa 3696 1562}%
\special{pa 3686 1586}%
\special{pa 3680 1598}%
\special{pa 3684 1600}%
\special{pa 3694 1592}%
\special{pa 3712 1574}%
\special{pa 3734 1550}%
\special{pa 3760 1520}%
\special{pa 3788 1484}%
\special{pa 3820 1448}%
\special{pa 3850 1410}%
\special{pa 3882 1372}%
\special{pa 3910 1338}%
\special{pa 3938 1308}%
\special{pa 3960 1284}%
\special{pa 3976 1266}%
\special{pa 3988 1260}%
\special{pa 3990 1262}%
\special{pa 3986 1278}%
\special{pa 3976 1302}%
\special{pa 3960 1332}%
\special{pa 3940 1368}%
\special{pa 3918 1408}%
\special{pa 3894 1448}%
\special{pa 3872 1488}%
\special{pa 3850 1526}%
\special{pa 3832 1556}%
\special{pa 3818 1582}%
\special{pa 3812 1596}%
\special{pa 3812 1600}%
\special{pa 3818 1592}%
\special{pa 3834 1574}%
\special{pa 3852 1548}%
\special{pa 3876 1516}%
\special{pa 3904 1478}%
\special{pa 3932 1440}%
\special{pa 3962 1400}%
\special{pa 3990 1362}%
\special{pa 4018 1328}%
\special{pa 4040 1300}%
\special{pa 4060 1280}%
\special{pa 4074 1270}%
\special{pa 4080 1270}%
\special{pa 4080 1284}%
\special{pa 4072 1310}%
\special{pa 4060 1342}%
\special{pa 4044 1382}%
\special{pa 4026 1424}%
\special{pa 4008 1464}%
\special{pa 3990 1504}%
\special{pa 3976 1538}%
\special{pa 3966 1564}%
\special{pa 3960 1578}%
\special{pa 3962 1580}%
\special{pa 3972 1568}%
\special{pa 3988 1544}%
\special{pa 4008 1512}%
\special{pa 4032 1476}%
\special{pa 4058 1436}%
\special{pa 4084 1394}%
\special{pa 4108 1356}%
\special{pa 4132 1320}%
\special{pa 4152 1294}%
\special{pa 4168 1276}%
\special{pa 4178 1270}%
\special{pa 4182 1278}%
\special{pa 4178 1300}%
\special{pa 4170 1332}%
\special{pa 4158 1372}%
\special{pa 4144 1414}%
\special{pa 4128 1458}%
\special{pa 4114 1500}%
\special{pa 4102 1536}%
\special{pa 4094 1562}%
\special{pa 4090 1578}%
\special{pa 4094 1580}%
\special{pa 4102 1566}%
\special{pa 4118 1540}%
\special{pa 4136 1506}%
\special{pa 4158 1468}%
\special{pa 4180 1424}%
\special{pa 4202 1382}%
\special{pa 4224 1344}%
\special{pa 4242 1310}%
\special{pa 4258 1284}%
\special{pa 4268 1272}%
\special{pa 4272 1272}%
\special{pa 4268 1288}%
\special{pa 4260 1316}%
\special{pa 4250 1354}%
\special{pa 4236 1396}%
\special{pa 4220 1440}%
\special{pa 4206 1482}%
\special{pa 4194 1522}%
\special{pa 4184 1552}%
\special{pa 4180 1574}%
\special{pa 4182 1582}%
\special{pa 4190 1574}%
\special{pa 4204 1554}%
\special{pa 4220 1524}%
\special{pa 4242 1488}%
\special{pa 4264 1446}%
\special{pa 4288 1404}%
\special{pa 4310 1362}%
\special{pa 4330 1324}%
\special{pa 4348 1294}%
\special{pa 4362 1272}%
\special{pa 4370 1260}%
\special{pa 4370 1264}%
\special{pa 4366 1282}%
\special{pa 4356 1310}%
\special{pa 4342 1346}%
\special{pa 4326 1388}%
\special{pa 4310 1430}%
\special{sp}%
\put(28.6000,-4.7000){\makebox(0,0)[lb]{polarizing}}%
\put(30.6000,-6.8000){\makebox(0,0)[lb]{plate}}%
%
\special{pn 13}%
\special{pa 240 1080}%
\special{pa 660 1080}%
\special{fp}%
\special{sh 1}%
\special{pa 660 1080}%
\special{pa 594 1060}%
\special{pa 608 1080}%
\special{pa 594 1100}%
\special{pa 660 1080}%
\special{fp}%
\special{pa 430 1430}%
\special{pa 430 560}%
\special{fp}%
\special{sh 1}%
\special{pa 430 560}%
\special{pa 410 628}%
\special{pa 430 614}%
\special{pa 450 628}%
\special{pa 430 560}%
\special{fp}%
\special{pa 350 910}%
\special{pa 500 1280}%
\special{fp}%
\special{sh 1}%
\special{pa 500 1280}%
\special{pa 494 1212}%
\special{pa 480 1232}%
\special{pa 456 1226}%
\special{pa 500 1280}%
\special{fp}%
\put(6.4000,-14.1000){\makebox(0,0)[rb]{x}}%
\put(3.5000,-5.7000){\makebox(0,0)[lb]{y}}%
\put(7.3000,-11.4000){\makebox(0,0)[lb]{z}}%
%
\special{pn 8}%
\special{pa 130 310}%
\special{pa 850 310}%
\special{pa 850 1640}%
\special{pa 130 1640}%
\special{pa 130 310}%
\special{da 0.070}%
\end{picture}%

		\end{center}
		\caption[The polarization of a photon]{This figure depicts how to prepare 
		photons with the polarization with $\cos \theta \, \ket{\leftrightarrow} + \sin \theta \, \ket{\updownarrow}$. }%
		\label{fig:photon}
	\end{figure}
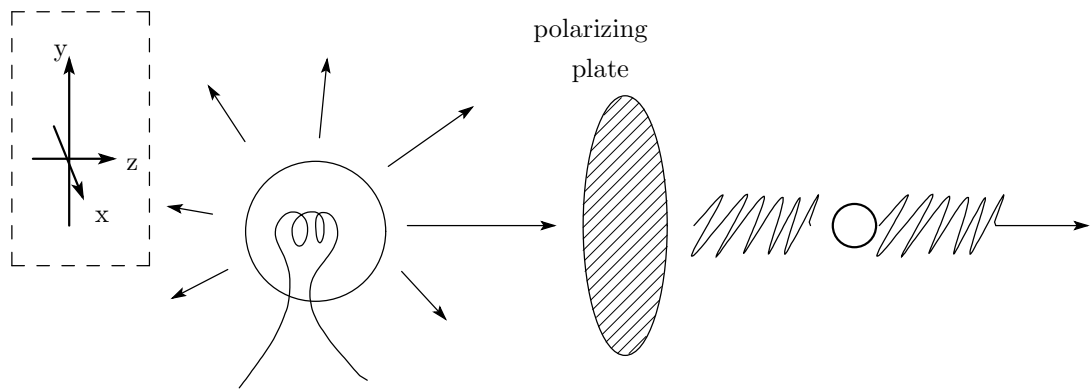
	The polarization of a photon
	flying in $z$-direction is conceptually represented 
	in a two-dimensional space with bases $\ket{\leftrightarrow}$ and $\ket{\updownarrow}$,
	that represent $x$-directional and $y$-directional polarization, respectively. 
	If the polarization angle is $\theta$ on the $xy$-plane, the state of the photon is represented as
	$\ket\theta = \cos \theta \, \ket{\leftrightarrow} + \sin \theta \, \ket{\updownarrow}$
	that is the {\it superposition} of $\ket{\leftrightarrow}$
	and $\ket{\updownarrow}$ with coefficients $\cos\theta$ and 
	$\sin\theta$.\index{superposition} 
	This vector $\ket\theta$ is represented as
	 $\left(\begin{smallmatrix}\cos\theta\\ \sin\theta\end{smallmatrix}\right)$.
	See Fig.\ref{fig:photon}.
	Note that the coefficients can be any complex number as long as their  squared summation is one. 
	$ \cos \theta \, \ket{\leftrightarrow} + e^{\sqrt{-1}\gamma}\sin \theta \, \ket{\updownarrow}$
	is such an example, which can be considered to have the phase shift of $\gamma$ in the factor of $\updownarrow$,
	and it is called to have {\it circular} or {\it elliptical} polarization.
	\index{circular polarization}\index{polarization!circular ---}
	\index{elliptical polarization}\index{polarization!elliptical ---}

	As a physical aspect, the dimensionality of two is enough for the vector space of the polarization 
	of a photon. In principle, the most possible measurement of the polarization of each single photon is equivalent
	 to check whether the photon passes through a prepared polarizing plate, \index{polarizing plate}\index{polarization!polarizing plate}
	and the passed photon loses the information of the polarization. Thus all the information the observer
	 can get about the polarization of the photon is  as few as two possibilities, which is related to the dimensionality 
	 of two here.

	\medskip\noindent
	{\bf The magnetic moment of a single silver atom:} 
	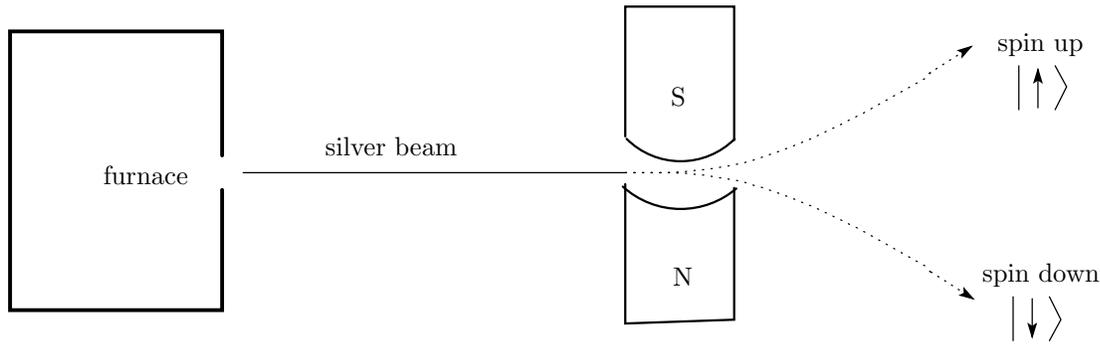
\begin{figure}[ht]
\unitlength 0.1in
\begin{picture}( 55.3000, 17.5000)(  0.9000,-23.6000)
%
\special{pn 13}%
\special{ar 3600 1020 400 400  0.7853982 2.3561945}%
%
\special{pn 13}%
\special{ar 3600 1220 450 450  0.8671825 2.3086114}%
%
\special{pn 13}%
\special{pa 3310 1290}%
\special{pa 3310 610}%
\special{pa 3880 610}%
\special{pa 3880 1290}%
\special{pa 3880 1310}%
\special{pa 3880 1310}%
\special{fp}%
%
\special{pn 13}%
\special{pa 3310 1540}%
\special{pa 3310 2270}%
\special{pa 3880 2250}%
\special{pa 3880 1570}%
\special{pa 3880 1570}%
\special{pa 3880 1560}%
\special{pa 3880 1560}%
\special{fp}%
%
\special{pn 8}%
\special{pa 3310 1480}%
\special{pa 3342 1482}%
\special{pa 3374 1482}%
\special{pa 3406 1482}%
\special{pa 3438 1482}%
\special{pa 3470 1482}%
\special{pa 3502 1482}%
\special{pa 3534 1480}%
\special{pa 3566 1478}%
\special{pa 3598 1476}%
\special{pa 3630 1472}%
\special{pa 3662 1468}%
\special{pa 3694 1464}%
\special{pa 3726 1460}%
\special{pa 3758 1454}%
\special{pa 3788 1450}%
\special{pa 3820 1442}%
\special{pa 3852 1436}%
\special{pa 3882 1428}%
\special{pa 3914 1420}%
\special{pa 3944 1412}%
\special{pa 3974 1404}%
\special{pa 4004 1394}%
\special{pa 4034 1384}%
\special{pa 4066 1374}%
\special{pa 4094 1364}%
\special{pa 4124 1352}%
\special{pa 4154 1340}%
\special{pa 4184 1330}%
\special{pa 4214 1316}%
\special{pa 4242 1304}%
\special{pa 4272 1292}%
\special{pa 4302 1278}%
\special{pa 4330 1264}%
\special{pa 4360 1250}%
\special{pa 4388 1236}%
\special{pa 4416 1222}%
\special{pa 4446 1208}%
\special{pa 4474 1194}%
\special{pa 4502 1178}%
\special{pa 4530 1162}%
\special{pa 4560 1148}%
\special{pa 4588 1132}%
\special{pa 4616 1116}%
\special{pa 4644 1100}%
\special{pa 4672 1084}%
\special{pa 4700 1068}%
\special{pa 4728 1050}%
\special{pa 4756 1034}%
\special{pa 4784 1018}%
\special{pa 4812 1002}%
\special{pa 4840 984}%
\special{pa 4868 968}%
\special{pa 4896 950}%
\special{pa 4924 934}%
\special{pa 4930 930}%
\special{sp -0.045}%
%
\special{pn 8}%
\special{pa 3310 1480}%
\special{pa 3342 1480}%
\special{pa 3374 1480}%
\special{pa 3406 1478}%
\special{pa 3438 1478}%
\special{pa 3470 1480}%
\special{pa 3502 1480}%
\special{pa 3534 1482}%
\special{pa 3566 1482}%
\special{pa 3598 1486}%
\special{pa 3630 1488}%
\special{pa 3662 1492}%
\special{pa 3694 1496}%
\special{pa 3726 1500}%
\special{pa 3758 1506}%
\special{pa 3788 1512}%
\special{pa 3820 1518}%
\special{pa 3852 1526}%
\special{pa 3882 1532}%
\special{pa 3914 1540}%
\special{pa 3944 1548}%
\special{pa 3974 1558}%
\special{pa 4004 1568}%
\special{pa 4034 1576}%
\special{pa 4066 1588}%
\special{pa 4094 1598}%
\special{pa 4124 1608}%
\special{pa 4154 1620}%
\special{pa 4184 1632}%
\special{pa 4214 1644}%
\special{pa 4242 1656}%
\special{pa 4272 1670}%
\special{pa 4302 1682}%
\special{pa 4330 1696}%
\special{pa 4360 1710}%
\special{pa 4388 1724}%
\special{pa 4416 1738}%
\special{pa 4446 1754}%
\special{pa 4474 1768}%
\special{pa 4502 1784}%
\special{pa 4530 1798}%
\special{pa 4560 1814}%
\special{pa 4588 1830}%
\special{pa 4616 1846}%
\special{pa 4644 1862}%
\special{pa 4672 1878}%
\special{pa 4700 1894}%
\special{pa 4728 1910}%
\special{pa 4756 1926}%
\special{pa 4784 1944}%
\special{pa 4812 1960}%
\special{pa 4840 1976}%
\special{pa 4868 1994}%
\special{pa 4896 2010}%
\special{pa 4924 2026}%
\special{pa 4930 2030}%
\special{sp -0.045}%
%
\special{pn 8}%
\special{pa 4900 960}%
\special{pa 5120 820}%
\special{dt 0.045}%
\special{sh 1}%
\special{pa 5120 820}%
\special{pa 5054 840}%
\special{pa 5076 850}%
\special{pa 5074 874}%
\special{pa 5120 820}%
\special{fp}%
%
\special{pn 8}%
\special{pa 4910 2000}%
\special{pa 5130 2140}%
\special{dt 0.045}%
\special{sh 1}%
\special{pa 5130 2140}%
\special{pa 5084 2088}%
\special{pa 5086 2112}%
\special{pa 5064 2122}%
\special{pa 5130 2140}%
\special{fp}%
%
\special{pn 8}%
\special{pa 3310 1480}%
\special{pa 1310 1480}%
\special{fp}%
%
\special{pn 20}%
\special{pa 1200 1390}%
\special{pa 1200 740}%
\special{pa 90 740}%
\special{pa 90 2200}%
\special{pa 1200 2200}%
\special{pa 1200 1570}%
\special{pa 1200 1570}%
\special{fp}%
\put(10.2000,-15.4000){\makebox(0,0)[rb]{furnace}}%
\put(17.4000,-13.9000){\makebox(0,0)[lb]{silver beam}}%
\put(52.6000,-8.7000){\makebox(0,0)[lb]{spin up}}%
\put(51.8000,-20.8000){\makebox(0,0)[lb]{spin down}}%
%
\special{pn 8}%
\special{pa 5370 920}%
\special{pa 5370 1150}%
\special{pa 5370 1150}%
\special{fp}%
%
\special{pn 8}%
\special{pa 5560 920}%
\special{pa 5620 1030}%
\special{pa 5560 1150}%
\special{pa 5560 1150}%
\special{pa 5560 1150}%
\special{fp}%
%
\special{pn 8}%
\special{pa 5470 1140}%
\special{pa 5470 940}%
\special{fp}%
\special{sh 1}%
\special{pa 5470 940}%
\special{pa 5450 1008}%
\special{pa 5470 994}%
\special{pa 5490 1008}%
\special{pa 5470 940}%
\special{fp}%
%
\special{pn 8}%
\special{pa 5340 2360}%
\special{pa 5340 2130}%
\special{pa 5340 2130}%
\special{fp}%
%
\special{pn 8}%
\special{pa 5530 2360}%
\special{pa 5590 2250}%
\special{pa 5530 2130}%
\special{pa 5530 2130}%
\special{pa 5530 2130}%
\special{fp}%
%
\special{pn 8}%
\special{pa 5440 2140}%
\special{pa 5440 2340}%
\special{fp}%
\special{sh 1}%
\special{pa 5440 2340}%
\special{pa 5460 2274}%
\special{pa 5440 2288}%
\special{pa 5420 2274}%
\special{pa 5440 2340}%
\special{fp}%
\put(35.5000,-11.3000){\makebox(0,0)[lb]{S}}%
\put(35.6000,-20.7000){\makebox(0,0)[lb]{N}}%
\end{picture}%
\index{furnace}\index{spin}
		\caption[Stern-Gerlach experiment]{Stern-Gerlach experiment}
		\label{fig:sg}
	\end{figure}
	Each silver atom has its momentum state
	on the vector space spanned by $\ket\uparrow$ and $\ket\downarrow$.
	The state is written as $\alpha\ket\uparrow+\beta\ket\downarrow$
	that is a superposition of $\ket\uparrow$ and $\ket\downarrow$.
	Physically, the momentum can be measured by making the atom flying through 
	nonuniform magnetic field, and the atoms curve in two possible directions.
	The measurement is done by observing which of the two directions the atom
	has curved. 
	If the magnetic field is set up to detect whether the atom has the moment of $\ket\uparrow$
	or of $\ket\downarrow$, then the atom turns to become $\ket\uparrow$ or $\ket\downarrow$
	by curving into the corresponding directions with the possibility $|\alpha|^2$ or $|\beta|^2$, respectively.
	This experiment is called {\it Stern-Gerlach} experiment \index{Stern-Gerlach experiment}, which
	is a typical experiment to show the physical phenomenon of quantization of the magnetic moment of atoms.

	A physical system is called a $d$-level system when the system is a space for  $d$-dimensional vectors. 
	\index{level}

	\bigskip
	The physical system to be observed the polarization of a single photon or the magnetic momentum of a 
	single silver atom is a typical $2$-level system. 
	\index{magnetic momentum of a single silver atom}

	\subsubsection{Ket and bra}
		As a mathematical convenience, $\ket{\psi}$ is called a {\it ket} vector. $\bra{\psi}$ is the Hermitian transpose, 
		or the conjugate transpose of $\ket{\psi}$, which is
		called a {\it bra} vector. \index{ket} \index{bra}
		The former is considered a column vector, and the latter is considered a row vector
		on matrix arithmetic. Thus, one can consider $\ketbra{\psi}$ to be  
		a square matrix.

	\subsubsection{Tensor product of pure states}

	\index{tensor product!--- of pure states}
	You might like to consider the state of a system which contains two particles and more. 
	Here, we give the concepts of {\it tensor} product for two pure states. 
	The tensor product $\mathcal{H}\otimes\mathcal{H}'$ of two spaces $\mathcal{H}$ and $\mathcal{H}'$
	is the $d \times d'$-dimensional vector space when $\mathcal{H}$ and $\mathcal{H}'$ are 
	$d$- and $d'$- dimensional, respectively. 
	The tensor product 
	of two vectors 
	of $\mathcal{H}$ and $\mathcal{H}'$
	is defined as follows:  
	\begin{equation}
		\left(
			\begin{array}{c}
				x_1  \\
				x_2  \\
				\vdots \\
				x_d \\
			\end{array}
		\right) 
		\otimes
		\left(
			\begin{array}{c}
				y_1  \\
				y_2  \\
				\vdots \\
				y_{d'} \\
			\end{array}
		\right) 
		=
		\left(
			\begin{array}{c}
				x_1 y_1 \\
				x_1 y_2 \\
				\vdots \\
				x_1 y_{d'}\\
				x_2 y_1\\
				 \vdots \\
				x_2 y_{d'}\\
				 \vdots\\ \vdots \\
				x_d y_{d'}\\
			\end{array}
		\right) . 
	\end{equation}

	For example, $\left(\begin{smallmatrix}\sqrt{0.3} \\ \sqrt{0.7} \end{smallmatrix}\right) \otimes \left(\begin{smallmatrix}\sqrt{0.4} \\ \sqrt{0.6} \end{smallmatrix}\right) 
	= \left(\begin{smallmatrix}\sqrt{0.12} \\ \sqrt{0.18} \\ \sqrt{0.28} \\ \sqrt{0.42} \end{smallmatrix}\right) $.
	By convention, $\ket{\phi_1}\otimes \ket{\phi_2}$ is sometimes abbreviated as $\ket{\phi_1}\ket{\phi_2}$ or  $\ket{\phi_1 \, \phi_2}$.
	One can consider that tensor product of states of two particles is considered to be the state of the whole two particles.

	\bigskip 

	\subsection{Mixed states and tensor product}

	\index{mixed state}\index{state!mixed ---}
	A mixed state is represented by a 
	semi-positive
	 Hermitian matrix%
	\footnote{Semi-positive Hermitian matrix: 
	A matrix $\rho$ is {\it Hermitian} as long as it satisfies $\rho =\rho^\dag$. 
	In this case, all its eigenvalues are real numbers. 
	It is {\it semi-positive} when all its eigenvalues are equal to or larger than zero. 
	\index{semi-positive}
	\index{positive!semi- ---}
	\index{Hermitian!--- matrix}
	\index{matrix!Hermitian ---}
	}
	of which the 
	trace,
	or the summation of the diagonal elements, is $1$.  
	It is regarded as a {\it stochastic} mixture of multiple pure states $\{\ket{\psi_i}\}_i$, as $\rho = \sum_i p_i \ketbra{\psi_i}$
	with $\{p_i\}_i$ a probability distribution.
	\index{stochastic!--- mixture}
	\index{mixture!stochastic ---}
	Note that {\it stochastic mixture} is a different notion from {\it superposition} of pure states. The semi-positive Hermitian matrix of a mixed state is 
	called the {\it density matrix}. \index{density matrix}\index{matrix!density ---}
	This dissertation employs the conventional phrase  ``a state $\rho$ {\it on} $\mathcal{H}$''
	or
	``a state $\rho$ {\it on} $\mathbb{C}^d$''
	 if $\rho$ is a $d\times d$ semi-positive matrix of which the 
	trace is one. 
	Notation $\mathcal{B}(\mathcal{H})$ is employed to explicitly specify the set of states on a Hilbert space $\mathcal{H}$. 

	\begin{Notation}[mixed states]

		$\BB(\HH)$ is the set of any mixed states $\rho$ on a Hilbert space $\HH$. 
		When $\HH$ is $d$-dimensional, $\BB(\HH)$ is equal to the set of 
		$d\times d $ semi-positive matrix of which the trace is one. 

	\end{Notation}

	Note that a pure state $\ket{\phi}$ is regarded to be equal to a mixed state $\ketbra{\phi}$ as 
	a stochastic mixture of a single state $\ket{\phi}$ with 100\% weight.
	\index{stochastic!--- mixture}
	\index{mixture!stochastic ---}
	Mixed states other than pure states may be called {\it non-pure} states. 

	\index{tensor product!--- of mixed states}
	The tensor product of mixed states is an expansion to the tensor product of pure states, as follows: 
	\begin{equation}
		\begin{split}
			&
			\left(
				\begin{array}{ccc}
					a_{11} & \cdots &a_{1d} \\ 
					\vdots &&\vdots \\ 
					a_{d1}& \dots & a_{dd} 
				\end{array} 
			\right)
			\otimes 
			\left(
				\begin{array}{ccc} 
					b_{11} & \cdots &b_{1d'} \\ 
					\vdots &&\vdots \\ 
					b_{d'1}& \dots & b_{d'd'} 
				\end{array} 
			\right)
			\hfill
			\\
			&\qquad\hfill
			= 
			\left(
				\begin{array}{ccc}
					\begin{array}{ccc} 
						a_{11}b_{11} & \cdots &a_{11}b_{1d'} \\ 
						\vdots &&\vdots \\ 
						a_{11}b_{d'1}& \dots & a_{11}b_{d'd'} 
					\end{array}
					 & \cdots &
					\begin{array}{ccc}
						a_{1d}b_{11} & \cdots &a_{1d}b_{1d'} \\ 
						\vdots &&\vdots \\ 
						a_{1d}b_{d'1}& \dots & a_{1d}b_{d'd'}
					\end{array} \\ 
					\vdots &&\vdots \\
					\begin{array}{ccc} 
						a_{d1}b_{11} & \cdots &a_{d1}b_{1d'} \\ 
						\vdots &&\vdots \\
						a_{d1}b_{d'1}& \dots & a_{d1}b_{d'd'} 
					\end{array}& \dots & 
					\begin{array}{ccc} 
						a_{dd}b_{11} & \cdots &a_{dd}b_{1d'} \\ 
						\vdots &&\vdots \\ a_{dd}b_{d'1}& \dots & a_{dd}b_{d'd'} 
					\end{array}  
				\end{array} 
			\right) 
			.
		\end{split}
	\end{equation}

	\bigskip
	\subsection{Arithmetic on states}

	This subsection deals with some arithmetic on quantum states,
	such as, tracing out, von Neumann entropy and  quantum divergence.

	\index{trace out}
	\begin{Notation}[trace out, tracing out]
		For a  state $\rho = \sum_i p_i \, \rho_i^A \otimes \rho_i^B $ on $\mathcal{H}_A \otimes \mathcal{H}_B $, 
		tracing out operations are defined as follows.  
		Tracing out the space of $\mathcal{H}_B$ from $\rho$ is defined as  
		\begin{equation}
			\Tr_B \rho = \sum_i p_i \, (\Tr \rho_i^B) \cdot \rho_i^A, 
		\end{equation}
		resultingly  a state on $\HH_A$. 
		Similarly, tracing out the space of $\mathcal{H}_A$ is,
		\begin{equation}
			\Tr_A \rho = \sum_i p_i \, (\Tr \rho_i^A) \cdot \rho_i^B \text{\quad on \quad} \mathcal{H}_B.
		\end{equation}
	\end{Notation}

	$\Tr_B$ is an operation tracing out the space of $B$, leaving the space of $A$. 
	Conversely,
	$\Tr_A$ is an operation tracing out the space of $A$, leaving the space of $B$. 
	Concrete examples of tracing out appear at pp.\pageref{pagetrout}.

	\index{logarithm!--- of a matrix}
	\index{matrix!logarithm of a ---}
	\index{square root of a matrix}
	\index{matrix!square root of a ---}
	\begin{Notation}[$\log\rho,\sqrt{\rho}$] 
	For a semi-positive Hermite matrix $\rho$, $\log \rho$ and $\sqrt{\rho}$ are defined as follows: 
	\begin{equation}
		\begin{split}
			&\log \rho = U \, {\rm diag}\,[\log t_1,\dots,\log t_d  ] \, U^\dag \\ 
			&\qquad\qquad\qquad\text{ for }  \rho = U \, {\rm diag}\,[t_1,\dots, t_d] \, U^\dag,
		\end{split}
	\end{equation}
	and
	\begin{equation}
		\begin{split}
			&\sqrt\rho = U \, {\rm diag}\,[\sqrt{t_1},\dots,\sqrt{t_d} ] \, U^\dag \\
			&\qquad\qquad\qquad
			\text{ for } \rho = U \, {\rm diag}\,[t_1,\dots, t_d] \, U^\dag,
		\end{split}
	\end{equation}
	where ${\rm diag}\,[t_1,\dots, t_d] $ is the diagonal matrix $\left(\begin{smallmatrix}t_1&&\\&\ddots&\\&&t_d\end{smallmatrix}\right)$,
	and $U$ is a unitary matrix.
	\end{Notation}

	To keep consistency and to avoid confusion the following conventions are employed here:  
	\begin{equation}
		\text{$ \log 0 $ is treated as $-\infty $, additionally, $0 \log 0$ is treated as $0$. }
	\end{equation}

	\index{von Neumann!--- entropy}\index{entropy!von Neumann ---}
	\begin{Notation}[von Neumann entropy, \cite{neu}]
		\begin{equation}
		 S(\rho) = - \Tr \rho \log \rho 
		\end{equation}
	\end{Notation}

	\index{Shannon!--- entropy}\index{entropy!Shannon ---}
	One can show that 
	$S(\rho)$ is equal to the Shannon entropy of the whole eigenvalues of $\rho$, 
	that is, $S(\rho)=\sum(-\lambda_i \log \lambda_i)$ with $\{\lambda_i\}$ being the eigenvalues of $\rho$.
	Note that for pure states $\ketbra{\phi}$, the von Neumann entropy is zero.  

	\index{divergence}
	\index{quantum!--- divergence}
	\begin{Notation}[quantum divergence]
		The quantum divergence for two mixed states from the same space are defined as follows: 
		\begin{equation}
			H(\rho||\sigma) = \Tr ( \rho \log \rho - \rho \log \sigma ).
			\label{eq:quantumdivergence}
		\end{equation}
	\end{Notation}

	Quantum divergence is also called {\it quantum relative entropy}.
	\index{quantum!--- relative entropy}
	\index{relative entropy!quantum ---}
	\index{entropy!quantum relative ---}
	Quantum divergence has following properties, 
	as it is similar to a measure of distance of two states,
	but it lacks the property of symmetry. 

	\begin{Proposition}[basic properties of $H(\cdot||\cdot)$]
		\label{pp:pdis}
		\begin{eqnarray}
			&H(\rho||\sigma) \ge 0, \\
			& H(\rho || \sigma ) = 0 \quad \Leftrightarrow \quad \rho = \sigma , \\
			&H(\cdot_1 || \cdot_2 ) \not\equiv H(\cdot_2 || \cdot_1) .
		\end{eqnarray}

	\end{Proposition}

\subsection{Mapping on quantum states} 
	This subsection deals with mapping on quantum states. 
	In this dissertation, a ``map'' or ``mapping'' means a linear map.
	One can associate a mapping as a time transition of a quantum state, 
	formalized as a quantum channel later. 
	\index{map|see{mapping}}
	\index{mapping}
	\index{linear mapping}\index{mapping!linear ---}

	\index{tensor product!--- of channels}
	\index{tensor product!--- of mapping}\index{mapping!tensor product of ---s}
	\begin{Notation}[tensor product of mapping]
		For two mappings $\T_1$ and $\T_2$, the tensor product $\T_1 \otimes \T_2 $ 
		is defined based on a restriction that $\T_1 \otimes \T_2 (\rho_1 \otimes \rho_2 ) 
		= \T_1 (\rho_1) \otimes \T_2 (\rho_2)$ for any input $\rho_1$ of $\T_1$ and 
		 any input $\rho_2$ of $\T_2$.
	\end{Notation}


	\index{CP}\index{mapping!CP ---}
	\index{complete positivity|see{CP}}
	\index{positive!complete ---}
	\begin{Notation}[complete positivity, CP-ness]
		A {\it complete positive} map $\T$ is a map such that 
		for any (finite) dimensional identity map $I$, $\T\otimes I$ 
		maps a semi-positive Hermitian matrix into a semi-positive Hermitian matrix.
		This complete positivity is also called {\it CP}-ness.
	\end{Notation}

	\index{TP}
	\index{trace preserving|see{TP}}
	\index{mapping!TP ---}
	\begin{Notation}[trace preserving, TP-ness]
		A {\it trace preserving}  map $\T$ is a map satisfying
		$\Tr \T \rho = \Tr  \rho $ for any $\rho$. 
		This trace preserving property is also called {\it TP}-ness.
	\end{Notation}

	\index{CPTP}
	\index{mapping!CPTP ---}
	\begin{Notation}[CPTP-ness]\label{notation:cptp}
		Combining the two terms above, a map of CP and TP is a {\it CPTP} map. 
	\end{Notation}

	\index{Bures distance}
	\index{fidelity}
	\begin{Notation}[Bures distance and the fidelity]
	The fidelity between given two states is defined as 
		\begin{equation}
			F(\rho,\sigma) = \Tr \sqrt{\,\sqrt{\rho} \, \sigma \, \sqrt{\rho}\,}.
		\end{equation}
		The Bures distance is defined as 
		\begin{equation}
			B(\rho,\sigma) = 2 \sqrt {1-F(\rho,\sigma)}. \label{eq:Bures}
		\end{equation}
	\end{Notation}

	\begin{Proposition}
	The fidelity has the properties such as: 
	\begin{eqnarray}
	F(\cdot_1,\cdot_2)& \le& 1 \\
	F(\cdot_1,\cdot_1) &\equiv &1 \\
	F(\cdot_1,\cdot_2)& \equiv& F(\cdot_2,\cdot_1) \\
	F(\ketbra{\psi},\ketbra{\phi})& =& \langle \psi| \phi \rangle \quad \text{ for the same dimensional } 
	\ket\psi,\ket\phi. \label{eq:fidelity3}
	\end{eqnarray}
	\end{Proposition}

	The fidelity is  considered to be how truly a quantum state is transmitted. 
	$100\%$ fidelity, or the fidelity being one is the perfect transmission, which means the 
	transmission is done without losing information the quantum
	states held. \index{perfect transmission}
	Note that in (\ref{eq:fidelity3}), the left hand side is determined 
	even if the dimension of $\ket\psi$ and $\ket\phi$ is 
	different while  the right hand side is not defined. 

	The Bures distance (\ref{eq:Bures}) is a derivation from the fidelity to take on the properties 
	of a distance. The nearer upward to 1 or 100\% 
	the fidelity between two states become, the smaller downward to 0 the distance between the 
	two states become.  

\section{A quantum channel and its capacity}\label{section:channelcapacity}


	There are various frameworks to deal with time transition of a quantum states as follows: \\
	\index{Schr\"odinger!--- equation}
	\qquad (1) Schr\"odinger equation --- \fbox{$H\ket\psi=\sqrt{-1}\hbar\frac{\partial}{\partial t}\ket\psi$}
		\index{quantum!--- circuit}
		\\
	\qquad (2) Quantum circuit --- \fbox{\parbox{0.6in}{
\unitlength 0.1in
\begin{picture}(  6.1000,  3.2100)(  0.0000, -3.2100)
\put(3.6000,-1.9000){\makebox(0,0)[lb]{H}}%
%
\special{pn 8}%
\special{pa 300 0}%
\special{pa 510 0}%
\special{pa 510 200}%
\special{pa 300 200}%
\special{pa 300 0}%
\special{fp}%
%
\special{pn 8}%
\special{pa 510 90}%
\special{pa 610 90}%
\special{fp}%
\special{pa 300 90}%
\special{pa 0 90}%
\special{fp}%
\special{pa 10 270}%
\special{pa 600 270}%
\special{fp}%
\special{pa 170 50}%
\special{pa 170 270}%
\special{fp}%
%
\special{pn 8}%
\special{ar 170 90 52 52  0.0000000 6.2831853}%
%
\special{pn 8}%
\special{sh 0.600}%
\special{ar 170 270 52 52  0.0000000 6.2831853}%
\end{picture}%
}} 
		and, \\
		\index{quantum!--- channel}
	\qquad (3) Quantum channel  --- \fbox{\parbox{1.35in}{
\unitlength 0.1in
\begin{picture}( 13.1600,  2.6400)(  3.0300, -4.4400)
%
\special{pn 8}%
\special{ar 596 312 30 132  0.0000000 6.2831853}%
%
\special{pn 8}%
\special{pa 596 180}%
\special{pa 1340 180}%
\special{fp}%
%
\special{pn 8}%
\special{pa 596 444}%
\special{pa 1340 444}%
\special{fp}%
%
\special{pn 8}%
\special{ar 1340 312 30 126  4.7123890 6.2831853}%
\special{ar 1340 312 30 126  0.0000000 1.5707963}%
%
\special{pn 20}%
\special{ar 350 312 46 52  0.0000000 6.2831853}%
%
\special{pn 20}%
\special{pa 488 312}%
\special{pa 602 312}%
\special{fp}%
%
\special{pn 20}%
\special{pa 1376 312}%
\special{pa 1460 312}%
\special{fp}%
\special{sh 1}%
\special{pa 1460 312}%
\special{pa 1392 292}%
\special{pa 1406 312}%
\special{pa 1392 332}%
\special{pa 1460 312}%
\special{fp}%
%
\special{pn 20}%
\special{ar 1574 306 46 52  0.0000000 6.2831853}%
\put(7.8700,-3.6600){\makebox(0,0)[lb]{channel}}%
\end{picture}%
}}.

		(1) and (2) treat only pure states, and (3) treats mixed states generally. To investigate quantum 
		information science, treating mixed states is necessary.

	\index{quantum!--- channel}
	\begin{Definition}[quantum channel]
	A quantum channel $\T$ is a CPTP map (see Notation \ref{notation:cptp}). Thus, 
	$ ( \T \rho ) \otimes \sigma $ is a quantum state for any quantum state $\sigma$. (See Fig.\ref{fig:cptp}.)

	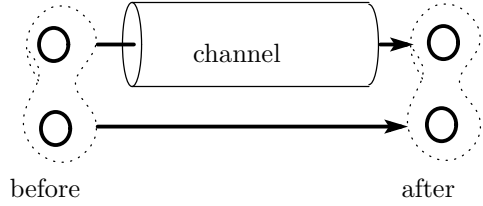
\begin{figure}
		\begin{center}
\unitlength 0.1in
\begin{picture}( 24.7200,  9.3000)(  1.5000,-11.0000)
%
\special{pn 8}%
\special{ar 790 400 50 220  0.0000000 6.2831853}%
%
\special{pn 8}%
\special{pa 790 180}%
\special{pa 2030 180}%
\special{fp}%
%
\special{pn 8}%
\special{pa 790 620}%
\special{pa 2030 620}%
\special{fp}%
%
\special{pn 8}%
\special{ar 2030 400 50 210  4.7123890 6.2831853}%
\special{ar 2030 400 50 210  0.0000000 1.5707963}%
%
\special{pn 20}%
\special{ar 380 400 78 88  0.0000000 6.2831853}%
%
\special{pn 20}%
\special{pa 610 400}%
\special{pa 800 400}%
\special{fp}%
%
\special{pn 20}%
\special{pa 2090 400}%
\special{pa 2230 400}%
\special{fp}%
\special{sh 1}%
\special{pa 2230 400}%
\special{pa 2164 380}%
\special{pa 2178 400}%
\special{pa 2164 420}%
\special{pa 2230 400}%
\special{fp}%
%
\special{pn 20}%
\special{pa 610 830}%
\special{pa 2200 830}%
\special{fp}%
\special{sh 1}%
\special{pa 2200 830}%
\special{pa 2134 810}%
\special{pa 2148 830}%
\special{pa 2134 850}%
\special{pa 2200 830}%
\special{fp}%
%
\special{pn 8}%
\special{pa 300 590}%
\special{pa 284 562}%
\special{pa 268 534}%
\special{pa 252 504}%
\special{pa 242 474}%
\special{pa 234 444}%
\special{pa 230 412}%
\special{pa 232 378}%
\special{pa 240 344}%
\special{pa 254 310}%
\special{pa 272 278}%
\special{pa 292 250}%
\special{pa 318 224}%
\special{pa 346 204}%
\special{pa 376 188}%
\special{pa 408 182}%
\special{pa 440 182}%
\special{pa 472 190}%
\special{pa 504 206}%
\special{pa 534 228}%
\special{pa 562 254}%
\special{pa 584 282}%
\special{pa 600 314}%
\special{pa 610 346}%
\special{pa 610 380}%
\special{pa 604 410}%
\special{pa 588 442}%
\special{pa 570 470}%
\special{pa 548 498}%
\special{pa 528 526}%
\special{pa 510 554}%
\special{pa 496 580}%
\special{pa 490 606}%
\special{pa 494 634}%
\special{pa 506 660}%
\special{pa 524 686}%
\special{pa 544 714}%
\special{pa 566 742}%
\special{pa 586 770}%
\special{pa 602 800}%
\special{pa 610 832}%
\special{pa 612 864}%
\special{pa 604 898}%
\special{pa 590 930}%
\special{pa 572 960}%
\special{pa 546 986}%
\special{pa 518 1010}%
\special{pa 488 1028}%
\special{pa 454 1038}%
\special{pa 420 1042}%
\special{pa 388 1036}%
\special{pa 354 1026}%
\special{pa 324 1008}%
\special{pa 294 988}%
\special{pa 268 962}%
\special{pa 246 932}%
\special{pa 228 902}%
\special{pa 216 870}%
\special{pa 210 838}%
\special{pa 210 808}%
\special{pa 216 776}%
\special{pa 226 746}%
\special{pa 238 716}%
\special{pa 254 686}%
\special{pa 268 658}%
\special{pa 284 628}%
\special{pa 296 596}%
\special{pa 302 566}%
\special{pa 300 536}%
\special{pa 288 508}%
\special{pa 272 478}%
\special{pa 252 450}%
\special{pa 230 420}%
\special{sp -0.045}%
%
\special{pn 8}%
\special{pa 2310 580}%
\special{pa 2294 552}%
\special{pa 2278 524}%
\special{pa 2262 494}%
\special{pa 2252 464}%
\special{pa 2244 434}%
\special{pa 2240 402}%
\special{pa 2242 368}%
\special{pa 2250 334}%
\special{pa 2264 300}%
\special{pa 2282 268}%
\special{pa 2302 240}%
\special{pa 2328 214}%
\special{pa 2356 194}%
\special{pa 2386 178}%
\special{pa 2418 172}%
\special{pa 2450 172}%
\special{pa 2482 180}%
\special{pa 2514 196}%
\special{pa 2544 218}%
\special{pa 2572 244}%
\special{pa 2594 272}%
\special{pa 2610 304}%
\special{pa 2620 336}%
\special{pa 2620 370}%
\special{pa 2614 400}%
\special{pa 2598 432}%
\special{pa 2580 460}%
\special{pa 2558 488}%
\special{pa 2538 516}%
\special{pa 2520 544}%
\special{pa 2506 570}%
\special{pa 2500 596}%
\special{pa 2504 624}%
\special{pa 2516 650}%
\special{pa 2534 676}%
\special{pa 2554 704}%
\special{pa 2576 732}%
\special{pa 2596 760}%
\special{pa 2612 790}%
\special{pa 2620 822}%
\special{pa 2622 854}%
\special{pa 2614 888}%
\special{pa 2600 920}%
\special{pa 2582 950}%
\special{pa 2556 976}%
\special{pa 2528 1000}%
\special{pa 2498 1018}%
\special{pa 2464 1028}%
\special{pa 2430 1032}%
\special{pa 2398 1026}%
\special{pa 2364 1016}%
\special{pa 2334 998}%
\special{pa 2304 978}%
\special{pa 2278 952}%
\special{pa 2256 922}%
\special{pa 2238 892}%
\special{pa 2226 860}%
\special{pa 2220 828}%
\special{pa 2220 798}%
\special{pa 2226 766}%
\special{pa 2236 736}%
\special{pa 2248 706}%
\special{pa 2264 676}%
\special{pa 2278 648}%
\special{pa 2294 618}%
\special{pa 2306 586}%
\special{pa 2312 556}%
\special{pa 2310 526}%
\special{pa 2298 498}%
\special{pa 2282 468}%
\special{pa 2262 440}%
\special{pa 2240 410}%
\special{sp -0.045}%
%
\special{pn 20}%
\special{ar 390 840 78 88  0.0000000 6.2831853}%
%
\special{pn 20}%
\special{ar 2420 390 78 88  0.0000000 6.2831853}%
%
\special{pn 20}%
\special{ar 2410 820 78 88  0.0000000 6.2831853}%
\put(11.1000,-4.9000){\makebox(0,0)[lb]{channel}}%
\put(1.5000,-11.0000){\makebox(0,0)[lt]{before}}%
\put(22.0000,-11.0000){\makebox(0,0)[lt]{after}}%
\end{picture}%

		\end{center}
		\caption[A quantum channel/CPTP. ]{A quantum channel must be a CPTP map in order to 
		keep the whole world a quantum state that is represented as a semi-positive Hermitian matrix. }
		\label{fig:cptp}
	\end{figure}

	\end{Definition}

	Because of the linearity
	of mappings of quantum channel, $\rho' = \T(\rho)$ is 
	abbreviated as $\rho' = \T \rho$. 

	\index{tensor product!--- of channels}
	\index{channel!tensor product of ---s}
	\begin{Definition}[Tensor product of channels]
		For channels $\T$ and $\T'$, the tensor product of these channels $\T\otimes \T'$ is 
		a linear map which maps $\rho \otimes \rho' $  to $ (\T \rho) \otimes (\T' \rho') $ where
		$\rho$ is any input of $\T$ and $\rho'$ is any input of $\T'$. 
	\end{Definition}

		The concept of tensor product is natural to consider multiple channels in the physical 
		world. 
	
	\index{Stinespring's dilation}
	\begin{Theorem}[Stinespring's dilation]
	\index{auxiliary!--- space}\index{auxiliary!--- state}
	For any quantum channel $\T:\rho\mapsto\rho'$, there exists some auxiliary space 
	$K$, some auxiliary state $\sigma$, and  a unitary matrix $U$ such that
	\begin{equation}
	\rho' = \Tr_K U \left(\rho \otimes \sigma \right) U^\dagger.   
	\end{equation}
	\end{Theorem}

	Now we have formalized what a quantum channel is. Then, how is the capacity of the 
	channel considered, {\it i.e.}, 
	how much information can be carried from the sender  to the receiver in a remote place
	through a quantum channel? One of the formalizations is the Holevo capacity. 
	The definition follows.

	\index{Holevo capacity}\index{capacity!Holevo ---}
	\begin{Definition}[Holevo capacity, \cite{Holevo73,Holevo98,SW97}]
	The Holevo capacity $C$ of a given quantum channel $\T$ is as follows: 
	\begin{equation}
	C(\T) = \max_{n ; \rho_1, \dots, \rho_n ; p_1, \dots, p_n } S(\sum_{i=1}^n p_i \, \T \rho_i) - \sum_{i=1}^n p_i \,  S( \T \rho_i) , \label{eq:00125}
	\end{equation}
	where $ n \in \mathbb{N}, p_i > 0 , \sum_{i=1}^n p_i = 1 $ and every $\rho_i$ s is an input of the channel $\T$. 
	\end{Definition}

	 The optimized $n$ in (\ref{eq:00125}) is regarded as the number of kinds of the input states of the channel $\T$,
	 when the communication capacity attains the Holevo capacity of $\T$ in an asymptotic sense. 

	\begin{Theorem}[\cite{ohya-petz-watanabe}]
	\begin{equation}
	C(\T) = \min_\sigma \max_\rho H( \T \rho \,|| \,\T \sigma)
	\label{eq:125}
	\end{equation}
	where $\sigma$ and $\rho$ are the inputs of the channel $\T$.
	\end{Theorem}

	Physically, the Holevo capacity is a classical information capacity of a given quantum channel
	at which input particles are not allowed to be entangled with each other, and the output particles are
	measured collectively. To fully utilize quantum aspects of a quantum channel, one 
	might consider the capacity at which the input particles are freely entangled with each other
	to send a message. 
	This capacity is called the full capacity.  

	\index{full capacity}\index{capacity!full ---}
	\begin{Definition}[Full capacity]
	\begin{equation}
	\bar{C}(\T) = \lim_{n\to\infty} C(\T^{\otimes n} ) / n 
	\end{equation}

	\end{Definition}
	 
	 
	\section{Entanglement and its quantification}
	\index{entanglement}
	This section defines  the {\it quantum entanglement}  in the case of a bipartite system.

	When considering bipartite quantum entanglement, we often think of two figures,
	named Alice and Bob. \index{Alice and Bob}\index{Bob|see{Alice and Bob}} 
	Their spaces, at which quantum states are considered, 
	are represented as Hilbert spaces
	$\mathcal{H}_A$ and $\mathcal{H}_B$, respectively.
	%
	%

	\index{entangled|see{entanglement}}
	\index{entanglement!--- of pure states}
	\index{entanglement!--- of mixed states}
	\begin{Definition}[entanglement]\label{def:entanglement}
	For a pure state $\ket{\phi} \in \HH_A\otimes\HH_B$, $\ket{\phi}$ is said to be entangled\\ 
	\qquad if $\ket{\phi}$ cannot be represented in a form $\ket{\phi_A}\otimes \ket{\phi_B}$ with 
	$\ket{\phi_A}\in\HH_A$ and $\ket{\phi_B}\in\HH_B$.\\
	For a mixed state $\rho$ on $\HH_A\otimes\HH_B$, $\rho$ is said to be entangled\\
	\qquad if 
	$\rho$ cannot be represented as $\sum_i p_i \ketbra{\phi_i}$ \\
	\qquad with unentangled pure states $\ket{\phi_i}$ and a probability distribution $\{p_i\}$.
	\end{Definition}
	An  unentangled state is called {\it separable}. \index{separable}
	 To explicitly refer to two system which are entangled, notation $\HH_A:\HH_B$ is 
	used in this dissertation. When considering the entanglement, our interest reside 
	at two sites (Alice and Bob) rather than their space $\HH_A$ and $\HH_B$. In such a 
	case we just denote {\tt Alice:Bob} or {\sf A:B}.

	\index{LOCC}
	\begin{Definition}[LOCC, cf. \cite{BDSW}]
	LOCC operations (standing for {\it Local Operation} plus {\it Classical Communication})
	for a bipartite system are defined to be comprised only of: 
	\\
	\qquad (1) local CPTP operations on each side, and \\
	\qquad (2) communications between the two sites by only classical means. \\
	{\it Classical communication} means the following: One side physically
	measures the state of its own system, and transmit the output of the measurement to 
	the other side. Based on this information,  the receiver may perform physical 
	operations of its own system. Physical measurements are formalized to be the concept 
	of {\it POVM}, which we do not take up in this dissertation. 
	%
	%
	\\

	\begin{figure}[hbt]
	\bigskip
	\begin{center}
	\includegraphics[width=11cm]{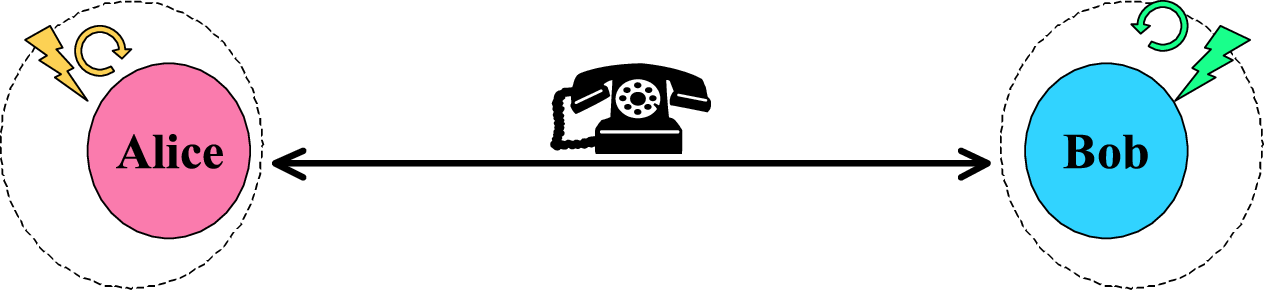}
	\end{center}
	\caption[LOCC]{LOCC operations: Only Local Operations and Classical Communications are allowed.
	Joint operations or sending quantum information is prohibited if it is beyond the allowed operations.}
	\end{figure}
	\end{Definition}

	There is a well-established entanglement measure for bipartite pure states.

	\index{reduced von Neumann entropy}
	\index{entropy!reduced von Neumann ---}
	\index{von Neumann!reduced --- entropy}
	\begin{Notation}[reduced von Neumann entropy]
	The {\it reduced} von Neumann entropy for a bipartite state  is defined  as
	\begin{multline}
	\hspace{4cm}
	E\AB(\ket\phi) = S ( \Tr_B \ketbra\phi )   \\
	\qquad\qquad\qquad \mbox{ for a pure state } \ket\phi \in  \HH_A \otimes \HH_B. 
	\hspace{4cm}
	\end{multline}
	One can easily confirm that $S(\Tr_B\ketbra\phi)= S ( \Tr_A \ketbra\phi )$ thus 
	$E\AB(\ketbra\phi)=S(\Tr_A\ketbra\phi)$. 

	\end{Notation}

	If a pure state is entangled, the reduced von Neumann entropy is larger than 0.

	\index{Bell!--- state}
	\index{state!Bell ---}
	\index{Bell!--- pair}
	\index{pair!Bell ---}
	\begin{Example}[Bell state, Bell pair]
	\begin{equation}
	\ket{\Phi^\pm}=\frac{\ket{00} \pm \ket{11} }{\sqrt{2}}, \ket{\Psi^\pm}=\frac{\ket{01} \pm \ket{10} }{\sqrt{2}}
	\end{equation}
	are called the Bell states.  These states $\ket{\Phi^+},\ket{\Phi^-},\ket{\Psi^+}$ and $\ket{\Psi^-}$ are often
	considered to be implicitly shared by Alice and Bob throughout this dissertation.   Their reduced von Neumann entropy is 1. 
	\end{Example}
	%

	Bell states are often considered to be units of quantum entanglement. 

	For bipartite mixed states, there are various candidates to measure 
	quantum entanglement as follows. 

	\index{entanglement!--- cost}
	\index{entanglement!--- distillation}
	\begin{Definition}[entanglement cost, entanglement distillation \cite{BDSW}]
	The {\it entanglement cost} is defined as
	\begin{equation}
	E_C\AB(\rho) = \inf \left\{ \> e  \>\big| \> ^\forall(\epsilon,\delta), ^\exists(m ,n, L) :
	|e - \frac{m}{n} | \le \delta , B \left( L(\ketbra{\Psi^-}^{\otimes m}) , \rho^{\otimes n} \right) \le \epsilon  
	\>\right\}, \label{eq:defec}
	\end{equation}
	and
	the {\it entanglement distillation } is defined as
	\begin{equation}
	E_D\AB(\rho) = \sup \left\{ \> e \> \big| \> ^\forall(\epsilon,\delta), ^\exists(m ,n, L) :
	|e - \frac{m}{n} | \le \delta , B \left( \ketbra{\Psi^-}^{\otimes m} , L(\rho^{\otimes n}) \right) \le \epsilon  
	\>\right\}, \label{eq:defed}
	\end{equation}
	where $e\ge 0, \; \epsilon,\delta > 0, \; m,n \in \mathbb{N}$ and $L(\cdot)$ is a LOCC operation.  $B(\cdot,\cdot)$ is the Bures distance defined in (\ref{eq:Bures}).
	\end{Definition}


	The entanglement distillation $E_D$ is the asymptotic quantity of  Bell states distilled 
	from $\rho$ with LOCC operations. Thus, one can consider $E_D$ as a measure to quantify the 
	entanglement as a resource to be used for quantum teleportation or quantum super dense coding. 

	The entanglement cost $E_C$ is the asymptotic quantity of Bell states necessary to produce 
	$\rho$ with LOCC operations. 

	\begin{figure}[h]
	\bigskip
	\begin{center}
	\includegraphics[width=12cm]{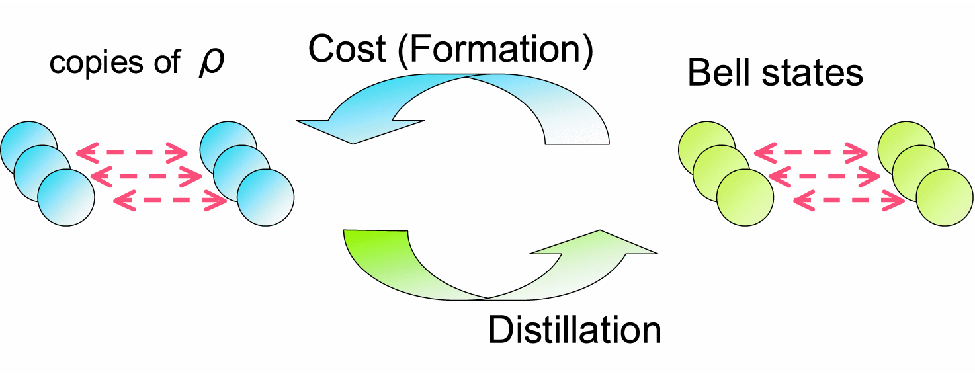}\vspace*{-1cm}
	\end{center}
	\caption{Difference between $E_C$ and $E_D$}
	\end{figure}

	\index{entanglement!--- of formation}
	\begin{Definition}[entanglement of formation \cite{BDSW}]\label{def:eof}
	The {\it entanglement of formation } is defined as
	\begin{equation}
	E_F\AB(\rho) = \min_{ n  ; \, \ket{\phi_1},\ldots, \ket{\phi_n};\, p_1,\ldots,p_n} 
	\sum_{i=1}^n p_i E\AB(\ketbra{\phi_i} ), \label{eq:defef}
	\end{equation}
	with  $p_i>0, \sum_{i=1}^n p_i = 1 , \sum_{i=1}^n p_i \ketbra{\phi_i} = \rho $. 
	\end{Definition}

	We may use the denotations $E(\cdot), E_C(\cdot), E_D(\cdot), E_F(\cdot)$ which omit the superscription $\sf A:B$ when the two sites are clear to consider entanglement. 

	\begin{Proposition}[ \cite{HHT}]
	The following holds: 
	\begin{equation}
	E_C (\rho ) = \lim_{n\to\infty} \frac{E_F(\rho^\oN)}{n} \label{eq:hht01}
	\end{equation}
	\end{Proposition}

	This equality (\ref{eq:hht01}) is significant in that the conceptually defined $E_C$ 
	represented as (\ref{eq:defec}) has become mathematically defined 
	by substituting (\ref{eq:defef}) into (\ref{eq:hht01}). 
	The calculation is, however, not simple, which leads to the main
	subject of this dissertation because  (\ref{eq:hht01}) being substituted with (\ref{eq:defef}) is in a 
	limitation form over infinitely many optimization forms
	$\{E_F(\rho^{\otimes n}) / n \}_{n=1}^{\infty}$. 

\section{Example: a two-level system}

	 This section presents  properties of the Bloch sphere representing states on a {\it qubit}, or a 2-level system.%
	\footnote{``Qubit'' can be considered as a unit of 
	quantum information as the counterpart of the concept of 
	``bit'' of the current information theory. Qubit is named by Benjamin Schumacher.}
	\index{qubit}

	\subsection{Bloch sphere and the Stokes parameterization} 

	\index{Stokes parameterization}
	\index{Bloch sphere}
	Assume 
	\begin{equation}
		\rho(x,y,z)=
		\frac{1}{2} 
		\left(
			\begin{array}{cc}
				1+z  &x-\I y \\ 
				x+\I y & 1-z 
			\end{array}
		\right)
		\label{def:stokes}
	\end{equation}
	 is a 
	2-dimensional mixed state. One can easily confirm that the semi-positivity is equivalent to
	$x^2 + y^2 + z^2 \le 1 $. The representation of a 2-dimensional mixed state under this condition
	is called the {\it Stokes parameterization}. The sphere $\{(x,y,z)  \,| \,x^2+ y^2+z^2 \le 1\}$, in which each 
	point is associated with
	the mixed state $\rho(x,y,z)$, is called the {\it Bloch sphere}. The surface of the Bloch sphere corresponds to
	pure states. 

	\begin{figure}[htbp]
	\begin{center}
	\includegraphics[width=8cm]{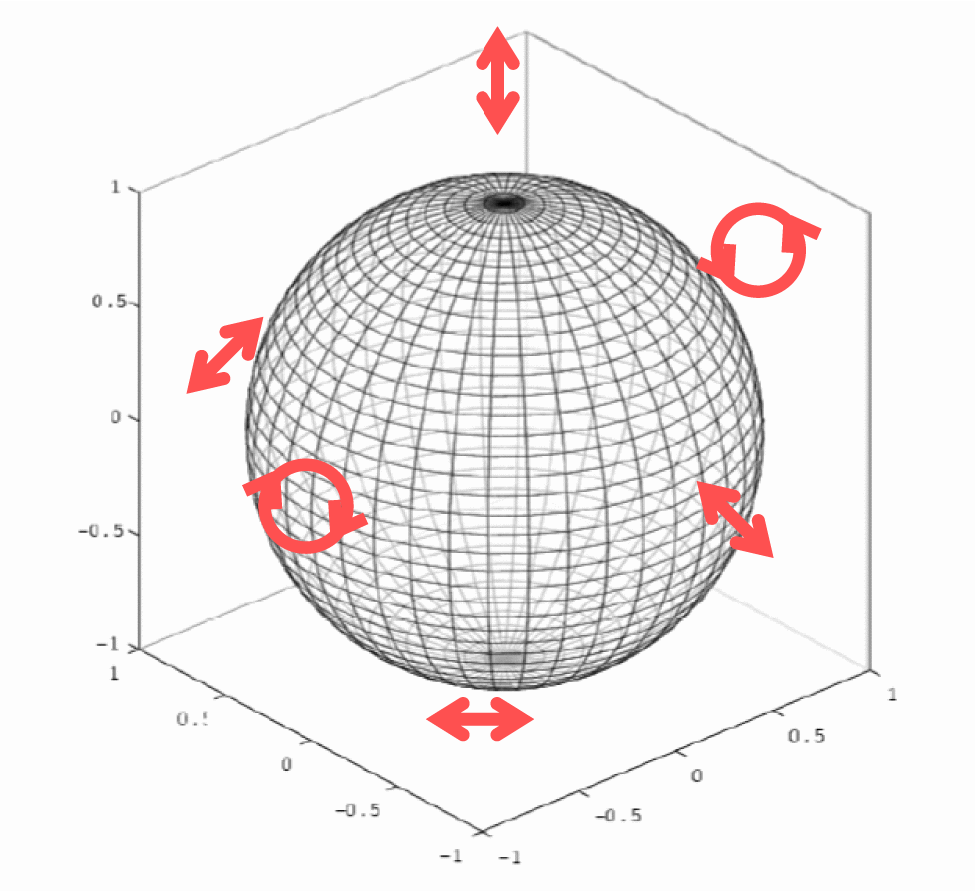}
	\caption[A photon's polarization on the Bloch sphere]{The polarizations of a photon on the Bloch sphere}
	\end{center}
	\end{figure}
	\index{polarization!--- of a single photon}

	\index{qubit!--- channel}\index{qubit}
	\subsection{A qubit channel}\label{subsec:aqc}
\begin{figure}[htbp]
	\begin{center}
	\includegraphics[width=8cm]{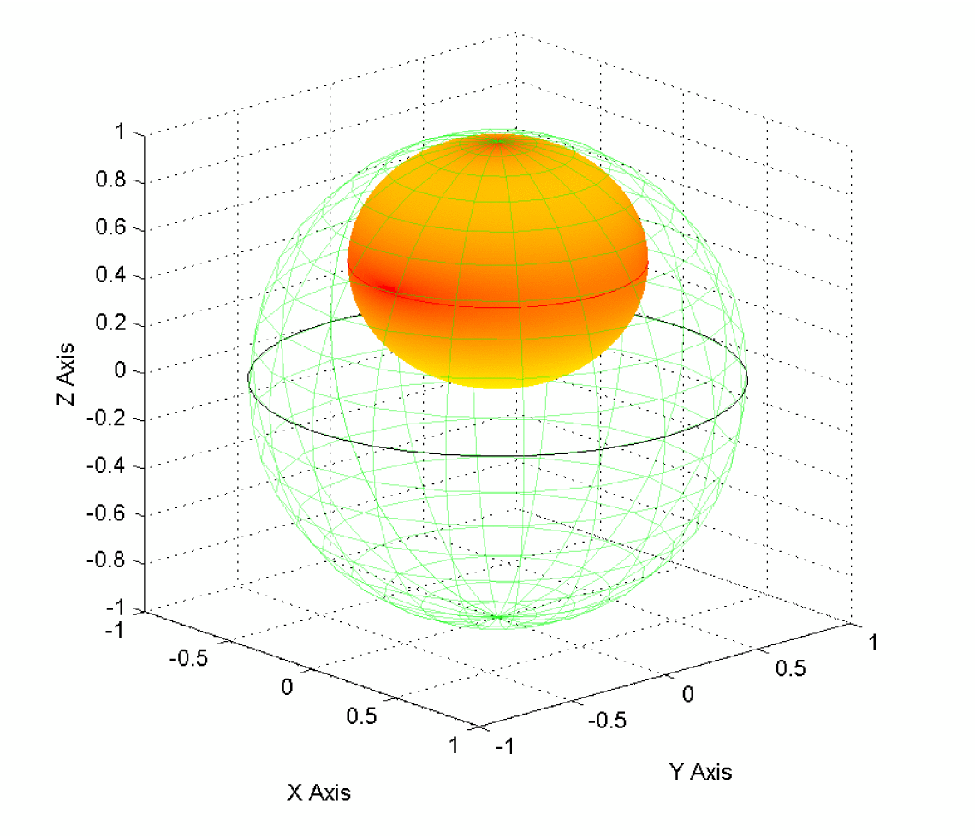}
	\end{center}
	\caption{A qubit channel's output ellipsoid}%
	\label{lemon}
	\end{figure}

	A qubit channel, which maps a qubit space to a qubit space, can be 
	regarded as an affine transformation on the Bloch sphere. Thus a qubit 
	channel is characterized by the {\it output ellipsoid} inside the Bloch sphere. See
	Fig.~\ref{lemon}.
	\index{output ellipsoid}

	\subsection{The von Neumann entropy on the Bloch sphere}

	In order to consider the Holevo capacity of a qubit channel, 
	here we consider the von Neumann entropy on the Bloch sphere.
	The eigenvalues of $\rho(x,y,z)$ (see \ref{def:stokes}) are 
	$\frac{1+\sqrt{x^2+y^2+z^2}}{2}$ and $\frac{1-\sqrt{x^2+y^2+z^2}}{2}$,
	so the von Neumann entropy is the binary entropy of 
	them. The von Neumann entropy on the Bloch sphere
	is spherically-symmetric and concave function.
	\index{binary entropy|(}
	\index{entropy!binary ---|(}

	\begin{figure}[htbp]
	\begin{center}
	\begin{minipage}{8cm} \includegraphics[width=10cm]{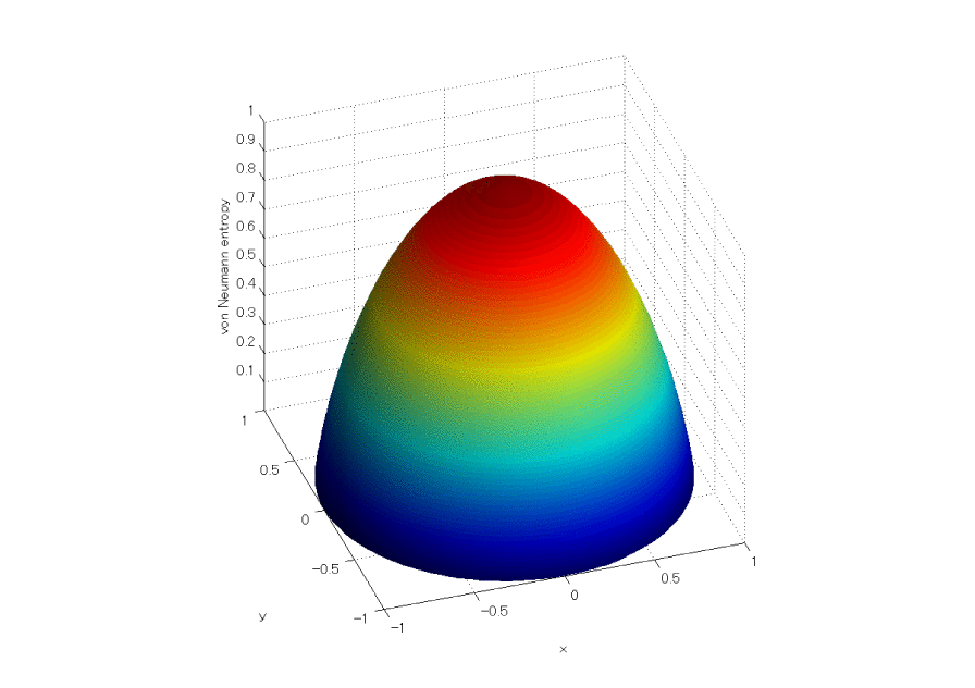} \end{minipage}\quad
	\begin{minipage}{7cm} \includegraphics[width=5.5cm]{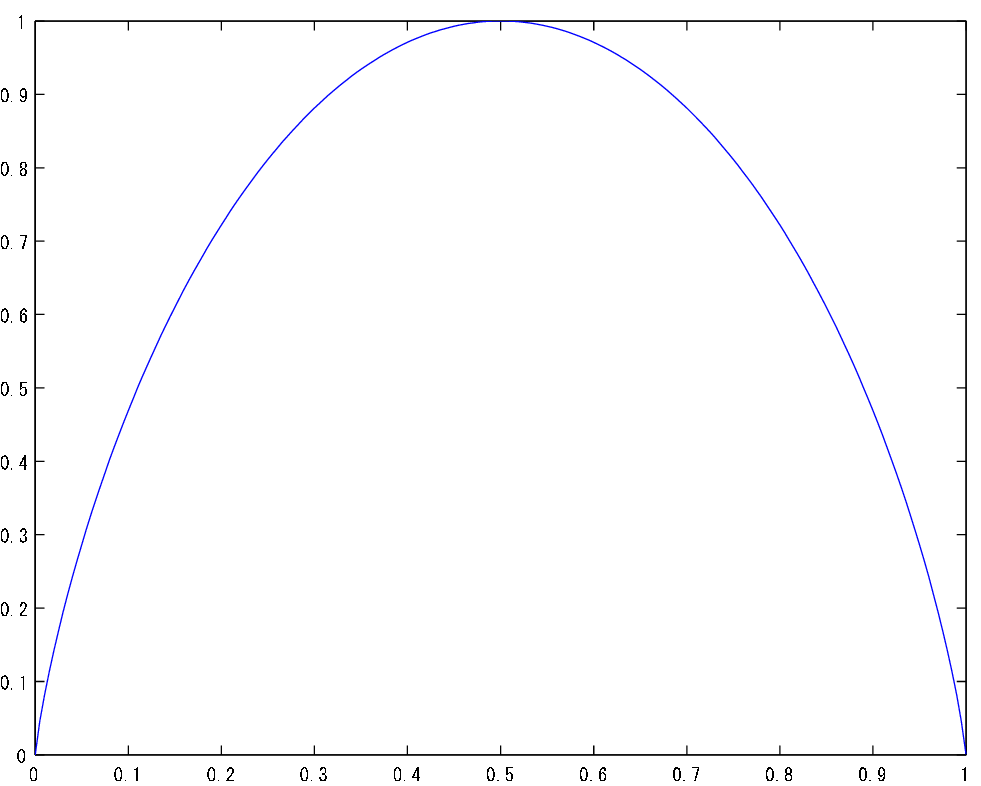} \end{minipage}
	\index{binary entropy|)}
	\index{entropy!binary ---|)}
	\index{von Neumann!--- entropy}
	\index{entropy!von Neumann ---}
	\caption[$-\Tr \rho\log\rho$ on a Bloch sphere's section, ---{\it cf.} the binary entropy]{
	{\sc Left:} The von Neumann entropy on the $xy$-section of the Bloch sphere. Note that
	the domain of this function is originally three-dimensional,
	and it is reduced to two-dimensional $xy$-plane in this figure. 
	\quad{\sc Right:}~ The binary entropy. 
	}
	\index{Alice and Bob}
	\end{center}
	\end{figure}

\section{Relation between $E_F$ and the Holevo capacity}
	\label{sec:stine}
	Due to Stinespring's dilation~\cite{Stinespring} any
	 CPTP map $\T:\BB(\HH_\mathtt{in}) \to \BB(\HH_\mathtt{out})$ can be represented as
	the composition of an isometric embedding\index{isometric!--- embedding}%
	\footnote{ A linear operator or map $U$ is called an {\it isometric embedding} if $U^\dag U $ is 
	an identity operator. The name {\it isometric embedding} emerged from the following two reasons: 
	\begin{itemize}
	\item Any vector $\ket{\phi}$ does not change its length after the mapping $U$, 
	as $ \left\|\, \ket{\phi} \,\right\| = \|\, U\ket{\phi} \,\| $. This is  because
	 $ \|\, \ket{\phi} \,\| ^2 = \langle \phi | {\phi} \rangle    = \bra{\phi} U^\dag U  \ket{\phi}   
	 =  (  U \ket{\phi} ) ^\dag  ( U \ket{\phi} ) = \|\, U\ket{\phi} |\,\| ^2 $ 
	 and $ \|\, \ket{\phi} \,\| \ge 0 , \|\, U\ket{\phi} \,\| \ge 0 $. 
	\item Any vectors do not change their distance after the mapping $ U $
	as $ \|\, \ket{\phi} - \ket{\psi} \,\| = \|\, U  \ket{\phi} - U \ket{\psi} \,\| $.
	\end{itemize}
	Note that isometric embedding is not always a unitary transformation, even though 
	any unitary transformation is always an isometric embedding.
	 } 
	of ${\BB(\HH_\mathtt{in})}$ into a bipartite
	system $\BB(\HH_\mathtt{out})\otimes \BB(\HH_\mathtt{aux})$ 
	followed by the operation tracing out $\BB(\HH_\mathtt{aux})$ 
	leaving  the output space $\BB(\HH_\mathtt{out})$, represented as,
	\begin{equation}
	  \label{eq:embedding}
	  \T: \quad 
	  \begin{aligned}
	  \BB({\HH_\mathtt{in}})
	   &  \>  \stackrel{U}  {\hookrightarrow}\> &\> \BB(\HH_\mathtt{out} \otimes \HH_\mathtt{aux})
	                 & \stackrel{\tr_{\HH_\mathtt{aux}}}  {\longrightarrow} & \BB(\HH_\mathtt{out})\quad \\
	   \rho\quad  & \>\mapsto &  U \rho \> U^\dag  \quad &  \longmapsto &  \quad \tr_{\HH_\mathtt{out}} U \rho \> U^\dag
	  \end{aligned}
	\end{equation}
	by choosing $\HH_\mathtt{aux}$ of which the dimension large enough
	and the isometric embedding $U$. 
	\begin{Notation}[$\KK_\lambda$, image vectors of the isometric embedding]
		Denote ${\cal K}_\T = U{\HH_\mathtt{in}}$, a subspace of $\HH_\mathtt{out}\otimes \HH_\mathtt{aux}$, the image
		subspace of $U$ where $U$ appeared in (\ref{eq:embedding}).
	\end{Notation}
	We can say that the channel $\T$ is equivalent to a 
	tracing out operation from ${\cal K}_\T$ to the output space $\HH_\mathtt{out}$,
	with an isometric embedding $U$.

	\begin{Theorem}[see \cite{MSW}]
		\label{thm:hcap:entf}
		\begin{equation}
			\label{eq:hcap:alt}
			C(\T) = \sup\{ S\bigl(\tr_{{\cal H}_\mathtt{aux}}\rho\bigr)-E_F\OA(\rho) :
			                                                 \rho\text{  on }{\cal K}_\T\}.
		\end{equation}
		where $C(\cdot)$ is the Holevo capacity defined in (\ref{eq:125}). 
	\end{Theorem}

	\begin{Notation}[$\OO_\lambda(\subset \BB( \KK_\lambda) )$, preimage matrices of the average output.]
		For a given channel, denote as follows: 
		\begin{equation}
			{\cal O}_\T = \mathop{\arg\max}\limits_{\rho\in\BB({\cal K}_\T)} 
			\{ S\bigl(\tr_{\HH_\mathtt{aux}}\rho\bigr)-E_F\OA(\rho) : \rho\text{  on }{\cal K}_\T\}.
		\end{equation}

	\end{Notation}

	\begin{Theorem}
		\label{thm:CC:implies:EE}
		If $C(\T\otimes \T')=C(\T)+C(\T')$ holds, then following holds: 
		\begin{equation}
			\forall\rho\in{\cal O}_\T,\forall\rho'\in{\cal O}_{\T'}\quad
			                            E_F\OA(\rho\otimes\rho')=E_F\OA(\rho)+E_F\OA(\rho').
		\end{equation}
		\qed
	\end{Theorem}
	\par
	It is interesting whether $C(\T^{\otimes n})=n C(\T)$ holds for a given/every channel $\T$,
	because one can conclude whether $E_F(\rho^{\otimes n})=n E_F(\rho)$ holds
	by the theorem above.
	As partial results, 
	following channels are known to satisfy this additivity property: 
	\begin{itemize}\setlength{\itemsep}{0pt}\setlength{\parsep}{0pt}
	\item Unital qubit channels%
	\footnote{A unital channel is a quantum channel $\T:\BB(\HH_\mathtt{in})\to\BB(\HH_\mathtt{out})$
	which maps the maximally mixed state of $\HH_\mathtt{in}$ 
	to the maximally mixed state of $\HH_\mathtt{out}$. \index{maximally mixed state}
	\index{state!maximally mixed ---}
	The {\it} maximally mixed state is defined as 
	 $\frac{1}{d}\sum_{i=1}^d \ketbra{i} 	$
	 for  $\BB(\mathbb{C}^d)$. 
	}%
	, cf. \cite{king:unital:add,king:unital:add:2}\index{unital channel}\index{channel!unital ---}
	\item Entanglement--breaking channels\footnote{%
	An entanglement-breaking channel $\T:\BHi\to\BHo$ is an {\it entanglement breaking} channel 
	if it satisfies the following condition. 
	Any output of $\T\otimes \tt{1}:\BHi\otimes\BHa\to\BHo\otimes\BHa$ is not entangled 
	between $\Ho$ and $\Ha$ for any auxiliary space $\Ha$. Here $\tt{1}$ is the identity 
	map of $\BHa$. 
	%
	}%
	, cf. \cite{Shor02}\index{entanglement-breaking channel}\index{channel!entanglement-breaking ---}
	\item Arbitrary depolarizing channels King\footnote{%
	A channel $\T:\BB(\HH)\to\BB(\HH)$ is an depolarizing channel 
	if it is a unital channel and there exist $s \in \mathbb{R}$ such that 
	$\T(\sigma)-\T(\rho) = s \,(\sigma - \rho)$, namely, $\T$ is a proportionally 
	shrinking mapping toward the maximally mixed state.
	}%
	, cf. \cite{king:depol}\index{depolarizing channel}\index{channel!depolarizing ---}
	\end{itemize}
	\index{additivity!--- of the Holevo capacity}

	\section{Additivity problems}\label{sec:addp}
	\index{additivity}
	\index{additivity!--- of entanglement of formation}

	  This section gives issues related to the additivity of $E_F$ and to the Holevo capacity. 

\begin{figure}[h]%
		\begin{center}
			\framebox[15cm][c]{ 
				\includegraphics[width=7cm]{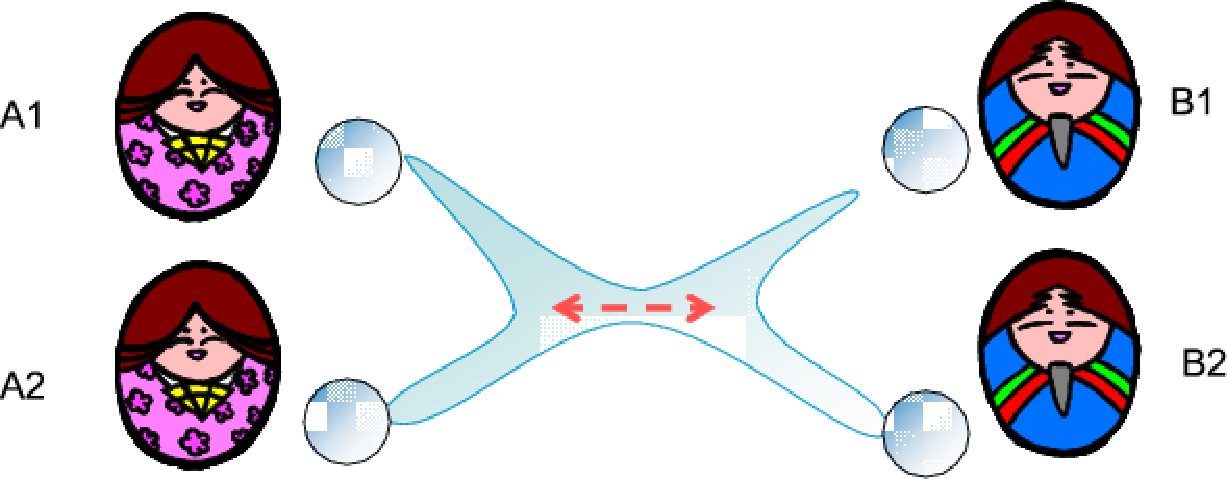}\qquad
				\includegraphics[width=7cm]{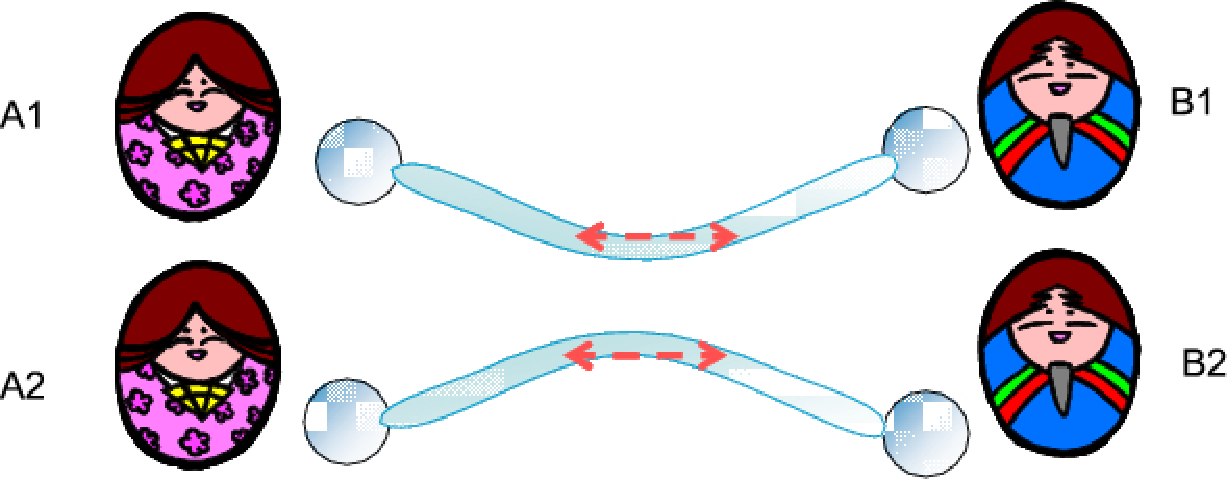}
			}
		\end{center}%
		\index{Alice and Bob}
		\caption[Strong superadditivity]
		{\small Illustration of the strong superadditivity. Entanglement is measured ({\sc left}) $A1+A2\;:\;B1+B2$, 
		({\sc right}) $A1\;:\;B1$ and  $A2\;:\;B2$ separately.
		The inequality (\ref{ineq:ssadditivity}) means the quantity of the left is larger than 
		the sum of the quantities of the right.}
		\label{fig:ssadditivity}
	\end{figure}

	\index{strong superadditivity}
	\index{additivity! strong super---}
	\index{superadditivity|see{strong superadditivity}}
	\begin{Definition}[strong superadditivity of $E_F$]
	For a state $\rho$ on $\HH_{A1}\otimes \HH_{A2} \otimes \HH_{B1}\otimes \HH_{B2}$, 
	the inequality of {\it strong superadditivity} of $E_F$ is defined as,
	\begin{equation}\label{ineq:ssadditivity}
		E_F\AB(\rho) \ge E_F\AB(\Tr_2{\rho} )  + E_F\AB(\Tr_1{\rho} ).
	\end{equation}
	Here $\Tr_2$ means tracing out the space of ${\HH_{A1}\otimes \HH_{B1}}$  and 
	$\Tr_1$ means tracing out the space of ${\HH_{A2}\otimes \HH_{B2}}$.
	See Fig.~\ref{fig:ssadditivity}.
	Note that whether this inequality holds for every case has not yet solved.
	\end{Definition}

	\begin{Theorem}[\cite{MSW,shor2}]
		The following four propositions are equivalent:
		 \begin{itemize}
		 \item The additivity in $E_F$ : $ ^\forall{(\rho,\sigma)} \>:\> E_F(\rho \otimes \sigma) = E_F(\rho) + E_F(\sigma) $ 
		 \item The additivity in the Holevo capacity  : $ ^\forall{(\T_1,\T_2)} \>:\> C(\T_1 \otimes \T_2) = C(\T_1) + C(\T_2) $ 
		 \item The strong superadditivity of $E_F$ for any mixed states
		 \item The strong superadditivity of $E_F$ for any pure states  
		 \end{itemize}
	\end{Theorem}

\section{Miscellaneous}
	Here, we fix some conventions, which might otherwise cause confusion. 

\subsection{Base of logarithm}
	\index{logarithm!base of ---}
	\index{base of logarithm}
	Throughout this dissertation, the base of logarithm is  fixed to two (2) 
	unless the base is specifically indicated. Therefore, $\log 2  =1 $. 
	Note that this rule is applied to the logarithm of matrices like $\log \rho$ as well,
	which effects von Neumann entropy and the quantum divergence and so on. 
\subsection{Convex and concave}
	\index{convex!--- function}
	\index{concave function}
	{\it Convexity} of a function: a function $f$ is {\it convex} when it bends {\it downward}, as 
	\begin{equation}
		f(\lambda x + (1-\lambda ) y ) \le \lambda f(x) + (1-\lambda) f(y)  \quad\mbox{ for }\quad 0\le \lambda \le 1.
	\end{equation}
	{\it Concavity} of a function: a function $f$ is {\it concave} when it bends {\it upward}, as 
	\begin{equation}
		f(\lambda x + (1-\lambda ) y ) \ge \lambda f(x) + (1-\lambda) f(y)  \quad\mbox{ for }\quad 0\le \lambda \le 1.
	\end{equation}

	\begin{figure}[h]
	\begin{minipage}{8cm}
	\includegraphics[width=7cm]{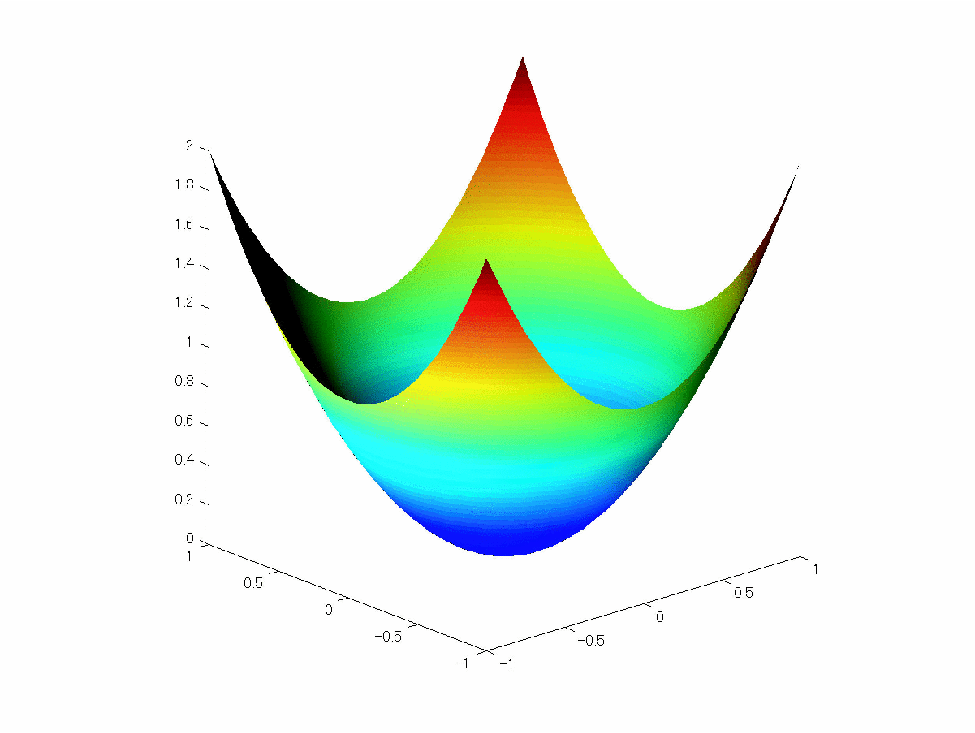}\\
	\hspace*{3cm} convex 
	\end{minipage}
	\begin{minipage}{8cm}
	\includegraphics[width=7cm]{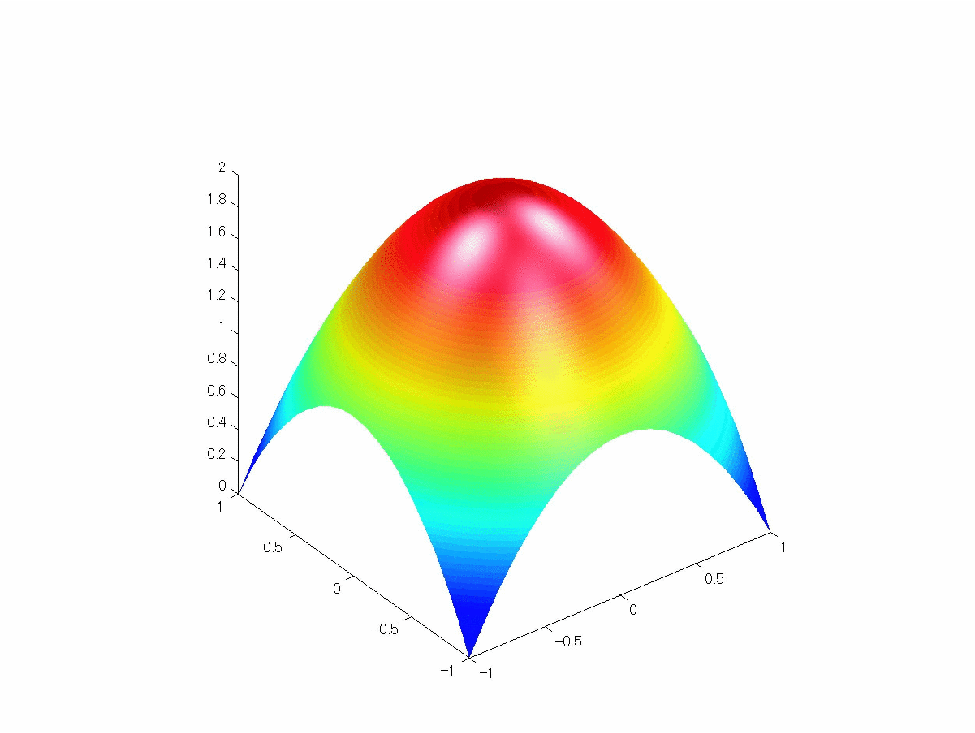}
	\\
	\hspace*{3cm}concave
	\end{minipage}
	\label{fig:vex}\caption[Convex/concave]{A convex function ({\sc Left}) and a concave function ({\sc Right}) }
	\end{figure}

	\subsection{Tensor products}
		In this chapter, or in this dissertation, the {\it tensor product} is 
		defined for 
		\begin{itemize}
		\item pure states, 
		\item mixed states, 
		\item mapping or quantum channels
		\end{itemize}

	\subsection{The terms level and dimension}
		In quantum informatics, the terms {\it level} and {\it dimension} may not be different in nature
		as one can say a state comes from 2-dimensional space or 2-level system. 
		In this dissertation, however, they may be distinguished in some context, as the level refers to the quantum space 
		and the dimension refers to technical derivation to perform the optimizing calculation 
		for the Holevo capacity.

\section*{Endnotes of Chapter \ref{ch:01} }

	\theendnotes

	\begin{table}[ht]
	\bigskip
	\begin{center}
	\begin{tabular}{cl}
	\hline
	$\ket{\phi} $ & ket \\
	$\bra{\phi} $ & bra \\
	$\BB(\HH)  $ & the set of mixed states of a given Hilbert space \\
	$\square^T   $ & transpose of a given matrix     \\
	$\square^\dagger$ & conjugate transpose or Hermitian transpose of a given matrix  \\
	$\rho(\cdot_x,\cdot_y,\cdot_z)$ &Stokes parameterization in the Bloch sphere \\
	\hline
	\\
	\multicolumn{2}{l}{$\rho,\sigma$ : mixed quantum states }\\
	\hline
	$B(\rho,\sigma) $    &  Bures distance between two states\\
	$F(\rho,\sigma)$   & fidelity between two states \\
	$H(\rho\, || \,\sigma)$ &  quantum divergence, quantum relative entropy \\
	$S(\rho)$  &  von Neumann entropy of a state \\
	\hline
	\\
	\multicolumn{2}{l}{$\cdot,\square$ : a mixed quantum state share by the two sites {\sf A} and {\sf B}}\\
	\hline
	$E\AB (\cdot)$   &  reduced von Neumann entropy \\
	$E_C\AB(\cdot)$ & entanglement cost \\
	$E_D\AB(\cdot)$ &  entanglement distillation \\
	$E_F\AB(\cdot)$ &  entanglement of formation \\
	$E_N\AB(\cdot)$  &  logarithm of negativity \\
	$E_R\AB(\cdot)$ &  relative entropy of entanglement \\
	$\square^\Gamma $ & partial transpose of a given matrix \\
	\hline
	\\
	\multicolumn{2}{l}{$\T$ : a quantum channel}\\
	\hline
	$C(\T)$    &  Holevo capacity of a quantum channel\\
	$\bar C(\T) $ &  full capacity of a quantum channel \\
	$\KK_\T$ & image vectors induced by an isometric embedding $U$ of a channel $\T$ \\
	$\OO_\T$ & preimage matrices of the average output of $\T$ in $\BB(\KK_\T)$\\
	\hline
	\end{tabular}
	\end{center}
	\caption[Denotations used in this dissertation]{Denotations used in this dissertation.}
	\end{table}

	\index{ @$\ket{\cdot}$|see{ket}}
	\index{ @$\bra{\cdot}$|see{bra}}
	\index{B@$B(\cdot,\cdot)$|see{Bures distance}}
	\index{C@$C(\cdot)$|see{Holevo capacity}}
	\index{C@$\bar C(\cdot)$|see{full capacity}}
	\index{E@$E(\cdot)$|see{reduced von Neumann entropy}}
	\index{E@$E_C(\cdot)$|see{entanglement cost}}
	\index{E@$E_D(\cdot)$|see{entanglement distillation}}
	\index{E@$E_F(\cdot)$|see{entanglement of formation}}
	\index{E@$E_N(\cdot)$|see{logarithmic negativity}}
	\index{E@$E_R(\cdot)$|see{relative entropy}}
	\index{F@$F(\cdot,\cdot)$|see{fidelity}}
	\index{H@$H(\cdot,\cdot)$|see{quantum divergence}}
	\index{S@$S(\cdot)$|see{von Neumann entropy}}
	\index{Hermitian!--- transpose}\index{transpose!Hermitian ---}
	\index{conjugate transpose}\index{transpose!conjugate ---}
	\index{partial transpose}
	\index{transpose!partial ---}
	\index{logarithm!---ic negativity}

	\index{state!quantum ---|see{pure state, mixed state}}
	\index{quantum!--- state|see{pure state, mixed state}}
	\index{Wootters' concurrence|see{concurrence}}

\makeendnotes
	%
	%
\chapter{Introduction to the dissertation}\label{ch:intro}
	The main body of this dissertation is 
	the author's researches
	in quantum information science with a main interest in quantum entanglement,
	and methodologies of conducting research as well.

	\medskip
	Before entering the chapters of the research performed,
	we are going to depict (i) the background of this dissertation, (ii) how 
	the research was conducted, and (iii) this dissertation's organization. 

\section{Background}

	\subsection{Quantum information science}
	Quantum information science,  
	the consolidation of information science and 
	quantum physics, both of which date back to the first half of the twentieth century%
	, has attracted the interest of many researchers 
	over the last ten years as the invention of efficient algorithms for 
	number factorization%
	\footnote{%
		Shor's algorithm : it solves the factorization in $O(n^3)$ with $n$ being the number of digits,
		even though the fastest classical algorithm does it in approximately $O(\exp(n^{2/3}))$.
		Note that the former is polynomial and the latter is superpolynomial.
	}\cite{shor1994}, 
	and database search%
	\footnote{%
		Grover's algorithm: it finds the target in a database 
		in size $n$ with the time proportional to $O(n^{1/2})$.
		This means the time cost grows only 10 times 
		if the database size grows as much as 100 times. 
	}%
	\cite{grover}%
	. %
	There is a characteristic phenomena in quantum physics %
	 that is called {\it quantum entanglement}, which
	plays an essential role in quantum computation \cite{JL}. 
	Various quantum protocols were devised using quantum entanglement,
	such as quantum teleportation%
	\footnote{%
		Quantum teleportation: 
		it teleports quantum states to a remote place with classical communication 
		and quantum entanglement between the two sites beforehand.
	}%
	\cite{BBC}
	and quantum superdense coding%
	\footnote{
		Quantum superdense coding: it transmits 2 bits while sending 1 qubit with prepared quantum entanglement.
	}%
	\cite{BW}.

	\index{factorization}
	\index{number factorization}
	\index{Shor's algorithm}
	\index{algorithm!Shor's ---}
	\index{database search}%
	\index{Grover's algorithm}%
	\index{algorithm!Grover's ---} \index{quantum!--- teleportation}
	\index{teleportation}
	\index{superdense coding}

	The author is concerned with the entanglement cost 
	\cite{BDSW}, 
	and also the Holevo capacity 
	\cite{Holevo73}.
		The entanglement cost  is one way to quantify the entanglement of  a bipartite quantum state, 
	and the Holevo capacity is a  classical communication capacity of a given quantum channel. 
	These two different measures are  related by \cite{MSW} through Stinespring's dilation theorem
	as presented in section \ref{sec:stine}.

	\bigskip 
	\subsection{Quantifying entanglement}

	How to quantify the entanglement of quantum states?  Quantum entanglement 
	raise up useful effects on quantum computation/communication mysteriously, 
	thus quantification of entanglement is significant and natural methodology 
	to clarify the mystery. 

	To measure the usefulness of entanglement, there is a simple way is to compare two 
	entangled pairs. There may be a situation that a pair particles shared by Alice and 
	Bob \index{Alice and Bob}is more entangled than 
	another pair particles also shared by the two figures
	because the former pair is more useful to be used in quantum 
	communication, say, sending a message in quantum superdense coding.
	Then how to compare?  An established way is to see the possibility of 
	transformation from a pair to the other through LOCC operation, as LOCC
	cannot create additional entanglement. If a pair {\#}1 can be transformed 
	into another pair {\#}2 by LOCC, then one can say that {\#}1 is more 
	useful than {\#}2. 

	\cite{vidal1999} shows a good theorem for this question for pure states. 
	Two pairs of pure states can be compared by the concept of {\it majorization}%
	\footnote{%
	Majorization: two pairs of pure states $\ket{\psi_1}$ and $\ket{\psi_2}$ are compared as follows: 
	\begin{list}{\hspace{8mm}-- 1.}{} 
	\item Let $\{\lambda_i\}$ and $\{\mu_i\}$ be each of the eigenvalues in descending order
	of $\Tr_A \ketbra{\psi_1}$ and $\Tr_A \ketbra{\psi_2}$, respectively.
	\item Compare the sequences 
	$\{\lambda_1, \lambda_1+\lambda_2, \lambda_1+\lambda_2+\lambda_3,\ldots\}$ 
	and 
	$\{\mu_1, \mu_1+\mu_2, \mu_1+\mu_2+\mu_3,\ldots\}$.
	\end{list} 
	They are equivalent 
	\begin{list}{\hspace*{8mm} --}{} 
	\item when $\ket{\psi_1}$ can be transformed to $\ket{\psi_2}$ by LOCC and 
	\item when  
	all of $\lambda_1\le\mu_1 , \lambda_1+\lambda_2 \le  \mu_1+\mu_2 ,
	\lambda_1+\lambda_2+\lambda_3 \le \mu_1 + \mu_2 +\mu_3 , \ldots$ hold. 
	 \end{list} 
	 This method to compare is called {\it majorization}.
	}
	\index{majorization}
	This majorization was the first significant step to measure quantum entanglement. 

	To quantify quantum entanglement literally in quantity, an {\it asymptotic} comparison 
	is conceived. The idea is, for a state $\rho$, to investigate the possibility to transform 
	$\rho^{\otimes N}$ from/into $\ket{\Psi^-}^{\otimes M}$ by LOCC with large $N$ and $M$.
	$\ket{\Psi^-}^{\otimes M}$ is considered to have $M$ {\it ebit}s. 
	The idea is, if $\rho^{\otimes N_1}$ can be transformed into $\ket{\Psi^-}^{\otimes M_1}$
	then $\rho$ is more than $N_1/M_1$ ebits, and if $\rho^{\otimes N_2}$ can be transformed from 
	$\ket{\Psi^-}^{\otimes M_2}$ then $\rho$ is less than $N_2/M_2$ ebits. The former 
	idea is formalized in {\it entanglement distillation} ($E_D$) and the latter is formalized in 
	{\it entanglement cost} ($E_C$) (see (\ref{eq:defec}) and (\ref{eq:defed}) ).

	\begin{figure}[hbt]
		\bigskip
		\begin{center}
			\includegraphics[width=12cm]{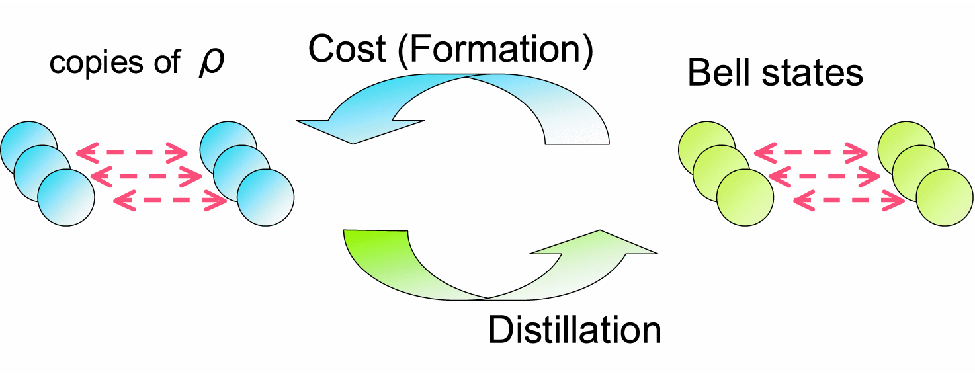}
		\end{center}
		\caption[$E_C$ and $E_D$]%
			{The depiction of 
			the entanglement cost and the entanglement of distillation. The transformations are
			LOCC. $\rho$ is compared with copies of a Bell state, say,  $\ket{\Psi^-}=\frac{\ket{01} \pm \ket{10} 
			}{\sqrt{2}}$.}
	\end{figure}

	For pure states %
	\cite{popescu:rohrlich,vidal2000,nielsen2000,donald:horodecki:rudolph},
	 the entanglement cost and the entanglement distillation coincide 
	each other.
	For mixed states, they are, however, different. 
	That is, irreversibility occurs in asymptotic transformation through LOCC for mixed bipartite state.
	Thus quantifying entanglement is 
	not simple for mixed states. Another idea is to put a restrict such as a measure should 
	be {\it additive}. There has been devised many measures and the {\it entanglement of 
	formation} ($E_F$) is also a good candidate to measure entanglement. It is originally
	defined as a formula (\ref{eq:defef}), and the relation (\ref{eq:hht01})  connects it 
	to the entanglement cost. Even though the entanglement distillation seems to break the 
	additivity \cite{evidence},  the entanglement cost and the entanglement of formation seem not
	break the additivity, or it is hard to find such a state $\rho$ that breaks the additivity.


	\subsection{Channel capacity beyond classical physics}
	\label{ss:ccbcp}

	\index{stochastic!--- mapping}\index{mapping!stochastic ---}
	For a classical channel, that is a stochastic mapping from the input alphabets to the output alphabets, 
	there is a sole capacity formalized by Shannon, which gives the supremum quantity of information 
	that is transmitted asymptotically per channel use \cite{Shannon48}. To determine the capacity for 
	classical case, there are efficient algorithms \cite{Arimoto72,Blahut72}.

	For a quantum channel, there could be many capacities for a channel. 
	An interesting quantity is the Holevo capacity, which was  upper-bounded by the Holevo quantity \cite{Holevo73}, 
	and proved to be equal to the Holevo quantity \cite{Holevo98,SW97}. 

	There were proposals to calculate
	the Holevo capacity for general quantum channels \cite{osawa-nagaoka, Shor03}. \cite{osawa-nagaoka} expands the method 
	of the Arimoto-Blahut algorithm. 
	\index{Arimoto-Blahut algorithm}
	\index{algorithm! Arimoto-Blahut ---}
	\cite{Shor03} made a 
	comment utilizing the column generation method. This dissertation presents seemingly the first  
	algorithm to calculate the Holevo capacity of qubit channels \cite{HIMRS04}.

	\index{Shannon!--- capacity}
	\index{capacity!Shannon ---}
	The Holevo capacity is a  measure which does not allow entanglement among input particles. 
	There is a possibility that allowing entanglement among the input particles might increase the capacity,
	even though the Shannon capacity keeps the additivity. 
	This problem is formalized 
	as whether $C(\T^{\otimes n} ) = n \, C (\T)$ for $n=2,3,\dots$, or $C(\T_1\otimes \T_2)=C(\T_1) + C(\T_2) $
	holds for arbitrary channels. 
	This additivity was related to the additivity of $E_F$ \cite{MSW}. Verifying the additivity of the 
	Holevo capacity is an interesting problem, thus numerically calculating the Holevo capacity for 
	high-dimensional case is expected.



	%
	\section{Research flow of this dissertation}
	This dissertation is concerned with the problems of quantification of quantum entanglement 
	and the computation of the Holevo capacity. As is written above, quantifying the entanglement
	of bipartite pure states is almost settled,  
	in that the reduced von Neumann
	entropy is the sole measure with the appropriate conditions. 
	 \index{reduced von Neumann entropy}\index{von Neumann!reduced --- entropy}
	\index{entropy!reduced von Neumann ---}
	The appropriate properties here 
	 are the reducing property against any non-quantum physical operation (formalized in LOCC), additivity property, and so on. 

	\subsection{Quantifying Entanglement}
	\index{entanglement!--- of formation}
	\index{entanglement!--- cost}
	Quantum states are generally mixed states, as each of them presented in the stochastic mixture of \index{stochastic!--- mixture}
	\index{mixture!stochastic ---}
	some pure states.  Various ways of quantifying the entanglement of a given mixed state
	are proposed \cite{BDSW}, such as the entanglement cost ($E_C$), the entanglement of formation ($E_F$), 
	 the entanglement distillation ($E_D$), and the relative entropy of entanglement ($E_R$)%
	\footnote{The relative entropy of entanglement $E_R$: 
	$E_R(\rho)$ is defined as $\min\limits_{\sigma:\mathrm{separable}} H(\rho||\sigma)$.}
	.  
	 \index{relative entropy!--- of entanglement}\index{entanglement!relative entropy of ---}
	 \index{entropy!relative --- of entanglement}
	For a given $\rho$, $E_C(\rho)$ is the measure of how many the Bell states are needed to construct $\rho$ in the asymptotic 
	sense.\footnote{The asymptotic sense here means that $\rho$ is not single to be compared with the Bell states. Rather,
	$\rho^{\otimes n}$ for large $n$ is compared with $m$ copies of the Bell states, and $\frac{m}{n}$ is considered
	in the limit of $n,m\to\infty$.}
	 $E_F$ is similarly defined, but rather technically. $E_D$ of $\rho$
	is defined as how many the Bell states can be distilled from $\rho$.  

	Naive questions might arise: ``Which is the best
	 [in quantifying the entanglement of bipartite mixed states]?'',
	 ``What is the appropriateness in measures?'',  ``Is there only one proper measure?''

	A relation  
	\begin{equation}
	E_C(\rho)= \lim_{n\to\infty} \frac{E_F(\rho^{\otimes n})}{n}
	\end{equation}
	 was given by
	\cite{HHT}, thus
	the following expectation were brought about: 
	\begin{itemize}
	\item $E_C$ could be calculated analytically and/or numerically.
	\item The additivity of $E_F$, that is
	\begin{equation}\label{nadd}
	E_F( \rho^{\otimes n } )   =^{?} n \, E_F(\rho) ,
	\end{equation}
	 might holds thus $E_C=E_F$ holds and $E_C$ is calculated rather easily. 
	\end{itemize}

	If the additivity does not hold, calculating $E_C$ is a formidable problem.  
	Calculating 
	\begin{equation}\label{eq:efrho1}
	E_F(\rho) = \min_{n; p_1,\dots,p_n ; \ket{\phi_1},\dots,\ket{\phi_n}}
	  \left\{
	 \sum_{i=1}^n p_i \, E(\ketbra{\phi_i} )
	   \Bigg| \;
	   {
	    \begin{matrix}
	    n \in\mathbb{N}, \hfill \\ \, p_1,\ldots,p_n >0, \, \sum_{i=1}^n p_i =1 , \\
	     \ket{\phi_1},\ldots, \ket{\phi_n} \text{ are pure states} , \\
	           \sum_{i=1}^n p_i \, \ketbra{\phi_i} = \rho \hfill
	    \end{matrix}       
	   }
	  \right\},
	\end{equation}
	where $E$ is the {\it reduced} von Neumann entropy, 
	\index{reduced von Neumann entropy}\index{von Neumann!reduced --- entropy}
	 is 
	quite an high-dimensional optimization problem, even if the dimension of $\rho$ is 
	quite low such 
	as two or three%
	\footnote{%
	For a bipartite $d$-dimensional system, $n$ in the formula (\ref{eq:efrho1}) could reach $d^2$, and 
	each of $\{\phi_i\}_{i=1}^{d^2}$ is a $d^2$-dimensional vector on the bipartite system on $\mathbb{C}$,
	represented in $2 d^2 -1 $ parameters in $\mathbb{R}$, as well as  $\{p_i\}_{i=1}^{d^2} $ 
	being considered $(d^2-1)$-dimensional parameters in $\mathbb{R}$. One can remove one parameter of a 
	global phase of a quantum system. As a whole, the number of parameters 
	is $ (2 d^2 -1) \times d^2  + (d^2-1) -1 = 2d^4  -2 $.  Then, how to optimize such a high-dimensional
	problems!? It seems very difficult to calculate $E_F$ even for $d=2$. (See the table below.)
	Fortunately, \cite{wootters1997, wootters1998} gave an analytical formula for $d=2$ to calculate $E_F$. 
	\begin{table}[hhh]
	\begin{center}
	\begin{tabular}{|c|cccccccc|}
	\hline
	$d$         & (1) &  2 &  3  &   4  &    5 &    6 & 7   & $\dots$  \\
	$2d^4  -2$  &  (0) & 30 & 160 & 510  & 1248 & 2590 & 4800& $\dots$\\
	\hline
	\end{tabular}
	\caption[The dimensional explosion in computing $E_F$]{The dimensional explosion in computing $E_F$}
	\end{center}
	\end{table}
	}
	. Still more, $E_C$ is a limitation form of this problem. 

	The first trial of calculating $E_C$ for nontrivial case was done for three-level antisymmetric states
	\cite{ViDC01}. 
	With the expectation of calculating the first case of $E_C$ for nontrivial case, the author gave 
	lower bounds of antisymmetric states \cite{Shimono}, which appear in Chapter 3. 
	More general form of (\ref{nadd}) is 
	\begin{equation}
	E_F(\rho\otimes\sigma) =^? E_F(\rho) + E_F(\sigma).
	\end{equation}
	The author seem to gave the first specific example \cite{Shimono03}, presented in Chapter 4. 
	Conclusively, the first examples to calculate $E_C$ for nontrivial case was done by \cite{VDC},
	for the mixture of Bell states. $E_C$ for three-dimensional antisymmetric states was calculated 
	by \cite{yura}, and the result was further expanded to more general antisymmetric case \cite{M:Y}.

	As of the days when  $E_C$ and $E_D$ were proposed \cite{BDSW}, whether they can be different or not
	was unknown%
	\footnote{The difference between $E_C(\rho)$ and $E_D(\rho)$ means some LOCC operations are
	irreversible (even) in an asymptotic sense.
	Namely, bulk of the Bell state $\ket{\Psi^-}$ in quantity $E_C(\rho)$ may be transformed 
	into bulk of $\rho$ in unit quantity by some LOCC operations, but any LOCC operation cannot 
	retrieve the original quantity of the Bell state (refer to the definition of $E_C$ and $E_D$.)
	}%
	.
	Such states were found by \cite{Horodecki98} for $3\otimes3$-level mixed states. 
	Attempts to find other examples was done by \cite{VDC}. The author of this dissertation
	also gave examples utilizing the following facts: 
	\begin{itemize}
	\item $E_F$ can be calculated through the Holevo capacity \cite{MSW}. 
	\item $E_D$ is bounded by another easily-calculable quantity called the logarithmic 
	negativity \cite{ViWe02} $E_N$, i.e. $E_D \le E_N$
	\end{itemize}
	This result is presented in Chapter 5.

	\subsection{Strong superadditivity}
	\index{strong superadditivity}
	\index{relative entropy!--- of entanglement}\index{entanglement!relative entropy of ---}
	\index{entropy!relative --- of entanglement}

	 There is another quantification called ``the relative entropy of entanglement''  \cite{VPRK} 
	 denoted $E_R$. It was believed to be additive for a while, however, negated later by finding 
	 counterexamples \cite{VW}. 

	It seems that determining only one measure is not appropriate because 
	$E_C(\rho) > E_D(\rho)$ holds for some state $\rho$ (examples is presented in Chapter 5),
	even though both $E_C$ and $E_D$ are good candidates to measure quantum entanglement.
	The additivity is the property $E(\rho \otimes \rho') = E(\rho) + E(\rho') $ or 
	$E(\rho^{\otimes n}) = n \, E(\rho)$. 
	It is of natural and expected requirement for any kind of ``measuring''.  As quantum entanglement is 
	the valuable peculiarity from an ordinal classical physical viewpoint, one might 
	like to quantify it like counting money. From a theoretical viewpoint of convenience, 
	if the additivity holds for $E_F$, then $E_F= E_C$, which leads to a difficult formula of $E_C$ in calculation 
	to a more feasible one calculating $E_F$. 

	Based on the concept written above, the value of 
	$E_F$ and $E_C$ for some specific states are tried calculated. For example, 
	 the antisymmetric states of bipartite three-level systems are taken up. 
	 (These states can be regarded as two particles of fermions in one system, \index{fermion}
	 thus consideration of this entanglement might cause practical applications.)
	In order to try to calculate $E_C(\rho)$ where $\rho$ is an antisymmetric 
	state, $E_F(\rho^{\otimes 2} ) $ is calculated. 
	(This work is completed by \cite{yura} as $E_C(\rho)=1$, which means $E_F(\rho^{\otimes n})$ for $\forall n\in\mathbb{N}$
	is calculated.
	Furthermore, it is extended to the space of $ \mathbb{C}^d\otimes (\mathbb{C}^d)^{\otimes d-2}$ \cite{M:Y}, as the value 
	is $\log_2 d-1$.)

	With the theorem of Stinespring dilation, 
	the article of \cite{MSW} related the Holevo capacity to $E_F$.  Utilizing this fact, 
	the author calculated $E_F$ for a $2\otimes 4$ - level system, which shows gaps between
	$E_F$ and $E_D$.  
	 To check the additivity of $E_F$ in the lowest nontrivial case,
	  the inequality of 
	the strong superadditivity for a twin $2\otimes 2$-level system\footnote{%
	\begin{list}{\hspace*{8mm}--}{} 
	\item Firstly, to consider quantum entanglement for a bipartite system, 
	$2 \otimes 2$ is the lowest-dimensional case; for an $n\otimes m$-level system,
	$n<2$ or $m<2$ is trivial. 
	\item Secondly, 
	 to check the additivity, twofold ({\it i.e.} $^{\otimes 2}$) of $2 \otimes 2 $ is the minimum. 
	\end{list} 
	 Therefore, to check the additivity, the nontrivial lowest case to calculate the 
	 quantity of entanglement is $(2\otimes 2)^{\otimes 2}$ dimensional, which becomes 16 dimensional.}. 
	Feasibility in calculation of the von Neumann entropy for a 4-level system and 
	calculation of $E_F$ for $2 \otimes 2$ \cite{wootters1997,wootters1998} is utilized, even though
	calculating $E_F$ seems formidable even for a $2^2\!\otimes\!2^2$-level system. 
	However, the inequality of the strong superadditivity is easy to check rather 
	than the additivity check, thus the attempts to check this is performed in 
	Chapter \ref{ch:6}.

	\subsection{The Holevo capacity}
	\index{Holevo capacity}
	\index{capacity!Holevo ---}

\begin{figure}[hbt]
		\begin{center}
			\includegraphics[width=12cm]{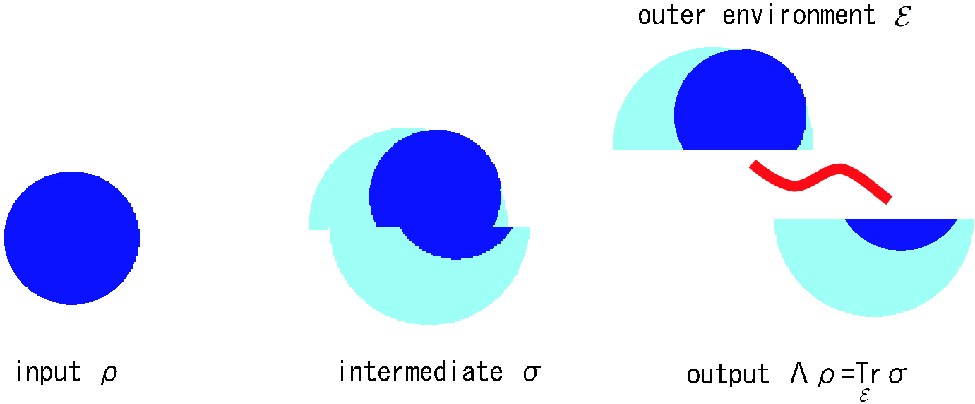}
		\end{center}
		\caption[Stinespring correspondese]%
			{This figure shows the Stinespring correspondense. Any quantum channel can be 
			represented as releasing some space $\mathcal{E}$ after
			a unitary operation over the input state and an additional space. }
		\label{fig:stinespring}
	\end{figure}

	 As the Stinespring correspondence introduces, the action of any quantum channels (CPTP 
	transformation) is equivalent to the result of the following procedure\cite{MSW} (see Fig. \ref{fig:stinespring}):\\
	\qquad -- making a tensor product from the input state and a certain auxiliary state,\\
	\qquad -- transforming it by a certain unitary operator, and then \\
	\qquad -- reducing into the space of the output. \\
	$E_F$ of the intermediate state is equal to \cite{MSW}:
	\begin{equation}\text{
	 the von Neumann entropy of the 
	average output \quad{\it minus}\quad the Holevo capacity. 
	}\end{equation}
	Therefore, if we can calculate the Holevo capacity, we can calculate the 
	$E_F$ for various states. In this dissertation, given examples are
	calculated 
	utilizing a 
	mathematical programming package NUOPT \cite{nuopt} of Mathematical Systems Inc., 
	which is capable of solving large optimization problems,
	and the convergence in calculation is analyzed. 

	The Holevo capacity is a classical information capacity of a given memoryless quantum channel. 
	Back to the history, it was formalized and upper-bounded by the Holevo quantity \cite{Holevo73}.  
	The Holevo capacity was proved to be equal to the Holevo quantity
	later \cite{Holevo98,SW97}, 
	which is defined as,
	\begin{equation}\label{defhsw} 
		 C( \T ) 
		 = \max_{
		 	\begin{smallmatrix}\\
		 	n; \,\quad\qquad \\ 
		 	\rho_1,\rho_2,\dots,\rho_n ; \, \\
		 	p_1, p_2, \dots , p_n \phantom{; \,}
		 	\end{smallmatrix}
		 	\hspace*{-2mm}
		 }
		  \objhsw
	\end{equation}
	for a quantum channel $\T:\BHi\to\BHo$, 
	where $S(\cdot)$ is  the von Neumann entropy.

	In 1998, another representation was provided as \cite{ohya-petz-watanabe}: 
	\begin{equation}\label{defopw}
	 C(\T) = \min_{\sigma\in\BHi} \max_{\rho\in\BHi} H (\T \rho \, || \,\T \sigma). 
	\end{equation} 
	where $H(\cdot||\cdot)$ is the quantum divergence.

	\subsubsection{Difficulty in calculation}

	\index{joint convexity}
	Both of (\ref{defhsw}) and (\ref{defopw}) contain the optimization operators such as {\it min} and {\it max}, and the functions to be optimized 
	seem challenging to optimize
	because:
	\begin{list}{\hspace{8mm}--}{}  
			\vspace*{-2mm}
		\item The objective function (\ref{defhsw}) is convex w.r.t.~$(\rho_i)_i$. 
	\end{list} 
	Therefore, it is hard to guarantee that a found local maximum is the global one.

\begin{figure}[hh]
		\begin{center}
			   \vspace*{0.5cm}
			  \fbox{
			   \vspace*{0.5cm}
			  \begin{minipage}{5cm}
			  \vspace*{1mm}
			  	\includegraphics[width=5cm,clip]{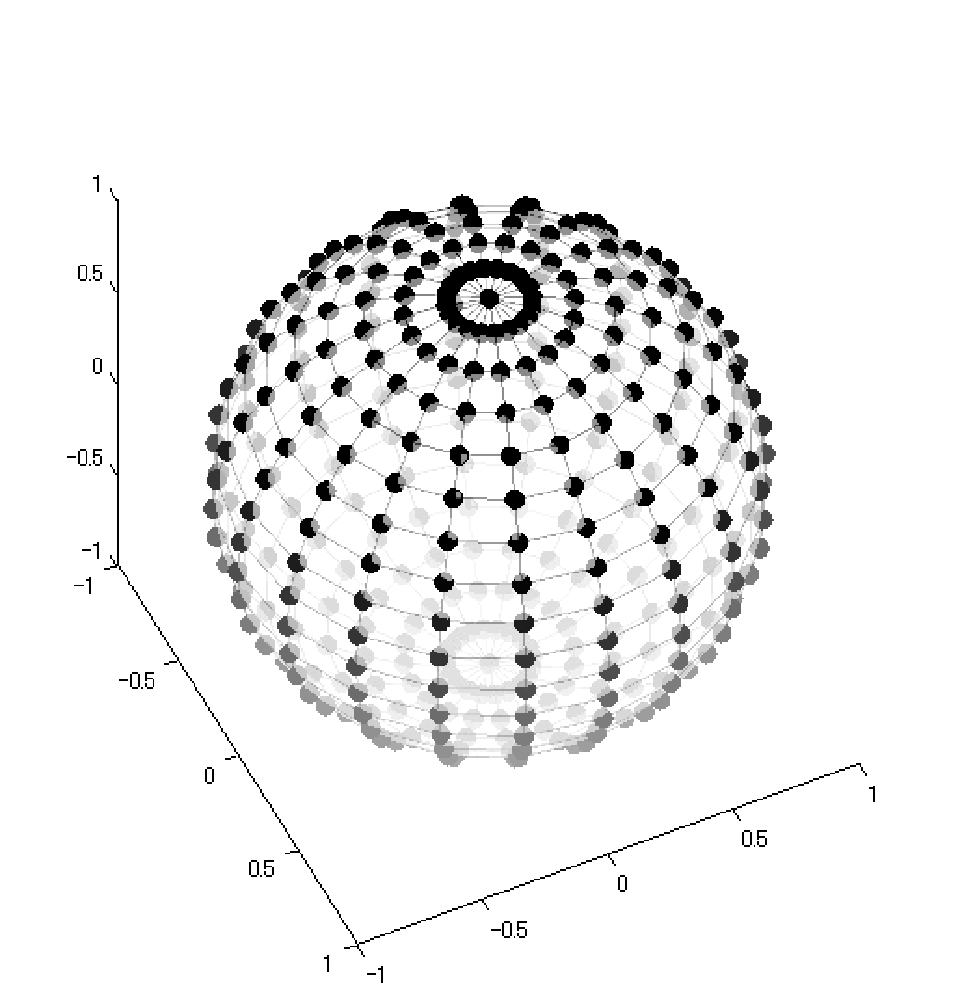}
			  	 \vspace*{-8mm}
			  	 \begin{center}
			  	 $ \ell_{20} $
			  	 \end{center}
			  \end{minipage}
			  \begin{minipage}{5cm}
			  \vspace*{1mm}
			    \vspace*{2mm}
			  	\includegraphics[width=5cm,clip]{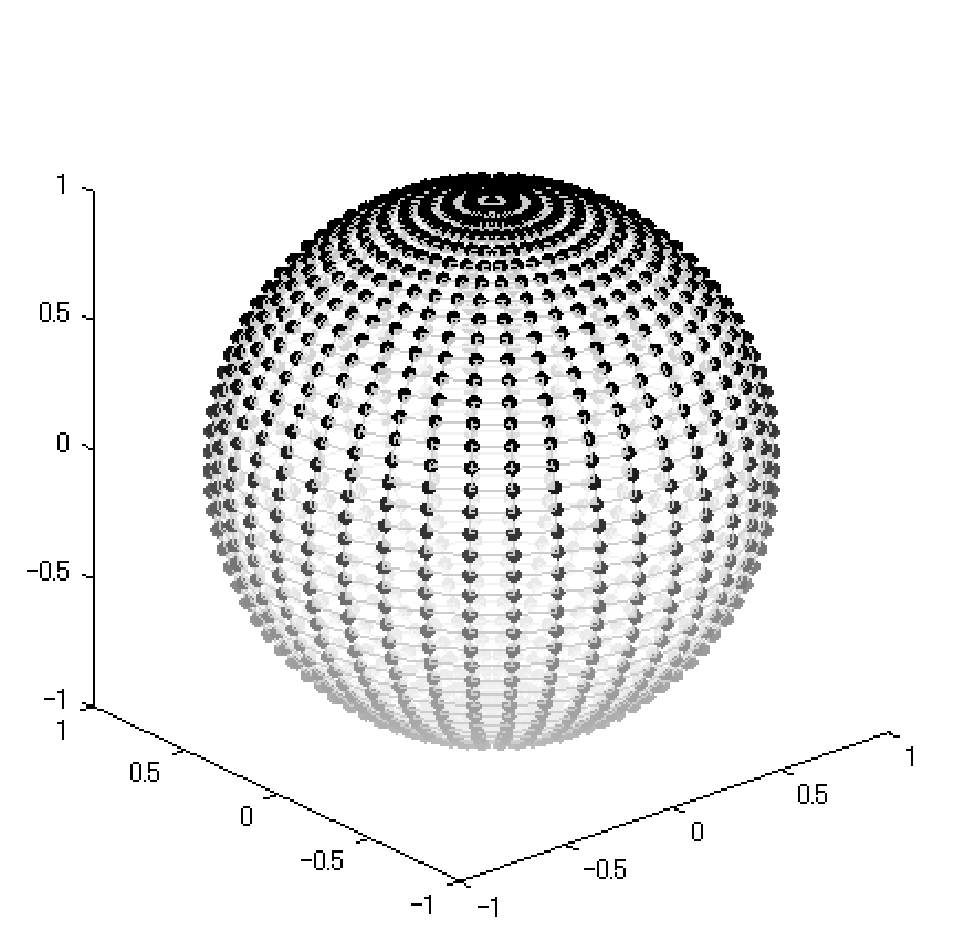}
			  	 \vspace*{-3mm}
			  	 \begin{center}
			  	 $ \ell_{40} $
			  	 \end{center}
			  \end{minipage}
			  \begin{minipage}{5cm}
			  \vspace*{-8mm}
			    \vspace*{9.5mm}
			    \includegraphics[width=5cm]{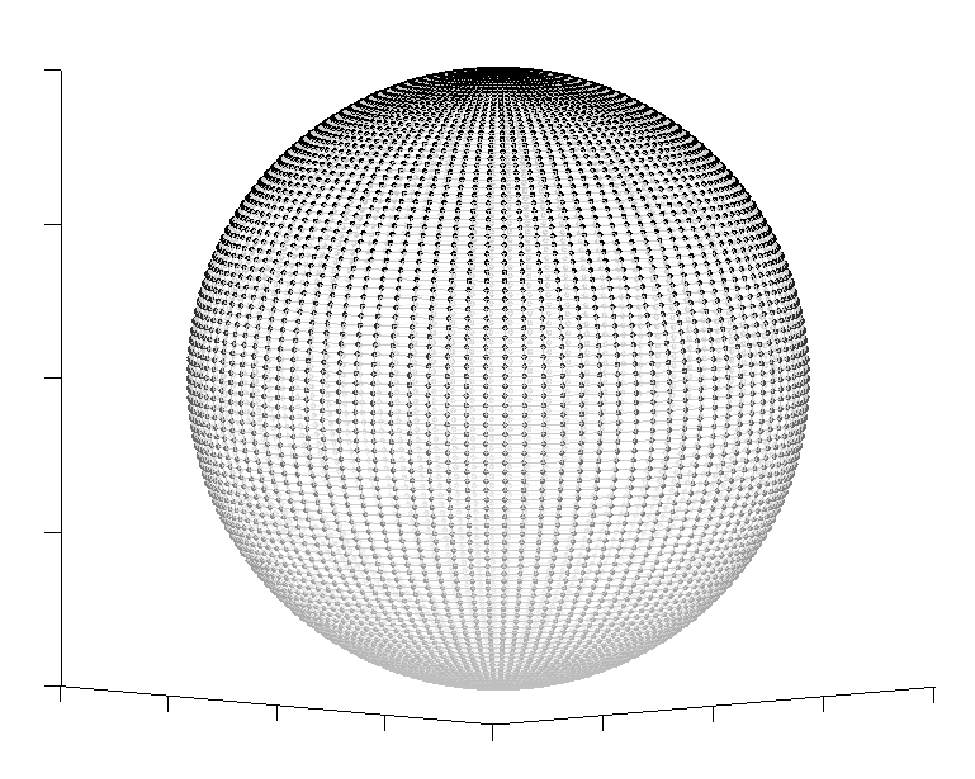}
			  	 \vspace*{-3mm}
			  	 \begin{center}
			  	 $ \ell_{100} $
			  	 \end{center}
			  \end{minipage}
			  }
			\caption
			[Lattices on the Bloch sphere]{%
			The Bloch spheres covered with lattices. 
			Each equator and meridian of them are divided into 20, 40 and 100 sections respectively.
			Specifying $(\rho_i)_i$ as the vertices of a lattice beforehand, 
			(\ref{defhsw}) is feasible to solve because it becomes a problem of maximizing 
			a concave function with a convex search space w.r.t.~$(p_i)_i$.
			The finer the lattice becomes, the closer the solution with the lattice becomes 
			to the actual Holevo capacity. 
			\index{Bloch sphere}
			}
			\label{fig:1bscm}
		\end{center}
	\end{figure}

	 There are, however, feasible ways to calculate the Holevo capacity numerically.%
	 \footnote{
		 The calculation was done in 2004 using an existing calculator; 
		  quantum computers are not yet available!
	  }%
	  The objective function of (\ref{defhsw}) is concave w.r.t.~each $p_i$
	  because the first term is concave as $S$ is concave and the second term is linear w.r.t.~each $p_i$. 
	  Therefore, once every $\rho_i$ is fixed, it is easy to find the global maximum
	  by gradual descending.  The author employs the methodology of covering with a fine lattice
	  the whole space to specify  $(\rho_i)_i$ beforehand for qubit channels (see Fig.\ref{fig:1bscm}).
	  The finer the lattice becomes, the closer the approximative solution becomes to the actual value. 
\subsubsection{Results}
	The results in terms of the Holevo capacity of this dissertation are as follows: 
	\begin{list}{\hspace*{8mm}--}{} 
		\item The Holevo capacity of qubit channels became possible to calculate 31 years after the formula was proposed. 
		\begin{list}{--}{} 
			\item $ \objhsw $ is maximized by NUOPT \cite{nuopt}.
			\begin{list}{--}{} 
				\item Qubit channels which require four inputs to achieve their Holevo capacity are found.
			\end{list} 
			\item The convergence of the optimization calculation is analyzed for 
			\begin{list}{--}{} 
				\item the algorithm maximizing $\objhsw$.
			\end{list} 
		\end{list} 
	\end{list} 


\subsection{Additivity issues}
	Here in this subsection, the additivity issues related with quantum entanglement is described. 
	
	\cite{MSW} revealed the relation between $E_F$ (the entanglement of formation) and
	the Holevo capacity which shows the additivity dependence between them. This relations became 
	to be called the MSW correspondence, and using this relation, \cite{shor2} found additivity equivalence 
	among various problems, including the strong superadditivity of $E_F$.

	As aforementined, additivity of measures is important; additivity means that 
	the quantity measured for two objects equals the sum of the quantities for the two measured separately,
	and this additivity property depends on the measure how to quantify. 
	Back to the topic, 
	whether $E_F$ is additive or not is important 
	because the additivity of $E_F$ leads the additivity of $E_C$ and no one knows whether $E_F$ is additive 
	or not. Additionary, other measures of entanglement such as $E_D$ (entanglement distillation) and $E_R$ 
	(the relative entropy of entanglement) were shown to be non-additive, and researchers may like to find out
	the measure for quantum entanglement that satisfies the additivity. Parallel to this, they like to 
	find out the capacity of quantum channel that satisfies the additivity, and the Holevo capacity is the 
	strong candidate, even though entanglement between input particles may increase the communication 
	efficienty more than the sum of the communication efficiency using non-entangled particles, 
	but nobody knows whether this capacity holds or not. 
	
	The importance of the equivalence properties\cite{MSW,shor2} is as follows.  
	\begin{quote}
		{\it If the additivity of $E_F$ holds}: the strong superadditivity of $E_F$ holds, and the 
		additivity of the Holevo capacity holds. 
	\end{quote}
	\begin{quote}
		{\it If the additivty of $E_F$ does not hold}, which means there  some two quantum states
		$\rho$ and $\sigma$ satisfying $E_F(\rho\otimes\sigma) < E_F(\rho)+E_F(\sigma)$: 
		the strong superadditivyt of $E_F$ breaks for 
		some quantum states, and the additivity of the Holevo capacity breaks
		(for some tow channels $\Lambda_1$ and $\Lambda_2$, $C(\Lambda_1\otimes \Lambda_2)
		> C(\Lambda_1) + C(\Lambda_2)$ with the aide of entanglement between input particles of 
		the two channels). 
	\end{quote}
	
	From the viewpoint of this additivity, this dissertation can be seen as follows: 
	Chapter 3 and 4 tried to check whether $E_F$ is additive for the limited, highly symmetric states.
	Chapter 6 and 7 tried to find the witness against the additivty, through trying to find the counterexample 
	against the superadditivity of $E_F$ and trying to find the breakage of the Holevo capacity of 
	a certain unordinary channel, respectively. 
\section{Organization of this dissertation}

	 The organization of this dissertation is as follows: 
	\smallskip
	\begin{quote}
	 Part I consists of Chapter 1 and this Chapter 2,
	 mentioning the dissertation's introductory matters.
	In Chapter 1, we review the basic notations and the definitions 
	in quantum information theory to be used in subsequent chapters. 
	This Chapter 2 is the introduction for the following main body of 
	this dissertation. 
	\smallskip
	  
	 Part II, consisting of Chapters 3, 4, 5 and 6, deals with the topics 
	related to entanglement quantification and its additivity problems. 
	Chapters 3 and 4 deal with the entanglement cost of antisymmetric states.
	Chapter 5 deals with the gap between $E_C$ and $E_D$. Chapter 6 deals with the strong superadditivity. 
	\smallskip

	Part III, consisting of Chapter 7, deals with the topics of calculating 
	the Holevo capacity. The Holevo capacity is a significant quantity 
	defined more than thirty years ago, the numerical calculation is, however,
	difficult. We present the algorithms to calculate it and application of it. 

	Part IV, the final part, consist of Chapter 8, summarize this dissertation. 
	\end{quote}

	 Plenty parts of this dissertation are derivation from journal articles and
	 conference presentations, such as : 

	\begin{quote}
	Chapter 3 from \cite{Shimono}, Chapter 4 from \cite{Shimono03}, Chapter 5 from \cite{MSW}, Chapter 7 from \cite{HIMRS04}.
	\end{quote}

	\section*{Endnotes of Chapter \ref{ch:intro} }
	{\small \theendnotes}
    \newpage

	\begin{figure}[tbph]
		\begin{center}
			\includegraphics[width=16cm]{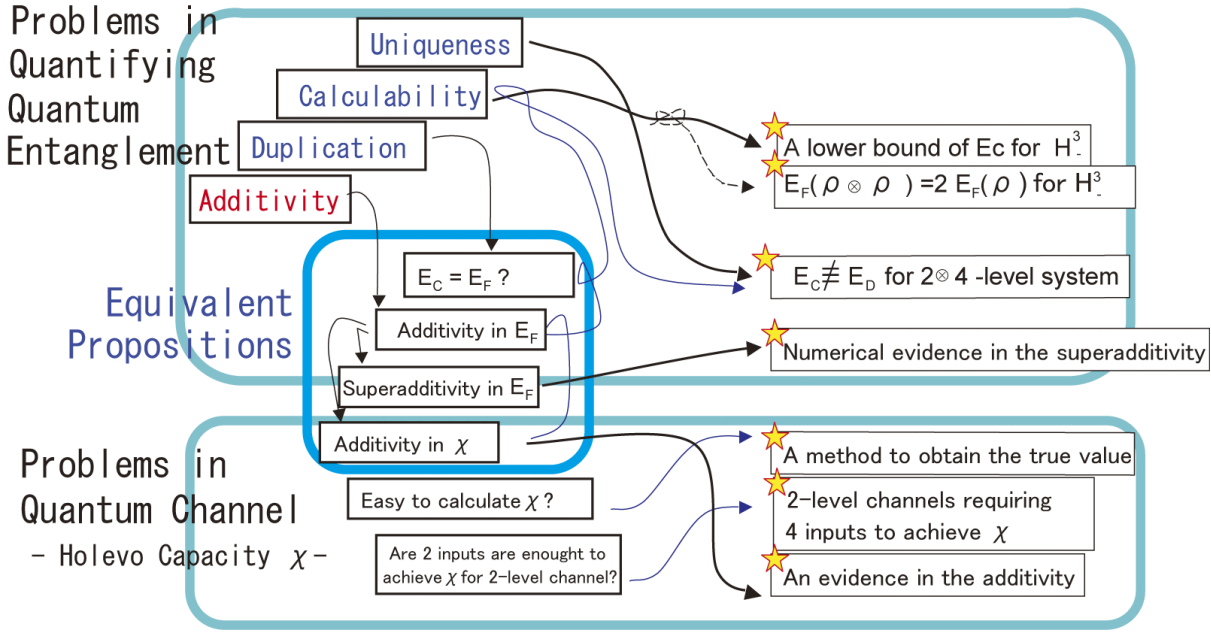}
		\end{center}
		\vspace*{-0.8cm}
		\caption[Contributions of this dissertation]%
			{The contributions of this dissertation.}
	\vspace*{5mm}
	\end{figure}
	\begin{figure}[btph]
		\begin{center}
			\includegraphics[width=16cm]{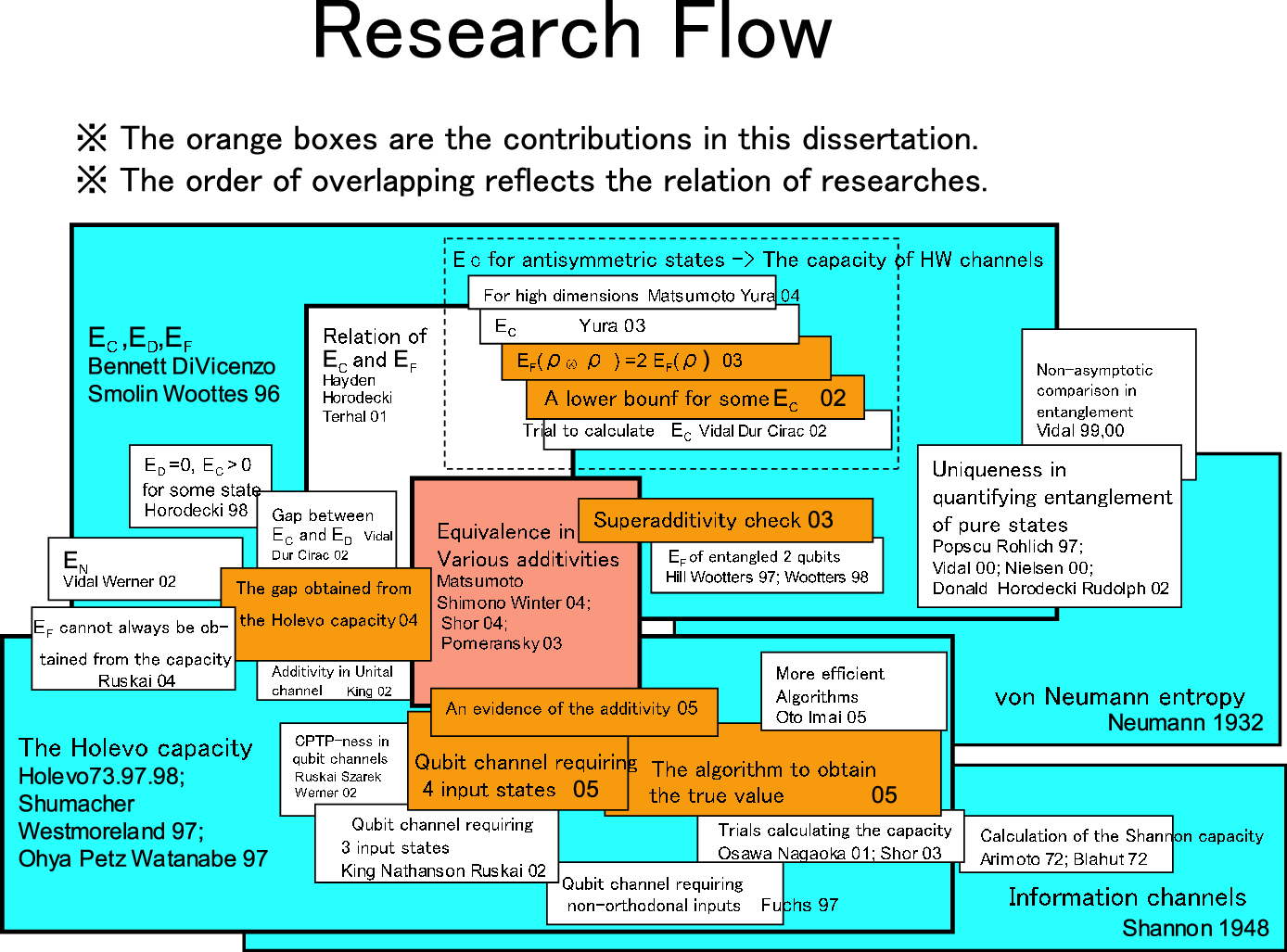}
		\end{center}
		\vspace*{-0.8cm}
		\caption[Research Flow]%
			{The contributions among the neighborhood.}
	\vspace*{-55mm}
	\end{figure}

\part{Entanglement}

  Problems concerning quantification of quantum entanglement are
 integrated in this part. The first two chapters deal with the problem of
 the entanglement of formation of antisymmetric states in a
  $3\otimes3$-level system. Next, irreversibility of entanglement is shown through presenting
 the difference in $E_F$ and $E_D$ for a mixed $2\otimes 4$-level system. 
 Lastly, the strong superadditivity for 
 a $(2\otimes 2)^{\otimes 2}$-level system is numerically verified, which gave circumstantial evidences for 
 the additivity of the entanglement of formation. 

\begin{center}
{\bf Chapters in Part II}
\begin{tabular}{cl}
\hline 
3 & $\forall \rho \mbox{ on } \mathcal{H}_-^3 :\; E_C(\rho) \ge  0.585$ , or $E_F(\rho^{\otimes n}) \ge 0.585 n $. \\
 & \qquad -- Attempts for calculating the $E_C$. \\
4 & $\forall \rho \mbox{ on } \mathcal{H}_-^3 :\; E_F(\rho)=1, E_F(\rho^{\otimes 2}) = 2 $. \\ 
 & \qquad -- Checking the additivity in $E_F$. \\
5 & $ \exists \rho \mbox{ on } \mathbb{C}^2\otimes \mathbb{C}^4:\; E_D(\rho) < E_C(\rho) $. \\ 
 & \qquad -- The gap shows the irreversibility in LOCC. \\
6 & $ \exists \rho \mbox{ on } \left(\mathbb{C}^2\otimes \mathbb{C}^2 \right)^{\otimes 2} : \;
E_F(\rho) \not\ge E_F(\Tr_1 \rho) + E_F(\Tr_2 \rho) ? $ \\
 & \qquad -- Attempts for seeking any counterexample against the strong superadditivity \\
\hline
\end{tabular}
\end{center}

\vspace*{5cm}
\begin{center}

\includegraphics[width=12cm]{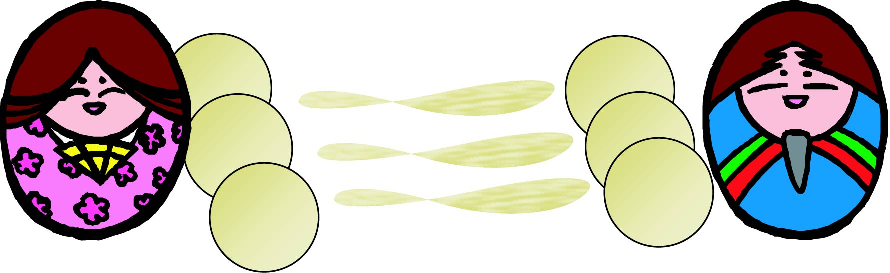}
\end{center}
 
\chapter{Lower Bounds for Entanglement Cost of Antisymmetric States}

{
Calculating $E_C$ is quite difficult. A breakthrough has been desired. We paid attention
to antisymmetric states. Lower bounds were found even though the exact values were unknown.
}
\medskip

This chapter gives a lower bound of the entanglement cost for 
antisymmetric states of a bipartite $d$-level system to be $\log_2 \frac{d}{d-1}$. 
\index{antisymmetric!--- state}
\index{state!antisymmetric ---}
\medskip
This chapter is a derivation from \cite{Shimono}.

\section{First step toward calculating $E_C$}
The entanglement cost can be determined by asymptotic behavior of 
the entanglement formation \cite{HHT}, but it is regarded to be very difficult to 
calculate.

This chapter gives  a lower bound of the entanglement cost of 
antisymmetric states for a bipartite $d$-level system.
We prove that all of the eigenvalues of each reduced matrix of any pure
state affiliating to $\HH_-^{\otimes n}$ for general $d$ is not greater than 
$\left(\frac{d-1}{d}\right)^n$, where $\HH_-$ is the set of antisymmetric states of a bipartite 
$d$-level systems, defined in the following section.
This is proved by investigating a CP map $\tM$ defined in (\ref{note:lambda2}). %

\section{Problem setup}\label{subsection:2.1}
The author makes the following assumptions: 
$\HH_A$ and $\HH_B$ are $d$-dimensional Hilbert spaces with the 
basis $D=\{|i\rangle\}_{i=1\ldots d}$. $\HH_{AB}=\HH_A\otimes\HH_B$.
 For $1\le i < j \le d$, 
\[ |(i,j)\rangle =
{\displaystyle \frac{
|i\rangle_A\otimes|j\rangle_B-|j\rangle_A\otimes|i\rangle_B
}{\sqrt{2}} 
}
\quad \in\quad \HH_{AB} \quad . 
\]
$D' = \{ |(i,j)\rangle \}_{1\le i \lneqq j \le d } $.
The antisymmetric space $\HH_- = \mathop{\mathrm{span}}D' \subset H_{AB}$ .
\begin{Notation}[matrices]
For a positive integer $m$, $\fM_m$ is a set of
$m\times m$-dimensional matrices with each entry a complex number
$\mathbb{C}$. 
For a set $\fX$,
$\left[a_{ij}\right]_{i,j\in \fX}$
is a matrix of which any 
$(i,j)$-component specified $a_{ij}$,
and 
$\fM(\fX)=\left\{[a_{ij}]_{i,j\in \fX} | \{a_{ij}\}\subset\mathbb{C}\right\}$
is a collection of matrices with each row and column  labeled with elements of
$\fX$.
\end{Notation}
\index{partial order between matrices}
\index{matrix!partial order between matrices}
\index{matrices!partial order between ---} 
\begin{Notation}[partial order between matrices]
The partial order $\le$ in $\fM(\fX)$ is employed as follows:
For $X_1,X_2\in\fM(\fX)$ , 
$X_1\ge X_2 \Leftrightarrow X_2 \le X_1 \Leftrightarrow X_1-X_2\ge 0
 \Leftrightarrow X_1-X_2$ 
is a semi-positive Hermitian matrix. 
\end{Notation}
\begin{Definition}[$\MM:\fM(D')\rightarrow\fM(D)$]
The map $\MM:\fM(D')\rightarrow\fM(D)$ is defined as follows:
First, $X\in\fM(D') $, is regarded as an antisymmetric state
$\rho_1 :=\sum_{I,J \in D} X_{IJ}|I\rangle\langle J| \in \HH_-$, Then 
$\rho_1$ is reduced into $\HH_A$ by the operation
$\rho_2 := \Tr\limits_B \rho_1\in \HH_A$, and is converted into the matrix 
representation $Y\in\fM(D)$ with the basis $D$ 
satisfying $\rho_2=\sum_{i,j\in D}Y_{ij}|i\rangle \langle j|$.
This transformation $X\mapsto Y$ is the map $\MM$.
\end{Definition}
%
The derivations of this map $\MM$ are investigated in section \ref{subsection:2.2}.
%
%
%
%
\begin{Notation}[$\fE_{ij}^\fX$]
For a set $\fX$ and every $i,j\in \fX$,
$\fE_{ij}^\fX\in\fM(\fX)$
is a matrix with entry $1$ only at the  $(i,j)$-component and $0$ 
elsewhere. For example,
for $\fX=\{1,2,3\}$,
 $\fE_{1,2}^\fX = \left(\begin{smallmatrix} 0&1&0\\0&0&0\\0&0&0
 \end{smallmatrix}\right)$.
\end{Notation}
Thus, 
$
[\fE^\fX_{IJ}]_{I,J\in \fX}
$ is 
\[
\begin{pmatrix} 
\fE^\fX_{11}&\fE^\fX_{12}&\fE^\fX_{13}\\\fE^\fX_{21}&\fE^\fX_{22}&\fE^\fX_{23}\\\fE^\fX_{31}&\fE^\fX_{32}&\fE^\fX_{33}
 \end{pmatrix}
=
\left(\begin{smallmatrix} 
1&0&0&&0&1&0&&0&0&1\\0&0&0&&0&0&0&&0&0&0\\0&0&0&&0&0&0&&0&0&0\\\\
0&0&0&&0&0&0&&0&0&0\\1&0&0&&0&1&0&&0&0&1\\0&0&0&&0&0&0&&0&0&0\\\\
0&0&0&&0&0&0&&0&0&0\\0&0&0&&0&0&0&&0&0&0\\1&0&0&&0&1&0&&0&0&1
 \end{smallmatrix}\right) 
\]
when $\fX$ is $\{1,2,3\}$ , which will be used in this chapter. This example indicates a $3\times 3$ block matrix with $3\times 3$ matrices, 
thus it is a $9\times 9$ matrix. 

%
%
\begin{Notation}[$\dM$]
$\dM$ is defined as a map
$X\mapsto \MM(X^\dagger)$,
{\it i.e.}  a compound transformation for the map 
$\MM$ after the matrix adjoint operation(Hermitian transpose).
Note that $\dM$ operates on Hermitian matrices, as $\MM$ operates,
i.e. for a Hermitian matrix $X$, $\dM(X)=\MM(X)$ because $X^\dagger=X$.
\end{Notation}

\noindent In this chapter ``map'' is a map between matrices.
\begin{Notation}[identities]
\noindent 
Let us assume each of $\fM,\fM'$ is either of $\fM_m$ or $\fM(\fX)$.
Then $\id\limits_\fM,\Id\limits_\fM,\oId\limits_{\fM,\fM'}$ are denoted as follows:
$\id\limits_\fM$  is the identity matrix of $\fM$, 
$\Id\limits_\fM$  is the identity map on $\fM$, 
$\oId\limits_{\fM,\fM'} $  is the  linear map $\fM \ni X \mapsto (\Tr X) \cdot\id\limits_\fM \in \fM' $.
$\fM,\fM'$ will sometimes be dropped, as in
 $\id,\Id$ and $\oId$.
\end{Notation}
%
%
\section{Propositions and theorems}\label{subsection:2.2}
%
%
%
\begin{Lemma}\label{lem:1}
For scalars $x$ and $y$, eigenvalues of 
$ \displaystyle 
  \left(
    \Id\limits_{\fM(D')}   \otimes \Big( x\MM + y\dM \Big)
  \right)
 \left[\fE^{D'}_{IJ}\right]_{I,J\in D'}
$
are\\ $ -y,\frac{1}{2}y,\frac{d-1}{2}x+\frac{1}{2}y $ .

%
%
\begin{proof}
The matrix considered above is equal to 
$\Xi := 
 \left[
\Big( x\MM + y\dM \Big)
\fE^{D'}_{IJ}
\right]_{I,J\in D'}
$
.
 For $(i,j),(k,l)\in D'$,
$
\MM \left( \fE^{D'}_{(i,j)(k,l)} \right)
= \Tr\limits_B |(i,j)\rangle\langle(k,l)|
= \frac{1}{2} 
\begin{smallmatrix}
     &
{{}_k \qquad {}_l} 
    \\
  \begin{smallmatrix} {\phantom |}^i \\{\phantom |}^j\end{smallmatrix}\!\!
    &
  \left(\begin{smallmatrix}
    \delta_{jl}&-\delta_{jk}\\-\delta_{il}&\delta_{ik}
  \end{smallmatrix}\right)
    \\
 &   \phantom{    {{}_i \qquad {}_l}  }
\end{smallmatrix}
$
where $\delta$ is the Kronecker's delta, and
$
\dM \left( \fE^{D'}_{(i,j)(k,l)} \right)
= \frac{1}{2} 
\begin{smallmatrix}
     &
{{}_i \qquad {}_j} 
    \\
  \begin{smallmatrix} {\phantom |}^k \\{\phantom |}^l\end{smallmatrix}\!\!
    &
  \left(\begin{smallmatrix}
    \delta_{jl}&-\delta_{il}\\-\delta_{jk}&\delta_{ik}
  \end{smallmatrix}\right)
    \\
 &   \phantom{    {{}_i \qquad {}_l}  }
\end{smallmatrix}
$
.
Observing the whole matrix $\Xi$, it is decomposed 
into the form of the direct sum 
$\displaystyle 
\Xi = \frac{y}{2}\Xi_1 \oplus \left(\frac{x}{2}\Xi_2+\frac{y}{2}\Xi_3\right)$ where \\ 
\begin{eqnarray*}
\Xi_1 
&=& \mathop{\bigoplus}\limits^{i,j,k}_{1\le i < j < k \le d	}
\begin{smallmatrix}
     &
{{}_{(i,j)\otimes k} \, {}_{(i,k)\otimes j}\, {}_{(j,k)\otimes i}} 
    \\
  \begin{smallmatrix}
 {\phantom |}^{(i,j)\otimes k} \\
 {\phantom |}^{(i,k)\otimes j} \\
 {\phantom |}^{(j,k)\otimes i}
\end{smallmatrix}\!\!
    &
  \left(\begin{matrix}    0&1&-1\\1&0&1\\-1&1&0   \end{matrix}\right)
    \\
 &   \phantom{    {{}_i \qquad {}_l}  }
\end{smallmatrix}
, \\
\Xi_2 
&=& \mathop{\bigoplus}\limits^{i}_{1\le i  \le d	}
\bordermatrix{
                               &       &        &        &        &        &        \cr
{\phantom x}^{(1,i)\otimes i}  &1      & \cdots & 1      & -1     & \cdots & -1     \cr
\quad          \vdots          &\vdots &        & \vdots & \vdots &        & \vdots \cr
{\phantom x}^{(i-1,i)\otimes i}&1      & \cdots & 1      & -1     & \cdots & -1     \cr
{\phantom x}^{(i,i+1)\otimes i}&-1     & \cdots & -1     & 1      & \cdots & 1      \cr
   \quad        \vdots         &\vdots &        & \vdots & \vdots &        & \vdots \cr
{\phantom x}^{(i,d)\otimes i}  &-1     & \cdots & -1     & 1      & \cdots & 1   
}
, \\
\Xi_3 
&=& \mathop{\bigoplus}\limits^{i}_{1\le i  \le d	}
\bordermatrix{
                               &       &        &        &        &        &        \cr
{\phantom x}^{(1,i)\otimes i}  &1      &        &        &        &        &        \cr
\quad          \vdots          &       & \ddots &        &        &   0    &        \cr
{\phantom x}^{(i-1,i)\otimes i}&       &        & 1      &        &        &        \cr
{\phantom x}^{(i,i+1)\otimes i}&       &        &        & 1      &        &        \cr
   \quad        \vdots         &       &   0    &        &        & \ddots &        \cr
{\phantom x}^{(i,d)\otimes i}  &       &        &        &        &        & 1   
}
.
\end{eqnarray*}
$ \frac{y}{2}\Xi_1 $ has eigenvalues $-y$ and $\frac{y}{2}$.
$ \left(\frac{x}{2}\Xi_2+\frac{y}{2}\Xi_3\right)$  has eigenvalues 
$\frac{1}{2}y$ and $\frac{d-1}{2}x+\frac{1}{2}y$ .

\end{proof}
\end{Lemma}

%
%
\begin{Lemma}
Let $\lambda({x,y}) = 
\max \{ |-y|,|\frac{1}{2}y|,|\frac{d-1}{2}x+\frac{1}{2}y| \}$  .
Then 
\begin{eqnarray}
\quad\mathop{\mathrm{arg\,min}}\limits^{(x,y)}_{x+y=1} \lambda({x,y}) &=& 
\left(\frac{1}{d},\frac{d-1}{d}\right)
 \quad \text{and} \\
\min\limits_{x+y=1} \lambda({x,y}) &=& 
\lambda\left(\frac{1}{d},\frac{d-1}{d}\right) = \frac{d-1}{d} 
.
\end{eqnarray} 
\end{Lemma}
\begin{Notation}[$\tilde\lambda,\tM$]
Denote that 
\begin{eqnarray}
\tilde\lambda &=&\frac{d-1}{d} \label{note:lambda1}
\quad \text{and} 
\\
\tM &=& \frac{1}{d} \MM + \frac{d-1}{d} \dM \label{note:lambda2}
.
\end{eqnarray} 
\end{Notation}
Note that due to the last two lemmas,
\begin{equation}\label{ineq:lambda}
  - \tilde\lambda \id 
                    \le 
  \left(
    \Id\limits_{\fM(D')}   \otimes \tM
  \right)
 \left[\fE^{D'}_{IJ}\right]_{I,J\in D'}
                    \le 
    \tilde\lambda \id ,
\end{equation}
i.e. the absolute values of all eigenvalues of the central side are not larger than 
(\ref{note:lambda1}). Note that (\ref{note:lambda2}) operates on Hermitian matrices, as $\MM$ does.
\begin{Notation}
${D'}^n$ indicates the bases index of $\HH_-^\oN$,  
if  $D'$ is the bases index of $\HH_-$, 
as it associates the direct sum of the set ${D'}$.

\end{Notation}
%
%
%
\begin{Lemma}\label{lem:2}
\begin{equation}\label{eq:2} 
\left( \Id\limits_{\fM(D')^\oN} \otimes 
\left(\tilde\lambda^n \oId - \tM^\oN  \right) \right)
\left[\fE^{{D'}^n}_{IJ}\right]_{I,J\in {D'}^n}
\ge 0 .
\end{equation}
Here, $\oId$ is a map from 
$ \fM(D')^\oN $ to $ \fM(D)^\oN $.
%

\begin{proof}
 The inequality (\ref{eq:2}) is equivalent to
\begin{equation}
\Big( \Id\otimes\tilde\lambda^n \oId \Big)
\left[\fE^{{D'}^n}_{IJ}\right]
\ge 
\Big( \Id\otimes\tM^\oN  \Big)
\left[\fE^{{D'}^n}_{IJ}\right] .
\end{equation}
The following is enough to show the above inequality.%
\begin{equation}
\left\{
\begin{array}{ll}
\mathit{LHS.} &=
\tilde\lambda^n \left[\oId 
\big(\fE^{{D'}^n}_{IJ}\big)\right]_{I,J}
=\tilde\lambda^n\left[\delta_{IJ}\id\right]_{I,J} 
= \tilde\lambda^n \id
\\
\mathit{RHS.}&=
\Big( \mathop{\Id^\oN}\limits_{\fM(D')}  \otimes \tM^\oN   \Big)
\Big(\big[\fE^{{D'}}_{IJ}\big]{}^\oN\Big)
=
\Big( \big(\Id\limits_{\fM(D')} \otimes \tM \big)
\big[\fE^{{D'}}_{IJ}\big]\Big)^\oN
\\
&\le
\left(\tilde\lambda \id \right)^\oN = 
\tilde\lambda^n \id
\qquad ( \le \mbox{  {\it due to }} (\ref{ineq:lambda}  ) \quad).
\end{array}
\right.
\end{equation}
\end{proof}
\end{Lemma}

\noindent The last lemma successively induces the next two propositions.
\begin{Proposition}
$\; \displaystyle 
\tilde\lambda^n
\oId - \tM^\oN   $  is a CP map.
\end{Proposition}
This is due to (\ref{eq:2}) and \cite{C}. (%
Any map $\Gamma:\fM_m\to\fM_n$ is CP if and only if\\
$\left(\Id_{\fM_m}\otimes\Gamma\right)\left[\fE_{ij}\right]_{i,j=1\ldots m}
=\left[\Gamma\left(\fE_{ij}\right)\right]_{i,j=1\ldots m}$ is a semi-positive matrix.)
\begin{Proposition}\label{prop:positivity}
$\displaystyle \; \MM^\oN(X) \le 
\tilde\lambda^n\id \;$ 
 for $X \in \fM(D')^{\oN}$ if
$ X \ge 0 \;, \;\Tr X=1$.
\end{Proposition}

\noindent
Proposition \ref{prop:positivity} is applied to the proof of Theorem \ref{Th:1}.
\begin{Theorem}\label{Th:1}
$E(|\Psi\rangle)\ge n \log_2\frac{d}{d-1}$
 for any pure state $|\Psi\rangle \in \HH_-^\oN$.
\end{Theorem}
This is because Proposition \ref{prop:positivity} indicates that all eigenvalues
of the reduced matrix from any antisymmetric states are less than or
equal to $\left(\frac{d}{d-1}\right)^{-n}$.

\begin{Lemma}
$E_F(\sigma)\ge n\log_2{\frac{d}{d-1}}$ 
for any density matrix $\sigma$ supported on 
$\HH_-^\oN$.
%

\begin{proof}
\noindent Note that the entanglement formation is defined as
\begin{equation}\label{eq:ef}
 \displaystyle E_F(\rho) =\min_{\bigl(p_i,|\Phi_i\rangle\bigr)_i \in \Delta(\rho)}\sum_i p_i E(\Phi_i) 
\end{equation}
where 
\begin{equation}
 \Delta(\rho) =\left\{\bigl(p_i,|\Phi_i\rangle\bigr)_i \Big|
(p_i>0, \|\Phi_i\|=1)\forall i,
\sum_i p_i = 1, 
\sum_i p_i |\Phi_i\rangle\langle \Phi_i|=\rho\right\}
\end{equation}
is the collection of all possible decompositions of $\rho$.
It is known that all of $|\Phi_i\rangle$ 
induced from $\Delta(\rho)$ satisfy 
$|\Phi_i\rangle\in\Range(\rho)$ , where
$\Range(\rho)$ is the 
image space of a matrix $\rho$, 
which is a collection of $\rho|\psi\rangle$ with $|\psi\rangle$ 
running over the domain of $\rho$.
Hence 
\begin{equation}
E_F(\rho)\ge \min \{ E(\Phi) |
 \Phi \in \Range(\rho),\|\Phi\|=1\}
.
\end{equation}
The condition of the lemma above implies $\mathop\mathrm{Range}(\rho)\subseteq H_-^\oN$ , 
therefore the last theorem implies $E_F(\sigma)\ge n$.
\end{proof}
\end{Lemma}
\hspace{0pt}\\
\noindent 
Due to 
$\displaystyle E_C(\rho)=\lim_{n\to\infty}\frac{E_F(\rho^{\otimes n})}{n}$ \cite{HHT}, 
the entanglement cost is given as follows:
\begin{Theorem}
$E_C(\sigma)\ge n
\log_2\frac{d}{d-1}
$ for any density matrix $\sigma$ supported on 
$\HH_-^\oN$.
\end{Theorem}
\begin{Corollary}[The lower bound of the entanglement cost for $\HH_-$]
 \label{Cor:1}
\begin{equation}\label{eq:main}
E_C(\sigma)\ge \log_2\frac{d}{d-1}
\end{equation} 
for any density matrix $\sigma$ supported on 
$\HH_-$.
\end{Corollary}
\section{Conclusion and discussion}
This chapter gave a lower bound of the entanglement cost of antisymmetric states 
for $d$-dimensional antisymmetric states as the inequality (\ref{eq:main}).

\noindent
\\

\chapter{Additivity of Entanglement of Formation of Two Three-Level Antisymmetric States}
\makeendnotes

The additivity of $E_F$ is an important problem. The feasible problems
to reach a hint of a breakthrough were sought. The author tried the problem
of $3\otimes 3$ plus $3\otimes 3$ for antisymmetric states.

\medskip
\index{antisymmetric!--- state}
\index{state!antisymmetric ---}
This chapter again focuses on antisymmetric states.
The author proves the entanglement of formation is additive 
for any tensor product of two three-dimensional bipartite 
antisymmetric states. 

%
\medskip
This chapter is a derivation from \cite{Shimono}.

\section{Antisymmetric states}
Let us start with an introduction of the notations and concepts used in this chapter.
$\HH_-$ denotes an antisymmetric Hilbert space, 
which is a subspace of 
a bipartite Hilbert space $ \HH_{AB}=\HH_A\otimes\HH_B$,
where both $\HH_A$ and $\HH_B$ are three-dimensional Hilbert spaces,
spanned by the vectors  $\{\ket{i}\}_{i=1}^3$.
$\HH_-$ is  a three-dimensional Hilbert space, spanned by states 
$\{\ket{i,j}\}_{ij=23,31,12}\;$,
where the state $\ket{i,j}$ is defined as $
\frac{\ket{i}\ket{j}-\ket{j}\ket{i}}  {\sqrt{2}}$ in this chapter.
 The space $\HH_-$ is called antisymmetric 
because by swapping the position of two particles of states $\ket{\psi}$ in $\HH_-$,
we get the state $-\ket{\psi}$. 
$\HH_-^\ox{n}$  is the tensor product of $n$ copies of $\HH_-$.
These copies will be distinguished by the upper index as
$\HH_-^{(j)}$, with $j=1,\ldots , n$. 
We assume $\HH_-^{(j)}$ is an antisymmetric subspace 
 of $\HH_A^{(j)}\otimes\HH_B^{(j)}$.
%
\section{The result and proof sketch}

It has been shown in \cite{ViDC01} that
 $E_F(\rho)=1$ for any mixed state $\rho\in\density{\HH_-}$. 
This result will play the key role in our proof.
Here we will prove the following theorem: 
\begin{Theorem}\label{th:1}
\begin{equation}\label{eq:1}
E_F(\rho_1\otimes\rho_2)= E_F(\rho_1) + E_F(\rho_2) \,\left(= 2\right) 
\end{equation}
for any ${\rho_1,\rho_2\in\density{\HH_-}}$. 

\begin{proof} 
To prove this theorem, it is sufficient to show that 
\begin{equation}\label{eq:ttl2}
 E_F(\rho_1\otimes\rho_2)\ge 2 
\end{equation}
since the subadditivity
 $E_F(\rho_1\otimes\rho_2)\le E_F(\rho_1)+ E_F(\rho_2) = 2$
  is trivial. 
{Indeed, it holds
\begin{eqnarray}
\!\!\!\!&E_F&\!\!\!\!(\rho_1\otimes\rho_2)=\inf\sum p_i E(\ketbra{\psi_i}) \nonumber\\
&\le&\!\!\!\! \inf\sum p_i^{(1)}p_i^{(2)} E(\ketbra{\psi_i^{(1)}}\otimes\ketbra{\psi_i^{(2)}})\nonumber\\
& = &\!\!\!\! \inf\sum p_i^{(1)} E(\ketbra{\psi_i^{(1)}})\nonumber\\
&&+\inf\sum p_i^{(2)} E(\ketbra{\psi_i^{(2)}})\nonumber\\
& = &\!\!\!\! E_F(\rho_1)+E_F(\rho_2)
\end{eqnarray} where 
$(p_i^{(j)},\ket{\psi_i^{(j)}})$ are subject to the condition
$\rho_j= \sum_i p_i^{(j)}\ketbra{\psi_i^{(j)}}$.
} 
To prove (\ref{eq:ttl2}), we raise the following proposition and prove it.
\begin{equation}\label{eq:3}
E(\ketbra{\psi})\ge 2, \text{ for any pure state }
\ignore{\forall} \ket{\psi}\in{\HH_-^\ox{2}}. 
\end{equation}	
Using the Schmidt decomposition%
\footnote{%
Any {\it pure} state $\ket{\psi}$ of a bipartite space $\HH_A\otimes\HH_B$
can be represented as $\ket{\psi} = \sum_i p_i \ket{\psi_i^A}\otimes \ket{\psi_i^B}
$ where $\ket{\psi_i^A}$ are orthogonal bases of $\HH_A$,
$\ket{\psi_i^B}$ are orthogonal bases of $\BB_B$.\index{Schmidt decomposition},
and $\{p_i\}_i$ is  a probability distribution.
}, the state $\ket{\psi}$ can be decomposed
as follows:
\begin{equation}
\ket{\psi}=\sum\limits_{i=1}^3\sqrt{p_i}\:\ket{\psi_i^{(1)}}\otimes\ket{\psi_i^{(2)}},
\end{equation}
where $p_1,p_2,p_3>0,\quad p_1+p_2+p_3=1$, and 
$\{\ket{\psi_i^{(j)}}\}_{i=1}^3$ is an orthonormal basis of 
the Hilbert space $\HH_-^{(j)}$ for $j=1,2$.
{
Note that this Schmidt decomposition is with respect to $\HH_-^{(1)} : \HH_-^{(2)} $,
or, 
 $\left(\HH_A^{(1)}\otimes\HH_B^{(1)}\right) : \left(\HH_A^{(2)}\otimes\HH_B^{(2)}\right)$.
It is not $\left(\HH_A^{(1)}\otimes\HH_A^{(2)}\right) : \left(\HH_B^{(1)}\otimes\HH_B^{(2)}\right)$.
Here ``:'' indicates how to separate the system into two subsystems for the decomposition.%
} 
\medskip


\bigskip\noindent
\begin{Fact}\label{lem:ttl1}
If $\{\ket{\psi_i}\}_{i=1}^3 $ is an orthonormal basis of $\HH_-$, 
then there exists a unitary operator $U$, acting on both $\HH_A$ and $\HH_B$, 
such that 
$U\otimes U $ maps the states $\ket{\psi_1},\ket{\psi_2},\ket{\psi_3}$ into
the states $ \ket{2,3},\ket{3,1},\ket{1,2}$, respectively.
\end{Fact}
The proof appears in the next section. 
\bigskip
Because of Fact \ref{lem:ttl1}, there exist unitary operators $U^{(1)},U^{(2)}$ such that
\begin{equation}\label{eq:46phi}
\begin{split}
\big( U^{(1)}&\otimes U^{(1)}\otimes U^{(2)}\otimes U^{(2)} \big)
\ket{\psi}\hfill\\
&=\sum\limits_{\begin{smallmatrix}{i,j}\\{ij=23,31,12}\end{smallmatrix}} 
 \sqrt{p_{ij}}\: \ket{i,j}\otimes\ket{i,j}
\; ,  
\end{split}
\end{equation}
where $p_{23}=p_1,\; p_{31}=p_2,\; p_{12}=p_3$. 
We denote $\ket{\phi'}$ with the value of (\ref{eq:46phi}). 

\bigskip
As is written in the following, we use the following fact: 

\medskip\noindent
\begin{Fact}\label{lem:ttl2}
\begin{equation}
E(\ketbra{\psi'})\ge 2 \quad\text{ if }\quad
\begin{cases}
\quad p_{23},p_{31},p_{12}\ge0\\
\phantom{p_1}p_{23}+p_{31}+p_{12}=1
\end{cases}
.
\end{equation}
\end{Fact}
The proof of this fact also appears in the next section. 
\medskip
Local unitary operators do not change
the von Neumann reduced entropy. 
Thus $E ( \ketbra{\psi}) = E(\ketbra{\psi'})\ge 2$.
Therefore the claim (\ref{eq:3}) is proven.

\medskip
The entanglement of formation is defined as 
\begin{equation}
E_F(\rho)
         =
\!\!  \inf_{  [(p_i,\psi_i)]_i \in{\Delta(\rho)}}  
 \sum_i p_i E(\ketbra{\psi_i})
\end{equation}
with 
$$\Delta(\rho)=   \left\{ \left[(p_i,\psi_i)\right]_i \Bigm|  \!\!
\begin{array}{l}
\sum_i p_i =1 , p_i >0 \forall i\\
\sum_i p_i \ketbra{\psi_i}=\rho, \braket{\psi_i}=1\forall i\!\!
\end{array}
\right\}  $$
and it is known that all $\ket{\psi_i} $ induced from $\Delta(\rho)$ satisfy 
$\ket{\psi_i}\in \Range(\rho)$, where $\Range(\rho)$ is 
is the set of $\rho\ket{\psi}$ with $\ket{\psi}$
running over the domain of $\rho$, called the image space of the matrix $\rho$. Hence 
\begin{equation}
E_F(\rho)\ge \inf \left\{ E(\ketbra{\psi}) \bigm| \ket{\psi}\in\Range(\rho), \braket{\psi}=1 \right\} .
\end{equation}
Since $\rho_1\otimes\rho_2\in\BB(\HH_-^\ox{2})$ and $\Range(\rho_1\otimes\rho_2)\subseteq \HH_-^\ox{2}$,
(\ref{eq:ttl2}) is proven. 
Therefore  (\ref{eq:1}) has been shown.  
\end{proof}
\end{Theorem}


{ 
\section{Proofs of lemmas in this chapter}
The proofs of two facts which appeared in the previous section are provided in this section. 
\begin{Lemma}
If $\{\ket{\psi_i}\}_{i=1}^3 \subset \HH_-$ is an orthonormal basis, 
there exists a unitary operator $U$, acting on both $\HH_A$ and $\HH_B$, 
such that 
$U\otimes U $ maps the states $\ket{\psi_1},\ket{\psi_2},\ket{\psi_3}$ into
the states $ \ket{2,3},\ket{3,1},\ket{1,2}$, respectively.
\begin{proof}
Let us start with some notational conventions. In the following, 
${}^{T\!\!}{\Box}$ stands for the transpose of a matrix, 
${\phantom{}^{\phantom{T}\!\!} }{\Box}^\dagger$ stands for taking the complex conjugate of
each element of a matrix, and 
${\phantom{}^{\phantom{T}\!\!} }{\Box}^\Theta$ denotes the transformation defined later.

\medskip
Let $U$ be represented as 
$\left(\begin{smallmatrix}u_{11}&u_{12}&u_{13}\\u_{21}&u_{22}&u_{23}\\u_{31}&u_{32}&u_{33}\end{smallmatrix}\right)$
with respect to the basis $\ket{1},\ket{2},\ket{3}$. %
An operator and its matrix representation might be different
objects, but we identify 
$U$ with 
$\left(\begin{smallmatrix}
u_{11}&u_{12}&u_{13}\\u_{21}&u_{22}&u_{23}\\u_{31}&u_{32}&u_{33}\end{smallmatrix}\right)$
here  for convenience. 
Lengthy calculations show that when a $9\times9$-dimensional matrix $U\otimes U$ is considered as a map 
from $\HH_-$ into $\HH_-$, it can be represented by the following $3\times 3$-dimensional 
matrix, with respect to the basis $\ket{2,3},\ket{3,1},\ket{1,2}$ : 
 \vspace*{0mm}
\begin{equation*}
 U^{\Theta} :=\begin{pmatrix}
u_{22}u_{33}-u_{23}u_{32} && u_{23}u_{31}-u_{21}u_{33} && u_{21}u_{32}-u_{22}u_{31} \\
u_{32}u_{13}-u_{33}u_{12} && u_{33}u_{11}-u_{31}u_{13} && u_{31}u_{12}-u_{32}u_{11} \\
u_{12}u_{23}-u_{13}u_{22} && u_{13}u_{21}-u_{11}u_{23} && u_{11}u_{22}-u_{12}u_{21} 
\end{pmatrix} . \vspace*{0mm}
\end{equation*} 
Then one can show that\vspace*{0mm}
$$U^{\Theta}\cdot ^{\:T\!\!\!\!}U = (\det U)
\left(\begin{smallmatrix}1&0&0\\0&1&0\\0&0&1\end{smallmatrix}\right),\vspace*{0mm}
$$
and multiplying  $U^\dagger$ from the right in the  equation above,
 one can obtain 
 $U^\Theta = (\det U)\cdot U^\dagger$,
since $U$ is a unitary matrix, and ${}^{T\!\!\!}U \cdot U^\dagger$ is equal to the identity matrix.
\\

Since  $\{\ket{\psi_i}\}_{i=1,2,3}  $
is an orthonormal basis of $\HH_-$,
there exists a unitary operator on $\HH_-$
such that 
$\ket{\psi_1}\mapsto\ket{2,3}, \ket{\psi_2}\mapsto\ket{3,1}, \ket{\psi_3} \mapsto\ket{1,2}$. 
Denote $\Theta_\psi$ the corresponding matrix with respect to the basis
$ \{\ket{i,j}\}_{ij=23,31,12}$.

Let 
 $ U_\psi:={({\det \Theta_\psi})^\frac{1}{2}}\cdot{\Theta_\psi^\dagger}$ 
.\footnote{In the definition above, it does not matter which of the two roots of 
$\det \Theta_\psi$ are taken. }
It holds that  $U_\psi^\Theta=\Theta_\psi$.%
\footnote{%
 Indeed,
$ U_\psi^\Theta  =(\det U_\psi){ U_\psi^\dagger}
= (\det \Theta_\psi)^{\frac{3}{2}} \det \Theta_\psi^\dagger
\cdot ({(\det \Theta_\psi)^{\frac{1}{2}} })^\dagger
\Theta_\psi 
= \Theta_\psi $ .
Note that $ \det U_\psi
= \det( \det(\Theta_\psi)^{\frac{1}{2}}\,  \Theta_\psi^\dagger )
=(\det \Theta_\psi)^{\frac{3}{2}} \det \Theta_\psi^\dagger $
because $\Theta_\psi^\dagger$ is a $3\times 3 $ matrix.
} Therefore $U_\psi\otimes U_\psi = U'_\psi$. 
The operator $U_\psi$ is the one needed to satisfy the statement of Lemma \ref{lem:ttl1}.
\end{proof}
\end{Lemma}

\begin{Lemma}
\begin{equation*}
E(\ketbra{\psi'})\ge 2 \quad\text{ if }\quad
\left\{
\begin{array}{l}
\ket{\psi'} = \sum\limits^{i,j}_{ij=23,31,12}\sqrt{p_{ij}}\:\ket{i,j}\ket{i,j}  \\
 p_{23},p_{31},p_{12}\ge0\\
p_{23}+p_{31}+p_{12}=1
\end{array}\right.
.
\end{equation*}

\begin{proof}
 Let $p_{32}:=p_{23} , p_{13}:=p_{31} , p_{21}:=p_{12}$.
Then it holds, \vspace*{0mm}
\begin{eqnarray*}
\ket{\psi'}  
&=& \sum^{i,j}_{1\le i < j\le 3}\sqrt{p_{ij}}\:\ket{i,j}\ket{i,j}  \vspace*{-6mm}
										\\
&=&\frac{1}{{2}}\sum^{i,j}_{1\le i < j\le 3}\sqrt{p_{ij}}\:     
\{\ket{ii;jj}-\ket{ij;ji}-\ket{ji;ij}+\ket{jj;ii}\}
                                                                                \vspace*{-6mm} \\
&=&\frac{1}{{2}}\sum^{i,j}_{1\le i\ne j\le 3}\sqrt{p_{ij}}\: 
\{\ket{ii;jj}-\ket{ij;ji}\} ,
                                                                       \vspace*{0mm}
\end{eqnarray*}                                                                                 
 where $\ket{i_1i_2;i_3i_4}$ denotes the tensor product 
$\ket{i_1}\otimes\ket{i_2}\otimes\ket{i_3}\otimes\ket{i_4}$ , 
where
 $\ket{i_1}\in\HH_A^{(1)}$, $\ket{i_2}\in\HH_A^{(2)}$, $\ket{i_3}\in\HH_B^{(1)}$  and  $\ket{i_4}\in\HH_B^{(2)}$ 
,
and the condition $1\le i\ne j\le 3$ actually means 
``$1\le i\le 3$, $1\le j\le 3$ and   $i\ne j$''. 

We are now going to calculate the reduced matrix of $\ketbra{\psi'}$, 
denoted $\Xi$ , and decomposed into the 
direct sum as follows:
\begin{eqnarray}
 \Xi &:=& \mathop{\mathrm{Tr}}\limits_{\HH_B^{(1)}\otimes\HH_B^{(2)} } \ket{\psi'}\bra{\psi'}  
 \nonumber          \\
 &=& \frac{1}{4}\sum^{i,j,k,l}_{\begin{smallmatrix}{1\le i\ne j\le 3}\\{1\le k\ne l\le 3}\end{smallmatrix}}
\sqrt{p_{ij}p_{kl}}\mathop{\mathrm{Tr}}\limits_{\HH_B^{(1)}\otimes\HH_B^{(2)} }
\left(
{\begin{matrix}
  {\ket{ii;jj}\bra{kk;ll}-\ket{ii;jj}\bra{kl;lk}\phantom{abc}}
   \\
   {\phantom{abc}-\ket{ij;ji}\bra{kk;ll}+\ket{ij;ji}\bra{kl;lk} }
\end{matrix} }
\right)
 \nonumber          \\
 &=& \frac{1}{4}\sum^{i,j,k,l}_{\begin{smallmatrix}{1\le i\ne j\le 3}\\{1\le k\ne l\le 3}\end{smallmatrix}}
\sqrt{p_{ij}p_{kl}}\mathop{\mathrm{Tr}}\limits_{\HH_B^{(1)}\otimes\HH_B^{(2)} }
\big(
{
  {\ket{ii;jj}\bra{kk;ll}+\ket{ij;ji}\bra{kl;lk} }
}
\big)
 \nonumber          \\
 &=&
 \frac{1}{4}\sum^{i,j,k}_{\begin{smallmatrix}{1\le i\ne j\le 3}\\{1\le k\ne l\le 3}\end{smallmatrix}}
 \sqrt{p_{ik}p_{jk}}\: \ket{ii}\bra{jj}
+
 \frac{1}{4}\sum^{i,j}_{{1\le i\ne j\le 3}}
 {p_{ij}}\: \ket{ij}\bra{ij}   
                       \nonumber          \\
   &   \cong &
\frac{1}{4}
\left ( 
\begin{smallmatrix} 
 p_{12}+p_{13} & \sqrt{p_{13}p_{23}} & \sqrt{p_{12}p_{23}} \\
 \sqrt{p_{13}p_{23}} & p_{12}+p_{23} & \sqrt{p_{12}p_{13}} \\
 \sqrt{p_{12}p_{23}}& \sqrt{p_{12}p_{13}} &p_{13}+p_{23}
\end{smallmatrix}
\right)  
\oplus
\frac{1}{4}(p_{12}) ^{\oplus 2} \oplus
\frac{1}{4}(p_{13}) ^{\oplus 2} \oplus
\frac{1}{4}(p_{23}) ^{\oplus 2} ,                       \label{eq:direct_sum}
\end{eqnarray}
where $\oplus$ denotes the direct sum of matrices,
and $\Box^{\oplus n}$ denotes the direct sum of $n$ copies of same matrices.

\medskip
We need to get the  eigenvalues of $\Xi$ 
in order to calculate the reduced von Neumann entropy 
$
E(\ketbra{\psi'})
 = -\Tr\left(\Xi\log_2\Xi\right)
 = -\!\!\sum\limits_{\lambda : \mathrm{e.v. of }\;\Xi}\!\!\lambda\,\log_2\lambda.$
In this case, fortunately, the whole eigenvalues can be determined explicitly
from  (\ref{eq:direct_sum}): 
\begin{equation}
\left\{
\frac{1-\cos\theta}{6}, 
\frac{1-\cos(\theta+\frac{2\pi}{3})}{6}, 
\frac{1-\cos(\theta+\frac{4\pi}{3})}{6}, 
\frac{p_{12}}{4},\frac{p_{12}}{4},
\frac{p_{13}}{4},\frac{p_{13}}{4},
\frac{p_{23}}{4},\frac{p_{23}}{4}
\right\}
\label{ex:explicit}
\end{equation}
 for a certain $-\frac{\pi}{3}<\theta\le\frac{\pi}{3}$.%
\footnote{%
The exact value of $\theta$ has no importance for us in the proof.} 
The eigenvalues above are denoted as 
$
\{\lambda_1, \lambda_2, \ldots
,\lambda_9\},
$ respectively. Although $\lambda_4,\dots,\lambda_9$ are trivial, 
$\lambda_1,\lambda_2,\lambda_3$ are the roots of the cubic polynomial
\begin{equation}\label{eq:theequation}
g(\lambda) := \lambda^3-\frac{1}{2}\lambda^2+\frac{1}{16}\lambda-\frac{p_{12}\:p_{13}\:p_{23}}{16},
\end{equation}
that is the characteristic polynomial function of the cubic matrix that appeared in
(\ref{eq:direct_sum}).
%
%
We must solve this cubic equation to obtain (\ref{ex:explicit}).
The cubic equation $g(\lambda)=0$ is a  Cardan's irreducible form,%
\footnote{A cubic equation is said to be in Cardan's irreducible form if 
its three roots are real numbers. }\index{Cardan's irreducible form}
 because $\Xi$ is a  density matrix.
 In such a case, the roots of the cubic equation are \vspace*{0mm}
\begin{equation}\label{eq:tri}
\alpha+\beta\cos\theta,
\alpha+\beta\cos(\theta+\frac{2\pi}{3}),
\alpha+\beta\cos(\theta+\frac{4\pi}{3}).\vspace*{0mm}
\end{equation}
One can easily show that
 $ \lambda_1+\lambda_2+\lambda_3=3\alpha,$ and $ 
\lambda_1^2+\lambda_2^2+\lambda_3^2=3\alpha^2+\frac{3}{2}\beta^2 $.
If $\lambda_1,\lambda_2,\lambda_3$ are equal to the roots of 
the cubic equation $\lambda^3+a_1\lambda^2+a_2\lambda+a_3=0$, then
$\lambda_1+\lambda_2+\lambda_3=-a_1$ and 
$\lambda_1^2+\lambda_2^2+\lambda_3^2=a_1^2-2a_2 $ hold.
Taking $a_1=-\frac{1}{2},a_2=\frac{1}{16}$ from  (\ref{eq:theequation}),
we get the simultaneous equations $\{ 3\alpha=\frac{1}{2},
3\alpha^2+\frac{3}{2}\beta^2 = \frac{1}{8} \}$
, with one of the solutions $(\alpha,\beta)=\left(\frac{1}{6},-\frac{1}{6}\right)$.
Applying this argument to (\ref{eq:tri}), we complete (\ref{ex:explicit}).

\bigskip  
The author's  idea is now to show that \vspace*{0mm}
\begin{equation}\label{eq:00}
E(\ketbra{\psi'}) = \sum_{i=1}^9(-\lambda_i\log_2\lambda_i)\ge 2 . 
\vspace*{0mm}
\end{equation}
 This will be shown if we prove that the following:  \vspace*{0mm}
\begin{equation}\label{eq:01}
\sum_{i=1}^3(-\lambda_i\log_2\lambda_i)\ge 1 \quad \text{ and } \quad 
\sum_{i=4}^9(-\lambda_i\log_2\lambda_i)\ge 1.\vspace{0mm}
\end{equation}
The second inequality is easy to verify by simple calculations.
Therefore, to finish the proof 
of the lemma
we  need to show that 
\begin{equation}\label{eq:last}
\sum_{i=1}^3(-\lambda_i\log_2\lambda_i)\ge 1.
\end{equation} 

\medskip
Without loss of generality, one can assume $\theta\in\left[0,\frac{\pi}{3}\right]$.%
\footnote{%
The set of $\{\lambda_i\}_{i=1}^3$ does not change if $\theta$ is 
replaced by $-\theta$. Thus we can change the assumption 
$\theta\in\left(-\frac{\pi}{3},\frac{\pi}{3}\right]$  into
$\theta\in\left[0,\frac{\pi}{3}\right]$. } 
Clearly,
$\lambda_1\in \left[0, \frac{1}{12}\right]$ and 
$
\lambda_2,\lambda_3=\frac{1}{4} -\frac{\lambda_1\pm\sqrt{\lambda_1-3\lambda_1^2}}{2}
 \in\left[\frac{1}{12}, \frac{1}{3} \right]
$.
($\lambda_2,\lambda_3$ can be regarded as the solution of the following simultaneous equations: 
$\lambda_1+\lambda_2+\lambda_3=\frac{1}{2},
\lambda_1^2+\lambda_2^2+\lambda_3^2=\frac{1}{8}$. ) 
One can also show that 
\begin{equation}\label{eq:2ineq}-z \log_2 z 
\ge
\begin{cases}
\phantom{ab}( \log_2 12)\:z 
  & 
\text{ if $ z\in\left[0,\frac{1}{12}\right]$} 
     \\
\phantom{ab} \frac{1}{2}+\frac{\log_{e} 4 -1}{\log_{e} 2}(z-\frac{1}{4})-4(z-\frac{1}{4})^2
  & 
\text{ if $z\in\left[\frac{1}{12},\frac{1}{3}\right]$}
\end{cases}
\end{equation}
(see Fig.1).
\begin{figure}\label{fig:ttl1}
\begin{center}\includegraphics[width=0.75\linewidth]{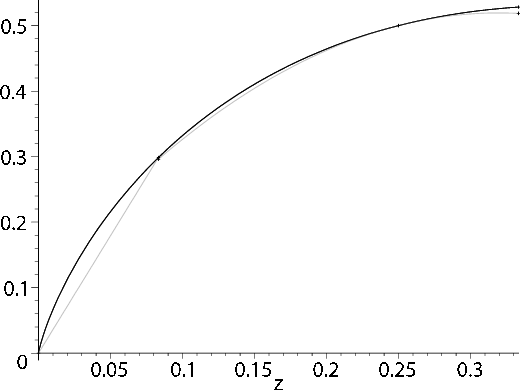}\end{center}
\caption[$-z\log_2 z$ : \quad lower-bounded with polynomial functions]{$-z\log_2 z $ is lower-bounded with the two polynomial functions as \newline 
$ \qquad\qquad\qquad\qquad\qquad
-z \log_2 z 
\ge
\begin{cases}
\phantom{ab}( \log_2 12)\:z 
  & 
\text{ if $ z\in\left[0,\frac{1}{12}\right]$} 
     \\
\phantom{ab} \frac{1}{2}+\frac{\log_{e} 4 -1}{\log_{e} 2}(z-\frac{1}{4})-4(z-\frac{1}{4})^2
  & 
\text{ if $z\in\left[\frac{1}{12},\frac{1}{3}\right]$}
\end{cases}
$
}
\end{figure}
The first inequality of (\ref{eq:2ineq}) is easily confirmed.
One way to prove  the second inequality is as follows: 
Let  $
f(z):=\bigl(-z\log_2 z\bigr) - 
\left(\frac{1}{2}+\frac{\log_{e} 4 -1}{\log_{e} 2}(z-\frac{1}{4})
-4(z-\frac{1}{4})^2\right)$.
Differentiating this expression with respect to $z$ once and twice respectively, we can get 
the increasing and decreasing table as Table.~\ref{tb:idt}.
\begin{table}[hh]
\begin{center}
\begin{tabular}{|c||c|c|c|c|c|c|c|} 
\hline
 $z$     & $\frac{1}{12}$ &    & $\frac{1}{8\log_{e} 2} $ &   &  $\frac{1}{4}$ &  
& $\frac{1}{3}$ \\ \hline\hline
$f(z)$ &   $+$  & $\curvearrowright$  &$+$ & $\searrow$ &$0$ &$\nearrow$ &  \\ \hline
$f'(z)$  &        &    &             & $-$  & $ 0 $  & $+$ &       \\ \hline
$f''(z)$ &        & $-$&     $0 $    & $+$  & $ + $  & $+$ &       \\ \hline
\end{tabular}
\caption{The increase/decrease table of $\bigl(-z\log_2 z\bigr) - 
\left(\frac{1}{2}+\frac{\log_{e} 4 -1}{\log_{e} 2}(z-\frac{1}{4})
-4(z-\frac{1}{4})^2\right)$}
\label{tb:idt} 
\end{center}
\end{table}
The table indicates 
 $ f(z)\ge 0 $  for $ z\in\left[\frac{1}{12},\frac{1}{3}\right]$. 
Now we indeed get the lower bounds with polynomial functions. 

Combining all of the above inequalities,
we get 
(\ref{eq:last}) as
\begin{equation*}
-\sum\limits_{i=1}^3 \lambda_i \log_2 \lambda_i \ge 1 + 
\left(\frac{\log_{e} 3 +2}{\log_{e} 2}-2\right)\lambda_1  + 4\lambda_1^2 \ge 1 \quad .
\end{equation*}
Therefore (\ref{eq:01}) and (\ref{eq:00}) are successively shown, and our proof is completed.
\end{proof}
\end{Lemma}
}%

\section{Conclusion and discussion}
The additivity of the entanglement of formation for two three-dimensional bipartite antisymmetric states 
has been proven in this paper. 


\section*{Endnotes}

\theendnotes
\chapter{A Gap Between $E_F$ and $E_D$}
\makeendnotes

\renewcommand{\emph}{}

Are there gaps between $E_D$ and $E_C$? That was an important problem. 
The answer was yes. We found such examples for a $2\otimes 4$-level system in 2004, 
after the bound entanglement for a $3\otimes 3$-level system was found \cite{Horodecki98} in Poland. 

\medskip
This chapter is a derivation from \cite{MSW}.

\section{Background}

\index{LOCC}
The problem is stated as follows: 
\begin{quote}
Is there a gap between the two important and not-yet-fully-analyzed quantities of 
quantum entanglement presented below? 
\begin{enumerate}
\item $E_C(\rho)$, the entanglement cost, \\
--- the asymptotic quantity of Bell states needed to produce $\rho$
under LOCC operation, \\
\item $E_D(\rho)$, the entanglement distillation, \\
--- the asymptotic quantity of Bell states  retrieved from
$\rho$ under LOCC operation. 
\end{enumerate}
\end{quote}

If there is no difference, meaning $E_C(\rho)=E_D(\rho)$ for any $\rho$, 
the problem of quantification of quantum entanglement may end up easy to handle. 
That the equality holds for pure states is already known \cite{popescu:rohrlich,vidal2000,nielsen2000,donald:horodecki:rudolph} thus the problem resides for 
$\rho$ being non-pure state. 
If there is a difference, which means $E_C(\rho) >E_D(\rho)$%
\footnote{$E_C(\rho)<E_D(\rho)$ does not occur for any $\rho$ in principle.
If it did, quantum entanglement would increase infinitely 
by LOCC operations. }, one can say the irreversibility 
occurred under LOCC operation and that the degradation of quantum entanglement is inevitable. 

Analyzing this problem is theoretically interesting when investigating the property of quantum 
entanglement.

\begin{figure}[hb]\begin{center}
\includegraphics[width=12cm]{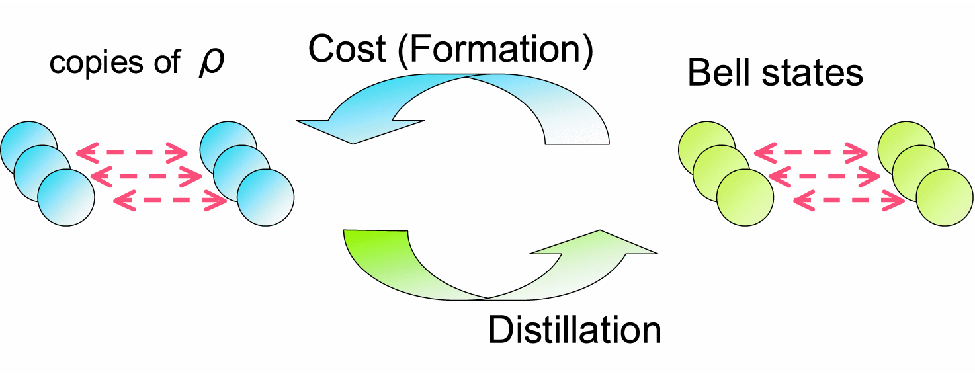}\end{center}
\caption[The irreversibility occurs?]{Does the irreversibility occurs during the production and the distillation?}
\end{figure}

\medskip 

  Here is a historical note. There are already answers in  that there exists such $\rho$ that satisfy
  $E_D(\rho) < E_C(\rho)$. 
  The first examples are 
such states $\rho$ that satisfy $E_C(\rho)> 0$ and $E_D(\rho)= 0$, 
 which are named
 {\it bound entanglement} \cite{Horodecki98}.
 \index{bound entanglement}\index{entanglement!bound ---}
 The bound entanglement appears in a $3\otimes 3 $-level system.\index{3$\otimes$3-level system}
  Other attempts are performed at \cite{ViDC01} for antisymmetric states. 
  \index{antisymmetric states}\index{state!antisymmetric ---}
This chapter gives the examples of a $2\otimes 4$-level system, with  mathematical software Maple,
\index{2$\otimes$4-level system}
through calculating the Holevo capacity of unital channels \cite{king:unital:add}. 

\medskip 


\section{Introduction}

 We need to estimate both the values of $E_C$ and $E_D$. 
$E_C$ can be calculated by the Holevo capacity 
if the full capacity $\bar{C}$ is determined through Stinespring's relation
because of the relation between $E_F$ and the Holevo capacity 
(see Sec. \ref{sec:stine}).
 To evaluate the value of $E_D$ there is a useful estimation theorem.
\index{logarithmic negativity}
\begin{Notation}[the logarithmic negativity]
The logarithmic negativity $E_N$ of $\rho$ where $\rho$ is shared by two sites is defined as follows: 
\begin{equation}
E_N(\rho) 
= \log \sum_i |\lambda_i| 
\end{equation}
 where $\{\lambda_i\}$ are the eigenvalues 
of $\rho^\Gamma$ where   
$\rho^\Gamma$ is the partial transpose of $\rho$. 
The partial transpose of a state for a bipartite system is defined as 
\index{partial transpose}
$(a_{i\otimes j, k\otimes l})_{(i\otimes j),(k\otimes l)}\mapsto (a_{k\otimes j, i\otimes l})_{(i\otimes j),(k\otimes l)}$. 
\end{Notation}
\begin{Theorem}[\cite{ViWe02}]%
\begin{equation}
E_D(\rho) \le E_N(\rho) 
\end{equation}
\end{Theorem}

Note that $E_N$ is simple to calculate even though the method to calculate $E_D$ for general cases is not known.

Therefore, 
the aforementioned gap $E_D(\rho)< E_C(\rho)$ is concluded, 
if 
\begin{list}{\hspace*{8mm}--}{} 
\item $E_C(\rho)$ is calculated and 
\item $E_N(\rho) < E_C(\rho)$ is determined, 
\end{list} 
by some method. 

\section{Unital channel and its associated state}

\index{depolarizing channel}
\index{channel!depolarizing ---}
\index{generalized depolarizing channel}
\index{channel!generalized depolarizing ---}
\index{output ellipsoid}
  We considered a generalized {\it depolarizing channel}%
  \footnote{%
		  A depolarizing channel and a generalized depolarizing channel:
			\begin{list}{\hspace*{8mm}--}{} 
				\item
				   A {\it depolarizing channel} for a qubit is a channel with its output ellipsoid
				   \index{depolarizing channel}\index{channel!depolarizing ---}
				  being a sphere and its center  located at the center of the Bloch sphere. 
				  ( The output ellipsoid is explained in Subsection 	 \ref{subsec:aqc}. ) 
				\item 
				  A {\it generalized} depolarizing channel here is defined as follows: 
				  a qubit channel is a generalized depolarizing channel as long as the center of 
				  its output ellipsoid is located at the center of the Bloch sphere. 
			\end{list} 
			  The output ellipsoid of the generalized depolarizing channel of (\ref{eq:gdct}) 
			  has three axes in $x,y,z$ directions with the radius of 
			  $p_0+p_x-p_y-p_z ,  p_0+p_y-p_x-p_z$ and $p_0+p_z-p_x-p_y$. 
		  \medskip
		  
		\begin{figure}[hhhhh]
			\begin{center}
				\includegraphics[width=3cm]{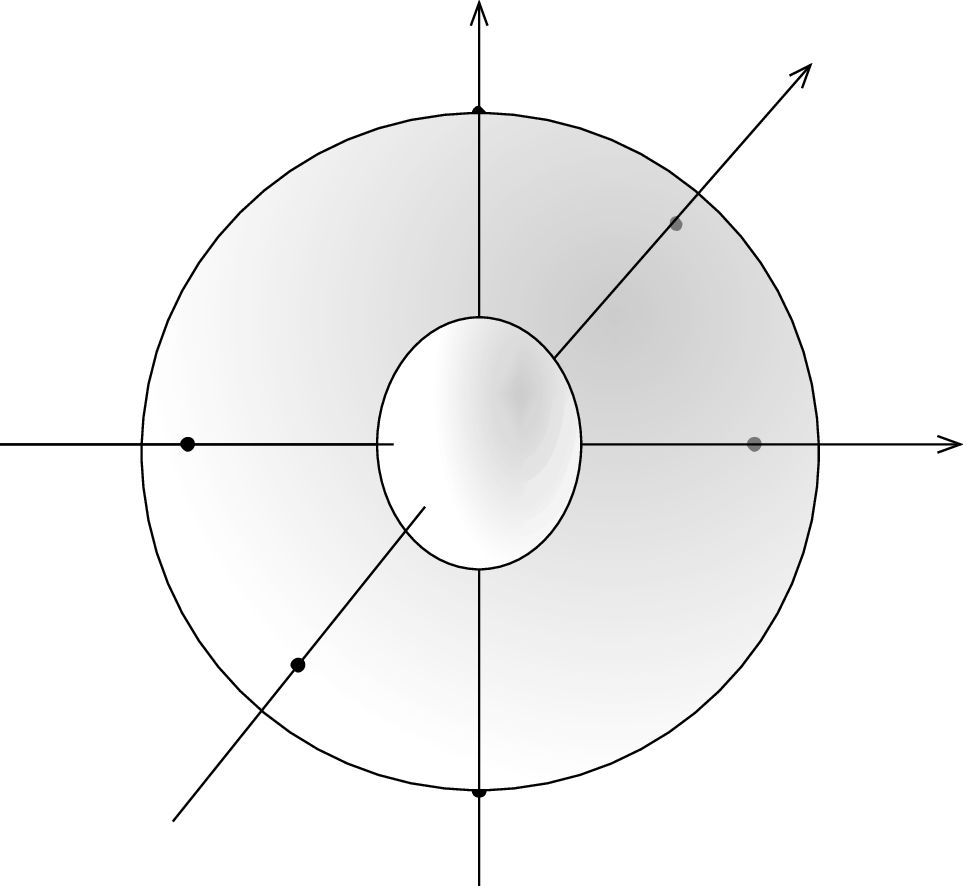}
			\end{center}
			\vspace*{-5mm}
			\begin{quote}
				\caption
				[
				A general depolarizing channel's output ellipsoid
				]
				{
					An illustration of the Bloch sphere and the output ellipsoid of a 
					general depolarizing channel. With the condition (\ref{eq:king-ruskai}), 
					the longest axis is $z$-directional.
					}
			\end{quote}
		\end{figure}
		} 
  on a qubit:
  \begin{equation}    
    \label{eq:gdct}
     \T: \rho\longmapsto
      \sum_{s=0,x,y,z} p_s\,\sigma_s\rho\sigma_s^\dagger, 
  \end{equation}
  with 
   the Pauli matrices\index{Pauli matrices}
\index{matrix!Pauli matrices}\index{matrices!Pauli ---}
  \begin{equation*}
    \sigma_0 =\left(\begin{array}{rr}
                       1 & 0 \\
                       0 & 1
                    \end{array}\right),\ 
    \sigma_x =\left(\begin{array}{rr}
                       0 & 1 \\
                       1 & 0
                    \end{array}\right),\ 
    \sigma_y =\left(\begin{array}{rr}
                      0 & -\I \\
                      \I &  0
                    \end{array}\right),\ 
    \sigma_z =\left(\begin{array}{rr}
                      1 &  0 \\
                      0 & -1
                    \end{array}\right),\ 
  \end{equation*}
  and a probability distribution $\{p_s\}_{s=0,x,y,z}$. 
  
  There is an interesting theorem about the additivity of the Holevo capacity for depolarizing channel:
  \index{additivity!--- of the Holevo capacity!generalized depolarizing channel}
  \begin{Theorem}[\cite{king:unital:add:2}]
  For  generalized depolarizing channels,
   additivity
  of the Holevo capacity in terms of  tensor product with an arbitrary channel holds.
  Namely,
  \begin{equation}
   C(\T\otimes \T') = C(\T) + C(\T') 
  \end{equation}
  for any generalized depolarizing channel $\T$ and any channel $\T'$.
  \end{Theorem}
  \par
  Note that up to unitary transformations on the input and the output systems, each unital
  qubit channel has this form 
  \cite{king-ruskai,fujiwara-algoet}. 

  We assume the condition
  \begin{equation}
    \label{eq:king-ruskai}
    p_0+p_z-p_x-p_y \geq |p_0+p_y-p_x-p_z|,|p_0+p_x-p_y-p_z|,
  \end{equation}
  which does not lose generality.
  Then for such a channel $\T$ the capacity is given by
  \begin{equation}
  C(\T)=1-S_{\rm min}(\T)
  \label{id:ch-en}
  \end{equation} 
  where $S_{\rm min}(\T)$ is  the {minimum output entropy} of $\T$, which is 
  achieved at either of the inputs  $\ket{0},\ket{1}$,  
   thus 
  $S_{\rm min}(\T)=S\bigl(\T(\ketbra{0})\bigr)=S\bigl(\T(\ketbra{1})\bigr)$.
  The optimal ensemble of input signals of the channel in order to achieve the Holevo capacity 
  is the uniform distribution $(1/2,1/2)$ on the two states $\ket{0},\ket{1}$.
  \par
  A Stinespring dilation for this map $\T:\BB(\HH_{\tt i})\to\BB(\HH_{\tt o})$ is given  by an isometric embedding
  $U:\HH_\mathtt{i}\longrightarrow\HH_\mathtt{o}\otimes\HH_\mathtt{a}$
  where $\HH_\mathtt{i},\HH_\mathtt{o}=\C^2$ and $\HH_\mathtt{a}=\C^4$, in the following $8\times 4$ block form:
  \begin{equation}
    U=\left(\begin{array}{r}
              \sqrt{p_0}\sigma_0 \\
              \sqrt{p_x}\sigma_x \\
              \sqrt{p_y}\sigma_y \\
              \sqrt{p_z}\sigma_z
            \end{array}\right),
  \end{equation}
  and the corresponding subspace ${\cal K}\subset\HH_\mathtt{o}\otimes\HH_\mathtt{a}$ is spanned by
  \begin{align}
    \ket{\psi_\T}       &=   \sqrt{p_0}\ket{0}\otimes\ket{0}
                          + \sqrt{p_x}\ket{1}\otimes\ket{x}
                          +\sqrt{-1}\sqrt{p_y}\ket{1}\otimes\ket{y}
                          + \sqrt{p_z}\ket{0}\otimes\ket{z}, \\
    \ket{\psi_\T^\perp} &=   \sqrt{p_0}\ket{1}\otimes\ket{0}
                          + \sqrt{p_x}\ket{0}\otimes\ket{x}
                          -\sqrt{-1}\sqrt{p_y}\ket{0}\otimes\ket{y}
                          - \sqrt{p_z}\ket{1}\otimes\ket{z}.
  \end{align}
  
  (You can easily verify that $\T$ is equal to the operation $\rho \mapsto \Tr_{\HH_\mathtt{a}} U\rho U^\dag$.)
  The average of the optimal input state\footnote{%
  Consider the definition of the Holevo capacity: 
  $$
  C(\T) = \max_{n ; \rho_1, \dots, \rho_n ; p_1, \dots, p_n } S(\sum_{i=1}^n p_i \, \T \rho_i) - \sum_{i=1}^n p_i \,  S( \T \rho) .
$$
  We call the optimized $\{\rho_i\}_i$ the {\it optimal input}. The {\it average input} is $\sum_i p_i \rho_i$.
  \index{optimal!--- input}\index{input!optimal ---}
  \index{average!--- input}\index{input!average ---}
  }
   is transformed to the equal mixture of the
  two pure states
  \begin{equation}
  \rho_\T = \frac{1}{2}\ketbra{\psi_\T}+\frac{1}{2}\ketbra{\psi_\T^\perp}
  \label{id:rhot}
  \end{equation}
   by $U$.
  \par
  With Theorem~\ref{thm:hcap:entf}, (\ref{id:ch-en}) leads to
  \begin{equation}
  E_F\oa(\rho_\T)=S_{\rm min}(\T)=H(p_0+p_z,p_x+p_y)
  \end{equation}
  where $H$ is the entropy function 
  and $E_F\oa(\rho_\T\otimes\sigma)=E_F\oa(\rho_\T)+E_F\oa(\sigma)$ for any
  $\sigma\in{\cal O}_{\T'}$, with an arbitrary channel $\T'$. 
  Theorem~\ref{thm:CC:implies:EE} leads to $E_F\oa(\rho_\T^{\otimes n})= n E_F\oa(\rho)$, thus 
  \begin{equation}
  E_C\oa(\rho_\T)=H(p_0+p_z,p_x+p_y).
  \end{equation}
  This leads to that  the decomposition of $\rho_\T^{\otimes n}$
  into the $2^n$ equally-weighted tensor products of $\ketbra{\psi_\T}$, 
  $\ketbra{\psi_\T^\perp}$ being is formation--optimal%
  \footnote{%
  The formation--optimal here means that the collection of states gives 
  the optimal value of (\ref{def:eof}), which is the formula of the 
  entanglement of formation.%
  }. By the convex
  roof property of $E_F$ this implies that \emph{any} convex combination
  of these states is a formation--optimal decomposition. (This argument was
  also used in~\cite{VW} to extend the domain of states with known entanglement
  of formation.) In particular, we can conclude that 
   \begin{equation}
     E_C\oa(\rho)=E_F\oa(\rho)=H(p_0+p_z,p_x+p_y)
   \label{eq:ec-qubit}
   \end{equation}
  for any mixture
  $\rho$ of $\ketbra{\psi_\T}$ and $\ketbra{\psi_\T^\perp}$.

 \medskip
 --- We may omit the superscription ($\oa\>$) from here.
 
\section{A gap between $E_C$ and $E_D$}
\label{sec:gap}
\begin{figure}[hhhhhhhhhhh]
	\vspace*{-0mm}
	\begin{center}
		\vspace*{-8mm}
		\includegraphics[height=4cm]{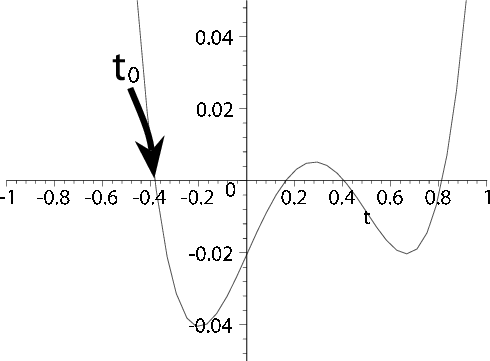}
	\end{center}
		\vspace*{-8mm}
	\caption
		[
		Plot: \quad $t^4-t^3+\f{19}{128}t-\f{21}{1024}$
		]
		{
		The graph of $f(t)$
		for $(p_0,p_x,p_y,p_z)=(\f{1}{16},\f{2}{16},\f{6}{16},\f{7}{16})$
		as an example
		}
	\label{figft416}
\end{figure}

  One can verify that the partial transpose $\rho_\T^{\Gamma}$ of the optimal state
  $\rho_\T$ is decomposed into a form of the direct sum of two $4\times 4$--matrices
  from (\ref{id:rhot}),
  which  have the same characteristic equation
  \begin{align}
    f(2t) &= 0,\ \text{ with} \\
    f(t)  &= t^4 - t^3 + 4(p_0p_xp_y+p_0p_xp_z+p_0p_yp_z+p_xp_yp_z)t - 16 p_0p_xp_yp_z.
    \label{eq:ft416}
  \end{align}
  This equation $f\left(2t\right)=0$ has only one negative root denoted $t_0$, and 
  $f$ is decreasing in a neighborhood of $t_0$. See Fig.~\ref{figft416}.

  Assume $\rho_\T^\Gamma$ has eigenvalues $\{t_0,t_0,t_1,t_1,t_2,t_2,t_3,t_3\}$. 
  Since $t_0 + t_0 + t_1+ t_1+ t_2+t_2+t_3+t_3 =1 $ , 
   $ \left\|\rho_\T^\Gamma\right\|_1 =
   | t_0 |+| t_0 |+| t_1|+| t_1|+| t_2|+|t_2|+|t_3|+|t_3 | 
     =
     |2 t_0 |+ |2 t_1+ 2 t_2+ 2 t_3 | 
    = (-2t_0) + (1 - 2t_0) = 1-4t_0
      $,
       where $\|\cdot\|_1$ is one-norm\index{one-norm}, or the sum of the abstract value of the eigenvalues.
    Hence $E_N(\rho_\T) = \log ( 1-4t_0)$.
   Thus 
  $E_N(\rho_\T) < E_C(\rho_\T)$ 
  is successively equivalent to
  $\log(1-4t_0) < H(p_0+p_z, p_x+p_y)$,
   $1-4t_0 < 2^{H(p_0+p_z, p_x+p_y)}$,
   $ \f{1-2^{H(p_0+p_z, p_x+p_y)}}{2} < 2t_0$,
   and 
  \begin{equation}
      f\left(-\frac{2^{H(p_0+p_z,p_x+p_y)}-1}{2}\right) > 0.
    \label{eq:ec>ed}
  \end{equation}
   \begin{figure}[thhbt]
    \centering
    \includegraphics[width=10cm]{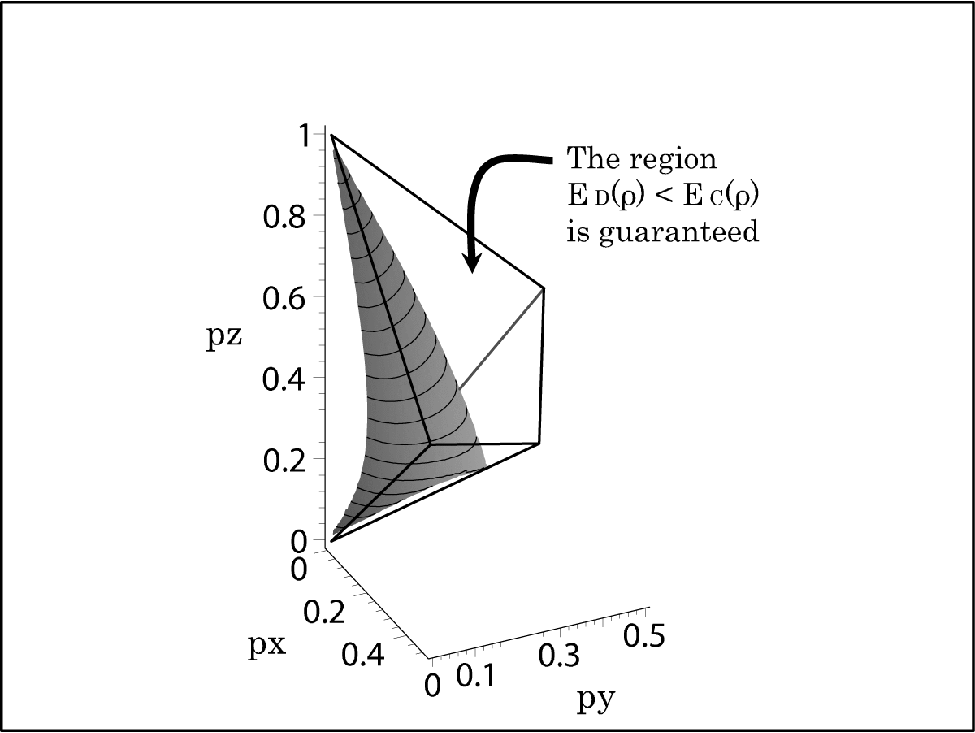}
    \caption[Pursuit of $E_D(\rho)<E_C(\rho)$ inside a pentahedoron]%
    {Pursuit of $E_D(\rho)<E_C(\rho)$: The issue resides whether there exist 
     points satisfying (\ref{eq:ec>ed}) with (\ref{eq:king-ruskai}). 
     The result is, at least all the points between the front of the curved 
     face inside the pentahedron are such points. 
     Here, the curved face corresponds to (\ref{eq:ec>ed})
    and the pentahedron corresponds to (\ref{eq:king-ruskai}).}
    \label{fig:1}
  \end{figure}
  \par
  This means if $p_0$, $p_x$, $p_y$, $p_z$ satisfy this inequality (\ref{eq:ec>ed}),
  there is a gap between the entanglement cost of $\rho_\T$,
  and its entanglement distillation, namely $E_D(\rho_\T)<E_C(\rho_\T)$; Fig.~\ref{fig:1} shows the plot
  of the region of these $(p_x,p_y,p_z)$. Due to this continuity, also for a mixture of 
  $\ketbra{\psi_\T}$ and $\ketbra{\psi_\T^{\perp}}$ which is sufficiently
  close to $\rho_\T$, we observe a similar gap.
  
  \begin{Example}
  Assume $(p_0,p_x,p_y,p_z)=(\frac{1}{2},\frac{1}{6},\frac{1}{6},\frac{1}{6})$. 
  Then $E_D(\rho_\T)<E_C(\rho_\T)$.

  \begin{proof}
  A short calculation reveals that $\left\|\rho_\T^{\Gamma}\right\|_1=5/3$,
  so $E_D(\rho_\T)\leq\log(5/3)\approx 0.737$, which is smaller
  than the entanglement cost $E_C(\rho_1)=H(1/3,2/3)\approx 0.918$. 
  
  You can also verify that LHS of (\ref{eq:ec>ed}) is $0.00784$ (which is larger than 0), thus $E_D(\rho_\T)<E_C(\rho_\T)$.
\end{proof}
  \end{Example}

  In this case, 
  
\newcommand{\3}{\sqrt{3}}

  \begin{equation}
   \rho_\T   = {
   \left(\ 
   \begin{matrix}
   \f{1}{4}   & 0 & 0 & \f{-\I}{4\3} &\vline& 0& \f{1}{4\3} & \f{1}{4\3} & 0 \\
   0 &  \f{1 }{12 } & \f{-\I}{4\3}&0 &\vline& \f{-\I}{12} &0 &0 & \f{\I}{12 }  \\
   0 & \f{\I}{4\3} & \f{1}{4}  & 0   &\vline& \f{1}{4\3} & 0  &0   &\f{-1}{4\3} \\
   \f{\I}{4\3} & 0 & 0 & \f{1 }{12 } &\vline& 0& \f{\I}{12 } & \f{\I}{12}  & 0 \\
   \hline 
   0& \f{\I}{12 } & \f{1}{4\3}& 0    &\vline& \f{1}{12}  & 0  &0   &\f{-1}{12 } \\
   \f{1}{4\3} & 0 & 0 & \f{-\I}{12 } &\vline& 0& \f{1}{12 } & \f{1}{12}  & 0 \\
   \f{1}{4\3} & 0 & 0 & \f{-\I}{12 } &\vline& 0& \f{1}{12 } & \f{1}{12}  & 0 \\
   0&  \f{-\I}{12 } &\f{-1}{4\3} &0  &\vline& \f{-1}{12} &0 &0 & \f{1}{12 }    \\
     \end{matrix}\ 
   \right)
   }
   \in \BB(\HH_{\sf o})\otimes \BB(\HH_{\sf a})
    \end{equation}
  %
  \begin{Example}
  Assume $(p_0,p_x,p_y,p_z)=(\frac{u}{2},\frac{1-v}{2},\frac{v}{2},\frac{1-u}{2})$ 
  with $u,v\in [0,\frac{1}{2}) \cup (\frac{1}{2},1]$.
  Then 
   $E_D(\rho_{\T,s})<E_C(\rho_{\T,s})$ holds
   for 
  $\rho_{\T,s}= s\ketbra{\psi_\T}+(1-s)\ketbra{\psi_\T^\perp}$  with $0\le s\le 1$.
  In this case, $\rho_{\T,s}$ is an $8\times 8$ matrix with variables $s, u,v$, 
  represented as 
\newcommand{\ut}{\sqrt{u}}
\newcommand{\vt}{\sqrt{v}}
\newcommand{\uu}{\sqrt{u'}}
\newcommand{\vv}{\sqrt{v'}}
\newcommand{\s}{\sqrt} 
  { \small
        \begin{equation}
          { 
           \f{1}{2} \left( \ 
           \begin{matrix}
             su    & 0 & 0 & s\s{uu'} & \vline &0& -s\s{-uv'}& s\s{uv} & 0 \\
           0 & s' v  & s' \s{-vv'} &0 &\vline&s'\s{uv}&0&0&-s'\s{u'v}  \\
           0 & -s'\s{-vv'} & s'v' &0 & \vline&-s'\s{-uv'}&0&0&s'\s{-u'v'} \\
           s\s{uu'} & 0&0& s u'&\vline& 0&-s \s{-u'v'}&s \s{u'v}&0 \\
           \hline 
           0 & s'\s{uv} & s'\s{-uv'} &0 &\vline& s'u &0 & 0& -s' \s{uu'}\\
           s\s{-uv'} & 0 & 0& s\s{-u'v'} &\vline &0& s v' & s \s{-vv'} &0 \\
           s\s{uv} &0 & 0 & s \s{u' v} &\vline & 0 & -s\s{-vv'} & s v & 0 \\ 
           0 & -s'\s{u'v} & -s'\s{-u'v'}&0 &\vline & -s'\s{uu'} & 0& 0& s'u'
           \end{matrix}\ 
           \right)
           }
        \end{equation}
    }
 where $s' = 1-s , u'=1-u, v'=1-v $.

  \begin{proof} 
   
  By eq.~(\ref{eq:ec-qubit}), $E_C(\rho_{\T,s}) = 1 $. 
 %
%
  The key observation is whether 
  $\log\|\rho_\T^{\Gamma}\|_{1} < E_C(\rho_{\T,s})$. This is true
  because the condition~(\ref{eq:ec>ed}) is always satisfied,
   as LHS 
  of (\ref{eq:ec>ed}) is 
 $\f{1}{2}(u-\f{1}{2})^2 + \f{1}{2}(v-\f{1}{2})^2 - (u-\f{1}{2})^2 (v-\f{1}{2})^2$ 
 which is zero at $(u,v)=(\f{1}{2},\f{1}{2})$; it is also an increasing function w.r.t.
 both $\tilde u =(u-\f{1}{2})^2$ and $\tilde v=(v-\f{1}{2})^2$ for $\tilde u, \tilde v \le \f{1}{4}$
 (See Fig.~\ref{fig:uv5}).
	\begin{figure}[hh]
		\begin{center}
			\includegraphics[width=4cm]{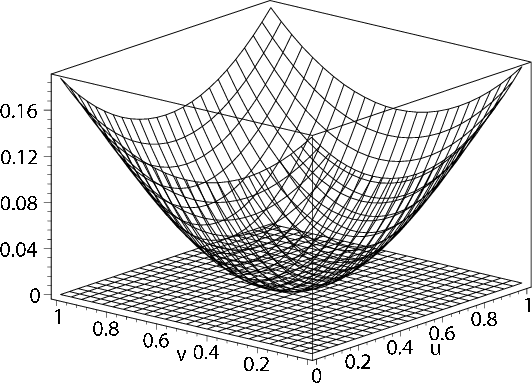}
			\caption
				[
				Plot:\quad$\f{1}{2}(u-\f{1}{2})^2 + \f{1}{2}(v-\f{1}{2})^2 - (u-\f{1}{2})^2 (v-\f{1}{2})^2$
				]
				{
				The graph of $\f{1}{2}(u-\f{1}{2})^2 + \f{1}{2}(v-\f{1}{2})^2 - (u-\f{1}{2})^2 (v-\f{1}{2})^2$ 
				}
			\label{fig:uv5}
		\end{center}
	\end{figure}
 \end{proof}
  \end{Example}

\section{Conclusion and discussion}
 We have found the states which show the gap as $E_D(\rho) <E_C(\rho) $ 
 for a $2\otimes 4$-level system. They were found by analyzing general depolarizing channels
 with Stinespring's dilation. This discovery also shows a usefulness of 
 the relationship between $E_F$ and the Holevo capacity (see Section \ref{sec:stine})
 found by \cite{MSW}.
 The examples shown in this chapter are the lowest dimensional cases among 
 those bipartite states $\rho$ being proved  $E_D(\rho) < E_C(\rho)$, 
 in that they are 8 dimensional ($2\otimes 4$), though the next lowest 
 cases are the bound entanglement which are 9 dimensional ($3\otimes 3$) 
 found in earlier in \cite{Horodecki98}. Our research may exemplify that 
 analyzing the entanglement even in the easiest cases would involve
 variant  ideas such as quantum channels that are tied to entanglement by Stinespring's relation.

\section*{Endnotes}
\theendnotes

\chapter{Numerical Verification of Superadditivity}\label{ch:6}
\makeendnotes

The strong superadditivity of $E_F$ is rather simple to check among the equivalent 
additivity problems. The author numerically checked the lowest-dimensional cases, 
which are described as ``$2\otimes 2 $ plus $2\otimes 2$''.
\medskip

In this chapter, the strong superadditivity of the entanglement of formation is numerically verified.
  This is a part of attempts to verify the expects 
that the whole additivity holds for the entanglement of formation (see Sec \ref{sec:addp}). 
It is interesting because if there were any single counterexample against the 
strong superadditivity, the expected relation $E_F \equiv E_C$ and the additivity of the Holevo capacities would collapse.
The additivity of the minimum entropy of channels would also collapse \cite{shor2}.

As a result, any single counterexample has been found from approximately a million randomly-chosen cases
and various directive searches.
Even though this is a negative result, it gave the knowledge that finding counterexamples is a challenging 
problem even if they exist.

\section{Introduction}

The strong superadditivity of the entanglement
of formation for four-partite qubits is numerically checked in this chapter.
The strong superadditivity is represented by the inequality defined in Section \ref{sec:addp}, 
as
\begin{multline}\label{eq:ineq}
 \qquad\qquad E_F(\rho) \ge E_F(\Tr_1 \rho ) + E_F(\Tr_2 \rho) \\
 \quad\mbox{ for }\quad\rho\quad  \mbox{ on } \quad 
 \HH_{A1}\otimes\HH_{A2}\otimes\HH_{B1}\otimes\HH_{B2} . \qquad\qquad 
\end{multline}

The additivity of 
entanglement of formation can be deduced directly
from this superadditivity if this inequality (\ref{eq:ineq}) holds for any pure case
\cite{MSW}.

\begin{figure}[htb]%
\begin{center}
\framebox[15cm][c]{ 
\includegraphics[width=7cm]{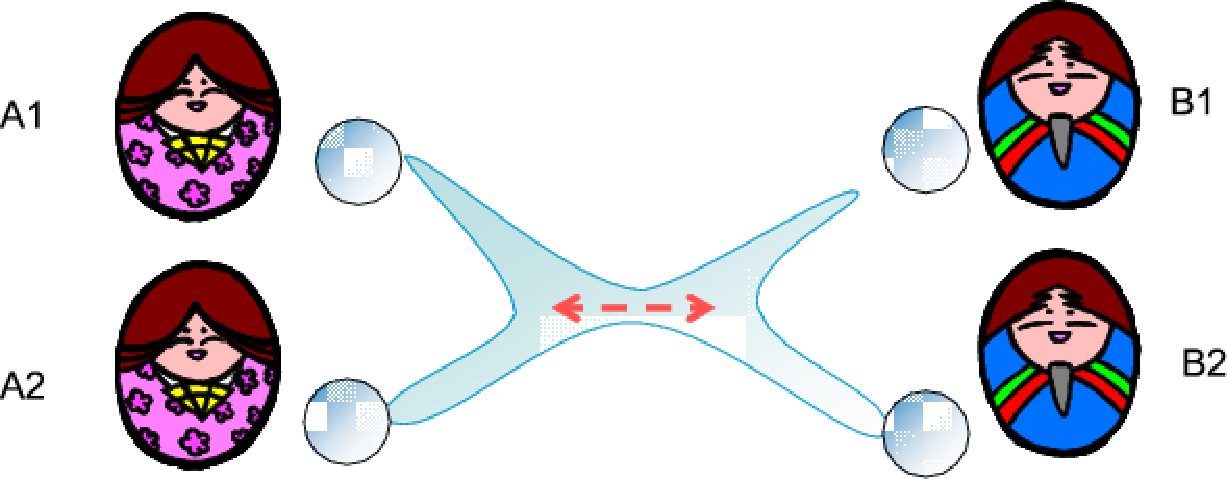}\qquad
\includegraphics[width=7cm]{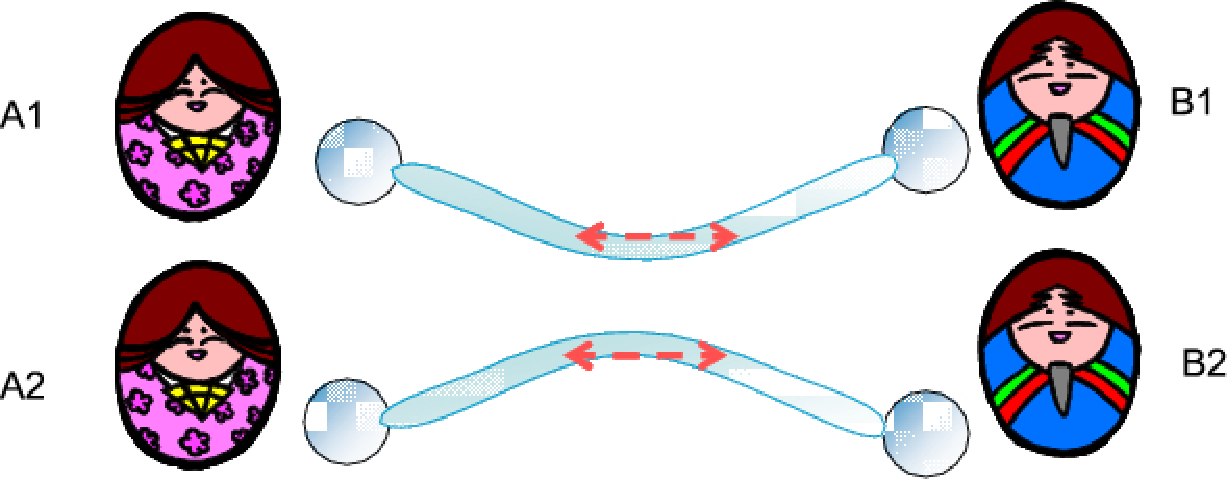}
}\end{center}%
\caption[Strong superadditivity]
{\small 
The left hand side of (\ref{eq:ineq}) corresponds to {\sc left} 
and the right hand side corresponds to {\sc right}.}
\label{fig:ssadd2}
\end{figure}

In this chapter $\HH_{A1}, \HH_{A2}, \HH_{B1}$ and $\HH_{B2}$ are assumed to be 
a two-level systems (four qubits). 
There are several reasons. First of all, this is the lowest-dimensional case. 
Second, there are feasible method to numerically calculate the values appeared in (\ref{eq:ineq}):
\begin{itemize}
\item The left hand side of (\ref{eq:ineq}) is just a von Neumann entropy when 
$\rho$ is a pure state. 
\item Each term of the right hand side of (\ref{eq:ineq}) is 
feasible to calculate by the method of the concurrence \cite{wootters1997,wootters1998}
when the argument is as low as two-dimensional. 
\end{itemize}

Note that  direct calculation of $E_F$ is formidable as it is defined as 
an optimization form on high-dimensional space as it is defined as (\ref{def:eof}). 

Furthermore, there is another reason to check the inequality
rather than to check the additivity. Technically, inequality check is 
easier than equality check. 
You cannot stop concerning about the calculation accuracy 
 when you check the equality numerically, 
whereas you can when you are checking the inequality numerically
unless the equality exactly holds. 


\section{Calculation procedure for the inequality}
In this section, we show the procedure how to numerically verify (\ref{eq:ineq}) 
for a chosen case. This is the base of the experiment executed millions times,
which works for $\HH_{A_1},\HH_{B_1},\HH_{A_2},\HH_{B_2}$ being 2-level systems.

\bigskip 

\begin{center}

\fbox{\begin{minipage}{14cm}\small
(1) Start from a four-partite qubit pure state that is represented as sixteen-dimensional vector
\begin{equation}
\begin{split}
|\Psi _{A_1B_1A_2B_2}\rangle
=\sum _{i,j,k,l=0}^1 \alpha _{ijkl}\, |i\rangle _{A_1}
|j\rangle _{B_1}|k\rangle _{A_2}|l\rangle _{B_2} \qquad \\
=(\alpha_{0000}\,\alpha_{0001}\,
\ldots\,\alpha_{1110}\,\alpha_{1111}\, )^T
\end{split}
\end{equation}
for a randomly chosen 16-tuple of complex numbers $\{\alpha_{ijkl}\}$
with the normalized restriction
$
\sum _{i,j,k,l=0}^1|\alpha _{ijkl} |^2=1.
$
\end{minipage}}
\\$\downarrow$\\
\fbox{\begin{minipage}{14cm}\small
(2) Calculate three reduced density operators,
$\rho _{A_1A_2}$, $\rho _{A_1B_1}$ and $\rho _{A_2B_2}$ from 
$\ketbra{\Psi _{A_1B_1A_2B_2} } $ , as 
\begin{eqnarray}
\rho_{A_1A_2} = \Tr_{B_1B_2} \ketbra{\Psi _{A_1B_1A_2B_2} }, \label{calc:a1a2} \\
\rho_{A_1B_1} = \Tr_{A_2B_2} \ketbra{\Psi _{A_1B_1A_2B_2} }, \label{calc:a1b1} \\
\rho_{A_2B_2} = \Tr_{A_1B_2} \ketbra{\Psi _{A_1B_1A_2B_2} }. \label{calc:a2b2} 
\end{eqnarray}
They are $4\times 4$ matrices.
Assume they are represented as
$\left[\begin{smallmatrix}0&1&2&3\\4&5&6&7\\8&9&A&B\\C&D&E&F\end{smallmatrix}\right]$.
Then each symbol $0,1,\ldots,F$ is the summation of the elements of 
$\rho_{A_2B_2} = \Tr_{A_1B_2} \ketbra{\Psi _{A_1B_1A_2B_2} }$
positioned by the same four letters as follows. 
(\ref{calc:a1a2}),(\ref{calc:a1b1}),(\ref{calc:a2b2})
for \\
{  \tiny $
\left[
\begin{smallmatrix} 
0&.&1&.&.&.&.&.&2&.&3&.&.&.&.&.\\
.&0&.&1&.&.&.&.&.&2&.&3&.&.&.&.\\
4&.&5&.&.&.&.&.&6&.&7&.&.&.&.&.\\
.&4&.&5&.&.&.&.&.&6&.&7&.&.&.&.\\
.&.&.&.&0&.&1&.&.&.&.&.&2&.&3&.\\
.&.&.&.&.&0&.&1&.&.&.&.&.&2&.&3\\
.&.&.&.&4&.&5&.&.&.&.&.&6&.&7&.\\
.&.&.&.&.&4&.&5&.&.&.&.&.&6&.&7\\
8&.&9&.&.&.&.&.&A&.&B&.&.&.&.&.\\
.&8&.&9&.&.&.&.&.&A&.&B&.&.&.&.\\
C&.&D&.&.&.&.&.&E&.&F&.&.&.&.&.\\
.&C&.&D&.&.&.&.&.&E&.&F&.&.&.&.\\
.&.&.&.&8&.&9&.&.&.&.&.&A&.&B&.\\
.&.&.&.&.&8&.&9&.&.&.&.&.&A&.&B\\
.&.&.&.&C&.&D&.&.&.&.&.&E&.&F&.\\
.&.&.&.&.&C&.&D&.&.&.&.&.&E&.&F\\
\end{smallmatrix}
\right]
\!\!,\!\!
\left[
\begin{smallmatrix}
0&.&.&.&1&.&.&.&2&.&.&.&3&.&.&.\\
.&0&.&.&.&1&.&.&.&2&.&.&.&3&.&.\\
.&.&0&.&.&.&1&.&.&.&2&.&.&.&3&.\\
.&.&.&0&.&.&.&1&.&.&.&2&.&.&.&3\\
4&.&.&.&5&.&.&.&6&.&.&.&7&.&.&.\\
.&4&.&.&.&5&.&.&.&6&.&.&.&7&.&.\\
.&.&4&.&.&.&5&.&.&.&6&.&.&.&7&.\\
.&.&.&4&.&.&.&5&.&.&.&6&.&.&.&7\\
8&.&.&.&9&.&.&.&A&.&.&.&B&.&.&.\\
.&8&.&.&.&9&.&.&.&A&.&.&.&B&.&.\\
.&.&8&.&.&.&9&.&.&.&A&.&.&.&B&.\\
.&.&.&8&.&.&.&9&.&.&.&A&.&.&.&B\\
C&.&.&.&D&.&.&.&E&.&.&.&F&.&.&.\\
.&C&.&.&.&D&.&.&.&E&.&.&.&F&.&.\\
.&.&C&.&.&.&D&.&.&.&E&.&.&.&F&.\\
.&.&.&C&.&.&.&D&.&.&.&E&.&.&.&F\\
\end{smallmatrix}
\right]
\!\!,\!\!
\left[
\begin{smallmatrix}
0&1&2&3&.&.&.&.&.&.&.&.&.&.&.&.\\
4&5&6&7&.&.&.&.&.&.&.&.&.&.&.&.\\
8&9&A&B&.&.&.&.&.&.&.&.&.&.&.&.\\
C&D&E&F&.&.&.&.&.&.&.&.&.&.&.&.\\
.&.&.&.&0&1&2&3&.&.&.&.&.&.&.&.\\
.&.&.&.&4&5&6&7&.&.&.&.&.&.&.&.\\
.&.&.&.&8&9&A&B&.&.&.&.&.&.&.&.\\
.&.&.&.&C&D&E&F&.&.&.&.&.&.&.&.\\
.&.&.&.&.&.&.&.&0&1&2&3&.&.&.&.\\
.&.&.&.&.&.&.&.&4&5&6&7&.&.&.&.\\
.&.&.&.&.&.&.&.&8&9&A&B&.&.&.&.\\
.&.&.&.&.&.&.&.&C&D&E&F&.&.&.&.\\
.&.&.&.&.&.&.&.&.&.&.&.&0&1&2&3\\
.&.&.&.&.&.&.&.&.&.&.&.&4&5&6&7\\
.&.&.&.&.&.&.&.&.&.&.&.&8&9&A&B\\
.&.&.&.&.&.&.&.&.&.&.&.&C&D&E&F\\
\end{smallmatrix}
\right]
$},\\ respectively. \label{pagetrout}
\end{minipage}}
\\$\downarrow$\\
\fbox{\begin{minipage}{14cm}\small
(3) Calculate the $E_F$ for a pure state $|\Psi _{A_1B_1A_2B_2}\rangle $, as 
\begin{eqnarray}
E_F(|\Psi _{A_1B_1A_2B_2}\rangle
\langle \Psi _{A_1B_1A_2B_2}|)
=S(\rho _{A_1A_2}),
\end{eqnarray}
where $S(\rho _{A_1A_2})$ is the von Neumann entropy that 
is calculated by the Shannon entropy $ -\sum \lambda_i\log \lambda_i$ with $\{\lambda_i\}_i$ 
the eigenvalues of the reduced density operator $\rho _{A_1A_2}$ that is a $4\times 4$ matrix. 
Thus one need to solve a quartic equation (a polynomial of order four) or utilize 
some mathematical package to calculate the eigenvalues of matrices.
\end{minipage}}
\\$\downarrow$\\
\fbox{\begin{minipage}{14cm}\small
(4) Calculate the 'spin-flip' transformations \cite{wootters1998} of the reduced density operators,
\index{spin-flip transformation}
$\tilde {\rho }_{A_1B_1}$ and $\tilde {\rho }_{A_2B_2}$
by calculating
\begin{eqnarray}
\tilde {\rho }_{A_1B_1}&=&
(\sigma _y\otimes \sigma _y)\rho ^T_{A_1B_1}
(\sigma _y\otimes \sigma _y),
\\
\tilde {\rho }_{A_2B_2}&=&
(\sigma _y\otimes \sigma _y)\rho ^T_{A_2B_2}
(\sigma _y\otimes \sigma _y).
\end{eqnarray}
$\sigma_y = \left(\begin{smallmatrix}0&-\sqrt{-1}\\\sqrt{-1}&0\end{smallmatrix}\right)$ 
is one of the Pauli matrices. Thus 
$\sigma_y \otimes \sigma_y = \left(\begin{smallmatrix}0&0&0&-1\\0&0&1&0\\0&1&0&0\\-1&0&0&0
	\end{smallmatrix}\right)$.
This spin-flip transformation $\rho\mapsto\tilde\rho$ is the operation
like 
$\left[\begin{smallmatrix}0&1&2&3\\4&5&6&7\\8&9&A&B\\C&D&E&F\end{smallmatrix}\right]
\mapsto
\left[\begin{smallmatrix}\ F&-E&-D&\ C\\-B&\ A&\ 9&-8\\-7&\ 6&\ 5&-4\\\ 3&-2&-1&\ 0\end{smallmatrix}\right]$.

\end{minipage}}
\\$\downarrow$\\
\fbox{\begin{minipage}{14cm}\small 
(5) Calculate the concurrence of $\rho _{A_1B_1}$ and \index{concurrence}
$\rho _{A_2B_2}$ by the following calculations:
first calculate the square roots of the eigenvalues of 
$\rho _{A_1B_1}\tilde {\rho }_{A_1B_1}$ and arrange them in
decreasing order $\lambda_1, \lambda_2, \lambda_3, \lambda_4$;
next, calculate the concurrence
\begin{eqnarray}
C(\rho_{A_1B_1} )=\max\{0, \lambda _1-\lambda _2-\lambda _3-\lambda _4\}.
\end{eqnarray}
Do the same for $\rho _{A_2B_2}$, as 
\begin{equation}
C(\rho_{A_2B_2} )= \max\{0, \mu_1-\mu_2-\mu_3-\mu_4\}
\end{equation} 
where $\mu_1,\mu_2,\mu_3,\mu_4$ are the eigenvalues of $\rho _{A_2B_2}\tilde {\rho }_{A_2B_2}$
arranged in decreasing order.
\end{minipage}}
\\$\downarrow$\\
\fbox{\begin{minipage}{14cm}\small 
(6) Calculate $E_F$ for $\rho _{A_1B_1}$ and
$\rho _{A_2B_2}$ by 
\begin{eqnarray}
E_F(\rho_{A_1B_1})= H_2\left(\frac{1-\sqrt{1-C(\rho_{A_1B_1})^2}}{2} \right), \\
E_F(\rho_{A_2B_2})= H_2\left(\frac{1-\sqrt{1-C(\rho_{A_2B_2})^2}}{2} \right),
\end{eqnarray}
where $H_2$ is the binary entropy function. \\
($E_F(\rho) =-\frac{1-\sqrt{1-C(\rho)^2}}{2}\log_2\frac{1-\sqrt{1-C(\rho)^2}}{2} 
-\frac{1+\sqrt{1-C(\rho)^2}}{2}\log_2\frac{1+\sqrt{1-C(\rho)^2}}{2} $)
\end{minipage}}
\\$\downarrow$\\
\fbox{\begin{minipage}{14cm}\small 
(7) Compare the values
\begin{eqnarray}
E_F(\ketbra{\Psi _{A_1B_1A_2B_2} })
 \quad\le^{??}\quad
E_F(\rho _{A_1B_1})+E_F(\rho _{A_2B_2}),
\end{eqnarray}
to check whether the superadditivity of $E_F$ holds or not.
\end{minipage}}

\end{center}

\vskip 2truecm

\section{Procedure to pick up random points and their results}

Three different methods are performed to pick up points, which are then
substituted into the procedure 
described in the section above.

\subsection{Random search}\label{ss:rs}
Sixteen parameters $z_{ijkl}\,(i,j,k,l=0,1)$ are chosen from uniformly
from the square 
whose four vertices are 
 $\pm 1 \pm\sqrt{-1}$  
on the Gaussian plane. Then they are normalized by $\alpha_{ijkl}=
\frac{z_{ijkl}}%
{\sqrt{\sum_{i,j,k,l=0}^1 |z_{ijkl}|^2}} $. These values are substituted 
into $\ket{\Psi_{A_1B_1A_2B_2}} = \sum_{i,j,k,l} \alpha_{i,j,k,l} \ket{i}\ket{j}\ket{k}\ket{l}$. 

\begin{figure}[htbp]
	\begin{center}
		\includegraphics[width=0.7\linewidth]{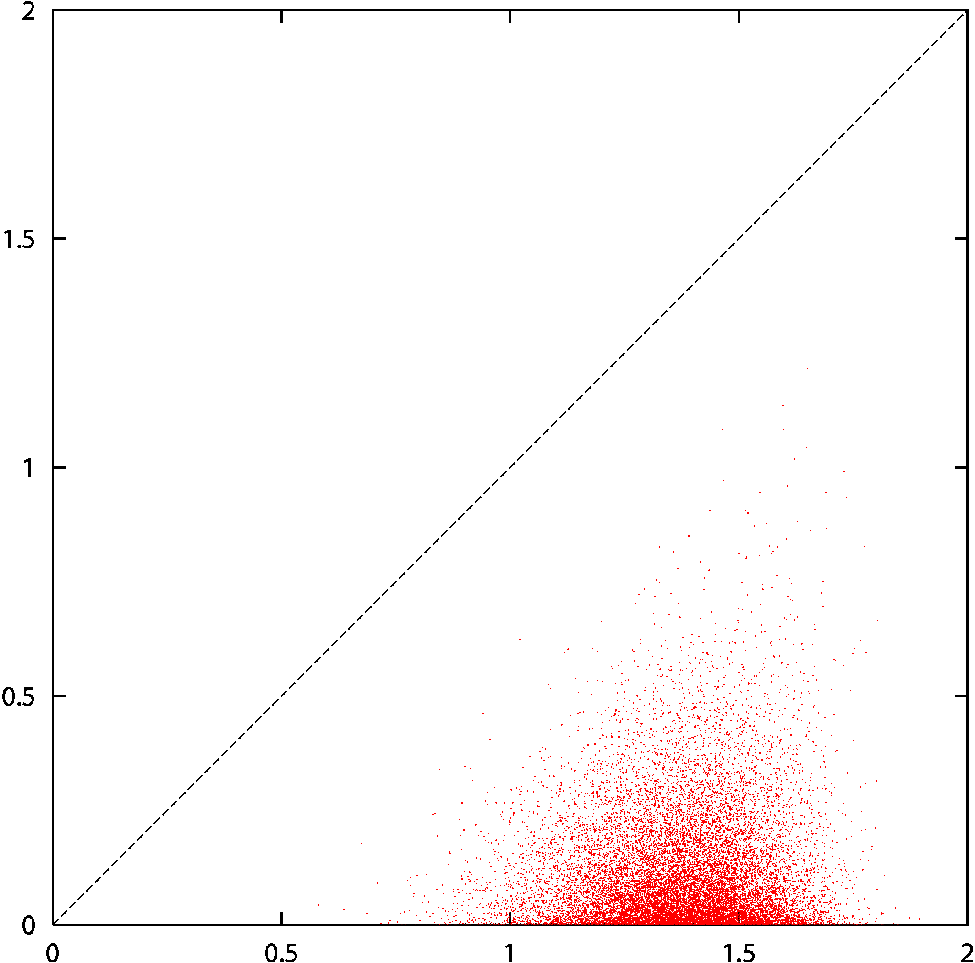}
		\caption
		[Superadditivity plot: { } 10,000 states from $(\mathbb{C}^2\otimes\mathbb{C}^2)^{\otimes 2 }$]%
		{\small Randomly chosen 10,000 states from the whole four-qubits space. 
		There is not  any exception against the inequality, as all the points in the figure satisfy the 
		condition $x \le y$.
		} \label{tenthousand}
	\end{center}
\end{figure}

Fig.~\ref{tenthousand} shows a typical case. Each of the dots is located at 
\begin{equation}
(x , y) = ( S(\rho_{A_1A_2}) ,  E_F(\rho_{A_1B_1})+E_F(\rho_{A_2B_2})),
\end{equation}
 
The result is that there has not been any counterexample for approximately million points.

\subsection{Pseudo zero-neighborhood search}

\begin{center}
\begin{figure}[htbp]
\begin{center}
\includegraphics[width=0.35\linewidth]{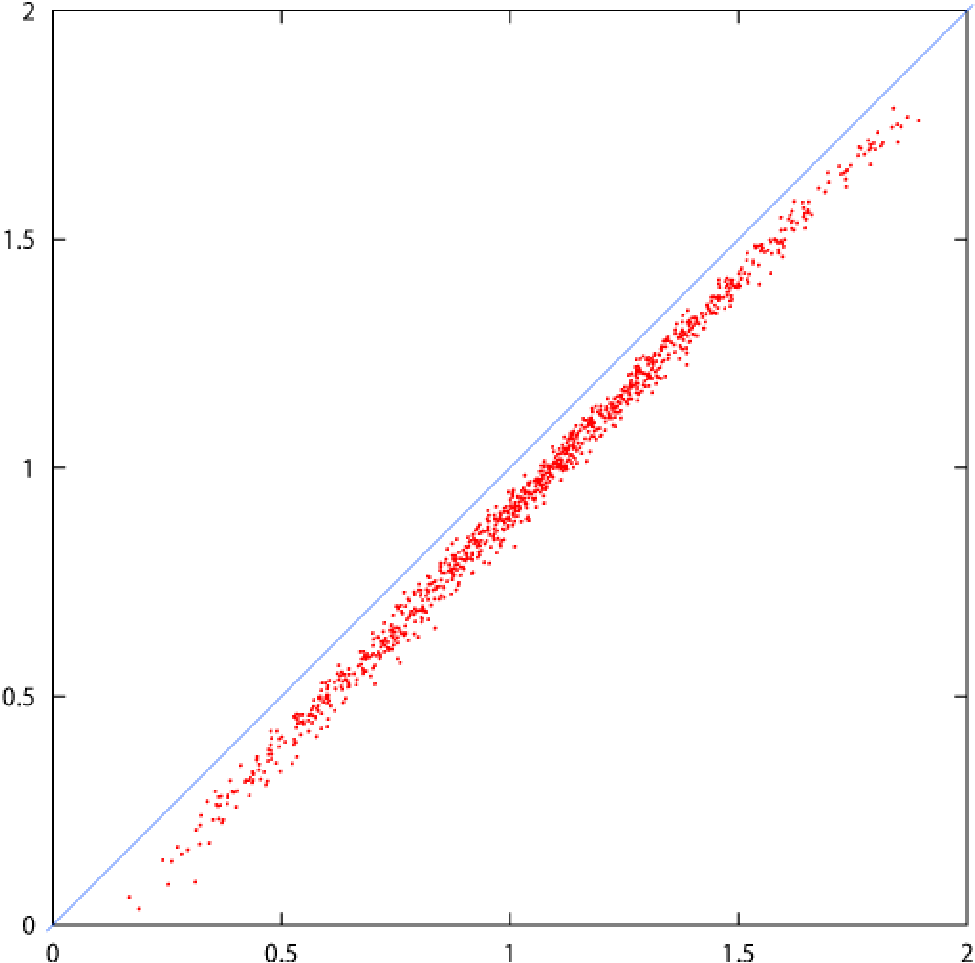} \hspace*{2cm}
\includegraphics[width=0.35\linewidth]{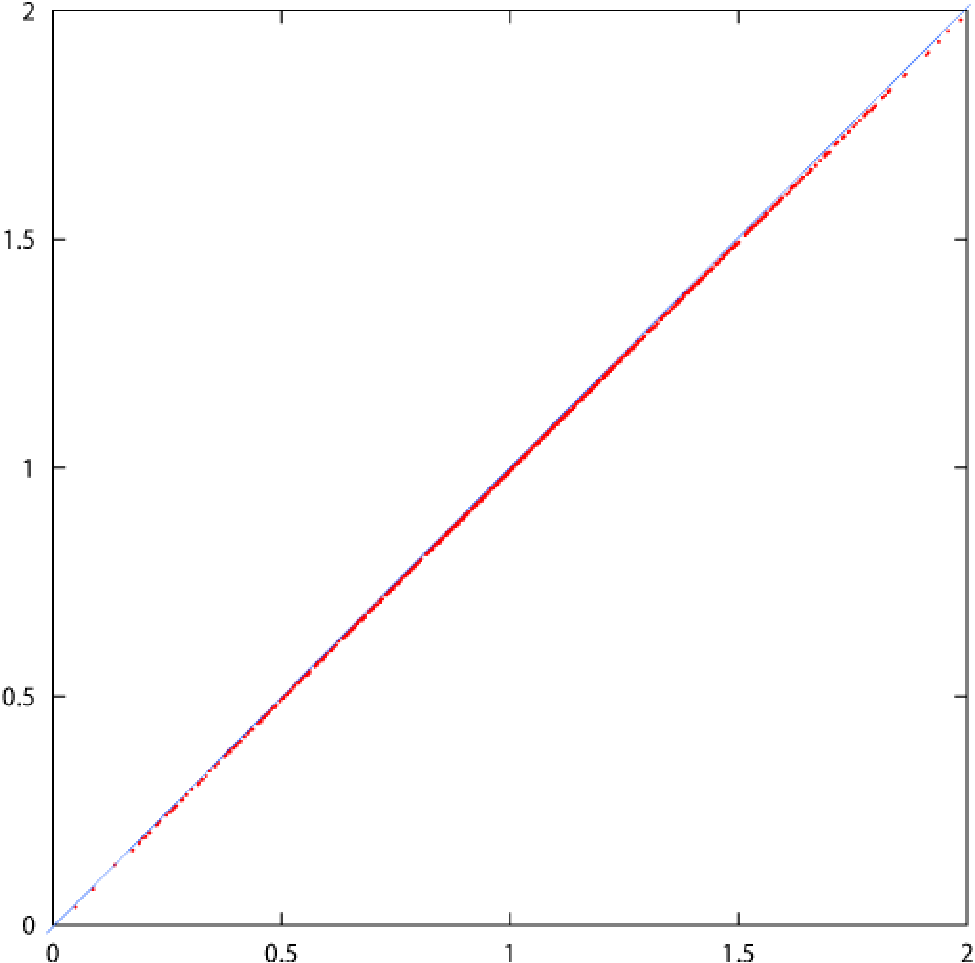}
\end{center}
\caption[Superadditivity plots: {} 1,000 points around the separable states.]{1,000 points are randomly chosen from each of certain neighborhoods of the separable states for each figure. {\sc left: }$\epsilon=0.2$,
{\sc right: }$\epsilon=0.05$. There is also no exception against the inequality,
as all the points in the figures satisfy the condition $x\le y$. }\label{thousand}
\end{figure}
\end{center}

If $\ket{\Psi_{A_1B_1A_2B_2}}$  is a separable state 
with respect to  one side $A_1B_1$ and the other side $A_2B_2$, equality holds 
for the inequality. Thus to find the counterexamples, 
it would be worth 
to check the neighborhood of these
separable states. 
We employed the following method to pick points around the separable states.
\begin{quote}
Set 
\begin{equation}
 \ket{\Psi}= 
 \sum_{i,j}\alpha_{ij}\ket{i}\ket{j} \otimes 
 \sum_{k,l}\alpha_{kl}\ket{k}\ket{l}
 + 
 \epsilon 
 \sum_{i,j,k,l} \alpha_{i,j,k,l} \ket{i}\ket{j}\ket{k}\ket{l}
 \label{def:epsilon}
\end{equation}
with 
where $\alpha_{ij}, \alpha_{kl}$ and $\alpha_{ijkl}$ are chosen by methods similar to the one 
in the previous subsection, and $\epsilon $ is some small scholar like 0.1 or 0.01.
This is normalized as 
\begin{equation}
 \ket{\Psi_{A_1B_1A_2B_2}} = \frac{\ket{\Psi}}{\sqrt{\braket\Psi}}.
\end{equation}
\end{quote}
Part of the results are shown in Fig.\ref{thousand}, for $\epsilon = 0.2 $ and $0.05$ 
with 1,000 points for each. The author has checked for many parameters $\epsilon$, 
but there have not been any of the aforementioned counterexamples. 

\subsection{Minimum search}

\begin{figure}[ht]
\begin{center}
\includegraphics[width=0.59\linewidth]{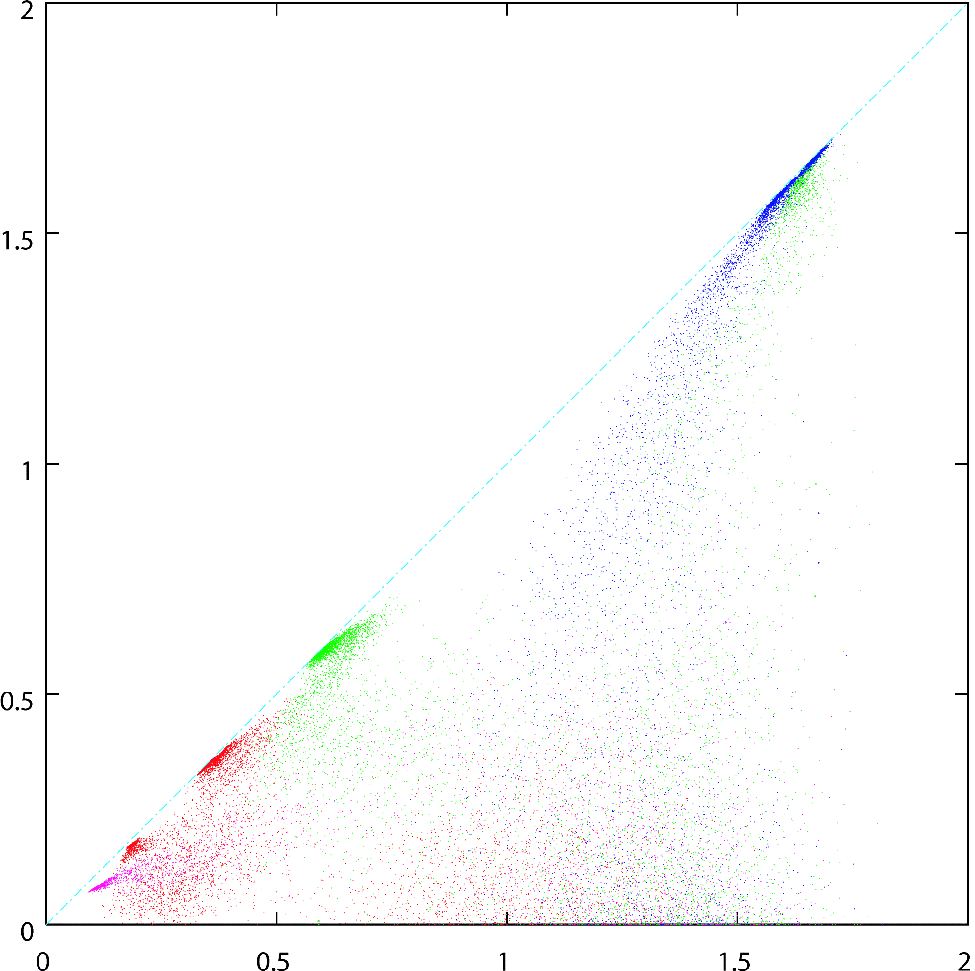}
\end{center}
\caption[Superadditivity plot: { } seeking the minimum]{The trails of seeking the minimum. Seeking the 
minimum of $x-y$ was tried tens of times. Each trial is distinguished by color. Every local minimum was zero. }\label{fig:colorful}
\end{figure}

Seeking the minimum value of  
\begin{equation}\label{terms:diff}
E_F(\rho_{A_1B_1}) +E_F(\rho_{A_2B_2}) - S(\rho_{A_1A_2})  
\end{equation}
 is also performed. 
 The issue is whether the value of (\ref{terms:diff}) can reach below zero. 
 The employed procedure is as follows:
\begin{itemize}
	\item First, pick a random point. 
	\item Next,
	repeat the following step:
	\begin{itemize}
	\item pick hundreds of points in a certain neighborhood of the previous points 
	and 
	\item the point giving the minimum value of (\ref{terms:diff}) is chosen to be the next point,
	\end{itemize}
	until the movement of the points is negligible. 
	The neighborhood shrinks as the steps progress.
\end{itemize}

The result is that there has not been any counterexample for the inequality 
for tens of trials (see Fig.~\ref{fig:colorful}). 
We observed that all of the minimum values 
of these experiments are zero with the convergence in the separable state represented by 
$\ketbra{\Psi_{A_1B_1}}\otimes\ketbra{\Psi_{A_2B_2}}$. 

%
\section{Results and Discussion}
\subsection{The experiments and the result}
More than one million cases  for the random search and tens of the directive search
described above were performed. As a result, no counterexample was been found. 

\subsection{Review of the search ways }

The experiments described in this chapter were a pilot test and there are many points 
to be improved. 

\medskip 

For the random search, sixteen parameters $\{z_{ijkl}\}$ would prefer to be chosen from 
the Gaussian distribution rather than squares of the Gaussian plain, to remove the 
bias on the pure states. 

At first glance, one million points seems almost myriad. The space those 
states reside on is, however, not two- or three- dimensional space.
The space $\ket{\Psi_{A_1B_1A_2B_2}}$ resides on is 
essentially fifteen dimensional in real%
\footnote%
{The pure states form a thirty-one dimensional manifold. Unitary
operations, which is four dimensional, on each four of qubits does 
not effect any difference on the inequality. Thus $31-4\times 4=15$ is the 
virtual dimension.}. 
Since $\sqrt[15]{1000000}\approx 2.5$, 
a figure with the diameter of a third ($\frac{1}{3}$) of the whole space would not 
be detected by a million random detections, in a rough estimation.
 Thus a numerical check performed here must be treated with care; 
 if the region of the counterexamples has a diameter less than a third of the collection 
 of the whole pure states, such a region is not detected.
 
 \medskip
For the (pseudo) zero-neighborhood search, we observed that 
on the plots (see Fig.\ref{thousand}), the points' distribution is 
strongly determined by the parameter $\epsilon$ of (\ref{def:epsilon}), 
as the possible locations of the points  on appearance do not overlap.
This fact may bring us some analytical knowledge. 

\medskip 
It would be better to employ a more refined way to seek the minimum values, 
as there are many methods for  seeking minimum values, though
we saved labor to implement the program. Then we can run a lot of trials 
to seek the minimum value. 
Since the result to seek the minimum value of (\ref{terms:diff})
shows that every local minimum is zero for tens of experiments, 
even if the minimum were less than zero, the minimum is hard to reach. 

\medskip
The experiments described above might some hints for the future analytical 
proof about whether the inequality holds or not. 

\subsection{For higher dimension}

The experiment performed above is for the lowest dimension that is two
for each of the four systems. This results in  turns to be the substituted pure state
being sixteen dimension ($2^4$). What would happen in higher dimension?
Even for the next lowest dimensional case, we do not know a feasible way
to calculate the entanglement of the formation, contrary to the case of 
two dimension. That makes the problem difficult.

\section*{Endnotes}
	\theendnotes
\clearpage
\section*{Appendix: Maple 6 program}
	The following is one of the  programs of Maple 6 used for the random search 
	(Subsection. \ref{ss:rs}).%
	\begin{center}
		\includegraphics[width=121mm]{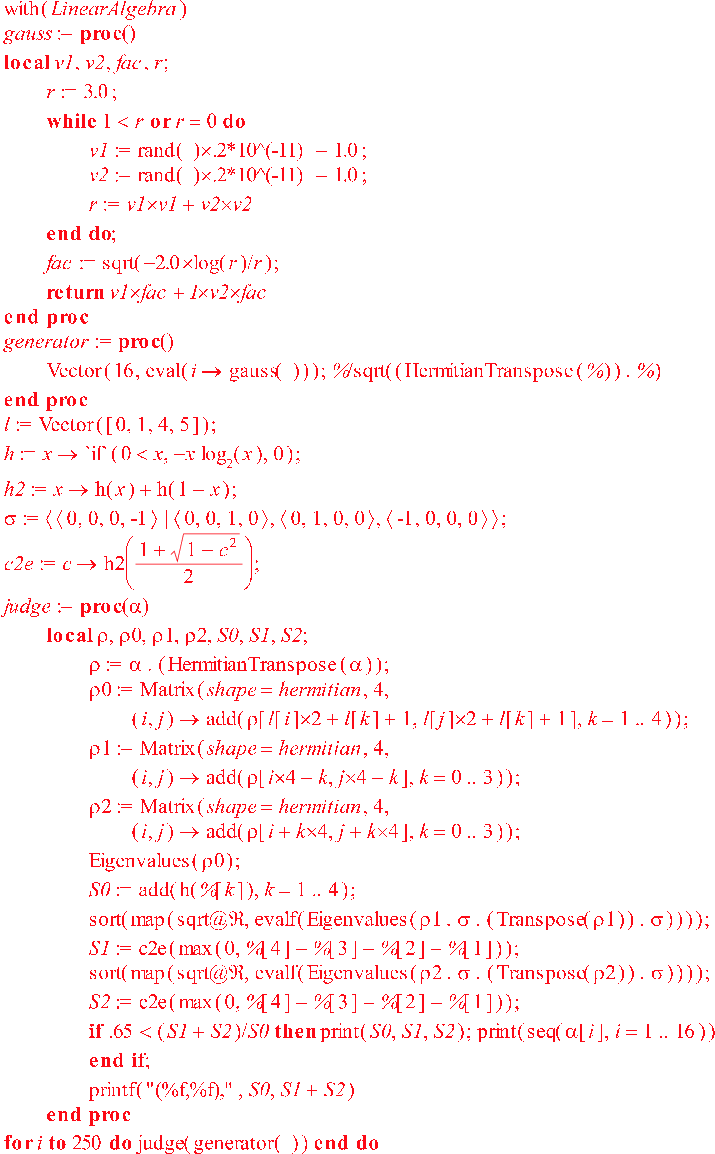}
	\end{center}
	The output of this program may be as follows: \begin{quote}
{\tt \small (1.423552,.077738), (1.241545,.057057), (1.227051,.029807), (1.301131,.075570), (1.366521,.004438), (1.235635,.175308), (1.446036,.026785), (1.299543,.082059), (1.480392,.178842), (1.010854,.224507), (1.515526,.209590), (1.424525,.195779), (1.478113,.011087), (1.161710,.040721), (1.416082,.311309), (1.120448,.072590), (1.528262,.000081), $\ldots$ }\end{quote}
\clearpage
\begin{minipage}{\linewidth}
\vspace*{7cm}
\part{Calculating the~Holevo~Capacity%
}
\end{minipage}
{
\begin{minipage}{\linewidth}
\vspace*{5cm}
\begin{center}
\includegraphics[width=12cm]{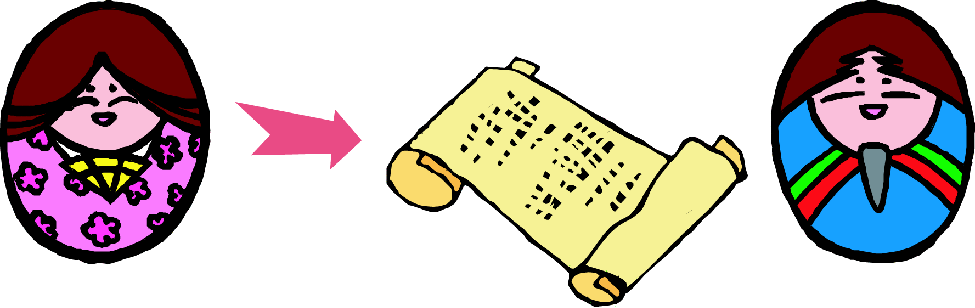}
\index{Alice and Bob}
\end{center}
\end{minipage}
}

\index{w.r.t.|see{contraction of ``{\it with respect to}''}}
\chapter{Calculating the Holevo Capacity}
\makeendnotes

The Holevo capacity, the classical information communication efficiency of a quantum channel, was
first formalized as the Holevo quantity in 1973 in Russia. 
The calculation method of the Holevo capacity of an arbitrary qubit channel 
with the guarantee of computation convergence 
was performed for the first time in 2004, which is presented in this chapter.

\medskip
This chapter is a derivation of \cite{HIMRS04} and \cite{OIIS04}.
%
%
\section{Significance in calculating the Holevo capacity} 
\subsection{The Definition of Holevo capacity} 
 Here we review concepts of the Holevo capacity of quantum channels.
 
 A {\it quantum channel} is a CPTP mapping from its input space 
 $\BHi$
 to its output space $\BHo$. 
 (One may supplement the 
 explanation, as this quantum channel is {\it memoryless}.%
 ) %
  This chapter mainly 
 deals with qubit channels, thus both their input and output spaces are 2-dimensional and
 each channel is characterized by its output ellipsoid presented as Fig.~\ref{figellipsoids}, 
 {\sc Upper Left}.

\medskip 
 The Holevo capacity  for a quantum 
 channel $\T:\BHi\to\BHo$ can be represented in two ways:

\medskip
\begin{enumerate}[\hspace*{8mm}1. ] 
	\item
		One way is \cite{Holevo73,Holevo77,SW97}
		\begin{equation}
			\begin{split}
				C(\T) = &
				\max_{n ;\, \rho_1, \dots, \rho_n ;\, p_1, \dots, p_n } 
				S(\sum_{i=1}^n p_i \, \T \rho_i) - \sum_{i=1}^n p_i \,  S( \T \rho_i) , 
			\\
				&\text{
				subject to $ n \in \mathbb{N}, p_i > 0 , \sum_{i=1}^n p_i = 1 $,
				$\rho_i  \in \BHi \,(i=1,\ldots,n) $, 
				} 		
			\end{split}
		\label{fm:eq}
		\end{equation}
		where $S(\cdot)$ is  the von Neumann entropy.
		We call each element $\rho_i$ of the optimized $(\rho_i)$ an 
		{\bf engaging input} of $\T$,
		and the number $n$ the {\bf engaging number}. 
		\index{engaging!--- input}\index{input!engaging ---}%
		\index{engaging!--- number}\index{number!engaging ---}%
		The engaging inputs are the input states which are used 
		to send messages through the quantum channel 
		when the communicating efficiency achieves the Holevo capacity. 
		
		Note that (\ref{fm:eq}) is the maximization form over
		%
		{\it the average entropy of the outputs} minus {\it the entropy of 
		the average of the outputs.}
		Therefore, it is interpreted as, 
			the maximum vertical distance between the following two surfaces: 
			\index{vertical distance}\index{distance!vertical ---}
			i) the epigraph of the von Neumann entropy function of which 
			the domain is constrained to
			the output image of the given quantum channel, and
\def\LL{{\sc Lower Left} figure}
\begin{figure}[p]
 	\fbox{   
 	\begin{minipage}{\linewidth}
		\begin{minipage}{0.95\linewidth}
			\begin{minipage}{4.5cm}
				\vspace*{3mm}
				\flushright{
					\includegraphics[height=4cm]{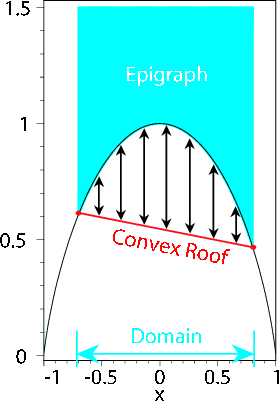}\quad\qquad						}
				\vspace*{-2mm}
			\end{minipage}\quad
			\begin{minipage}{105mm}
				\normalsize
				Given a function, its epigraph is the set of points 
				on and above the surface of its graph. The convex roof 
				is the function determined by the convex hull of the epigraph.
				The vertical distances are indicated by the arrows in the left
				figure.\\
			\end{minipage}
			\caption
			[Vertical distances between epigraph and convex roof]
			{The vertical distances between the epigraph and the convex roof}
			\label{fig:vdistance}
		\end{minipage}
	\end{minipage}
	}
%
%
	\vspace*{6mm}	
	
	\fbox{
	\begin{minipage}{\linewidth}
	\begin{minipage}{\linewidth}
		\begin{minipage}{\linewidth}
 		  \begin{minipage}{8cm}
			\begin{center}
				\includegraphics[width=8cm]{lemon.eps}\\
				\includegraphics[width=8cm]{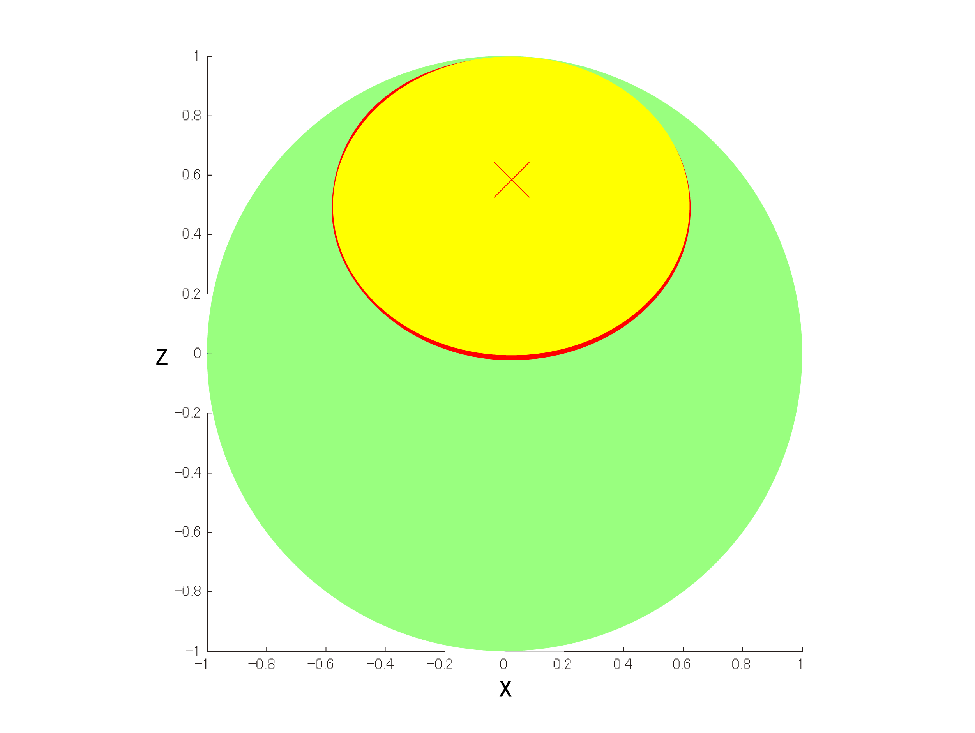}
			\end{center}
		  \end{minipage}
		  \begin{minipage}{7.5cm}
		  		\vspace*{1cm}
				\footnotesize
				{\sc \hspace*{-6mm}\underline{Upper Left}:}\\
					The output ellipsoid  inside the Bloch sphere\\
					representing a qubit channel of
					$$
						\Tq: 
						\rho(x,y,z) 
						\mapsto 
						\rho(0.6x+0.021,0.601y, 0.5z+0.495).
					$$
				\vspace*{0mm}
				{\sc \hspace*{-6mm}\underline{Lower Left}:}
						$xz$-section of the \LL \\
					\begin{tabular}{rl}
						Green circle: & the Bloch sphere
						\\
						Red oval:  &
						the minimum enclosing sphere
						\\
						& w.r.t.~the quantum divergence 
						\\
						Yellow ellipse: &
						the output ellipsoid
						\\
						Red cross: &
						the center of the red ``circle''
						\\
						& w.r.t.~the quantum divergence 
					\end{tabular}

			In the \LL, the Bloch sphere (green) and the output ellipsoid 
			 (yellow) are a sphere and an ellipsoid, respectively, w.r.t.~the Euclidean metric. 
			 The enclosing sphere is, however,  like an ellipsoid and is subtly distorted, 
			 even though it is a ``sphere'' w.r.t.~the quantum divergence.
				
				\hspace*{-8mm} 
				\begin{picture}(120,0)
					\put(0,0){\line(1,0){240}}
				\end{picture}
				\\ 
				The color of the surface of the output ellipsoid in the 
				{\sc Upper Left} figure corresponds to the quantum divergence 
				$\qdv{\rho'}{\sigma'}$  
				with the following color:\\
				\includegraphics[width=8cm]{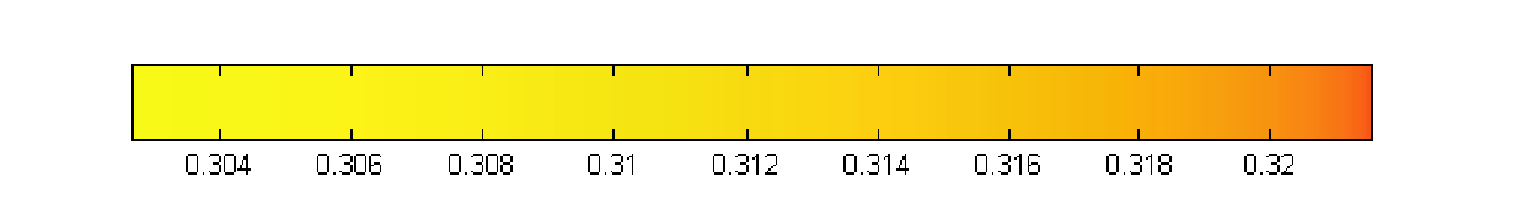}
				\\
				Here $\sigma'$ is the center of the minimum enclosing sphere 
				 (black cross of the \LL) and  $\rho'$ is substituted 
				 with 
				 points on the output ellipsoid. 
				Therefore, the most reddish points are the {\it engaging} 
				outputs.
				\vspace*{2mm}
			\end{minipage}
		\end{minipage}%
		\caption%
			[The Holevo capacity as the minimum radius]%
			{The Holevo capacity as the quantum divergence radius 
			 of the minimum enclosing sphere } 
		\label{figellipsoids}
	\end{minipage}
 	\end{minipage}	}
\end{figure}

			ii) the graph of the convex roof of this function.\index{convex!--- roof}
			See Fig.~\ref{fig:vdistance} for a conceptual illustration.	
		\medskip

	\item
		The other way is \cite{ohya-petz-watanabe}
		\begin{equation}
			C(\T) = \min_{\sigma\in\BHi} \max_{\rho\in\BHi} H( \T \rho \,|| \,\T \sigma)
			\label{fm:opw}
		\end{equation}
		where 
		$H(\cdot||\cdot)$ is the quantum divergence. 
		It is
		{\it the radius of the minimum enclosing  sphere
		containing  the output ellipsoid} 
		when the radius is measured by the quantum divergence.
		(See Prop.~\ref{pp:pdis} to refer to the pseudo distance properties of the quantum divergence.) 
\end{enumerate}

The physical meaning of the Holevo capacity is as follows: it is 
{\it a classical information capacity%
\footnote{%
	Classical information capacity:
	 it is {classical}, the antonym of {quantum}. 
	The Holevo capacity measures the efficiency of 
	transmitting the sequences of the alphabets of some message from the sender to the 
	receiver through a given quantum channel,
	while the quantum capacity may be defined as the 
	efficiency of transmitting the input quantum states themselves with high fidelity through 
	the channel.
	\begin{center}
		\noindent\includegraphics[width=7cm]{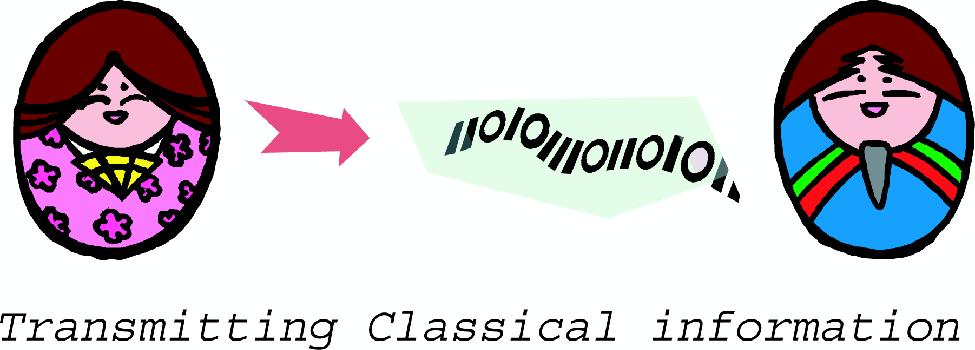}\qquad
		\noindent\includegraphics[width=7cm]{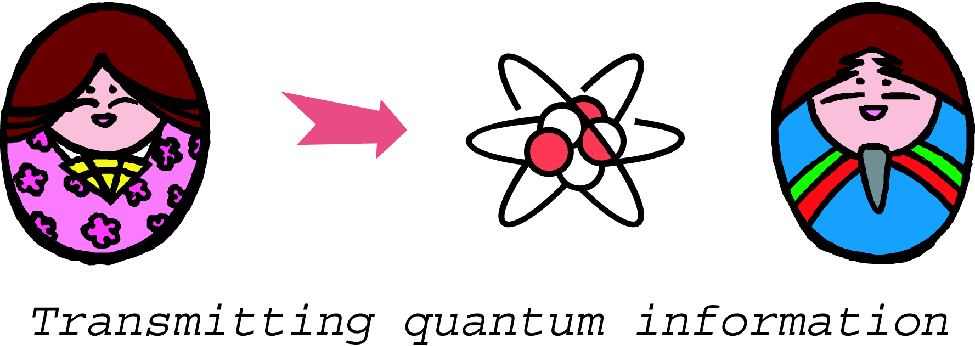}
		\index{Alice and Bob}
	\end{center}
}%
 of a given quantum channel}
 at which input particles are not allowed 
entangled with each other, and the output particles are measured collectively. 

To fully utilize quantum aspects of the quantum channel, one may consider 
the capacity at which the input particles are freely entangled with each other
to send messages 
in order to take advantage of the communication efficiency.
This capacity is called the {\it full capacity}, and it is represented as
\begin{equation}
	\bar{C}(\T)
	 =
	  \lim_{N\to\infty} 
	  \frac{C(\T^{\otimes N} )}{N} .
	  \label{obj:sptpst}
\end{equation}

\subsection{Why do we calculate the Holevo capacity?} 

For historical background, see Subsection~\ref{ss:ccbcp}.

 What if the Holevo capacity $C(\T)$ can be calculated?  The benefits are as follows: 

\begin{itemize}
	\item 
		The actual benefit would occur when the full capacity 
		$\bar{C}(\T)=\lim_{N\to\infty}C(\T^{\otimes N})/N$ 
		is calculated for arbitrary  channels.
		(The full capacity provides the theoretical upper-bound of the channel capacity in the 
		 age of quantum information commencing tens of  years from now, when each light particle, {\it i.e.} photon, carries 
		 a significant amount of information.)
		 Therefore, it is very desirable that 
		 $C(\T^{\otimes N})$ for $N=1,2,\ldots,\infty$ 
		 can be calculated. 	 
	\item
		 As a matter of theoretical contribution, theorists like to verify whether the additivity holds for 
		the Holevo capacity, that is, whether $C(\T_1\otimes \T_2) = C(\T_1) + C(\T_2) $ holds for arbitrary channels 
		$\T_1$ and $\T_2$. 
		(The author is interested in finding circumstantial evidence as to whether this additivity holds 
		or not. A piece of evidence which numerically suggests the additivity appears in Subsection 	\ref{ss:addcc}.)
\end{itemize}

\section{Algorithm to calculate the Holevo capacity}
\subsection{Difficulty in optimization}

\begin{figure}[bbhhtp]
	\fbox{
		\begin{minipage}{\linewidth}\vspace*{-8mm}
			\begin{center}
				\begin{minipage}{7cm}
					\begin{center}
                        \vspace*{13.33mm}
						\includegraphics[width=8cm,clip]{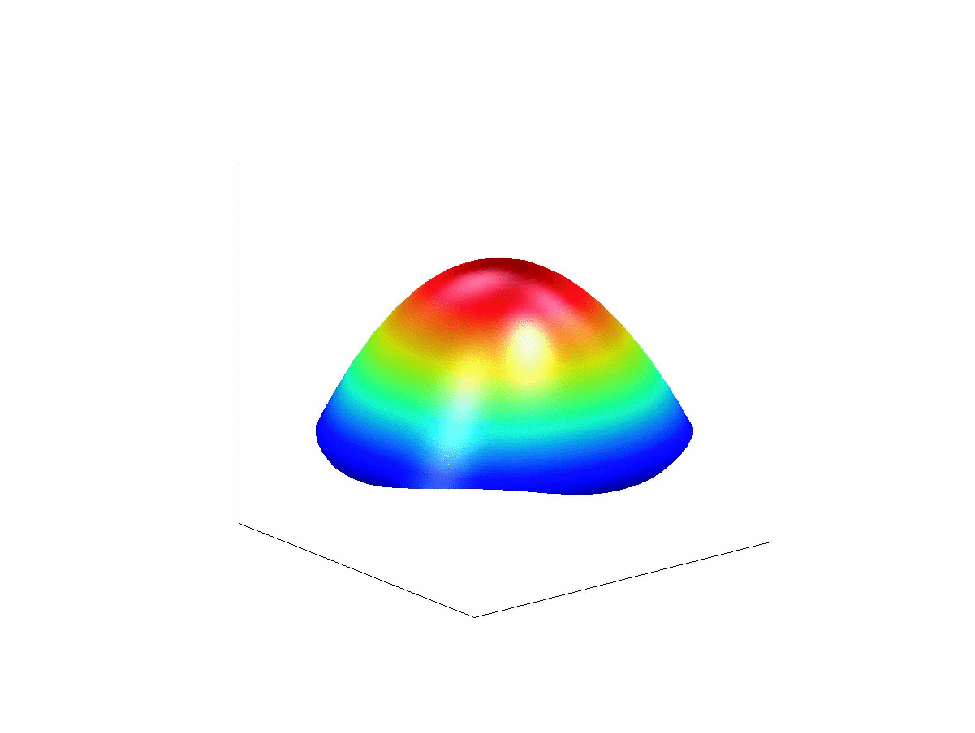}\\\vspace*{-0.7cm}
						{\footnotesize a concave~function~with~a convex~search~space}
					\end{center}
				\end{minipage}
				\begin{minipage}{7cm}
					\begin{center}
                        \vspace*{9.95mm}
						\includegraphics[width=8cm,clip]{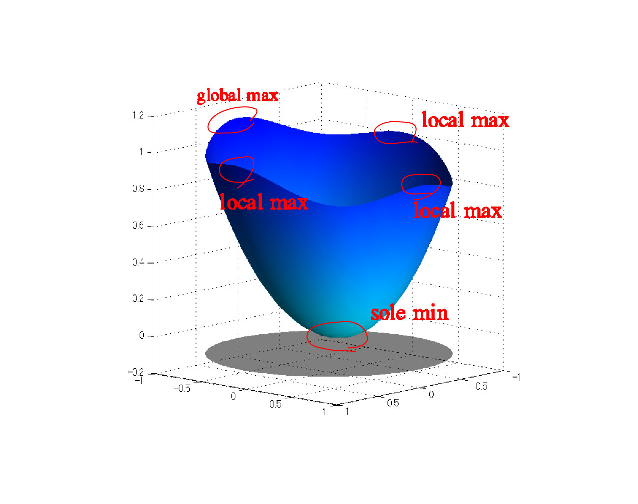}\\\vspace*{-1cm}
						{\footnotesize a convex~function~with~a convex~search~space}
					\end{center}
					\end{minipage}
			\end{center}
			\caption[Extrema of a convex/concave function with a concave domain]%
			{
			\small
			{\sc Right:} Calculating the maximum of a concave function with a convex domain is feasible 
			because the solution is obtained by simple gradual ascending or {\it hill climbing}.
			{\sc Left:} Calculating the maximum of a convex function 
			is, however, rather difficult.
			The maximum cannot be generally reached by gradual ascending 
			because there could be multiple local maximums, which
			appear at the boundary of the search space.
			}
			\label{fig:opts}
		\end{minipage}
		}
\end{figure}
\newcommand{\ssp}{\vspace*{-2mm}}
The objective function \ssp
\begin{equation}
	\objhsw \label{objhsw}
\end{equation}\ssp
 to be maximized has the following two properties: 
\begin{enumerate}[\hspace*{8mm}-- 1.]
	\item \underline{Convex} w.r.t.~$(\rho_i)_i$, the set of input states. 
	\ssp
	\begin{quote}
		Due to the joint convexity \cite{ohya-petz}, that is, \index{joint convexity}\index{convex!joint ---ity}
		\begin{equation}
			\begin{split}
				&
				\lambda \qdv{\rho}{\sigma} + (1-\lambda) \qdv{\rho'}{\sigma'} 
				\ge \qdv{\>\lambda \rho+ (1-\lambda)\rho'\>}{\>\lambda \sigma+ (1-\lambda)\sigma'\>}
			\end{split}
		\end{equation}
		for $0\le\lambda\le1$,
		the quantum divergence $H(\rho|| \sigma)$ is convex w.r.t.~$\rho	$. 
		Note that the objective function (\ref{objhsw}) is rewritten as 
		\begin{equation}
		 \sum_i
		  p_i \, H(\,{\T \rho_i}\,||\,{\sum_j p_j\, \T\rho_j} \,).
		\end{equation}
		Since $H(\rho||\sigma) \;(=\Tr(\rho\log\rho -\rho\log\sigma) \;)= -S(\rho)-\Tr \rho\log\sigma$, 
		as this first term is convex and the second term is linear w.r.t.~$\rho$, 
		the objective function
		is convex w.r.t.~$(\rho_i)_i$. 
	\end{quote}
	\ssp
	\item \underline{Concave} w.r.t.~$(p_i)_i$, the probability distribution.
	\ssp
	\begin{quote}
		The first term $S(\sum_{i=1}^n p_i \,\T \rho_i)$ is concave w.r.t.~$(p_i)_i$
		due to the convexity of the von Neumann function $S(\cdot)$ itself. 
		The second term $\sum_{i=1}^n p_i \,S(\T \rho_i)$ is linear w.r.t.~$(p_i)_i$.
	\end{quote}
	\ssp
\end{enumerate}

In such a function (\ref{obj:sptpst}), it is hard to identify the function's global maximum. 
When one employs a method of gradual ascending or {\it hill climbing}, there is no way to guarantee 
whether the obtained solution is the global maximum (see Fig.~\ref{fig:opts}).

Because of the convexity of the objective function (\ref{obj:sptpst})
w.r.t.~$(\rho_i)_i$, 
each $\rho_i$ of the solution of $(\rho_i)_i$ appears 
at the {\bf boundary} of $\BHi$. \index{boundary}
Hereafter, we denote the bound of $\BHi$ as 
\underline{$\partial\BHi$}. 

%
\subsection{Solution --- Computation with lattices}

The issue is ascribable to the difficulty in guaranteeing the calculation error.
In order to guarantee the convergence in calculation, 
the following method is effective, 
 as it simplifies the problem of maximizing a complicated function 
 into one  maximizing a concave function on a convex search space.

\medskip
\fbox{ 
	\begin{minipage}{0.92\textwidth}
		Solve the following {\bf approximated problem}:
		\vspace*{-3mm}
		\begin{equation}
		  \begin{split}
				c_\ell^\T =	\max_{ \{p_i\}_{i=1}^n } 
				&
				S(\sum_{i=1}^n p_i \, \T \rho_i) - \sum_{i=1}^n p_i \,  S( \T \rho_i) , 
			\\
				& \text{ subject to $ \{p_i\}_{i=1}^n $ being a probability distribution} , 
			\\
				& \text{ where $ \ell = \{\rho_i\}_{i=1}^n $ is a set of fixed points 
				covering $\partial\BHo$}
			\\
				&\quad \text{with a specified (small) coarseness $\delta$.}
		  \end{split}
		  \label{obj:hsw-m}
		\end{equation}
	\end{minipage}
}
\medskip

\begin{figure}[hhttt]
	\fbox{
	\begin{minipage}{\linewidth}	
		\begin{minipage}{8cm}
			\begin{center}
				\includegraphics[width=8cm]{budha20.eps}\\
				\footnotesize
				{\sc Left:} A lattice $\ell_{20}$ on the Bloch sphere
			\end{center}
		\end{minipage}
		\begin{minipage}{8cm}
			\begin{center}
				\includegraphics[height=8cm]{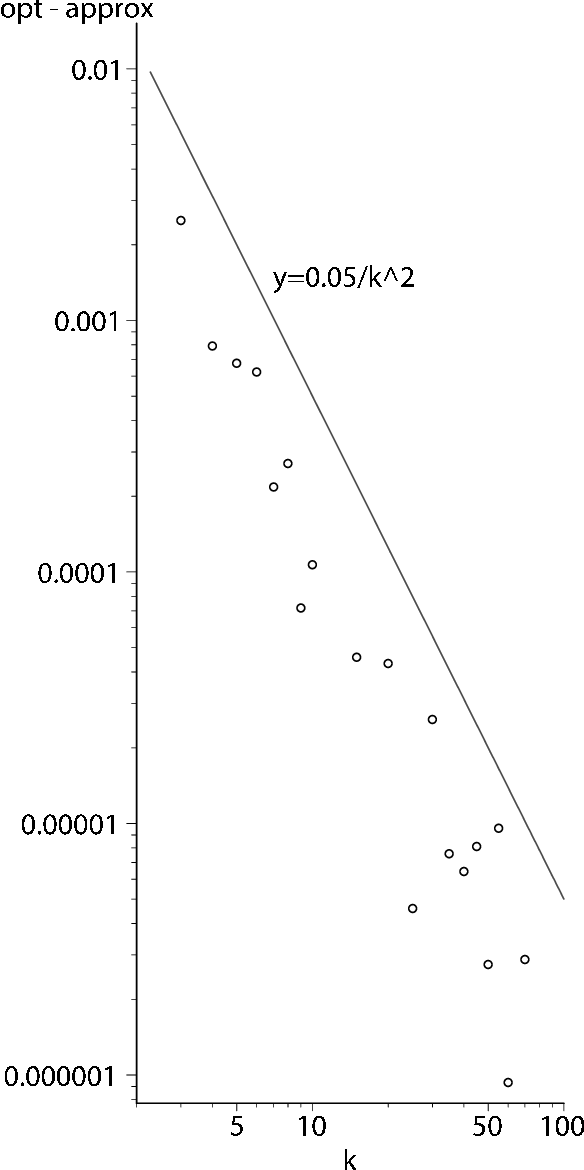}\\
				\footnotesize
				{\sc Left:} How the calculated values converges
			\end{center}
		\end{minipage}
		\caption
		[A covering of $\partial\BHo$]
		{{\sc Left:} A lattice covers $\partial\BHo$. 
		To solve the maximization problem 
		(\ref{obj:hsw-m}), 
		$\{\rho_i\}_i$ is specified beforehand, such as $\ell_{20}$. 
		(Other lattices such as $\ell_{40}$ and $\ell_{100}$ appear in Fig.~\ref{fig:1bscm}.)
		
		Because of
		$
		\big<
		  \begin{smallmatrix}
			\text{\small the concavity w.r.t.~$(p_i)_i$ of the objective function,}\\ 
			\text{\small and the convexity of the search space $(p_i)_i$ }
		  \end{smallmatrix}
		\big> 
		$,
		it is feasible to maximize the objective function.
		One can get a value close to the actual Holevo capacity, 
		when the coarseness on the lattice is small enough.
		\\
		{\sc Right:} The actual convergence for an example channel. 
		The horizontal axis is a log plot of $k$, and the vertical axis is a log plot 
		representing the 
		difference between the calculated value and the optimum value. A line $y=0.05/k^2$ is drawn for reference.
		}
		\label{figlatconv}
	\end{minipage}
	}
\end{figure}

Here the covering $\partial \BHo$ is a set of points, for example, 
the lattice shown in Fig.~\ref{figlatconv}, {\sc Left}.
The {\bf coarseness} $\delta$ is defined as \index{coarseness}
the maximum distance from arbitrary points on the sphere 
to its nearest vertices among $\{\rho_i\}_{i=1}^n$.

We employ the following {\bf lattice}, and we denote it \underline{$\ell_k$} for a positive integer $k$:
\medskip

\fbox{
	\begin{minipage}{0.92\textwidth}
		The lattice $\ell_k$ is defined as,
		\begin{quote} \vspace*{-3mm}
			$\{\rho_i\}_{i=1}^n$ being the set of vertices of the {lattice} which divides 
			both the equator and the meridian of the Bloch sphere, \index{lattice}
			as it is a set of 
			\end{quote}
		\begin{equation}
			\left\{
				 \left(
				 	\;
					 \sin \f{u\pi}{k} \cos \f{2v \pi}{k} \; , 
				 	\;
					 \sin \f{u\pi}{k} \sin \f{2v \pi}{k} \; , 
				 	\;
					 \cos \f{u\pi}{k} 
				 	\;
				 \right)
			\right\}%
			_{
			   \left\{
				   \begin{smallmatrix}
					   {{u=0,\ldots,k\phantom{-1}}}
					   \\
					   { v=0,\ldots,k-1}
				   \end{smallmatrix}
			   \right. 
			 }
		\end{equation}
		\begin{quote} \vspace*{-3mm}
			of which the total number is $n=k^2-k+2$.
		\end{quote}
	\end{minipage}
}

\vspace*{8mm}

One may be overwhelmed by the number of the elements of $\ell_k$
because solving the $(k^2-k+2)$-dimensional problems is required here.
There is, however, a way to solve this part of  problem.
The calculation of (\ref{obj:hsw-m}) is performed by the mathematical programming package 
NUOPT 
developed by  Mathematical Inc.\cite{nuopt}.
NUOPT is capable of solving 
large optimization problems
with an interior point method.	
The computation time is
approximately one hour for $n=1562\ (k=40)$ and 
approximately one day for $n=4831\ (k=70)$
on a Sun Microsystems workstation with a CPU UltraSPARC-II 360MHz and 
2048 Megabytes of RAM. It seems to take time proportional to $n^3$,
or $n$ to the power of three.

As  $k$ becomes larger, the finer the lattice becomes, 
and the smaller the coarseness $\delta$ becomes
due to $\delta = O(k^{-1})$.
Fig.~\ref{figlatconv}, {\sc Right} shows how each solution of the approximated problems 
(\ref{obj:hsw-m}) converges to the Holevo capacity of the qubit channel 
\begin{equation}
	\Tq: \rho(x,y,z) \mapsto \rho(0.6x+0.021,0.601y, 0.5z+0.495), 
	\label{ch4engaging}
\end{equation}
which is a distinct one,
as the author calculated the Holevo capacity with quite high precision,
of which the detail is described in Section~\ref{s:application}. 

\subsection{Error analysis}


The following theorem is significant in that 
it guarantees the convergence of the approximated solution to the actual 
Holevo capacity, {\it i.e.} $c_{\ell_k}^\T \to C(\T)$ as $k\to \infty$.
\begin{Theorem}
	 The error, or the difference, between the Holevo capacity $C(\T)$ and 
	 the solution of the approximated solution of (\ref{obj:hsw-m})
	 is upper-bounded by $O(-\delta^2 \log\delta)$,
	  which is equal to $O(k^{-2} \log k)$ when the lattice is $\ell_k$.
	Particularly, when the output ellipsoid does not touch the surface of 
	the Bloch sphere, it is bounded by $O(-\delta^2)$, which is  $O(k^{-2})$
	for $\ell_k$.

	\begin{proof}
	

		Denote $c_k$ as the solution of the approximated problem (\ref{obj:hsw-m})
		with the lattice $\ell_k$,
		and $c_\infty$ as the Holevo capacity $C(\T)$. Then the following conditions hold: 
		\begin{eqnarray}
			c_k      &\le& c_\infty 
			\label{eq:ck}
			\\
			c_k      &=  & \max \{ \qdv{\T \rho}{\T\sigma_k} ; \, \rho \in \ell_k \}
			\label{ineq:ck}
			\\
			c_\infty &\le& \max \{ \qdv{\T \rho}{\T\sigma_k} ; \, \rho \in \partial\BHi \}
			\label{ineq:ci}
		\end{eqnarray}
		where $\sigma_k = \sum_i p_i \rho_i$, the average of calculated engaging inputs
		of (\ref{obj:hsw-m}) with $\ell_k$.
		Denote 
		\begin{equation}
			f_k(\rho)=\qdv{\T\rho}{\T\sigma_k} \quad \text{ for }\quad  \rho \in \partial\BHi. 
		\end{equation}
	  	The above equality and inequalities (\ref{eq:ck}), (\ref{ineq:ck}) and (\ref{ineq:ci}) lead to
		\begin{equation}
			| c_\infty -c_k | 
			\le 
			\max_{ \rho \in \partial\BHi } f_k(\rho)
			-
			\max_{ \rho \in \ell_k       } f_k(\rho) 
			.
			\label{inq:maxmax}
		\end{equation}

		Note that $\ell_k \subset \partial\BHi$ and the coarseness $\delta$ 
		is the upper bound of the distance from any points on $\partial\BHi$
		to the nearest point among  $\ell_k$. 
		Therefore, estimating RHS of (\ref{inq:maxmax}) is reduced to:
		\begin{equation} 
			\text{
				the infinitesimal analysis of $f_k(\cdot)$ with 
				the domain $\partial\BHi$ around the maximum.
			}
		\end{equation}
		
		We need to consider the following two cases:  
		
		\begin{enumerate}[\hspace*{8mm}1.]
			\item 
				When the maximum is given at $\rho_0$ where $\T\rho_0\in\BHo \backslash \partial\BHo$,
				$f(\cdot)$ is as doubly continuously differentiable as $\qdv{\cdot}{\T\sigma_k}$ 
				on $\BHo \backslash \partial\BHo$.
				Note that the compactness of $\partial\BHi$ bounds the coefficient of the second derivative
				 of $f(\cdot)$. 
				These lead to	
				(\ref{inq:maxmax} , RHS) $=O(\delta^2)$. 
			\item 
				Otherwise (when the maximum is given at $\partial\BHo$),
				 one needs to consider the infinitesimal behavior  
				of $\qdv{\cdot}{\T\sigma_k}$ around $\partial\BHo$. 
				This infinitesimal behavior is as same as the infinitesimal behavior of the function $-z\log z$ 
				substituted by $z=t^2$ around $t=0$. 
				This leads to 
				(\ref{inq:maxmax} , RHS) 
				$=O(-\delta^2\log \delta^2) = O(-\delta^2 \log \delta)$.
		\end{enumerate} 
		Combining these two results above,
		RHS of (\ref{inq:maxmax}) is upper-bounded by 
		$\max\{O(\delta^2),O(-\delta^2\log\delta)\} = O(-\delta^2\log\delta) $. 
		Namely,
		\begin{equation}
			|c_\infty-c_k| = O(-\delta^2\log\delta) 
		\end{equation}
		Since $\delta = O(k^{-1})$, 	
		$|c_\infty-c_k| = O(k^{-2}\log k)$ holds. 
		
		\begin{figure}[th]
		 \begin{center}
			\includegraphics[width=6cm]{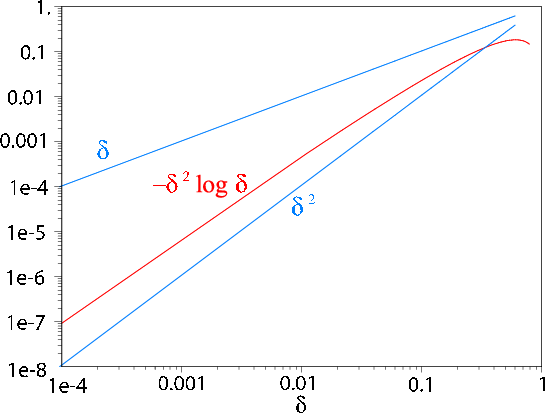}
			\qquad
			\includegraphics[width=6cm]{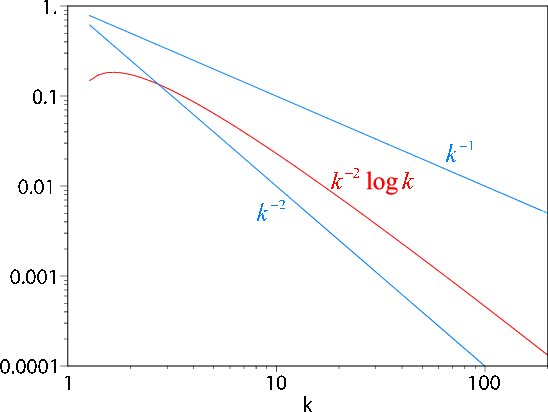}
		 \end{center}
		 \vspace*{-8mm}
			\caption[Plot:\quad$-\delta^2\log\delta,k^{-2}\log k$]
			{How $-\delta^2\log\delta$ and $k^{-2}\log k$ converge to 0 as $\delta\to 0 , k\to \infty$}
		\end{figure}
	\end{proof}
\end{Theorem}


%
%
\section{Application --- finding qubit channels with four engaging inputs} 
\label{s:application}

The channel
\begin{equation}
	\Tq: \rho(x,y,z) \mapsto \rho(0.6x+0.021,0.601y, 0.5z+0.495), 
\end{equation}
already appeared in (\ref{ch4engaging}) has a distinguished property:
The number of its engaging inputs is four.
As far as the author checked tens of times by the method presented in the previous 
section, the number of the engaging inputs of qubit channels are, 
mostly two, rarely three.  Four does not appear without elaborate preparation.

\subsection{When a qubit channel has four engaging inputs}
\label{ss:when}

When all the approximated positions of the engaging inputs of a given 
qubit channel $\T$ are located through the method described in the previous 
section, 
one can calculate its Holevo capacity $C(\T)$ much more precisely
by the Newton-Raphson method, even though the significant precision
seems limited to 5 -- 6 digits by the method in the previous section. 
We employed this method for the channel $\Tq$, and the capacity and
the engaging inputs are shown in Table.~\ref{tab:4st}. 
The calculation is performed by mathematical software Maple 6, and we 
actually obtained the precision of $C(\Tq)$ as much as more than 50 digits.

\begin{table}[p]
	\fbox{
		\begin{minipage}{1.0\linewidth}
		    $$
			    \Tq \rho(x,y,z) =  \rho(0.6x +0.021, 0.601y, 0.5z+0.495)  
			     ~~\qquad ~~
			     \text{capacity} =  0.3214851589 
		    $$
		    \vskip-0.5cm

\footnotesize
		     $$
			     \begin{array}{ccrc}
					 \text{probability}        &     \text{engaging input } (x,y,z)         &     \phi  ~~~&   \theta
					\\
					\hline
					 0.2322825705  & ( ~~0.2530759862,-0.0000000000,~~ 0.9674464043) 
					 & 14.66^\circ & ~~~0.00^\circ
					\\
					 0.2133220819  &  (~~ 0.9783950999, ~0.0000000000, ~~0.2067438718) 
					 & 78.07^\circ & ~~~0.00^\circ
					\\
					 0.2771976738  & (-0.4734087533, ~~0.8646461389,-0.1681404376) 
					 & 99.68^\circ & ~118.67^\circ
					\\
					 0.2771976738  &  (-0.4734087533,-0.8646461389,-0.1681404376)
					 & 99.68^\circ & -118.67^\circ
					\\
					 \text{average}     &  ( ~~0.0050428099, ~~0.0000000000, ~~0.1756076944) &   
					\\
					\hline
				 \end{array}  
			$$
		     \flushright {$\phi, \theta$ denote the angular coordinates of the engaging inputs.}  \vskip-0.2cm
			$$ 
				\begin{array}{ccc}
					\text{probability}       &     \text{engaging output } (x,y,z)      & S[\Tq(\rho)] \\
					\hline
					0.2322825705 &~( ~0.1728455917, ~0.0000000000, ~0.9787232022)   & 0.0300135405  
				  \\
					0.2133220819  &  ~ ( ~0.6080370599, ~0.0000000000,~ 0.5983719359)  & 0.3786915585  
				  \\
					0.2771976738  & ~(-0.2630452520, ~0.5196523295, ~0.4109297812)  &  0.5935800377   
				  \\
					0.2771976738   &~ (-0.2630452520,-0.5196523295, ~0.4109297812) & 0.5935800377  
				  \\
					\text{average}  & ~(~ 0.0240256859, ~0.0000000000,~ 0.5828038472)  & 0.7383180644 
				  \\
				    \hline
				\end{array}  
			$$

			\caption{Data for the qubit channel $\Tq$ with the engaging number four}
			\label{tab:4st}
		\end{minipage}
	}
\end{table}
\begin{figure}
	\fbox{
		\begin{minipage}{\linewidth}
%
		\begin{center}
		\includegraphics[width=14.5cm]{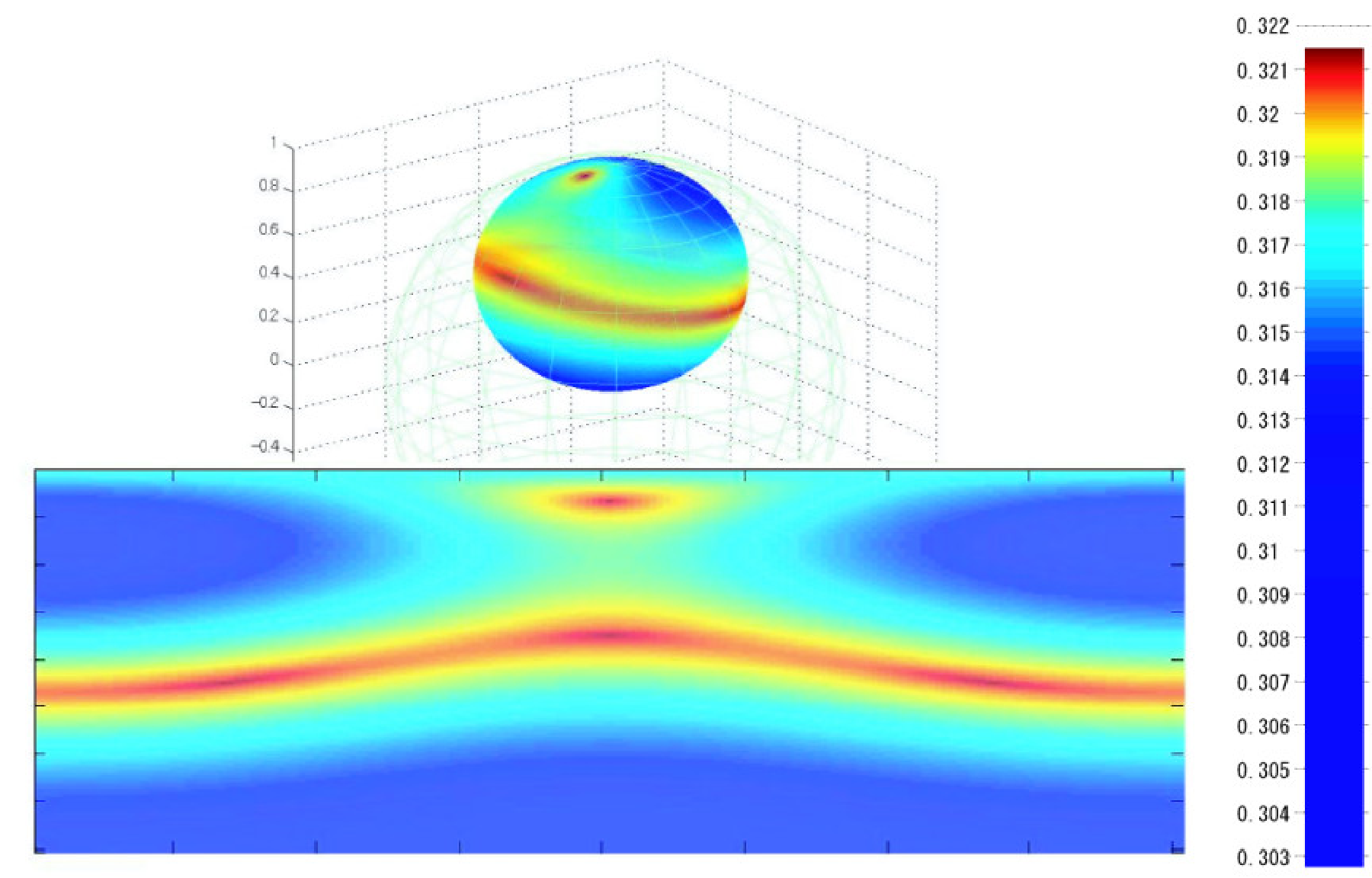}
		\end{center}
			\caption[$\qdv{\Tq \rho}{\Tq \sigma}$ on the ellipsoid]
			{The quantum divergence $\qdv{\Tq\rho}{\Tq\sigma}$ on the surface of the output ellipsoid of $\Tq$.
			$\rho$ runs on the surface of the Bloch sphere and the $\sigma$ is the optimal average input.
			The four most reddish points correspond to the engaging output because they are the farthest 
			points from $\Tq\sigma$ w.r.t.~the quantum divergence. The rectangular region is a projection
			from the ellipsoid surface. (The ellipsoid in the figure is drawn slightly transparent to
			see through to the opposite side.)}
		\end{minipage}
	}
\end{figure}
Incidentally, by Carath\'eodory's theorem, the theoretical upper bound 
of the engaging numbers %
 of qubit channels is four \cite{HIMRS04}.
Therefore, the discovery of this channel $\Tq$ settle the problem
of the actual maximum engaging number for qubit channels. 
It is the following discovery 
after \cite{king-nathanson-ruskai}
which found qubit channels which require three inputs to achieve 
the Holevo capacity. Moreover, it is a kind of the final goal 
that \cite{fuchs} attempted to dispel a prejudice that 
the engaging inputs should be orthogonal states,
{\it i.e.} two antipodal points on the Bloch sphere;
the idea is that a general quantum channel is noisy,
therefore the input should be kept to the most distinguishable 
states. On that prejudice, for any qubit channel, the engaging 
inputs should be two. 

One of the reasons that non-orthogonal inputs achieve 
greater capacity is that collective measurement 
on the output is allowed. 
This consideration may lead to the idea that the collective 
preparation of inputs, which are entangled input states, may
achieve further capacity than the Holevo capacity. This leads 
to the additivity check of the Holevo capacity, described in
Subsection~\ref{ss:actt}.

\subsection{How was $\Tq$ found? }

How the 
channel 
$
	\Tq: 
	\rho(x,y,z)
	\mapsto 
	\rho(0.6x +0.021, 0.601y, 0.5z+0.495)
$,
which has four engaging inputs, 
was found is explained here. 

\begin{enumerate}[\hspace*{8mm}-- 1.]
\item 
	The author perceived the tendency that the engaging outputs appear 
	near the two endpoints of the major axis of the output ellipsoid. Moreover,
	when the surface of the output ellipsoid very closely approaches the surface 
	of the Bloch sphere, then the third engaging output on the ellipsoid may appear
	near 
	the Bloch sphere.
		\begin{center}
		\includegraphics[width=8.5cm]{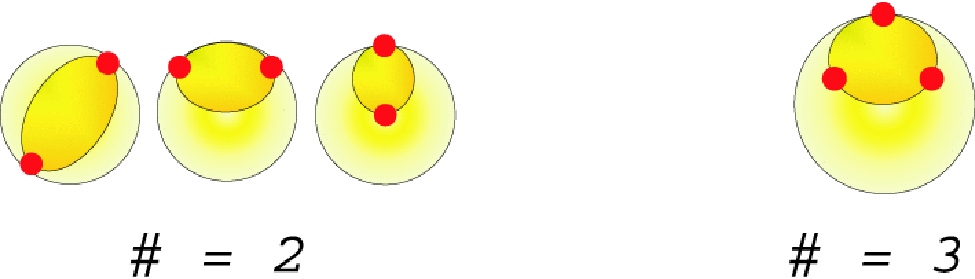}
		\end{center}
\item
	The output ellipsoid of an already-found three-engaging channel 
	$\T_{\tt 3}: \rho(x,y,z)\mapsto\rho(0.6x, 0.6y, \\0.5z +0.5)$
	is stretched along the $y$-direction very slightly to an adequate extent.
	%
	The channel became $\T_{\tt 3'}: \rho(x,y,z)\mapsto\rho(0.6x, 0.601y,0.5z +0.5)$.
	It is desired that one of the engaging 
	outputs splits into two (we denote this phenomenon {\it bifurcation}); however, 
	the excessive symmetry in terms of the location of the ellipsoid
	prevents bifurcation. \index{bifurcation}

\item 
	To reduce the symmetry, moving ahead in the $x$-direction is attempted, and to 
	keep the output ellipsoid inside the Bloch sphere, one must pull the ellipsoid to 
	the center slightly along the $z$-direction.
	The channel became $\T_{\tt 3'\epsilon}: \rho(x,y,z)\mapsto\rho(0.6x+\epsilon, 0.601y,0.5z +0.495)$.
	
\item 
	Various $\epsilon$ were tried, and bifurcation occurred when $\epsilon=0.021$. 
	
		\begin{center}
		\includegraphics[width=7cm]{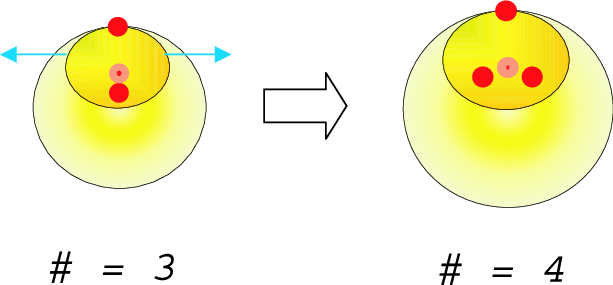}
		\end{center}

\item 
	CPTP-ness was checked by the method introduced by \cite{ruskai-szarek-werner}. 
	The method showed $|\epsilon| \le 0.05277..$  is equivalent to $\T_{\tt 3'\epsilon}$ being CPTP 
	\cite{HIMRS04}.

\end{enumerate}

Another qubit channel with four engaging inputs: 
\begin{equation}
	\T:\rho(x,y,z)\mapsto\rho(0.8x+0.022, 0.8015y, 0.75z+0.245)
\end{equation}
was found by the prescription above from a channel $\T:\rho(x,y,z)\mapsto\rho(0.8x, 0.8y, 0.75z+0.25)$.

\subsection{Additivity check for $\Tq\otimes\Tq$}\label{ss:addcc}
\label{ss:actt}

As mentioned in Subsection~\ref{ss:when}, a quantum channel $\T$ can get more communication efficiency 
if collective measurement by the receiver is allowed. This efficiency is 
called the Holevo capacity of this channel. One might like to consider whether ``collective preparation'' by the 
sender increases further efficiency, {\it i.e.} whether input particles with quantum entanglement 
increase the efficiency. This is formalized 
as whether $C(\T^{\otimes n} ) > n \, C(\T) $ for a 
channel $\T$ and a positive integer $n$. We checked this for $n=2$ with $\Tq$, which is an 
eccentric channel as already stated. 

There is a useful relation to check the additivity \cite{HIMRS04}
\begin{equation}
C(\T^\d ) > 2C(\T) 
\Leftrightarrow
\max_{\omega\in\partial\left(\BHi^\d\right)} \qdv{\T^\d \omega}{{\sigma'_\T}^\d} > 2C(\T),
\end{equation}
where $\sigma'_\T$ is the optimal average output of $\T$.
It is useful in that it greatly reduces the dimension of the search space.
Therefore, $\omega=\ketbra{\psi}$ is substituted by 
\begin{equation}
\ket{\psi} 
	= \sqrt{p}\, \ket{u}\otimes \ket{v} 
	+ e^{\I\nu}\sqrt{1-p}\,
		\ket{u^\bot}\otimes \ket{v^\bot}
 \label{fm:Psi}
\end{equation}
where 
$
\ket{u}=		\Big(\begin{smallmatrix} \cos \theta_u \\ e^{\sqrt{-1}\phi_u} \sin \theta_u\end{smallmatrix}\Big)
,\ket{v}=		\Big(\begin{smallmatrix} \cos \theta_v \\ e^{\sqrt{-1}\phi_v} \sin \theta_v\end{smallmatrix}\Big)
$,  
$\ket{u^\bot}= 		\Big(\begin{smallmatrix} e^{-\sqrt{-1}\phi_u} \sin \theta_u \\ -\cos \theta_u \end{smallmatrix}\Big) $
and 
$\ket{v^\bot}= 		\Big(\begin{smallmatrix} e^{-\sqrt{-1}\phi_v} \sin \theta_v \\  -\cos \theta_v \end{smallmatrix}\Big) $.
($\ \ket{u^\bot},\ket{v^\bot}$ are orthogonal vectors of $\ket{u}$ and $\ket{v}$, respectively.)

The range of the parameters of (\ref{fm:Psi}) are
\begin{equation}
0\le p \le 1 ,\; 0\le \theta_u,\theta_v,\nu \le 2\pi, \; 0\le\phi_u,\phi_v\le \f{\pi}{2}.
\end{equation}

\begin{figure}[ht]
	\fbox
	{
	\begin{minipage}{\textwidth} 
		\begin{minipage}{8cm}
			\begin{center}
				\vspace*{4mm}
				\includegraphics[width=6.5cm]{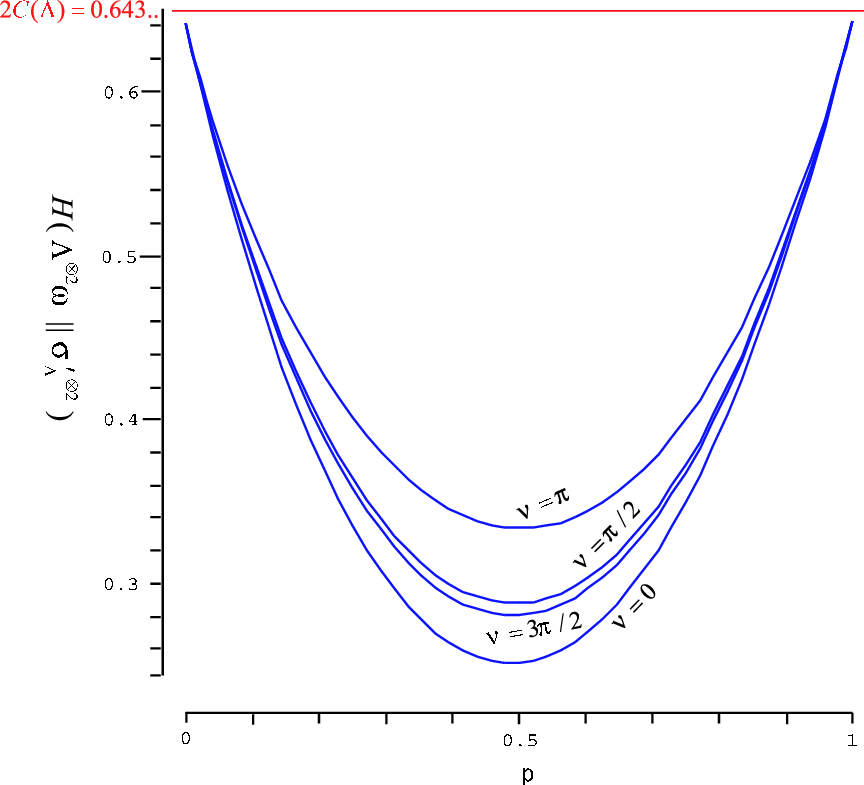}
			\end{center}
		\end{minipage}
		\begin{minipage}{7.5cm}
			\small
			If there is an extraordinary gain 
			in the communication capacity
			of $\Tq^\d$,
			there exists a pure state $\omega$
			shared by two particles
			which satisfies
				$
					\qdv{\Tq^\d \omega}{{\sigma'_\Tq}^\d} 
				> 2C(\Tq).
				$
			Our numerical experiments, however, show that 
			the more the two particles are entangled, 
			the more the communication capacity falls,
			as the figure shows. 
		\end{minipage}

		\caption
		[Convexity of \quad$\qdv{{\Tq}^\d \omega}{{\sigma'_{\Tq}}^\d}$]
		{Convexity of $\qdv{\Tq^\d \omega}{{\sigma'_\Tq}^\d}$}
		\label{fig:vexp}
	\end{minipage}
	}
	
	\medskip
	\fbox{
	\begin{minipage}{\textwidth}
		\begin{minipage}{8cm}
			\begin{center}
				\vspace{0mm}
				\includegraphics[width=6.5cm]{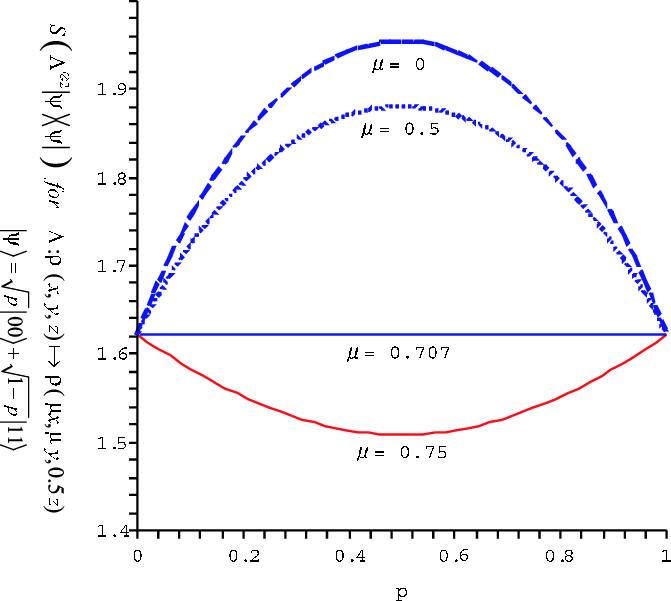}
				\vspace{-2mm}
			\end{center}
		\end{minipage}
		\begin{minipage}{7.5cm}
		\small
			One might like to conjecture that entanglement always causes
			convexity as illustrated in Fig.~\ref{fig:vexp}. 
			Note that the convexity 
			of $H(\T^\d\omega||{\sigma'_\T}^\d)$ causes 
			the concavity of $S(\T^\d\omega)$ 
			because $H(\T^\d\omega||{\sigma'_\T}^\d)+S(\T^\d\omega)$
			is linear as of $\omega$. The figure on the left,
			however, shows the opposite for $\T:\rho(x,y,z)\mapsto\rho(0.75x,0.75y,0.5z)$
			with $\ket{\psi}=\sqrt{p}\ket{00}+\sqrt{1-p}\ket{11}$,
			as of a function of $p$.
			
					\end{minipage}

		\caption
		[A counterexample of the concavity conjecture]
		{A counterexample of the concavity conjecture}
		\label{fig:cecj}

	\end{minipage}
	}
\end{figure}

(\ref{fm:Psi}) is simply an ordinal  Schmidt decomposition form of a bipartite pure state. 
We plot $\qdv{{\Tq}^\d \omega}{{\sigma'_{\Tq}}^\d}$ with various $u$ and $v$ as a function of $p$ and
$\nu$.
A typical plot is shown in Fig.~\ref{fig:vexp}. All the graphs of the numerical experiments show that 
$\qdv{{\Tq}^\d \omega}{{\sigma'_{\Tq}}^\d}$ is deeply convex w.r.t.~$p$, the entanglement parameter 
of the Schmidt decomposition. Therefore, the extra gain in the Holevo capacity of $\Tq^\d$ 
does not seem expected.

(As a result, such deep convexities  may lead to whether this convexity universally holds for any quantum channel. 
If it holds, the additivity holds. 
However, this convexity does not always hold. The convexity in $\qdv{\rho}{\sigma}$
as of $\rho$ leads to the concavity of $S(\rho)$; however, as Fig.~\ref{fig:cecj}
shows,
 $S(\T^\d \ketbra{\psi})$ of $\T:\rho(x,y,z)\mapsto\rho(0.75x,0.75y,0.5z)$ 
with $\ket{\psi}=\sqrt{p}\, \ket{00}+\sqrt{1-p}\, \ket{11}$ is convex as of $p$.
Therefore, the additivity problem is not yet solved.  )

%
%

\section{Conclusion and discussion} 

\subsection{The calculating method with universal applicability and convergence}
The calculations of the Holevo capacity were performed by the author 
for the first time in the world in the following sense: 
\begin{enumerate}[\hspace*{8mm}-- 1.]
	\vspace*{-2mm}
	\item The calculation method was applicable to any qubit channel.
	\vspace*{-2mm}
	\item The answer was truly close to the real answer; 
		convergence to the real answer was guaranteed.
	\vspace*{-6mm}
\end{enumerate}
The method involved a high-dimensional optimization and was 
actually performed for approximately a hundred channels. 

The convergence speed is, however, not so fast in that 
bounding the error within $\epsilon>0$ requires a lattice with the coarseness of 
$O(\epsilon^{1/2+0})$. 
If this lattice is $\ell_k$, $k=\Omega(\epsilon^{-\f{1}{2} -0})$, thus
$n=\Theta(\epsilon^{-1-0})$ where $n$ is the number of the points of $\ell_k$.
Assuming the calculation cost in time is $\Theta(n^3)$, it ends up to become equal to
$\Theta(\epsilon^{-3-0})$.
(Here $+0$ and  $-0$ mean arbitrary positive and negative numbers, respectively,
with the intention that they are non-zero and are close to zero.)

\subsection{An attempt to check the additivity in Holevo capacity}
To investigate a mysterious aspect of 
quantum physics, the additivity of the Holevo capacity was 
checked for a channel with four engaging inputs, 
which was discovered by the proposed calculation method. 
Since the communication efficiency through a quantum channel
 increases by collective measurement on the receiver side, 
one might expect it to also increase by the collective preparation,
{\it i.e.} entanglement among inputs may increase the efficiency.
The discovered channel was such a candidate, 
as it increases  efficiency by the collective measurement 
by the receiver, 
but the 
analysis presented in this chapter showed the collective preparation
of the sender cannot seem to increase the efficiency.

In order to refine this attempts, we may need to investigate: 
\begin{enumerate}[\hspace*{8mm}-- 1.]
\vspace*{-2mm}
\item how the input and the output of a quantum channel are reversed, 
and
\vspace*{-2mm}
\item how communication efficiency relates to the collectiveness 
of both the input and the output of a quantum channel. 
\end{enumerate}

\subsection{Beyond qubit channels}
One may like to calculate the Holevo capacity of a quantum channel which is not 
a qubit channel. Because of dimensional explosion, it is impossible to calculate 
the Holevo capacity even for three-level systems. The reason is as follows:
the corresponding space of 
the Bloch sphere for $d$-level system forms a $(d^2-1)$-dimensional manifold which 
is the convex hull of a $(2d-2)$-dimensional manifold. What will happen if we 
cover the $(2d-2)$-dimensional manifold with a mesh of $ \underbrace{40 \times \dots \times 40}_{2d-2}$?
A roughly approximated number of 
vertices for each $d$ is as shown in Table~\ref{tab:dimex}.
\begin{table}[h]
\begin{center}
\begin{tabular}{c|ccccc}
\hline
$ d  $       &  2   &   3   &   4   &  8   &  16   \\
\hline
$2d-2$       &  2   &   4   &   6   &  14  &  30\\
$40^{2d-2}$  & $1.6\times 10^3 $ & $2.6\times 10^6  $& $4.1 \times 10^{9}$ & $2.7\times 10^{22}$ & $1.2 \times 10^{48} $ \\

\hline
\end{tabular}
\caption[Number of vertices on the lattices]{\small Vertices number of the lattices for each quantum channel of $d$-level system.}
\label{tab:dimex}
\end{center}
\end{table}
\vspace*{-6mm}
Thus it is quite hard to calculate the Holevo capacity beyond qubit channels. 
A quite elaborate methodology may be 
required to realize this calculating, for example, an efficient configuration of the vertices of the mesh.

\section*{Endnotes} 
\theendnotes

\part{Epilogue}

\chapter{Conclusion}

\section{Summary of this dissertation}
\makeendnotes
In this dissertation, the following concepts in quantum information science have been 
dealt with: 
\begin{enumerate}[\hspace*{8mm}-- 1.]\setlength{\itemsep}{0pt}\setlength{\parsep}{0pt}\setlength{\baselineskip}{13pt}
	\item the {\bf entanglement cost} and 
	\item the {\bf Holevo capacity}.
\end{enumerate}
These two concepts are important quantification in the emerging science of quantum information.
To be slightly more precise, they are both quite important candidates to measure 
\begin{enumerate}[\hspace*{8mm}-- 1.]\setlength{\itemsep}{0pt}\setlength{\parsep}{0pt}\setlength{\baselineskip}{13pt}
	\item the anomalous correlation between multiple figures peculiar to quantum physics, and
	\item the ultimate communication efficiency of given channels in this physical world,
\end{enumerate}\vspace*{-2mm}
respectively. Note that 
the anomalous correlation, which is called {\bf quantum entanglement}, 
is an indispensable resource in quantum computation and quantum communication. 
Also note that the two concepts are related theoretically through Stinespring's dilation\cite{MSW}. 

From an objective overview, the desirable goal in investigating entanglement measures and 
capacity of quantum channels might be as follows: 
\begin{list}{\hspace*{8mm}-- }{}\setlength{\itemsep}{0pt}\setlength{\parsep}{0pt}\setlength{\baselineskip}{13pt} 
	\item identifying the most approvable entanglement measure if it exists,
	\item discovering how to calculate this measure,
	\item determining whether measures based on various concepts are equal to each other,
	\item determining the additivity of the measures 
		(perhaps $\epsilon(\rho\otimes \sigma)\neq
		\epsilon(\rho)+\epsilon(\sigma)$ would occur for some measure $\epsilon$ contrary to expectation), and  
	\item determining whether the Holevo capacity is the most approvable communication capacity 
	of quantum channels; this issue is equivalent to whether the additivity of the Holevo capacity holds.
	\item discovering how to calculate the communication efficiency of quantum channels.  
\end{list} 

With the purpose of settling the problems written above, 
this dissertation has attempted to solve the following problems:
\begin{list}{\hspace*{8mm}-- }{}\setlength{\itemsep}{0pt}\setlength{\parsep}{0pt}\setlength{\baselineskip}{13pt} 
	\item 	$E_C$ (the entanglement cost) and  	$E_F$ (the entanglement of formation) of antisymmetric states,
 	\item   the difference between $E_C$ and $E_D$(the entanglement distillation),  
	\item   the superadditivity of $E_F$,
	\item   the algorithm to calculate the Holevo capacity, and 
	\item 	the additivity of the Holevo capacity.
\end{list}\vspace*{-2mm} 


\section{Application in daily life \\--- A long, long time from now or
in a galaxy far, far away --- }
 How will the research of this dissertation contribute to the real world? 
One of the amusing applications is {\bf quantum teleportation} 
\cite{BBC}. \index{quantum!--- teleportation}\index{teleportation}\index{cricket}\index{galaxy}

\begin{figure}[bhtbp]
	\fbox{
		\vspace*{4mm}\\
		\begin{minipage}{0.97\linewidth}
			\begin{center}
			\includegraphics[width=13cm]{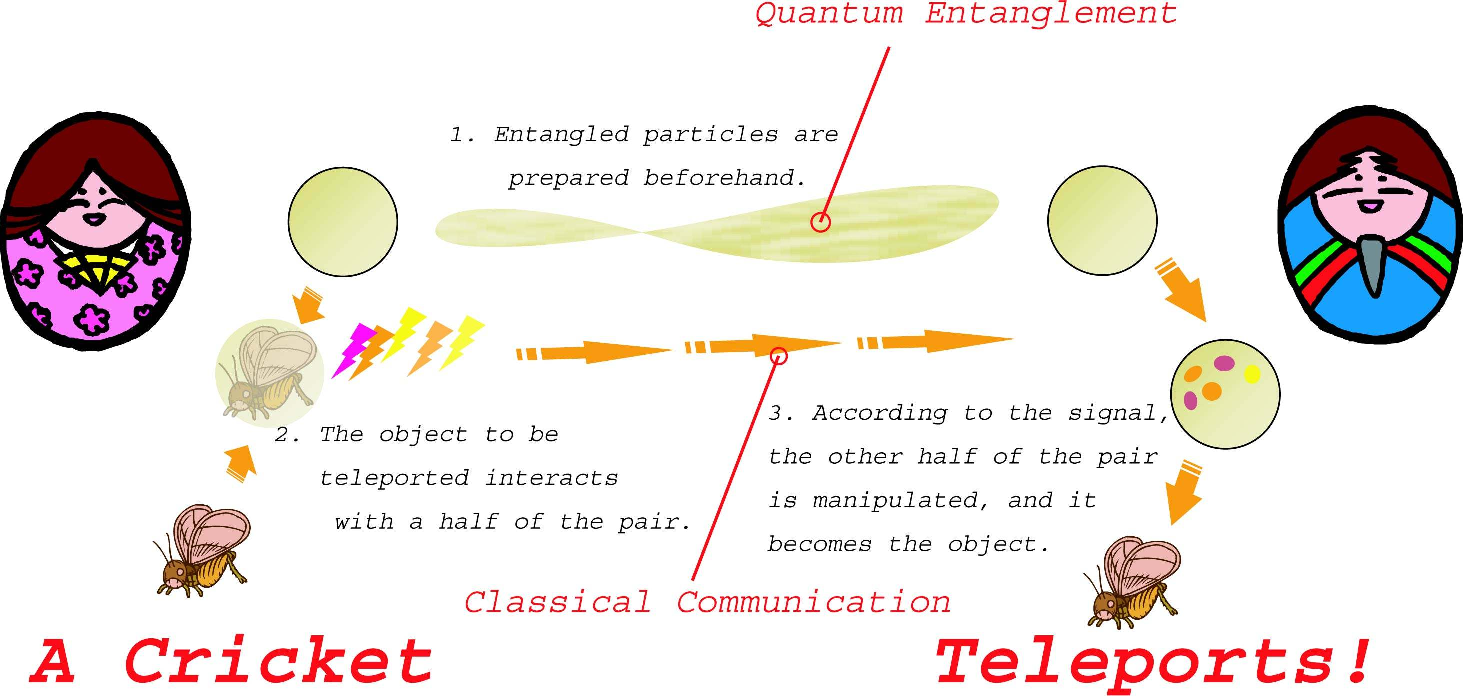}
			\vspace*{-8mm}
			\end{center}
			\caption[Sketch of quantum teleportation]
			{%
			A sketch of quantum teleportation. 
			
			\small
			A quantum state (a cricket in the sketch) in Alice's hand is teleported to Bob with 
			quantum entanglement and classical communication. 
			When the age of quantum teleportation comes, each of the amount of
			the quantum entanglement and the classical communication
			in quantum teleportation is important. 

			}%
			\label{fig:teleportation}
		\end{minipage}\\
		\vspace*{4mm}
	}
\end{figure}

\bigskip

Quantum teleportation is a method to teleport the quantum state in 
one place to a remote place with quantum entanglement shared by the two 
sites beforehand.
You can consider Alice desires to send a quantum state to Bob in a remote place. 
They need to have pairs of entangled particles divided between Alice's site and Bob's site. 
The quantum state to be teleported may be anything, but it is desirable to be a small object
like DNA molecules or mineral samples on asteroids.
Alice may teleport herself with a huge amount of entanglement 
and with a transmitter of huge information. 
A brief explanation of the method to teleport the object is as follows (See Fig.~\ref{fig:teleportation}):

\begin{quote}
Alice makes 
the object and the particles in her hand
interact with each other.
 After getting signals from the interaction, 
she sends the signals to Bob. 
According to the signals, 
he manipulates the particles in his hand 
which turn into the object Alice once had. 
\end{quote}

\newcommand{\sA}{_{\sf A}}
\newcommand{\sB}{_{\sf B}}
\newcommand{\sO}{_{\sf O}}
How are these operation performed? Here is a theoretical description 
how to teleport the object ${\sf O}$ of $d$-level system of which the basis is
a $\{\ket{i}\}_{i=0,1,\ldots,d}$. Assume the 
state of {\sf O} is  $\ket{\psi}\sO=\sum_{i=0}^{d-1} a_i\,\ket{i}\sO$, and  $\{a_i\}_{i=0}^{d-1}$ is not 
need to be known. 
In the following, $\omega = e^{\frac{2\sqrt{-1}\pi}{d}}$, and $+$ operation is done in modulo $d$.

%
\begin{enumerate}[\hspace*{2.5mm}-- 1.]
\item 
	The shared entangled particles {\sf AB} are prepared in 
	$ {\sum_{i=0}^{d-1} \frac{1}{\sqrt{d}}\;\ket{i}\sA\otimes\ket{i}\sB}$.
\item
	The interaction is done by `measuring' the particles in her hand to physically identify with one of 
	$\displaystyle \left\{
	\ket{\psi_{x,y}}
	 =  {\sum_{i=0}^{d-1} \frac{\omega^{ix}}{\sqrt{d}}\;\ket{i}\sO\otimes\ket{i+y}\sA}\right\}_{
        \begin{smallmatrix} x=0,\ldots, d-1 \\ y=0,\ldots, d-1 \end{smallmatrix} }$
	; 
		$\{
		\ket{\psi_{x,y}}
		\}$ 
		is an orthogonal basis of the space of {\sf OA}.
		
\item
Bob manipulates his particles {\sf B} with a 
unitary operator $U_{xy} = (u_{ij}  ) $ where $u_{ij}= \delta_{i,j+y	} \,\omega^{ix} $ and 
$\delta$ is the Kronecker delta ($\delta_{kk}=1$ and $\delta_{kl}=0$ when $k\neq l$). 

\end{enumerate}
(For $d=2$, the entangled state shared beforehand is the Bell state, the measurement is 
the Bell measurement, and the unitary operator is the Pauli operator.)

When Alice measures that the state of {\sf OA} is $\ket{\psi_{x,y}}_{\sf OA}$, 
the state of the particles in Bob's hand turns to be $\ket{\phi_{x,y}}\sB=\sum_{i=0}^{d-1}\omega^{-ix} \, a_i \,\ket{i+y}\sB$,
because the state of the whole {\sf OAB} before the measurement is 
$\sum_{x,y} \frac{1}{\sqrt{d}} \;\ket{\psi_{x,y}}_{\sf OA}\otimes \ket{\phi_{x,y}}\sB$.
He retrieves the state of $\sum_{i=0}^{d-1}a_i \,\ket{i}\sB$ from 
$\ket{\phi_{x,y}}\sB$
by the unitary operation with the information $x$ and $y$.
When the physical attributes, such as atomic configuration or the state of electrons, 
of $\ket{i}\sO$ and $\ket{i}\sB$ correspond, one observes the object {\sf O} in Alice's site teleports
to Bob's site.

\medskip
The technical difficulties are as follows: how to cope with the decay of the entanglement
(except photons, the quantum state is quite easily interacts with its outer environment within a millionth second,
which causes quite hard to preserve quantum entanglement), and how to transmit the huge information
(even a 1mm$^3$ (cubic millimeter) of water requires $d>10^{10^{20}}$, 
thus Alice needs to tell $x$ and $y$ to Bob with huge digits%
.
It takes much more than a trillion years by the present technology%
).
Finding the technology to preserve quantum entanglement for a long time 
and to load  information on  each light particle of ordinary brightness 
will solve these difficulties.

\medskip
When the age of quantum teleportation comes, measuring 
quantum entanglement and also the ultimate capacity of information-transmitting channels
will become important. 
They can be compared to the amount of petroleum and the road capacity for an automobile. 
The research of this dissertation will hopefully contribute to the furthering understanding these 
points. 

%

	\def\see#1#2{$\Rightarrow$ #1}
	\printindex%
\newpage
\pagestyle{empty}
\vspace*{-29mm}\hspace*{-32.3mm}
\noindent\includegraphics[width=210mm,height=297mm,bb=77 72 535 720]{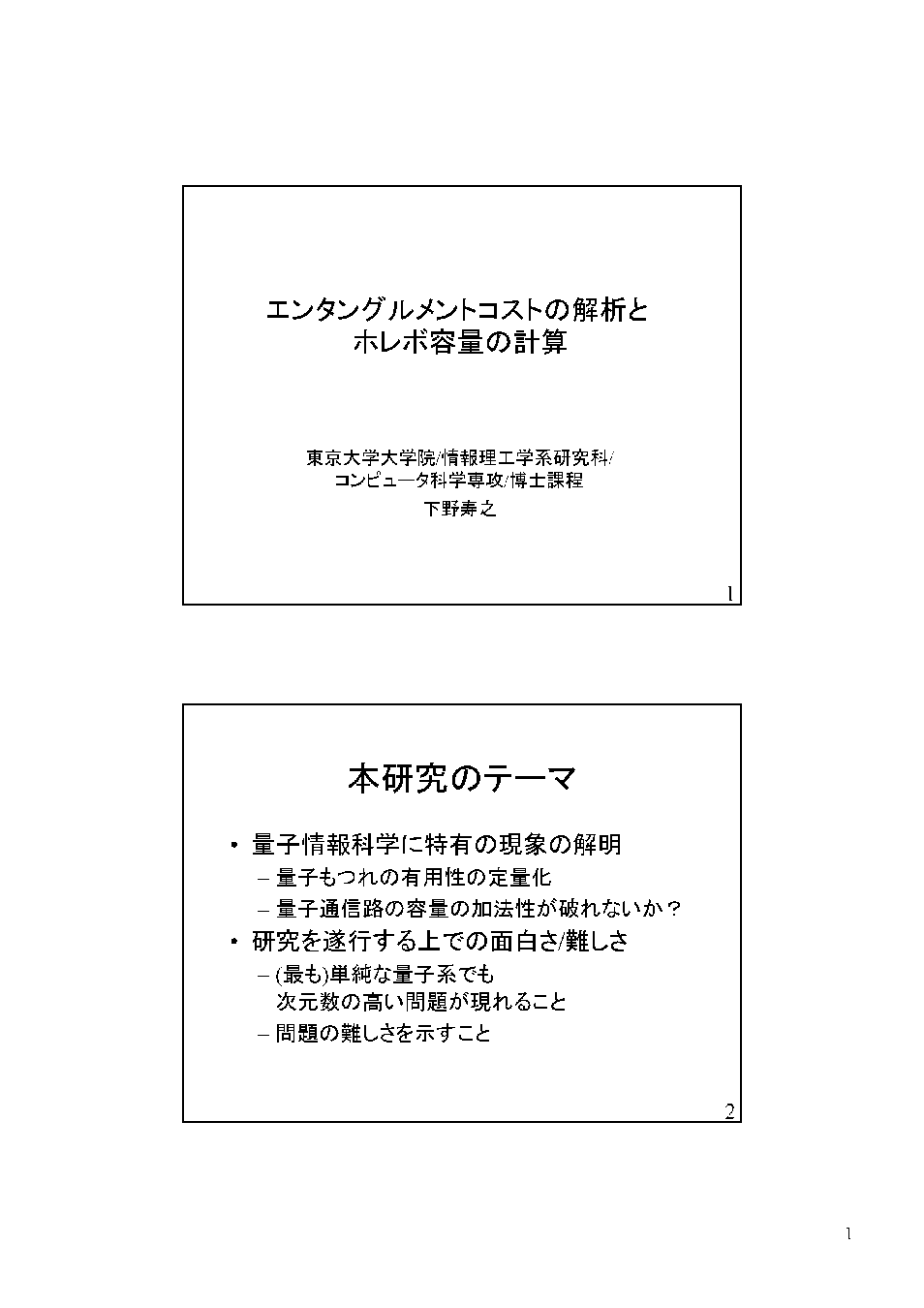}
\newpage
\thispagestyle{empty}
\vspace*{-29mm}\hspace*{-32.3mm}%
\noindent\includegraphics[width=210mm,height=297mm,bb=77 72 535 720]{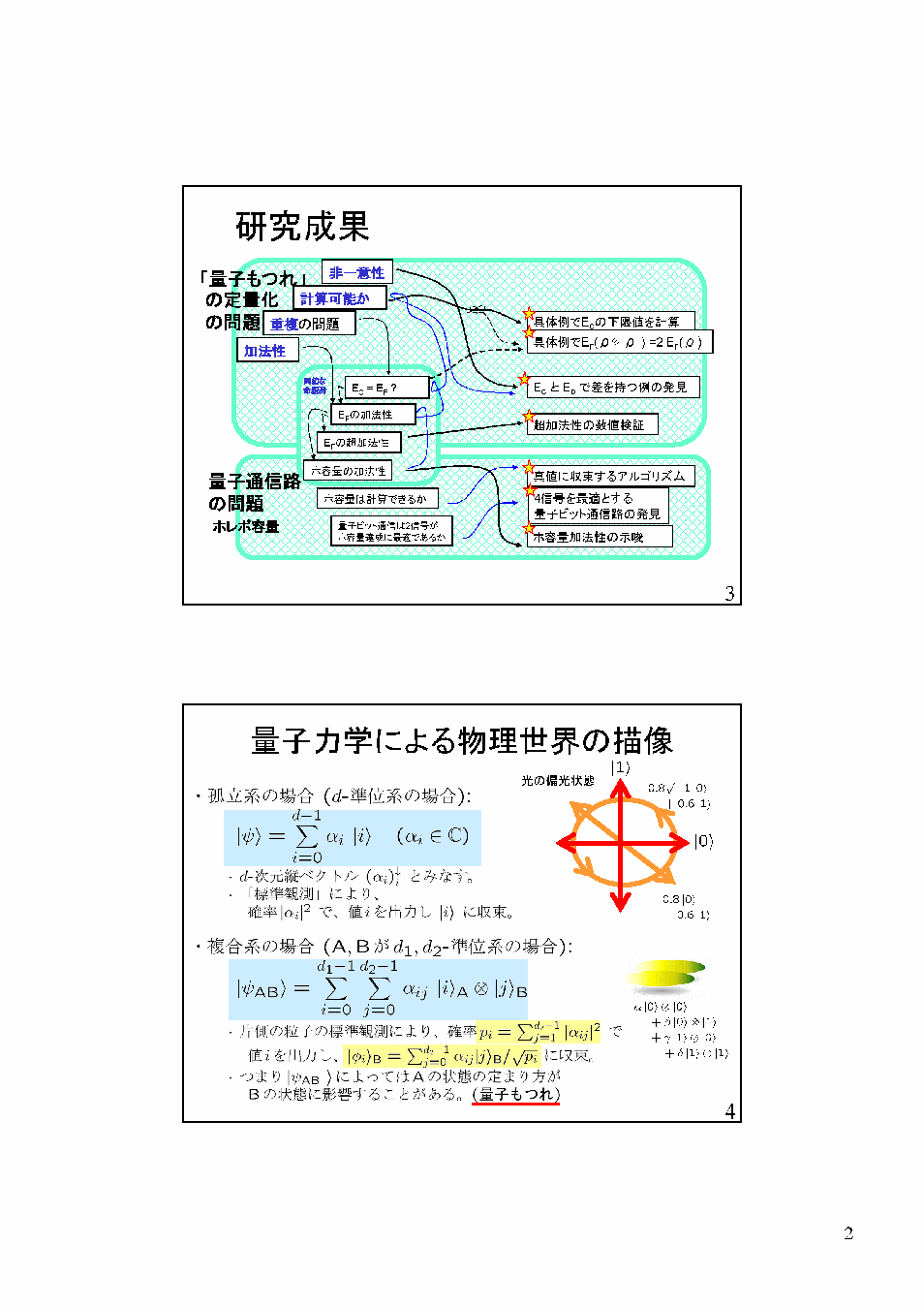}
\newpage
\thispagestyle{empty}
\vspace*{-29mm}\hspace*{-32.3mm}%
\noindent\includegraphics[width=210mm,height=297mm,bb=77 72 535 720]{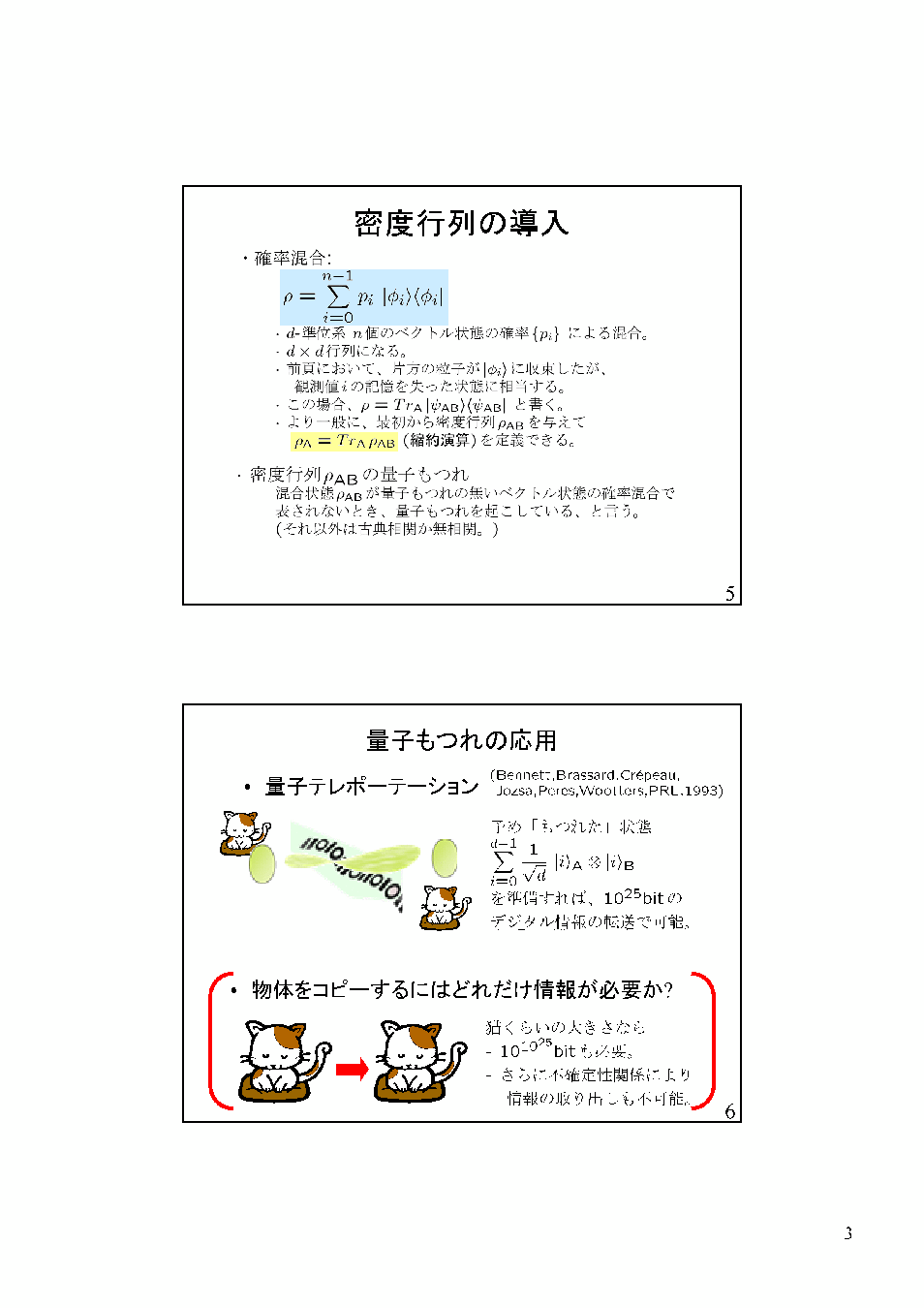}
\newpage
\thispagestyle{empty}
\vspace*{-29mm}\hspace*{-32.3mm}%
\noindent\includegraphics[width=210mm,height=297mm,bb=77 72 535 720]{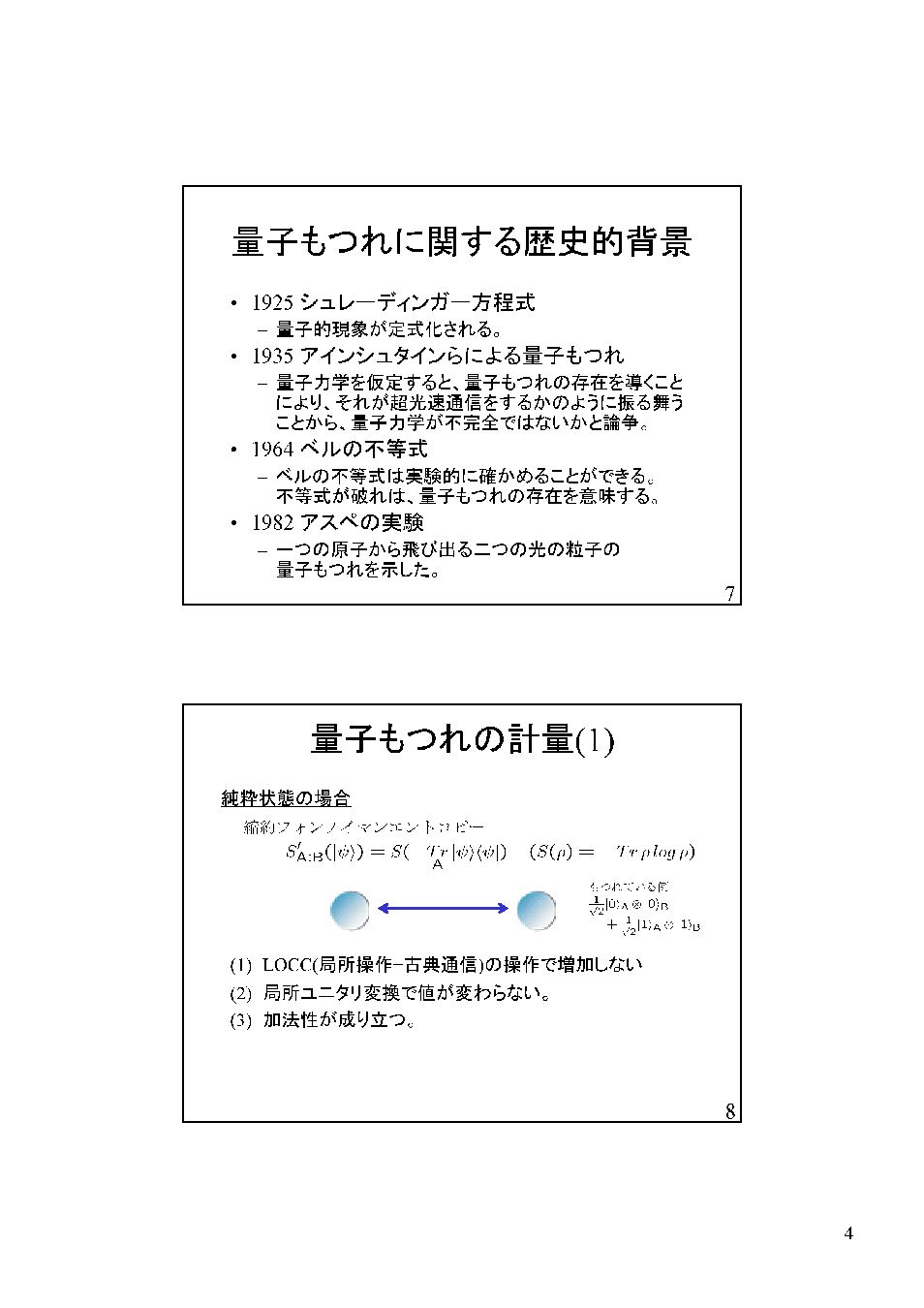}
\newpage
\thispagestyle{empty}
\vspace*{-29mm}\hspace*{-32.3mm}%
\noindent\includegraphics[width=210mm,height=297mm,bb=77 72 535 720]{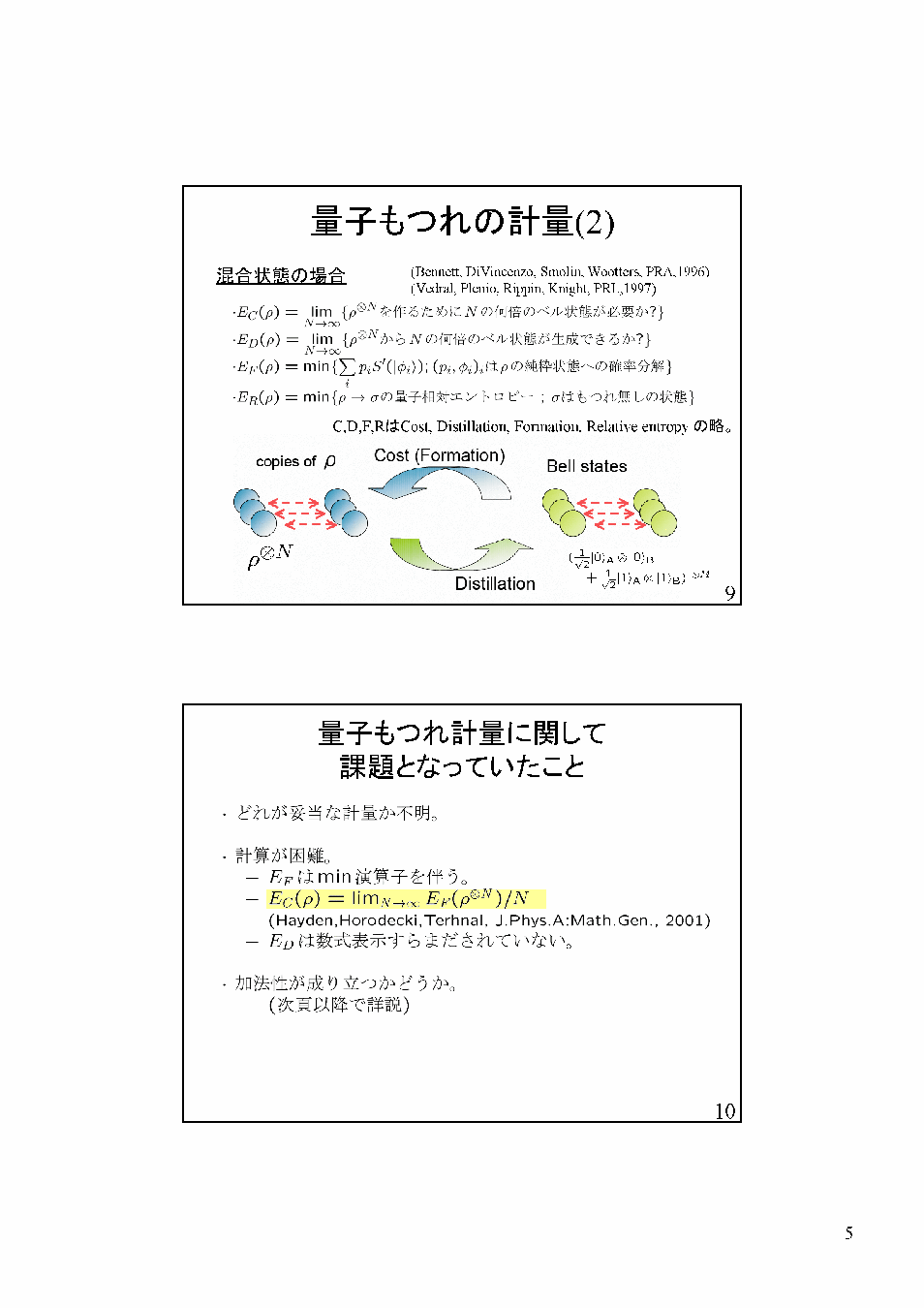}
\newpage
\thispagestyle{empty}
\vspace*{-29mm}\hspace*{-32.3mm}%
\noindent\includegraphics[width=210mm,height=297mm,bb=77 72 535 720]{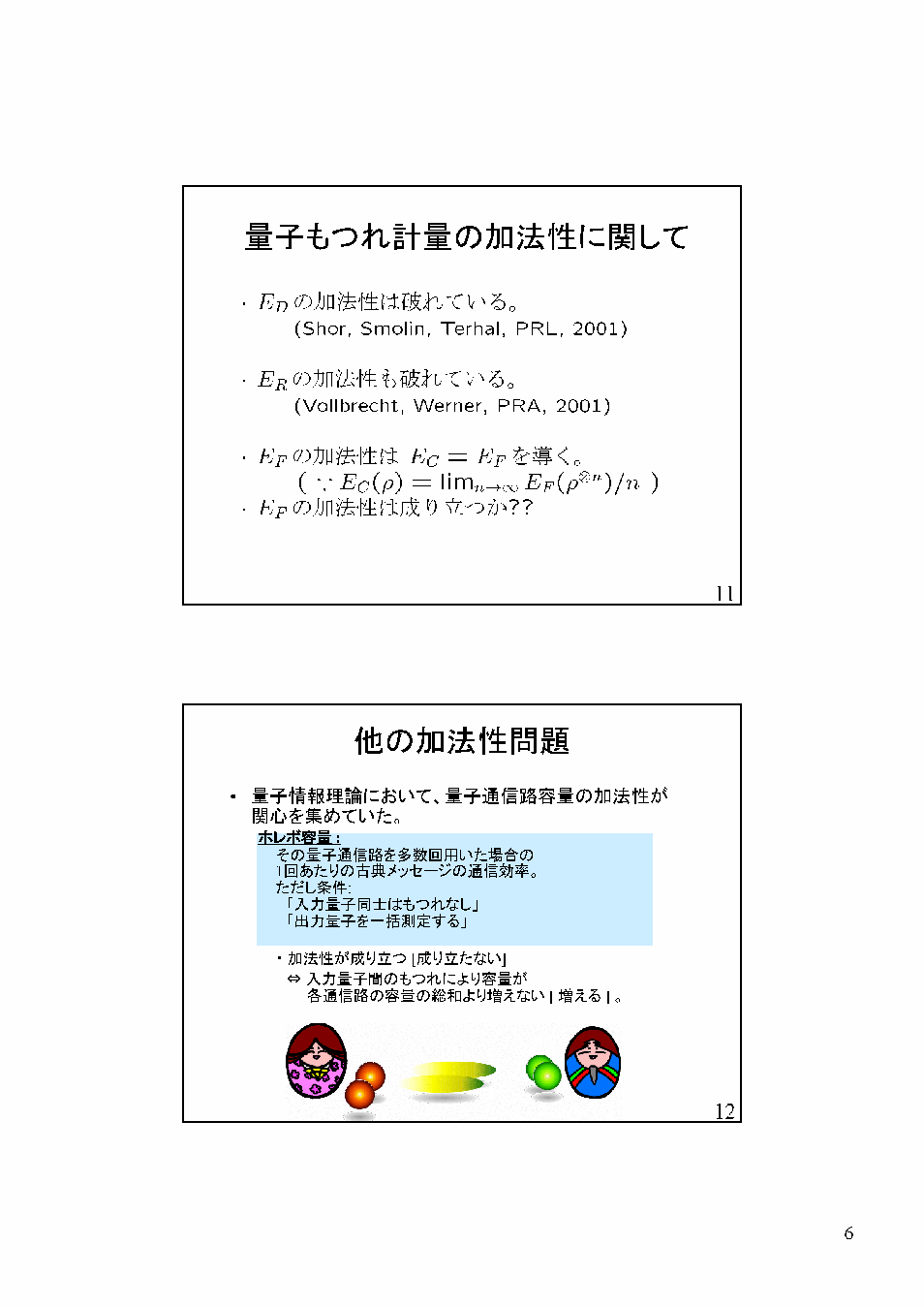}
\newpage
\thispagestyle{empty}
\vspace*{-29mm}\hspace*{-32.3mm}%
\noindent\includegraphics[width=210mm,height=297mm,bb=77 72 535 720]{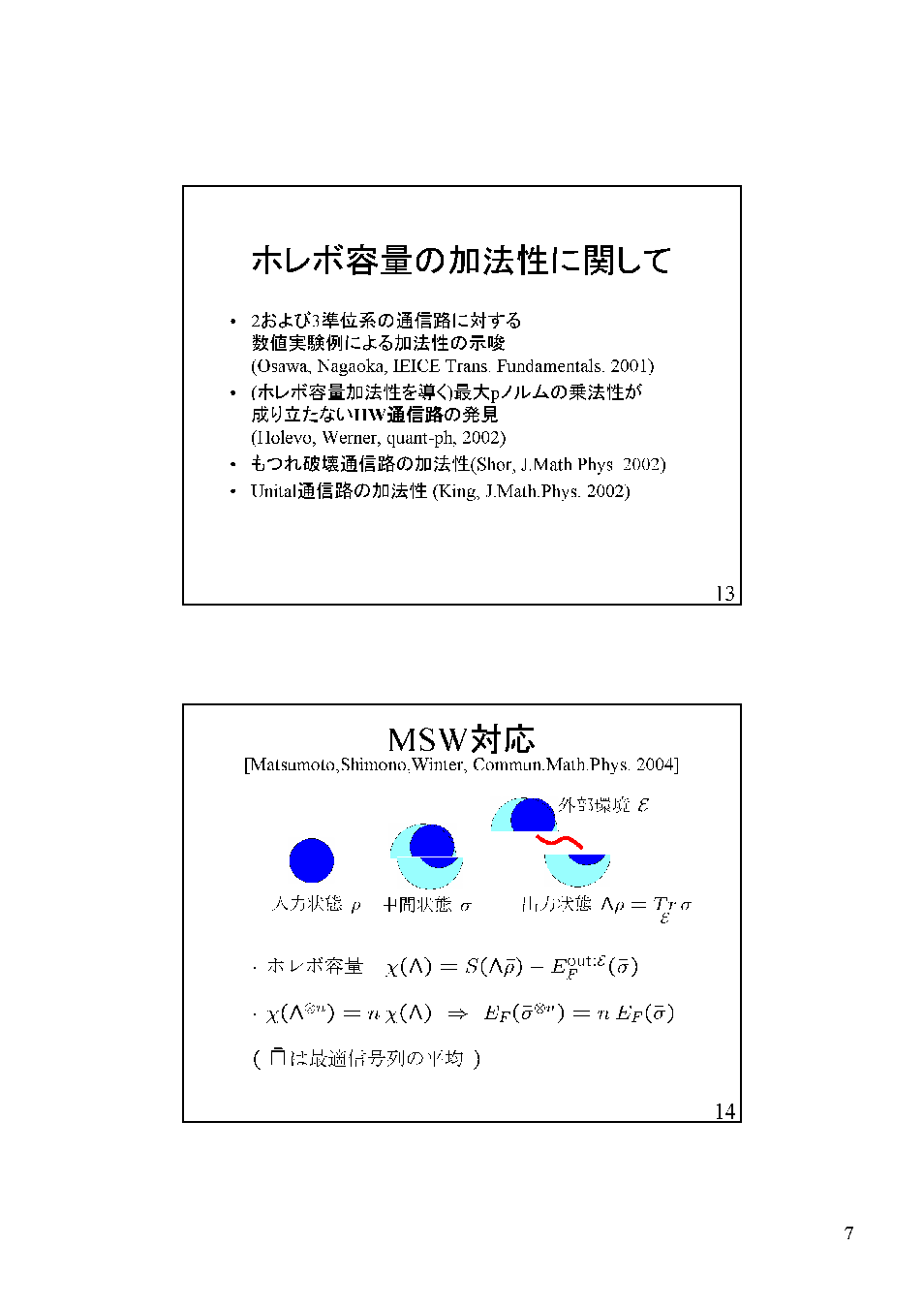}
\newpage
\thispagestyle{empty}
\vspace*{-29mm}\hspace*{-32.3mm}%
\noindent\includegraphics[width=210mm,height=297mm,bb=77 72 535 720]{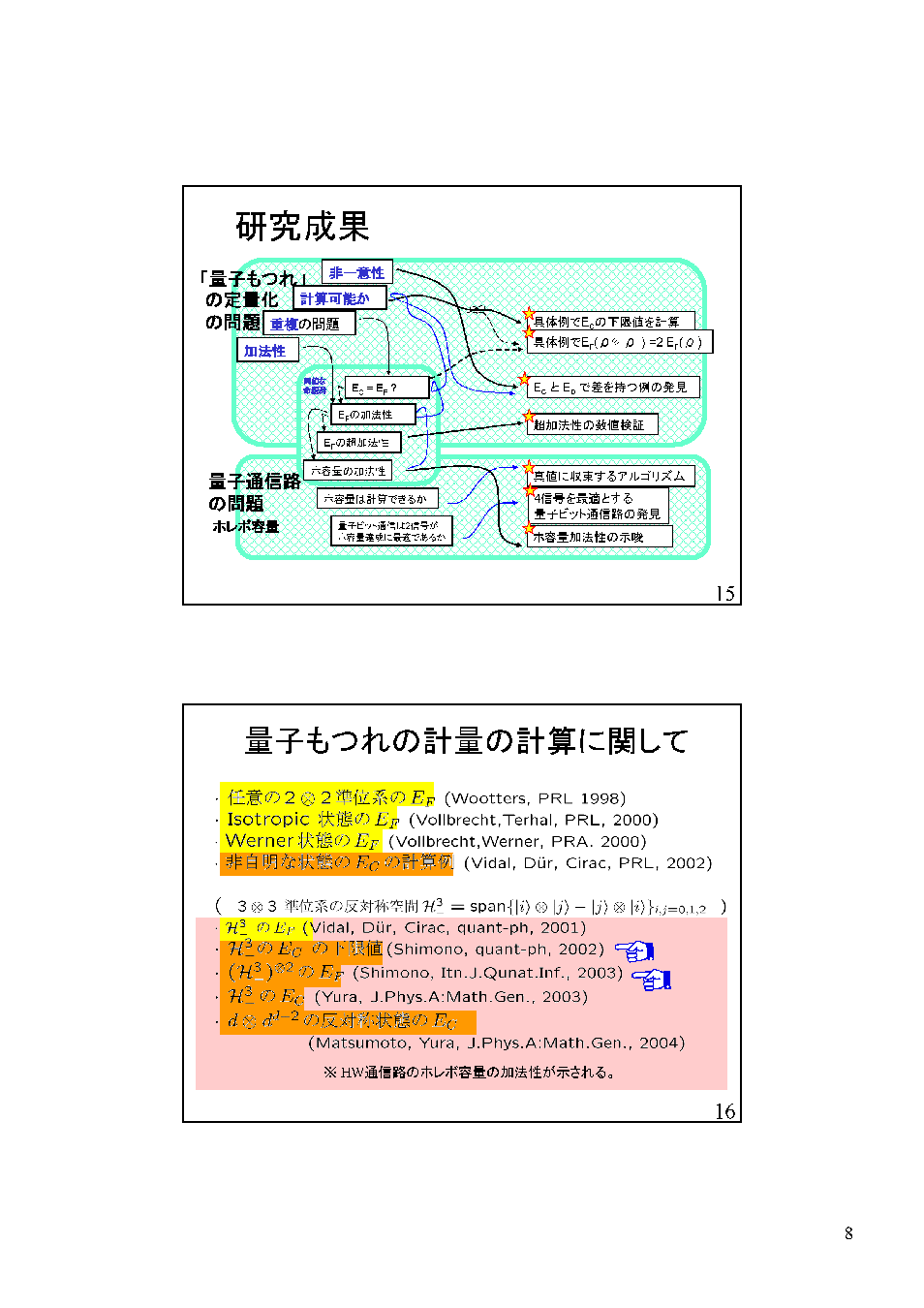}
\newpage
\thispagestyle{empty}
\vspace*{-29mm}\hspace*{-32.3mm}%
\noindent\includegraphics[width=210mm,height=297mm,bb=77 72 535 720]{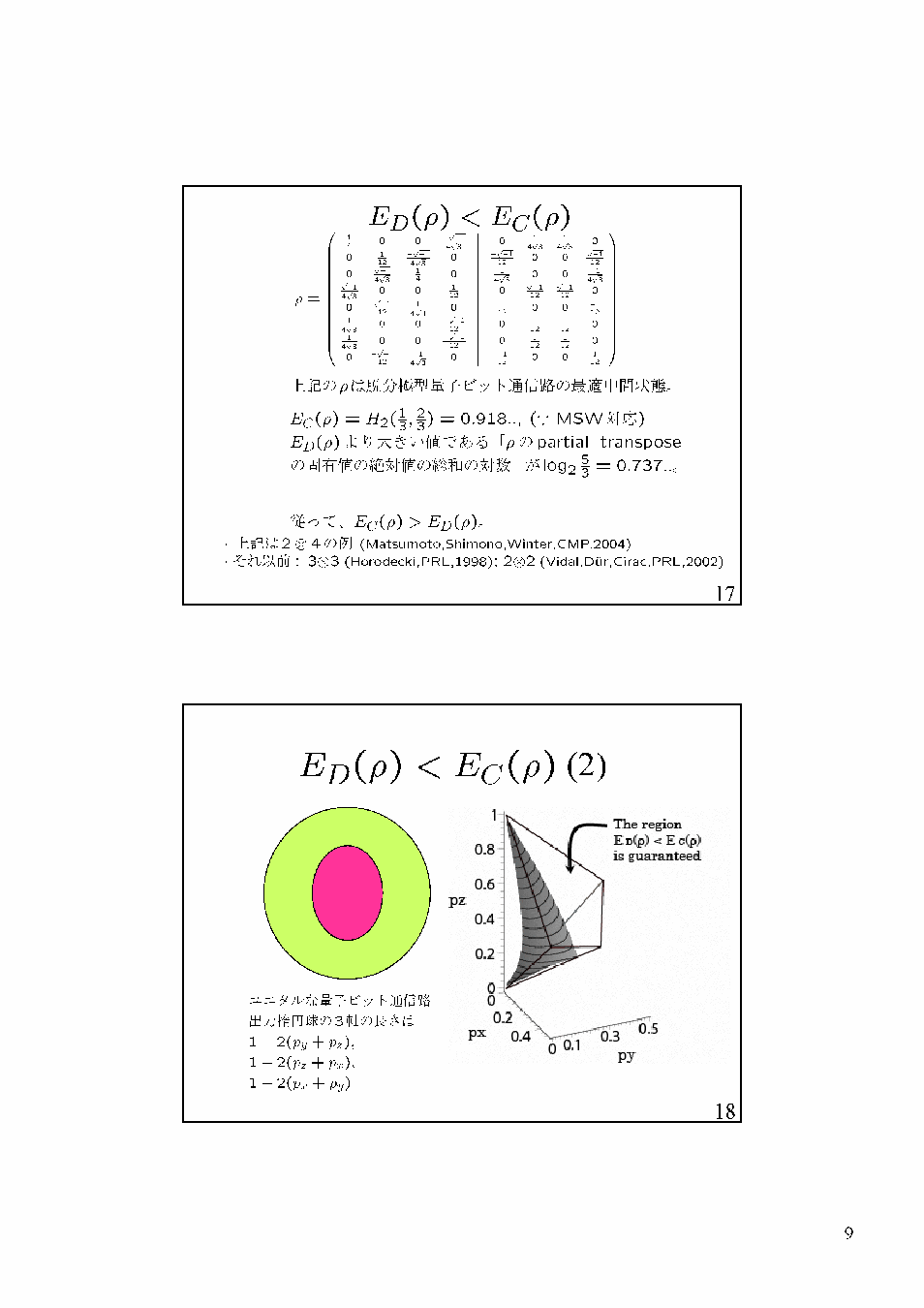}
\newpage
\thispagestyle{empty}
\vspace*{-29mm}\hspace*{-32.3mm}%
\noindent\includegraphics[width=210mm,height=297mm,bb=77 72 535 720]{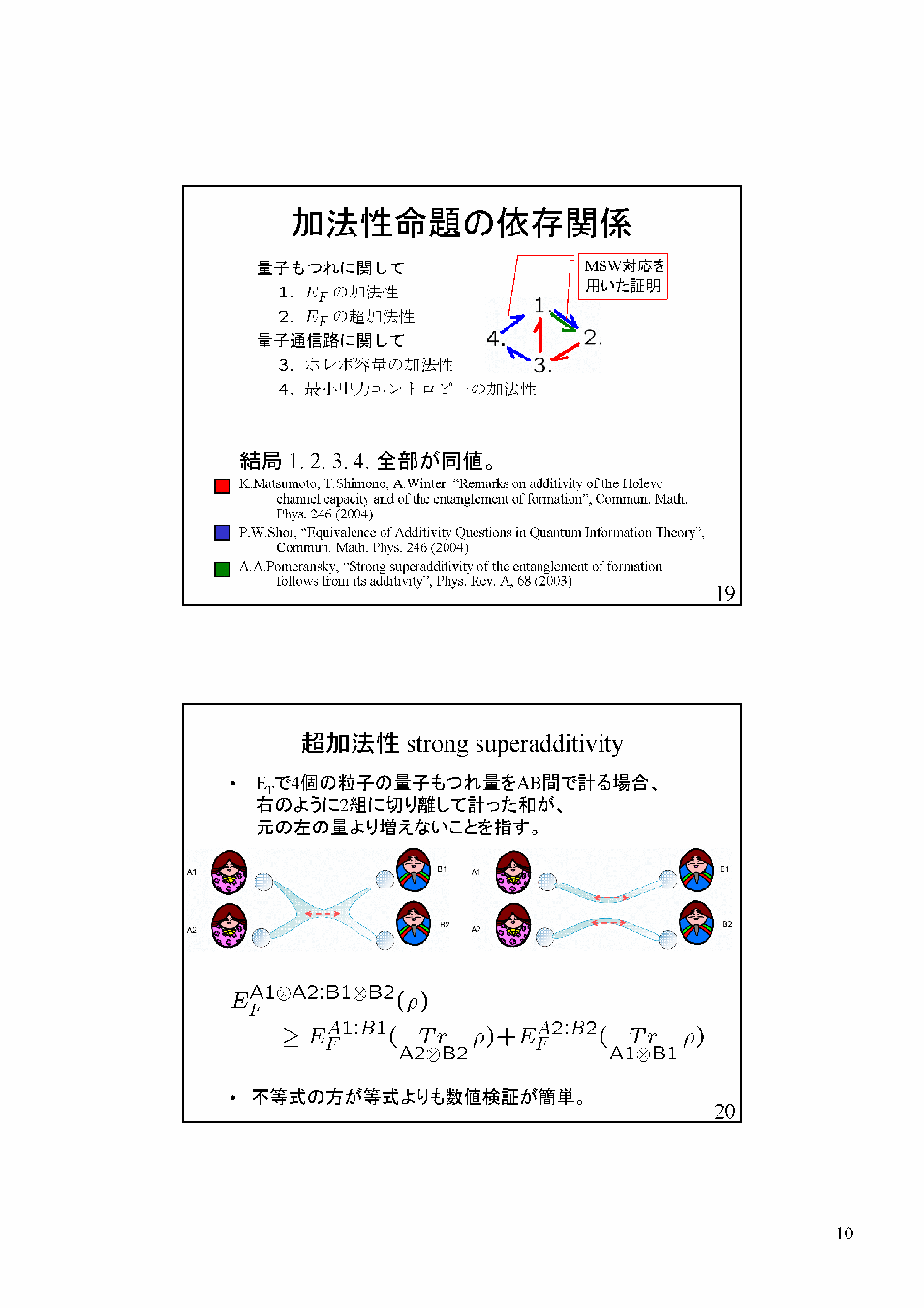}
\newpage
\thispagestyle{empty}
\vspace*{-29mm}\hspace*{-32.3mm}%
\noindent\includegraphics[width=210mm,height=297mm,bb=77 72 535 720]{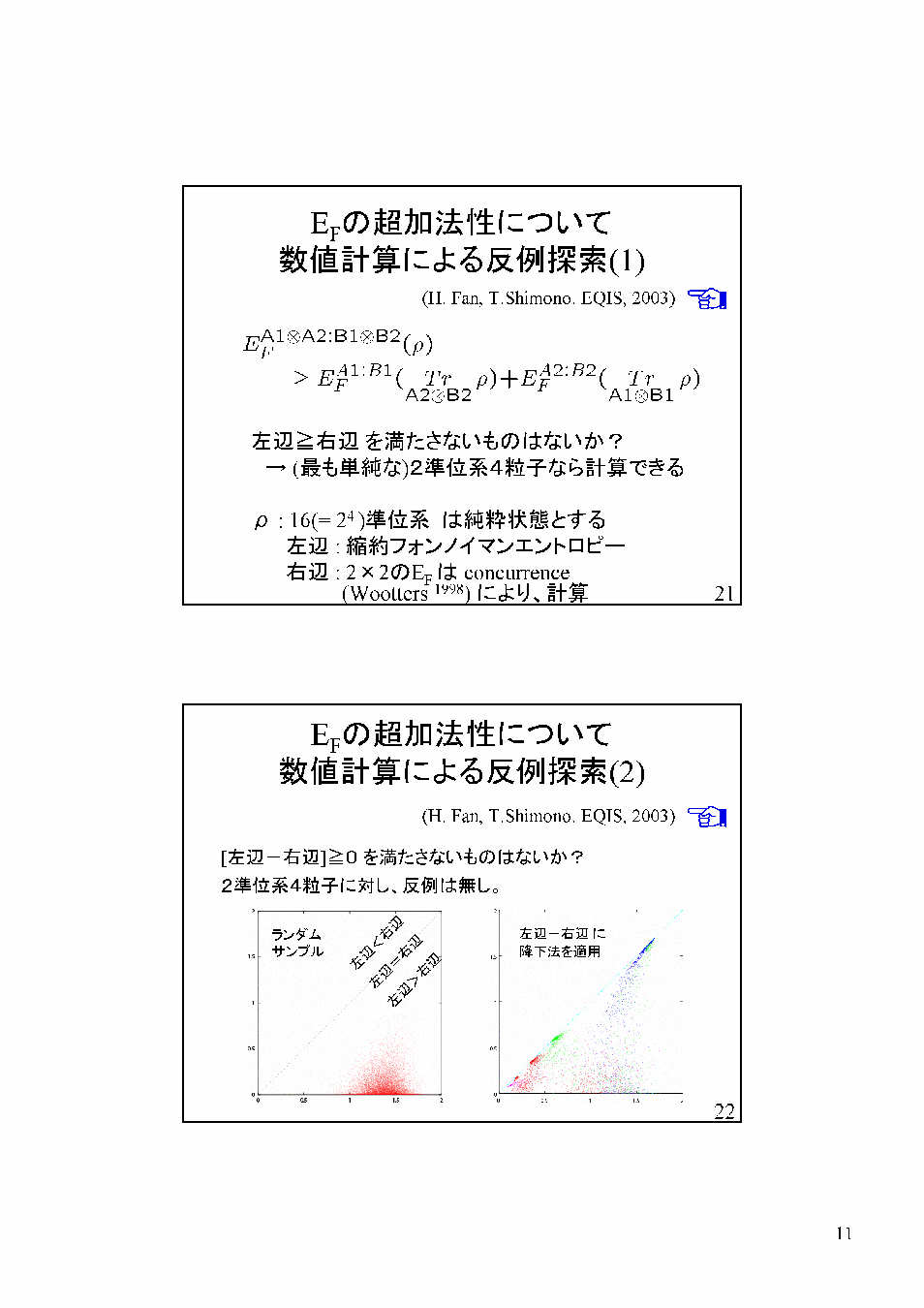}
\newpage
\thispagestyle{empty}
\vspace*{-29mm}\hspace*{-32.3mm}%
\noindent\includegraphics[width=210mm,height=297mm,bb=20 20 575 806]{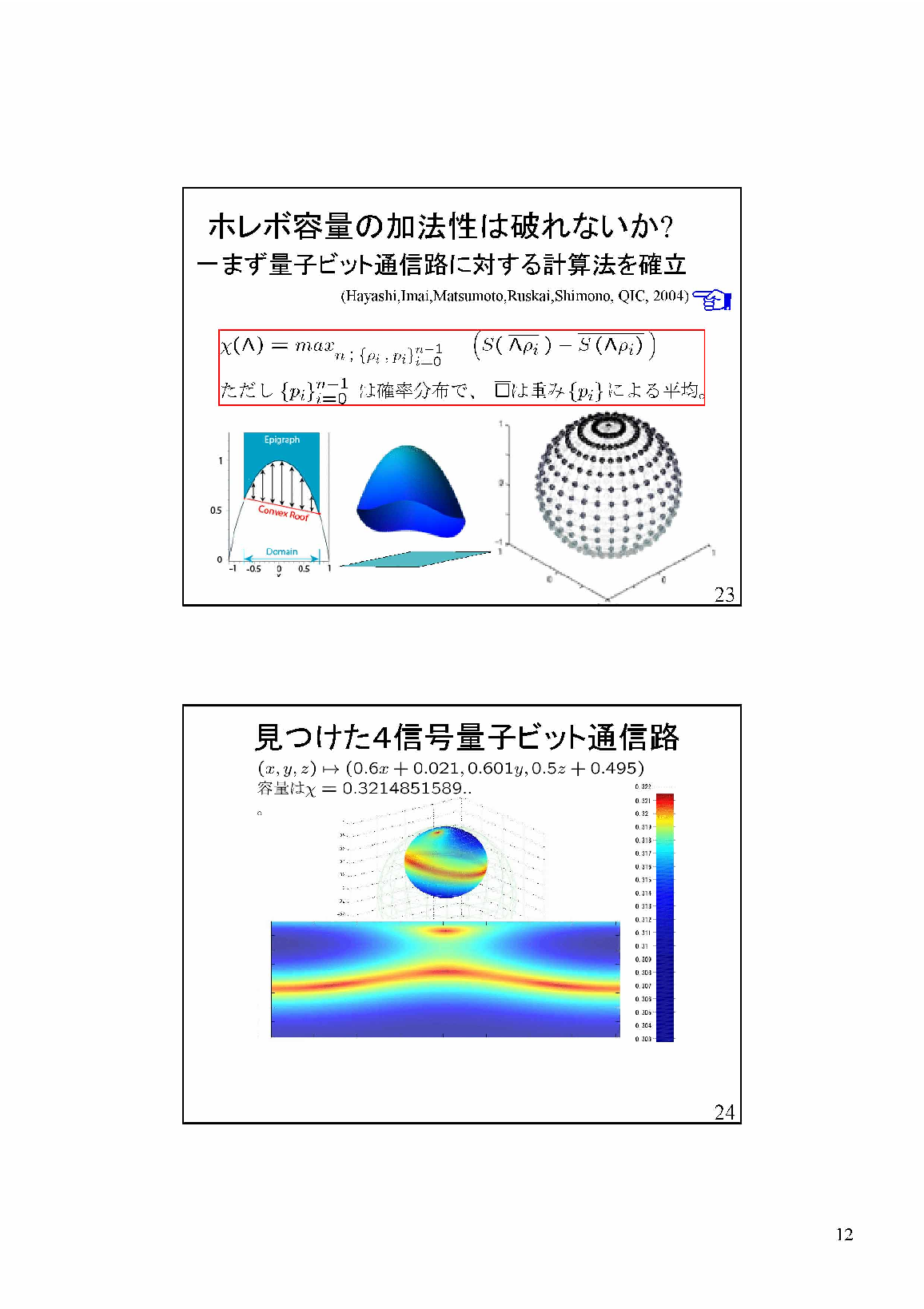}
\newpage
\thispagestyle{empty}
\vspace*{-29mm}\hspace*{-32.3mm}%
\noindent\includegraphics[width=210mm,height=297mm,bb=77 72 535 720]{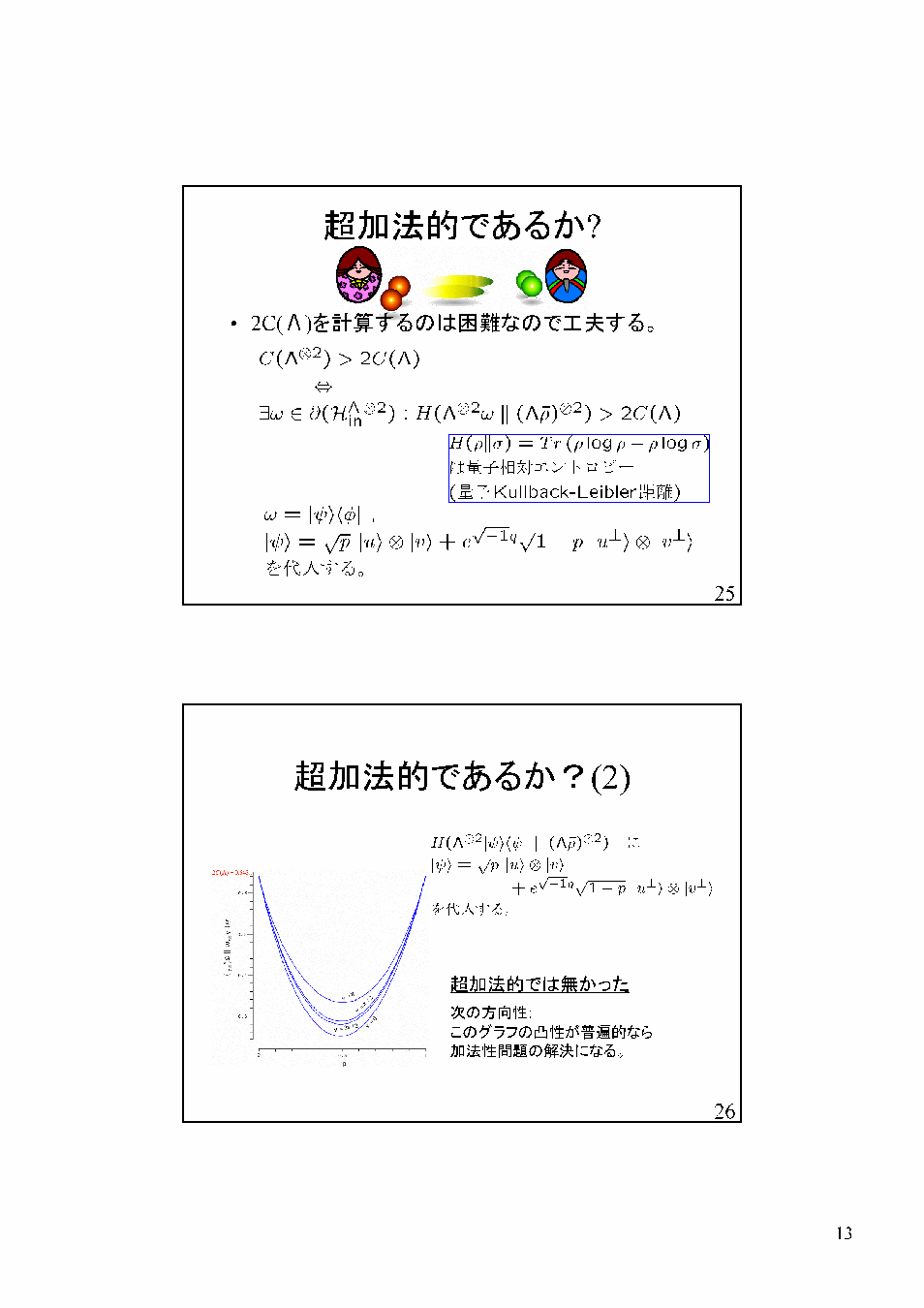}
\newpage
\thispagestyle{empty}
\vspace*{-29mm}\hspace*{-32.3mm}%
\noindent\includegraphics[width=210mm,height=297mm,bb=77 72 535 720]{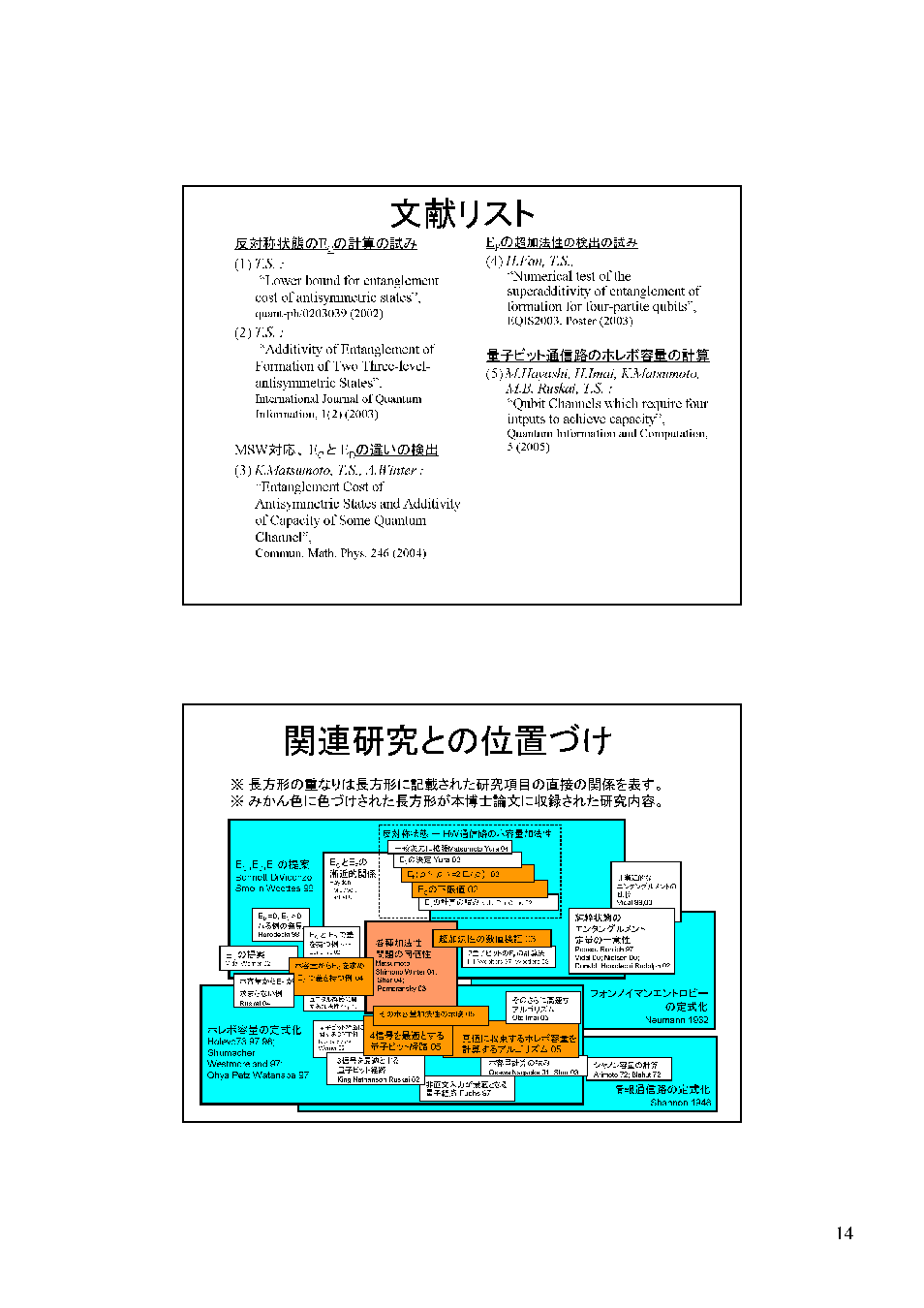}
\newpage
\thispagestyle{empty}
\vspace*{-5mm}\hspace*{-32.3mm}%
\noindent\includegraphics[width=210mm,bb=20 20 685 839]{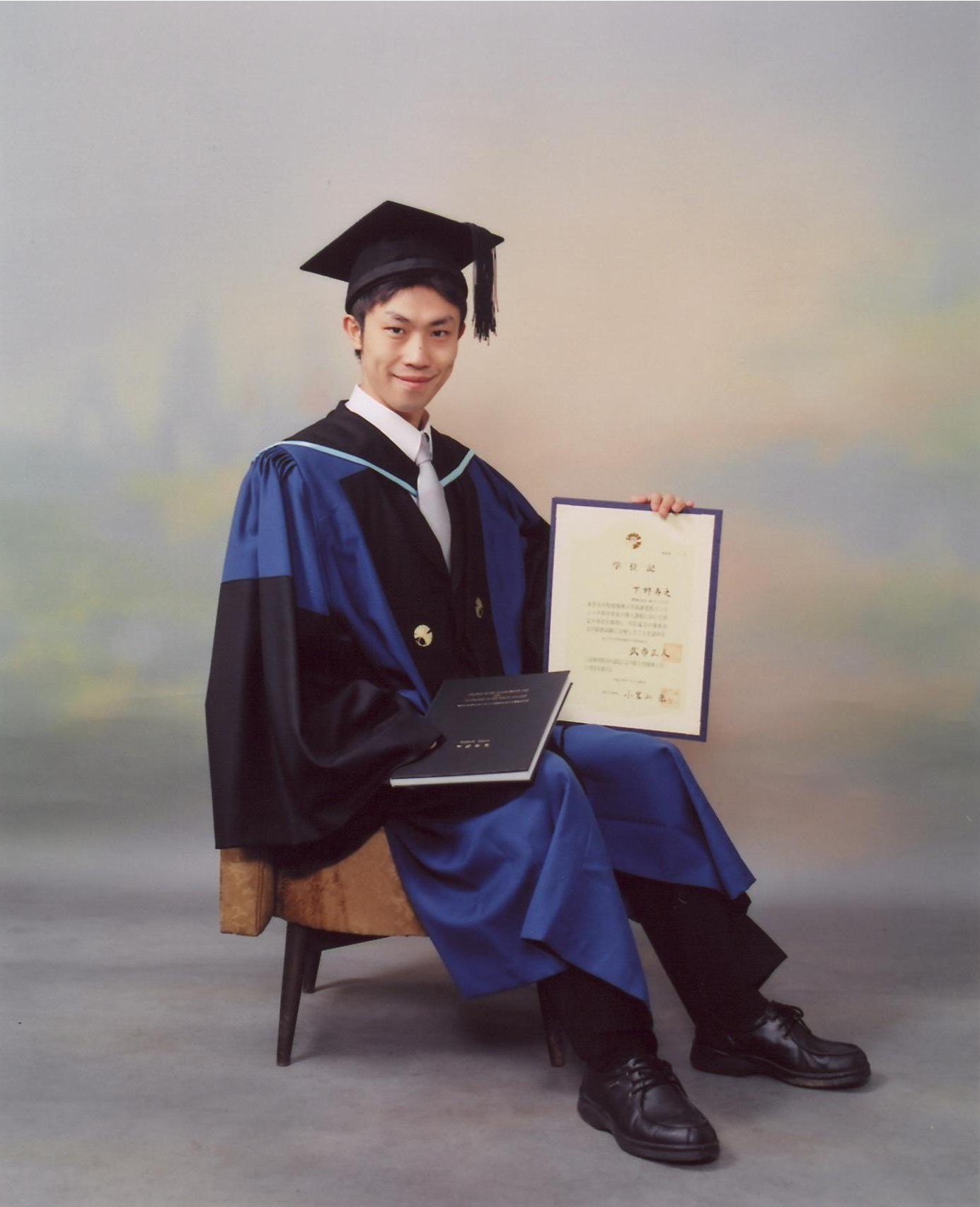}

\end{document}